\documentclass[aps,prd,preprintnumbers,groupedaddress,nofootinbib]{revtex4}

\pdfoutput=1

\usepackage{graphicx}
\usepackage{latexsym}
\usepackage{amsfonts}
\usepackage{amssymb}
\usepackage{amsmath}
\usepackage{slashed}
\usepackage{array}
\usepackage{feynmp}
\usepackage{url}
\usepackage{multirow}

\begin{document}
\title{Super-Razor and Searches for Sleptons and Charginos at the LHC}

\author{Matthew R.~Buckley$^{1,2}$, Joseph D.~Lykken$^3$, Christopher Rogan$^{4,5}$, and Maria Spiropulu$^4$}
\affiliation{$^1$Center for Particle Astrophysics, Fermi National Accelerator Laboratory, Batavia, IL 60510}
\affiliation{$^2$Department of Physics and Astronomy, Rutgers University, Piscataway, NJ 08954}
\affiliation{$^3$Theoretical Physics Department, Fermi National Accelerator Laboratory, Batavia, IL 60510}
\affiliation{$^4$Lauritsen Laboratory of Physics, California Institute of Technology, Pasadena, CA 91125}
\affiliation{$^5$Department of Physics, Harvard University, Cambridge, MA 02138}

\preprint{FERMILAB-PUB-13-456-A-T, CALT 68-2866}
\date{\today}

\begin{abstract}
Direct searches for electroweak pair production of new particles at the LHC are a difficult proposition, due to the large background and low signal cross sections. We demonstrate how these searches can be improved by a combination of new razor variables and shape analysis of signal and background kinematics. We assume that the pair-produced particles decay to charged leptons and missing energy, either directly or through a $W$ boson. In both cases the final state is a pair of opposite sign leptons plus missing transverse energy. We estimate exclusion reach in terms of sleptons and charginos as realized in minimal supersymmetry. We compare this super-razor approach in detail to analyses based on other kinematic variables, showing how the
super-razor uses more of the relevant kinematic information while achieving higher selection efficiency on
signals, including cases with compressed spectra.
\end{abstract}

\maketitle

\section{Introduction \label{sec:intro}}

Searches with the ATLAS and CMS detectors at the Large Hadron Collider have already placed strong lower bounds on the mass of pair-produced strongly-interacting gluinos or degenerate squarks decaying into final states with missing transverse energy \cite{Aad:2012naa,Chatrchyan:2012uea,Chatrchyan:2013fk,Aad:2013wta,CMS-PAS-SUS-13-004}. 
A  determination of the role of supersymmetry in electroweak symmetry breaking requires a much broader campaign of searches, many of which are already underway. 
Some of these searches present special challenges at a hadron machine, even when they involve the pair production of relatively light superpartners. Examples include light stops whose decays closely resemble those of top quarks \cite{Carena:2008mj,Bi:2011ha,
Bai:2012gs,Alves:2012ft,Han:2012fw,Bhattacherjee:2012mz,Carena:2013iba,
Delgado:2012eu,Dutta:2012kx,Evans:2012bf,Kilic:2012kw,Buckley:2013lpa,Bai:2013ema}, a variety of models with compressed spectra, $R$-parity violating models \cite{Barbier:2004ez,Csaki:2011ge,Berger:2013sir,Evans:2012bf,CMS:yut,CMS:swa}, and relatively long-lived superpartners with displaced decays \cite{Aad:2013txa,Chatrchyan:2012jwg}. 

Of particular importance to this program is the direct electroweak production of charginos, neutralinos, and sleptons at the LHC. Relatively light charginos and neutralinos have a possible connection to weakly-interacting dark matter in supersymmetry models with conserved $R$-parity. Light sleptons are motivated by the measured value of the anomalous magnetic moment $(g-2)$ of the muon \cite{Jegerlehner:2009ry,Miller:2007kk}, providing a thermal annihilation cross section for bino-like neutralino dark matter \cite{Buckley:2013sca}, and the possibility that the branching fraction of the newly-discovered Higgs boson into two photons is enhanced over the Standard Model prediction \cite{Carena:2012gp}. 
Charginos, neutralinos, and sleptons could also appear in cascade decays of heavier colored superpartners, but this prospect merely emphasizes the importance of being able to produce these lighter superpartners directly.

We will focus on electroweak pair production of charged particles that decay to charged leptons and a stable (or long-lived) neutral particle, appearing in the detector only as missing transverse energy ($\vec{E}^\text{miss}_T$). The decay to leptons can occur either directly or through the leptonic decay of a $W$ boson. We will consider two canonical examples: sleptons of the first or second generation ($\tilde{e}^-\tilde{e}^+$ or $\tilde{\mu}^-\tilde{\mu}^+$) with 100\% branching into leptons and the lightest supersymmetric particle (LSP) neutralino, and charginos ($\tilde{\chi}^+\tilde{\chi}^-$) decaying through an on- or off-shell $W$ boson and the neutralino LSP. In the latter case, we require the $W$ to decay leptonically. In both cases, we set all other superpartner masses heavy, including the other charginos and neutralinos. Though our study is performed assuming a supersymmetric model, it can easily be generalized to other scenarios that contain similar particles with the same broad characteristics. The pair production of tau partners ({\it e.g.}~staus) has different backgrounds and will be considered in a later work.

Searches at LEP have already set lower bounds on the masses of new charged particles, ranging between 90 and 105 GeV assuming supersymmetric-like cross sections \cite{Beringer:1900zz}. The ATLAS and CMS collaborations have performed model-independent dilepton searches for both the slepton and chargino pair production scenarios we consider in this paper. ATLAS, using 20.3 fb$^{-1}$ of integrated luminosity at 8 TeV places an upper bound of 300 GeV on left-handed sleptons (assuming massless neutralinos), and an upper limit of 450~GeV on charginos assuming a 100\% branching ratio to leptons and neutralinos \cite{:2012gg,ATLAS-CONF-2013-049}. CMS places a 300~GeV bound on pair production of degenerate selectrons and smuons using 19.5~fb$^{-1}$ at 8 TeV, and 550~GeV on chargino pair production decaying to neutralinos with 100\% branching ratio \cite{CMS-PAS-SUS-12-022,CMS-PAS-SUS-13-006}. Both experiments \cite{CMS-PAS-SUS-12-022,CMS-PAS-SUS-13-006,Aad:2012ku,Aad:2012hba,ATLAS-CONF-2013-036,ATLAS-CONF-2013-035} have also performed multilepton searches for production of heavier chargino/neutralino pairs (such as $\tilde{\chi}^0_2 \tilde{\chi}^0_2$ or $\tilde{\chi}^0_2 \tilde{\chi}^\pm_1$), followed by cascades of the form $\tilde{\chi}^0_2 \to W^\pm \tilde{\chi}^\mp_1 \to \ell^\pm \ell^\mp \nu \bar{\nu} \tilde{\chi}^0_1$ to obtain three or more leptons in the final state.


We propose several techniques that can increase the sensitivity of the LHC experiments to electroweak pair production in the dilepton channel, using the data currently available from the completed 8 TeV run. Starting from the CMS razor variables \cite{Rogan:2010kb,Chatrchyan:2011ek} (see Refs.~\cite{Fox:2012ee,CMS:2012dwa,Chatrchyan:2012gq} for applications), 
we develop an improved version that more accurately approximates the production frame and center-of-mass (CM) energy scale of the pair production event, compared to the original razor formulation. This ``super-razor'' results in a set of mass variables, $\sqrt{\hat{s}_R}$ and $M_\Delta^R$ that contain information about the mass differences involved in the pair production and subsequent decay, allowing for discrimination between signal and background. In addition, the derivation of these mass variables involves constructing the approximate boost to the pair production frame, followed by a boost to an approximation of the decay frame. From this boost direction and the momenta of the visible particles, we construct angular variables $\Delta\phi_R^{\beta}$ and $|\cos\theta_{R+1}|$ that also distinguish between the signal events and background. Using these super-razor variables, we develop a new set of selection criteria and apply a multi-dimensional shape analysis to maximize the sensitivity to signal over the dominant backgrounds (primarily $W^-W^+$ and Drell-Yan + jets).  Shape analyses have been implemented by experimental groups \cite{Rogan:2010kb,CMS:2012dwa} and have been used in theoretical proposals for new searches \cite{Alves:2012ft,Fox:2012ee}. As we show through direct comparison to ATLAS- and CMS-like searches, this technique is promising in difficult channels.

In the next section we review the construction of the standard razor variables, followed by a derivation of the improved super-razor and the associated angular variables of interest. The background and signal simulations are described in Section~\ref{sec:simulation}, along with comparisons to the alternative searches by the ATLAS and CMS experiments that employed the kinematic variables $M_{T2}$ \cite{Lester:1999tx,Barr:2003rg} and $M_{CT\perp}$ \cite{Matchev:2009ad,Tovey:2008ui}. The shape analysis techniques and statistical tools are described in Section~\ref{sec:shape}. Our expected exclusion limits for 20~fb$^{-1}$ of integrated luminosity at 8 TeV are presented in Section~\ref{sec:conclusion}.

\section{Kinematic Variables \label{sec:variables}}

We are interested in the pair production of particles that each decay either into a lepton and a massive undetected ``invisible'' particle, or into an invisible particle and a $W$ boson, followed by leptonic decays of the $W$'s. For specificity we consider slepton pair production and chargino pair production
as in the minimal supersymmetric standard model (MSSM):
\begin{eqnarray}
p p \to \tilde{\ell}^-\tilde{\ell}^+ & \to & (\ell^- \tilde{\chi}^0_1)(\ell^+ \tilde{\chi}^0_1)  \\
p p \to \tilde{\chi}_1^-\tilde{\chi}_1^+ & \to & (W^-\tilde{\chi}_1^0)(W^+\tilde{\chi}_1^0) \to (\ell^-\bar{\nu}\tilde{\chi}_1^0)(\ell^+\nu\tilde{\chi}_1^0).
\end{eqnarray}
In both cases, the observables at the LHC are the same: opposite-sign leptons (which may or may not be of the same flavor) and large missing transverse energy. Searching for these types of new particles is difficult for several reasons. The production cross sections are small, on the order of tens of femtobarns to a few picobarns before branching fractions. The background cross sections are large. The dilepton backgrounds (primarily $W^-W^+$ and Drell-Yan + jets production but also with contributions from $WZ$, $ZZ$, and top pair production) have kinematic distributions that
are similar to the signal, since most of these backgrounds have two charged leptons and real missing transverse momentum from neutrinos. Kinematic variables sensitive to the mass (or mass squared) differences between the parent and invisible particles are less effective in regions of the mass plane when the parent/daughter mass difference is close to or smaller than the $W$ mass. The $M_{CT\perp}$ \cite{Matchev:2009ad,Tovey:2008ui} and $M_{T2}$ \cite{Lester:1999tx,Barr:2003rg} variables  used by the CMS \cite{CMS-PAS-SUS-13-006} and ATLAS \cite{ATLAS-CONF-2013-049} (see also Refs.~\cite{Chatrchyan:2012jx,Aaltonen:2009rm} for other experimental applications of $M_{T2}$) searches have this drawback, as does the original formulation of the razor variable, as we will show.

Our new work is motivated by the razor variables $M_R$ and $R$, originally developed in Ref.~\cite{Rogan:2010kb,Chatrchyan:2011ek}  to distinguish between new massive strongly interacting particles ({\it e.g.}~squarks and gluinos) and QCD background, and implemented by CMS \cite{CMS:2012dwa,Chatrchyan:2012gq} in various searches. Razor variables have also been demonstrated to be of use in distinguishing signal and background in electro-weak channels \cite{Fox:2012ee}. Here, we describe the motivating principles behind the razor (for a full description, see Ref.~\cite{roganthesis}), and then propose a series of improvements that more accurately capture the relevant mass differences in events with final states relevant to electroweak production. We then introduce new kinematic variables, motivated by the construction of the improved razor, which contain information about the ratio of mass scales of the particles in the event.

\subsection{Principles of the razor}

The razor variables are intended for use in a very generic new physics scenario. Two massive particles, $S_1$ and $S_2$, with a common mass $m_S$, are produced at the LHC. Each then decays into a set of visible particles ($Q_1$ and $Q_2$, respectively) and an invisible particle ($\chi_1$ and $\chi_2$) with common mass $m_\chi$. For this paper, we will be assuming that the visible decays each consist of a single effectively massless particle (an electron or muon). In a more inclusive razor analysis decays may include more than one visible particle, in which case their four-momenta are summed to create two visible objects known as megajets.

If we could identify the rest frames of the $S_i$ decay, then in that frame the energies $E_i$ of the visible $Q_i$ would be 
\begin{equation}
2E_1 = 2E_2 = \frac{m_S^2-m_\chi^2}{m_S} \equiv M_\Delta.
\end{equation}
If this frame could be identified using the available visible momenta and the $E_{T}^\text{miss}$ of the invisible particles, then the momentum of the visible $Q_i$ in signal events would be easily distinguished from background, which does not inherit information about this scale (save in cases where $M_\Delta \sim m_W$).

However, as is well understood in hadron colliders, with $S_i$ both decaying into at least one invisible particle, we do not possess enough kinematic information to reconstruct the decay frames. The approach of the razor is to make a series of assumptions which, while not capable of reconstructing the precise decay frames event-by-event, approximate the relevant frames on average. In both simulations and data these approximations work well in the experimental environment of the LHC.

\begin{figure}[t]
\includegraphics[width=0.9\columnwidth]{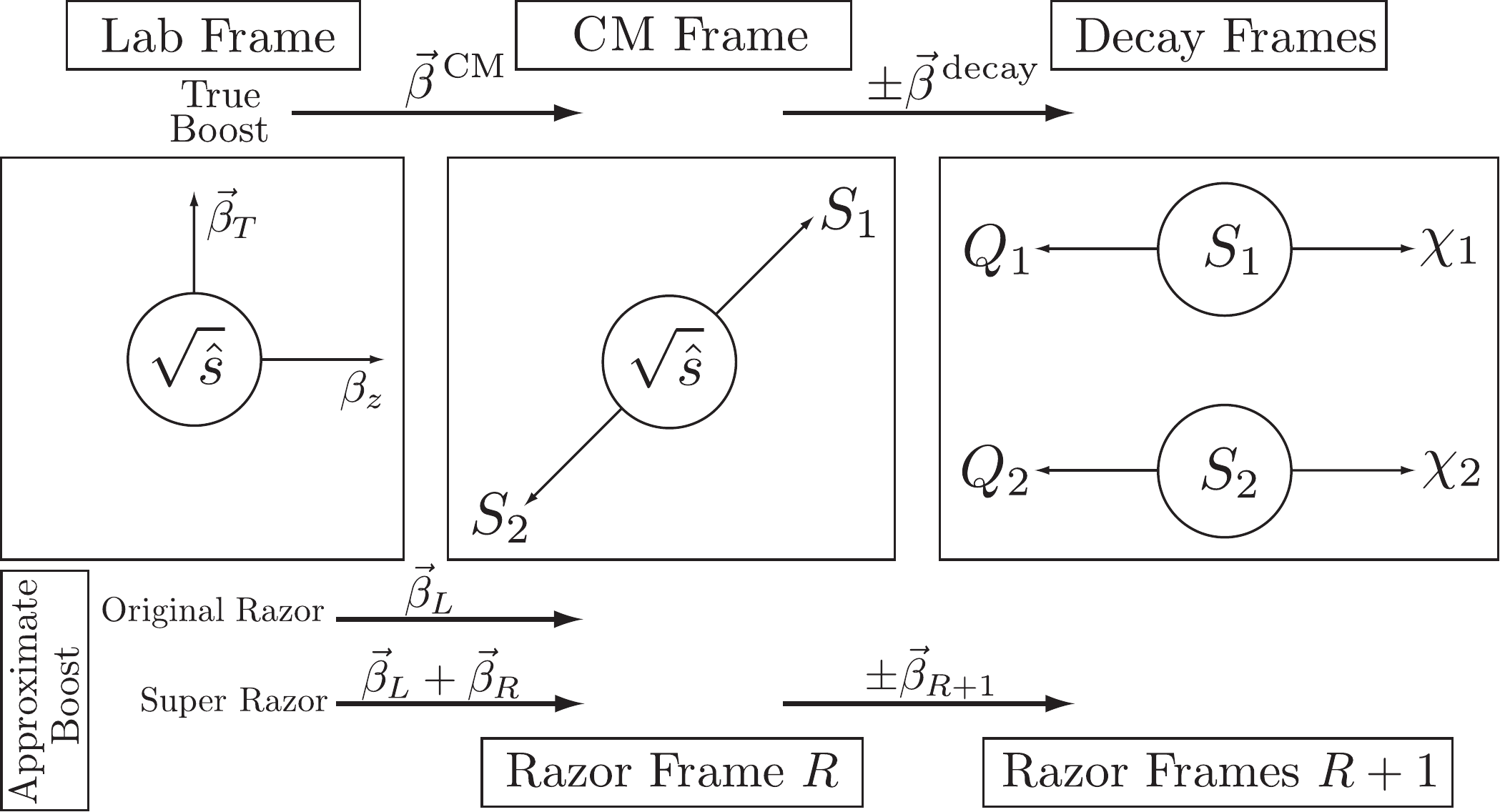}

\caption{Sketch of the three sets of frames relevant to the razor reconstruction: the lab frame, the pair production frame for $S_1$ and $S_2$, and the two decay frames of the particles $S_i$. The approximate razor frame identified with each physically relevant frame is also shown, along with the actual and approximate boosts from one frame to the next. By convention, we label each boost by the destination frame ({\it i.e.}~boost $\vec{\beta}^{\, \rm CM}$ takes you {\it from} the lab from {\it to} the pair production center of mass frame). \label{fig:frames}}
\end{figure}

There are three kinds of frames relevant to pair production at the LHC: the lab frame, the pair production center-of-mass (CM) frame, and the two decay frames (see Figure~\ref{fig:frames}). 
The initial assumption made by the original razor construction is that the heavy parent particles are generally produced near threshold, due to the fall-off of the parton distribution functions with CM energy $\sqrt{\hat{s}}$. If we could identify the boost $\vec{\beta}^{\, \rm CM}$ from the lab frame into the $S_1$ and $S_2$ production frame (the center of mass frame CM), then this could serve as an approximation to the decay frames. We approximate this frame by making a longitudinal boost $\vec{\beta}_L$ to the razor frame $R$, which is defined here as the frame where the two sets of visible decay products $Q_1$ and $Q_2$ have equal and opposite $z$-component of momentum. This boost has magnitude
\begin{equation}
\beta_L = \frac{q_1^z+q_2^z}{E_1+E_2}. \label{eq:betaL}
\end{equation}
Here, $E_i$ is the energy of decay product $Q_i$ and $q_i^z$ is the $z$-component of the momentum.

In this razor frame, we expect $2E_{R1} \approx 2E_{R2} \approx M_\Delta$. Writing the boosted momenta in terms of lab-frame observables, we define a longitudinally boost-invariant mass
\begin{equation}
M_R^2 = (E_1+E_2)^2 - (q_1^z+q_2^z)^2. \label{eq:MR}
\end{equation}
We expect that the distribution of $M_R$ for signal events will have a peak near $M_\Delta$, assuming that our approximations of near-threshold production and $q_1^z \approx - q_2^z$ are correct on a statistical basis. Background events will not, in general, have any special feature near $M_\Delta$. For example, events consisting only of visible particles and $E_{T}^\text{miss}$ from mismeasurement would be expected to have an $M_R$ distribution proportional to the distribution of CM energy $\sqrt{\hat{s}}$, as in the case of QCD backgrounds. 

We then define a second mass variable that inherits knowledge of the mass splitting $M_\Delta$, using the visible and invisible transverse momentum in the event. Note that this information was not used in the definition of $M_R$. Motivated by the fact that backgrounds with no invisible particles must have $Q_1$ and $Q_2$ back-to-back (a fact that mismeasurement dos not tend to greatly change), we define a transverse mass in terms of the visible transverse momenta, $q_{1T}$ and $q_{2T}$, and the missing transverse energy $E_T^\text{miss}$:
\begin{equation}
(M_T^R)^2 = \frac{1}{2}\left[E_T^\text{miss} (q_{1T}+q_{2T})-\vec{E}^\text{miss}_T \cdot (\vec{q}_{1T} +\vec{q}_{2T})\right]. \label{eq:MTR}
\end{equation}
Assuming pair production at threshold, $M_T^R \leq M_\Delta$ for signal events. Introducing the dimensionless ratio
\begin{equation}
R^2 = \left(\frac{M_T^R}{M_R}\right)^2, \label{eq:R}
\end{equation}
we expect $R^2 < 1$ for signal events, with a rough spread around $R^2 \sim \tfrac{1}{4}$, while
for background without real $E_T^\text{miss}$ we expect $R\sim 0$. 

The razor variables $M_R$ and $R^2$ were originally designed to separate QCD and other 
backgrounds from pair production of strongly-interacting heavy particles \cite{Rogan:2010kb,Chatrchyan:2011ek,CMS:2012dwa,Chatrchyan:2012gq}.\footnote{The initial use of the razor was in squark and gluino searches,
where a major background is QCD. As QCD is essentially scale-free at LHC energies, the QCD background in the razor variable $M_R$ falls exponentially. Requiring a minimum value of $R^2$, the background falls off
more and more steeply as the $R^2$ threshold is increased. This ``slicing away" of the background is the
origin of the name ``razor.''}
When used in these studies, all visible particles are assumed to fall into one of the decay chains of the parent particles $S_1$ or $S_2$. Therefore, all visible particles are assigned to a megajet $Q_1$ or $Q_2$ by a
simple algorithm, and their momenta summed. Calculation of $M_R$ and $R^2$ then proceeds as if there were only two visible objects.

\subsection{The super-razor}

Consider events that have both visible and invisible particles. Rather than splitting the visible particles into two objects $Q_1$ and $Q_2$, suppose we can divide them into three classes: particles (or groupings of particles) $Q_1$ and $Q_2$ that are assumed to come from the decay of the new physics particles $S_1$ and $S_2$, and a third class of particles that come from initial state radiation or something else extraneous to
the heavy particle decays. In electroweak production of non-colored particles, every jet in an event can be assigned to this third class. The sum of the momenta of all particles in this class is $\vec{J}$. By construction
\begin{equation}
\vec{J}_T = -\vec{E}^\text{miss}_T-\vec{q}_{1T}-\vec{q}_{2T}.
\end{equation}

The effect of $\vec{J}$ is to shift the production frame by an additional boost that was not taken into account by the original longitudinal razor boost of Eq.~\eqref{eq:betaL}. To correct for this, we want to make an additional transverse boost which takes us to the frame in which is recoiling against the jet contamination. The direction of this transverse boost is trivial: we must boost in the direction opposite to $\vec{J}$. However, there is insufficient information in the events at the LHC to unambiguously determine the magnitude of the boost. The correct boost from the lab frame to the pair production frame is
\begin{equation}
\vec{\beta}^{\, \rm CM} = \frac{\{-\vec{J}_T,p^{\, \rm CM}_z\}}{\sqrt{|\vec{J}_T|^2+(p^{\rm CM}_z)^2+\hat{s}}}, \label{eq:truelabbeta}
\end{equation}
where $p^{\rm CM}_z$ is the $z$-momentum of the center of mass frame relative to the lab frame. Neither $p^{\rm CM}_z$ or $\hat{s}$ can be determined from the available visible particle momenta at the LHC. 

We therefore must make new assumptions to build our approximate boost to the frame $R$, the razor frame that is our best guess to the pair production frame. To build this approximate boost $\vec{\beta}_R$, we make the longitudinal boost $\beta_L$, and then construct an additional boost $\vec{\beta}_R$ from approximate center of mass energy $\sqrt{\hat{s}}_R$, defining
\begin{equation}
\vec{\beta}_R = \frac{\{-\vec{J}_T,p^R_z\}}{\sqrt{|\vec{J}_T|^2+|p^R_z|^2+\hat{s}_R}}. \label{eq:razorbeta}
\end{equation}
There are two necessary assumptions to build $\hat{s}_R$. The first assumption is that the invariant mass of the visible system is equal to the invariant mass of the invisible system. This guess will result in $\hat{s}_R$ systematically lower than the actual $\hat{s}$ when the weakly interacting particles in the event are massive. Conveniently, this will actually turn out to be useful in our construction of further discriminating variables, which will be discussed shortly. The second assumption we must make is that the constructed variables (such as $\hat{s}_R$) do not depend on the unknown $p_z^R$. Clearly, this is not correct on an event-by-event basis, but allows for a determination of $\hat{s}_R$ to be made (up to a two-fold ambiguity, which we resolve by taking the positive solution). By requiring $\partial \sqrt{\hat{s}}_R/\partial p_z^R = 0$, we find (in terms of the razor variable $M_R$ of Eq.~\eqref{eq:MR})
\begin{equation}
\frac{\hat{s}_R}{4} = \frac{1}{2}\left( M_R^2 + \vec{J}_T\cdot(\vec{q}_1+\vec{q}_2) + M_R \sqrt{M_R^2+|\vec{J}_T|^2+2\vec{J}_T\cdot(\vec{q}_1+\vec{q}_2)}\right). \label{eq:hatsR}
\end{equation}
This new variable $\hat{s}_R$ can be thought of as a ``jet-corrected'' version of the original razor variable $M_R^2$ (up to a factor of four). Which is to say, it inherits information about the mass difference $M_\Delta$ and the overall pair-production energy scale $\sqrt{\hat{s}}$. 

\begin{figure}[ht]
\includegraphics[width=0.4\columnwidth]{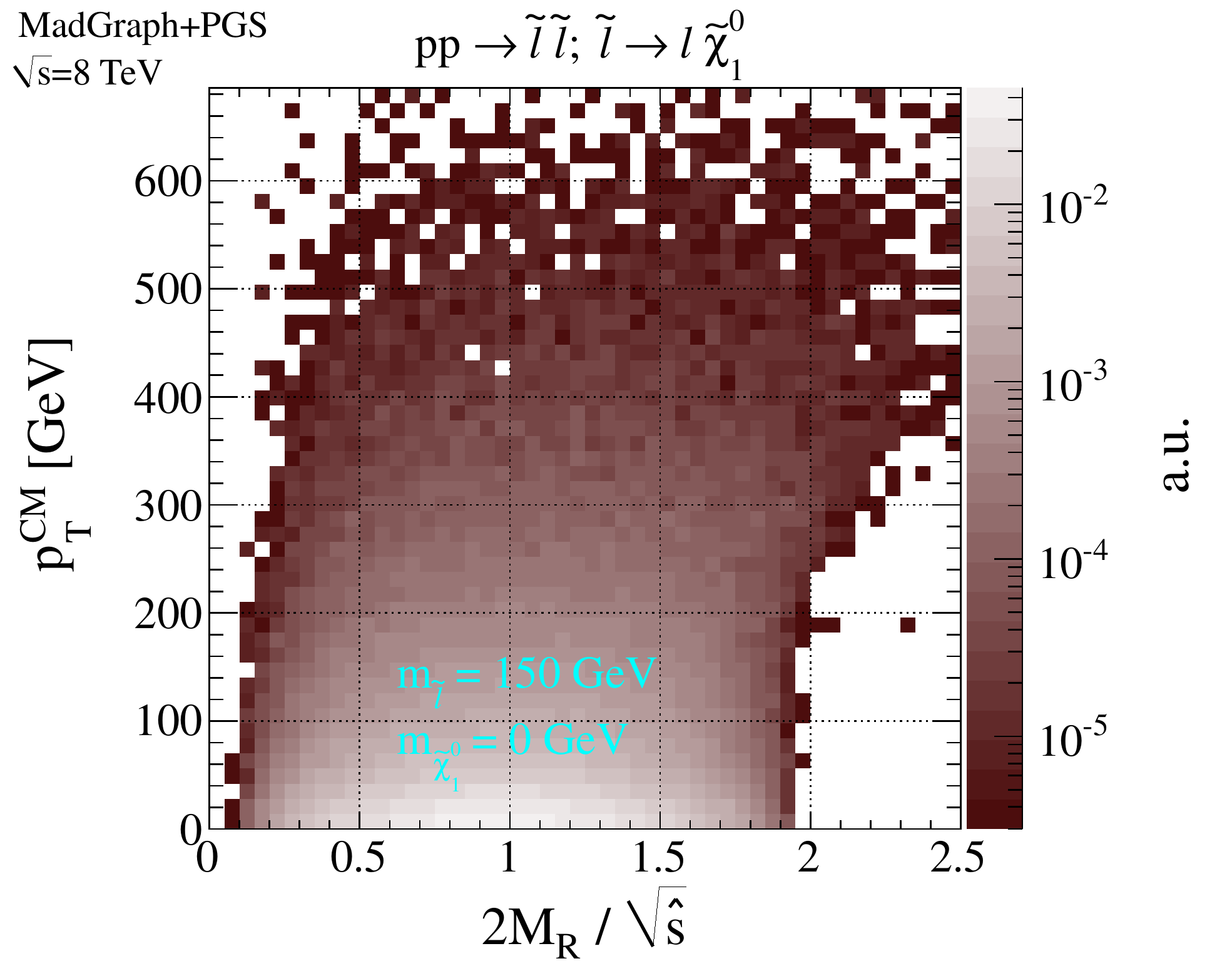}
\includegraphics[width=0.4\columnwidth]{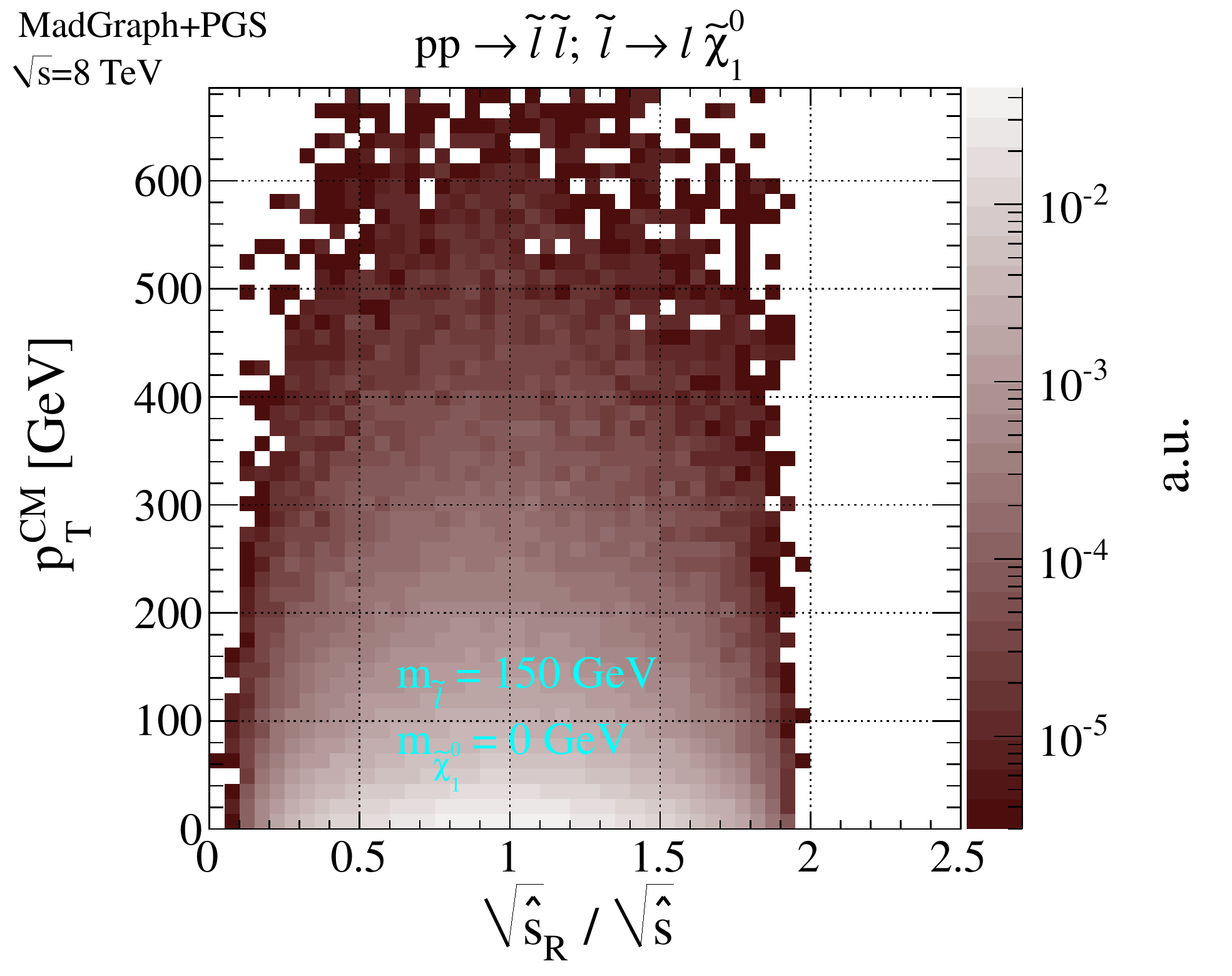}
\caption{Distributions of razor variable $M_R$ normalized to $\sqrt{\hat{s}}$ (left) and $\sqrt{\hat{s}}_R$ normalized to $\sqrt{\hat{s}}/2$ (right) versus CM $p_T$ compared for 150~GeV slepton pair production followed by decay to leptons and massless neutralinos,. See Section~\ref{sec:simulation} for details of the simulation. \label{fig:hatsMR}}
\end{figure}

In Figure \ref{fig:hatsMR}, we show the distributions of $M_R$ and $\sqrt{\hat{s}}_R$ (normalized to $\sqrt{\hat{s}}$) versus the $p_T$ of the CM frame, for representative slepton pair production decaying to leptons and neutralinos. As can be seen, while both $M_R$ and $\sqrt{\hat{s}}_R$ peak at the expected value given by the actual energy scale of the pair production ($\sqrt{\hat{s}}/2$ or $\sqrt{\hat{s}}$, respectively), when the center of mass is boosted to high $p_T$, the $M_R$ variable begins to show deviations from the smooth distribution. Boosting against the jets corrects for the high $p_T$ of the center of mass, as is seen in the distribution of $\sqrt{\hat{s}}_R$. The signal distributions are simulated using {\tt MadGraph5} \cite{Alwall:2011uj}, {\tt Pythia 6.4}  \cite{Sjostrand:2006za}, and {\tt PGS}; complete details of our simulations and cuts are discussed in the next section. 

Interestingly, this variable $\hat{s}_R$ was constructed in Ref.~\cite{Rainwater:1999sd}, 
using a separate line of reasoning. In the razor framework, interpreting this invariant mass as the energy associated with a boost to an approximation of the pair production frame allows us to reconstruct that boost. As we will show, this leads to additional variables that add to our ability to distinguish signal and background.

Now that we are in the razor frame $R$, we can attempt to build boosts to approximations of the two decay frames of the parent particles $S_i$. Given the incomplete information available for the event, our choices for boosts are constrained. As there are two decay frames, which must have equal and opposite boosts from the pair production frame, we approximate the boost $\vec{\beta}^{~\rm decay}$ by the boost
\begin{equation}
\vec{\beta}_{R+1} = \frac{\vec{q}_{R1}-\vec{q}_{R2}}{E_{R1}+E_{R2}}, \label{eq:razorbeta1}
\end{equation}
where $q_{R1}$ and $q_{R2}$ are the 4-momenta of the two visible particles $Q_1$ and $Q_2$ in the razor frame $R$. This boost has the correct symmetry property, in that the boost to the decay frame of $S_2$ is the negative of the boost to the decay frame of $S_1$. 

If we correctly identified the boost $\vec{\beta}^{\, \rm decay}$, then the invariant mass of the pair production frame would be related to the mass of the particles $S_i$ by
\begin{equation}
\sqrt{\hat{s}} = 2 \gamma^{\rm decay}m_S.
\end{equation}
We have constructed our boosts using information from the visible system $Q_1$ and $Q_2$, so our approximate boost $\vec{\beta}_{R+1}$ and approximate CM energy $\sqrt{\hat{s}}_R$ should be related not to the mass $m_S$, but the mass difference $M_\Delta$. We therefore define a second razor variable $M_\Delta^R$,
\begin{equation}
M_\Delta^R = \frac{\sqrt{\hat{s}}_R}{2\gamma_{R+1}}, \label{eq:mdeltaR}
\end{equation}
where $\gamma_{R+1}$ is the Lorentz factor associated with the boosts $\vec{\beta}_{R+1}$. This variable should approximate $M_\Delta$ for signal events.

\begin{figure}[ht]
\includegraphics[width=0.4\columnwidth]{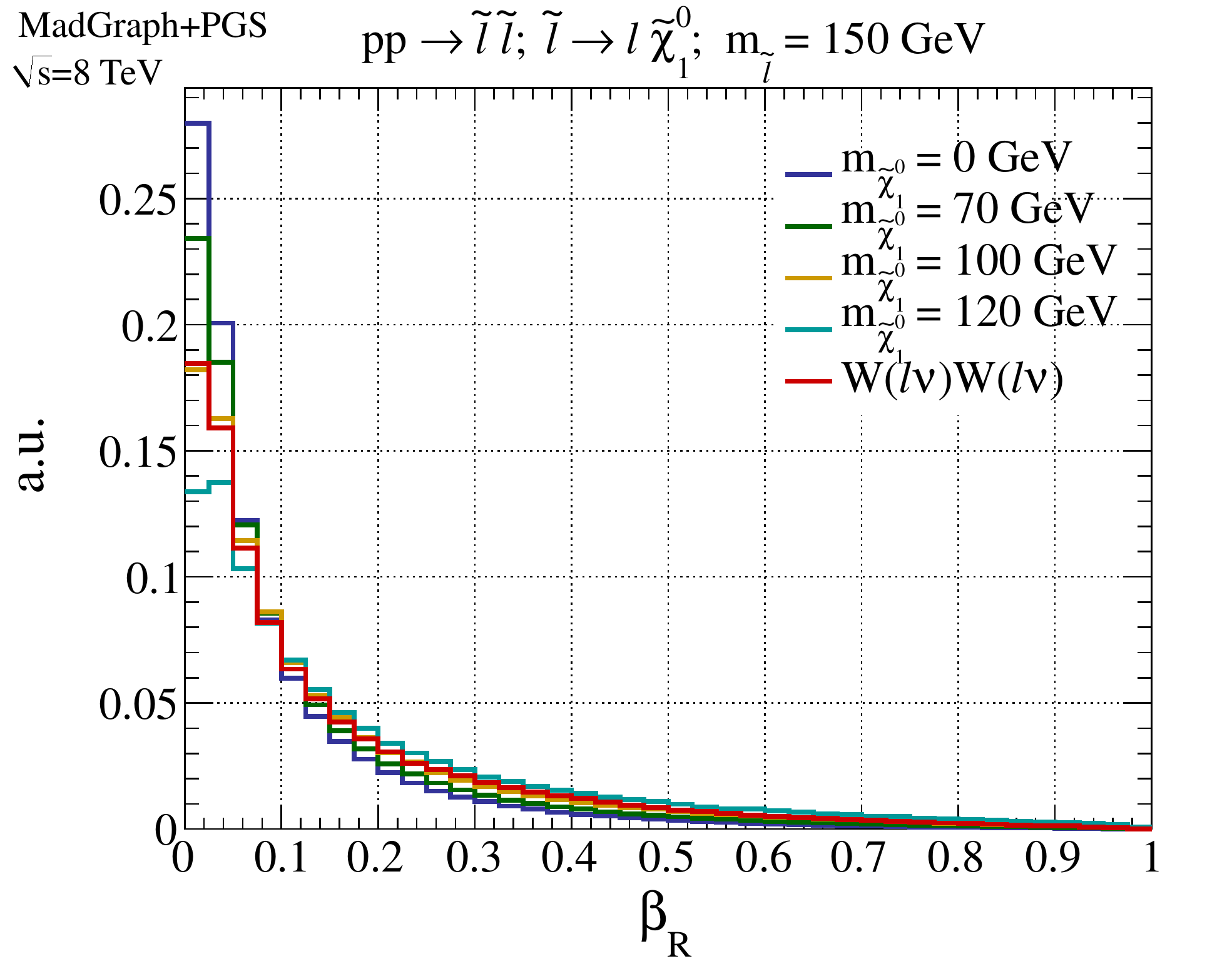}
\includegraphics[width=0.4\columnwidth]{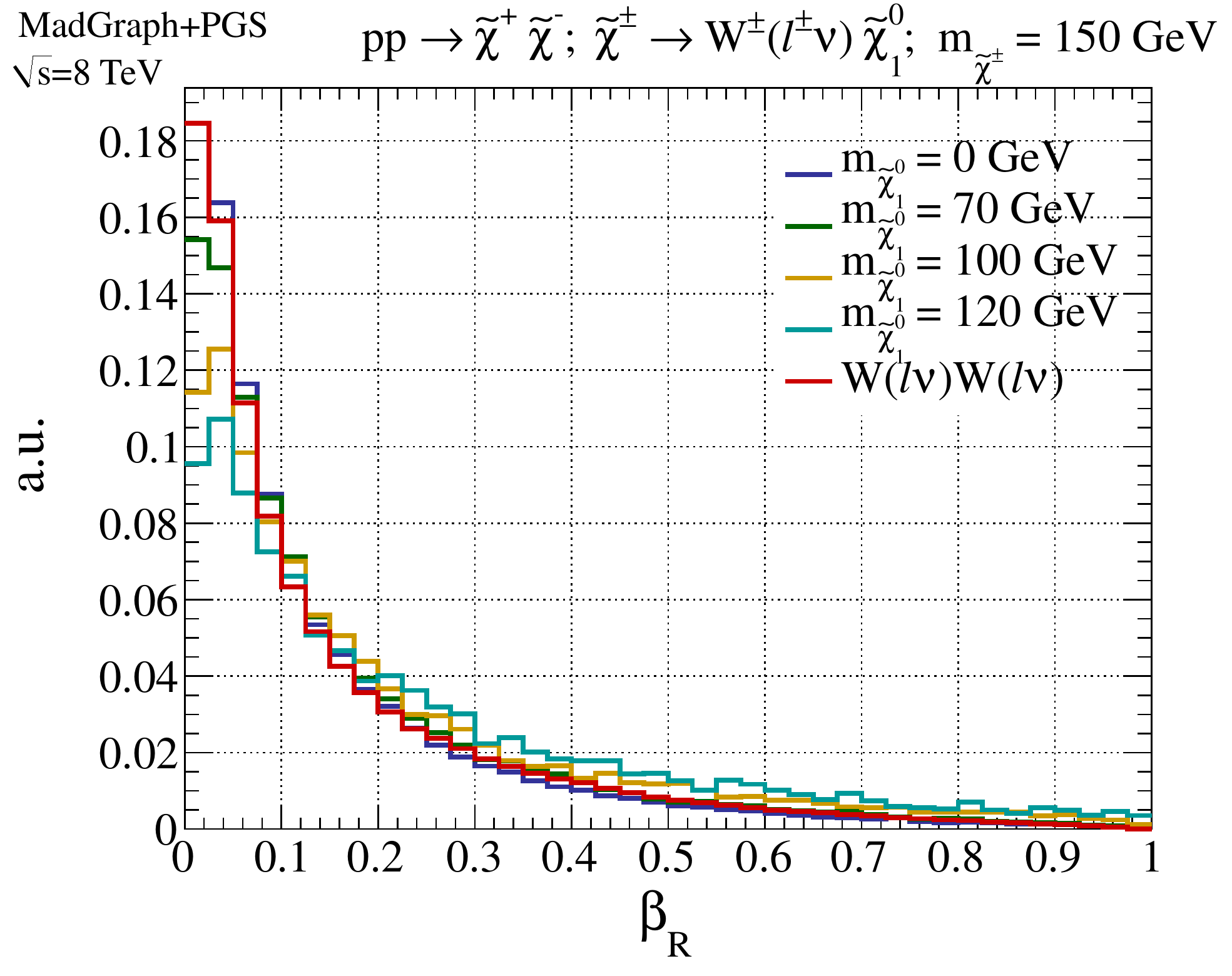}
\includegraphics[width=0.4\columnwidth]{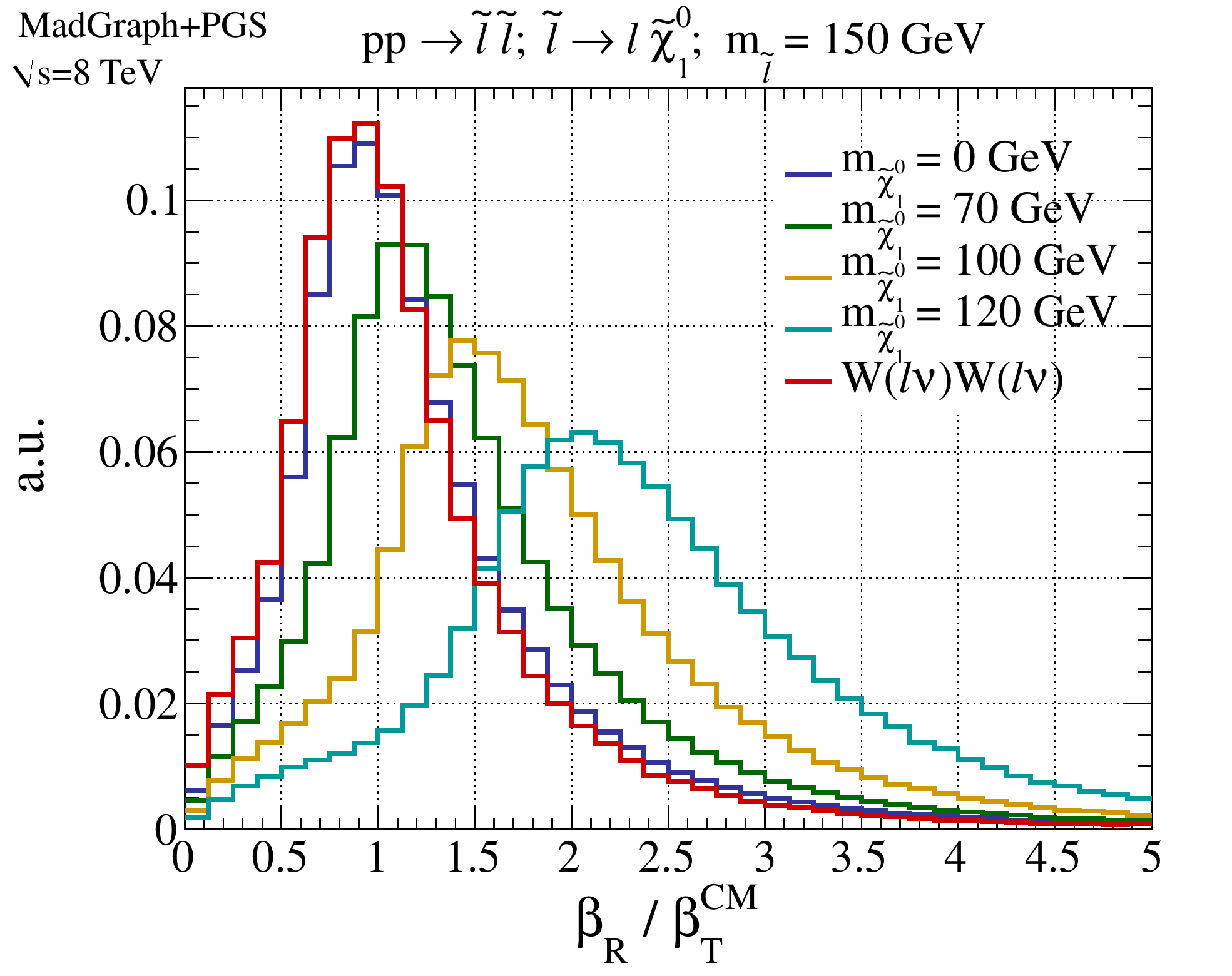}
\includegraphics[width=0.4\columnwidth]{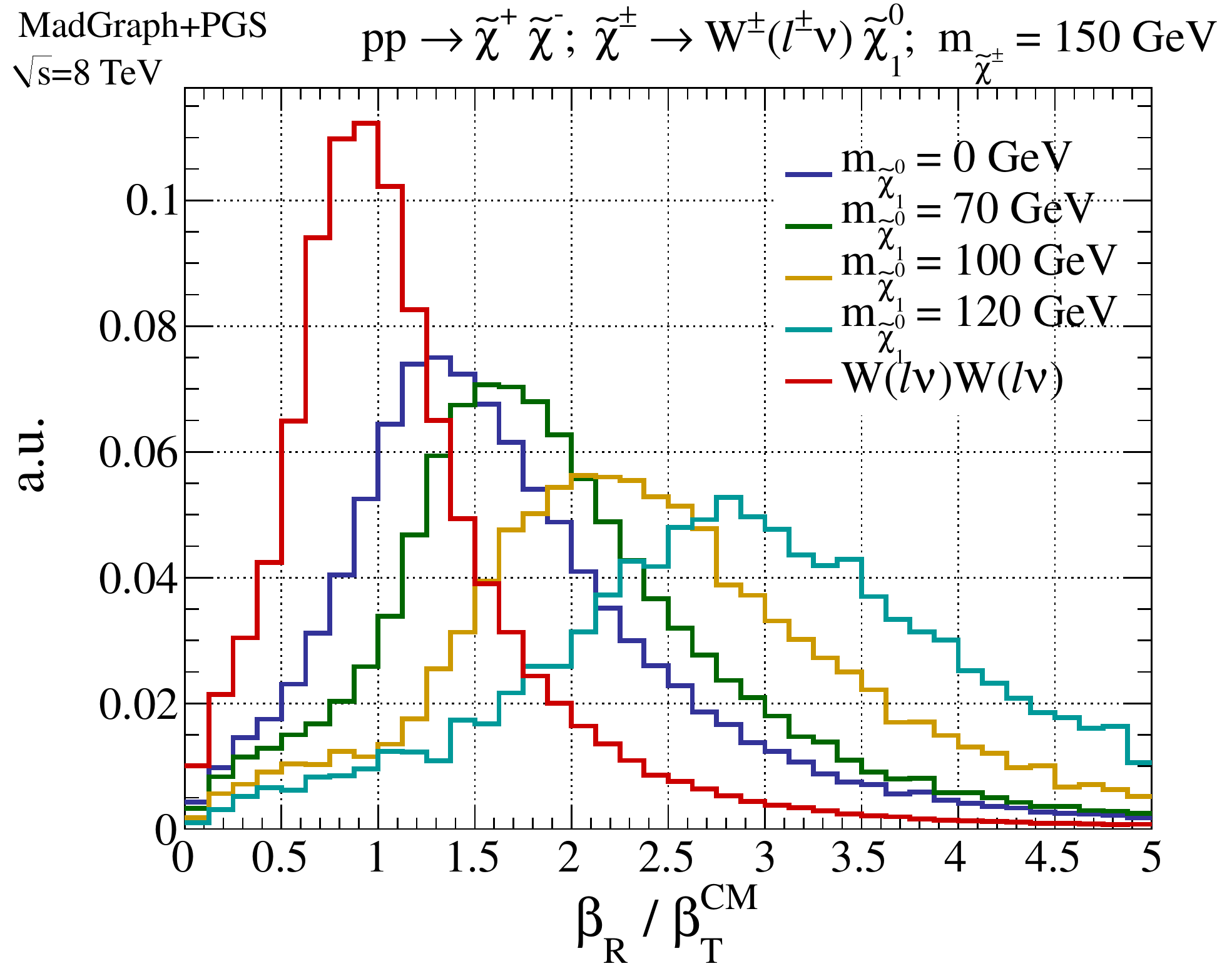}

\caption{Top Row: Distributions of $\beta_R$ for 150~GeV selectrons (left) or charginos (right) decaying into neutralinos and electrons, for a range of neutralino masses. Also shown is the distribution of the $W^-W^+$ background. Bottom Row: Distributions of normalized $\beta_R/\beta_T^{\, \rm CM}$ (right) for 150~GeV selectrons (left) or charginos (right) decaying into neutralinos, again for a range of neutralino masses. \label{fig:beta}}
\end{figure}

\begin{figure}[ht]
\includegraphics[width=0.4\columnwidth]{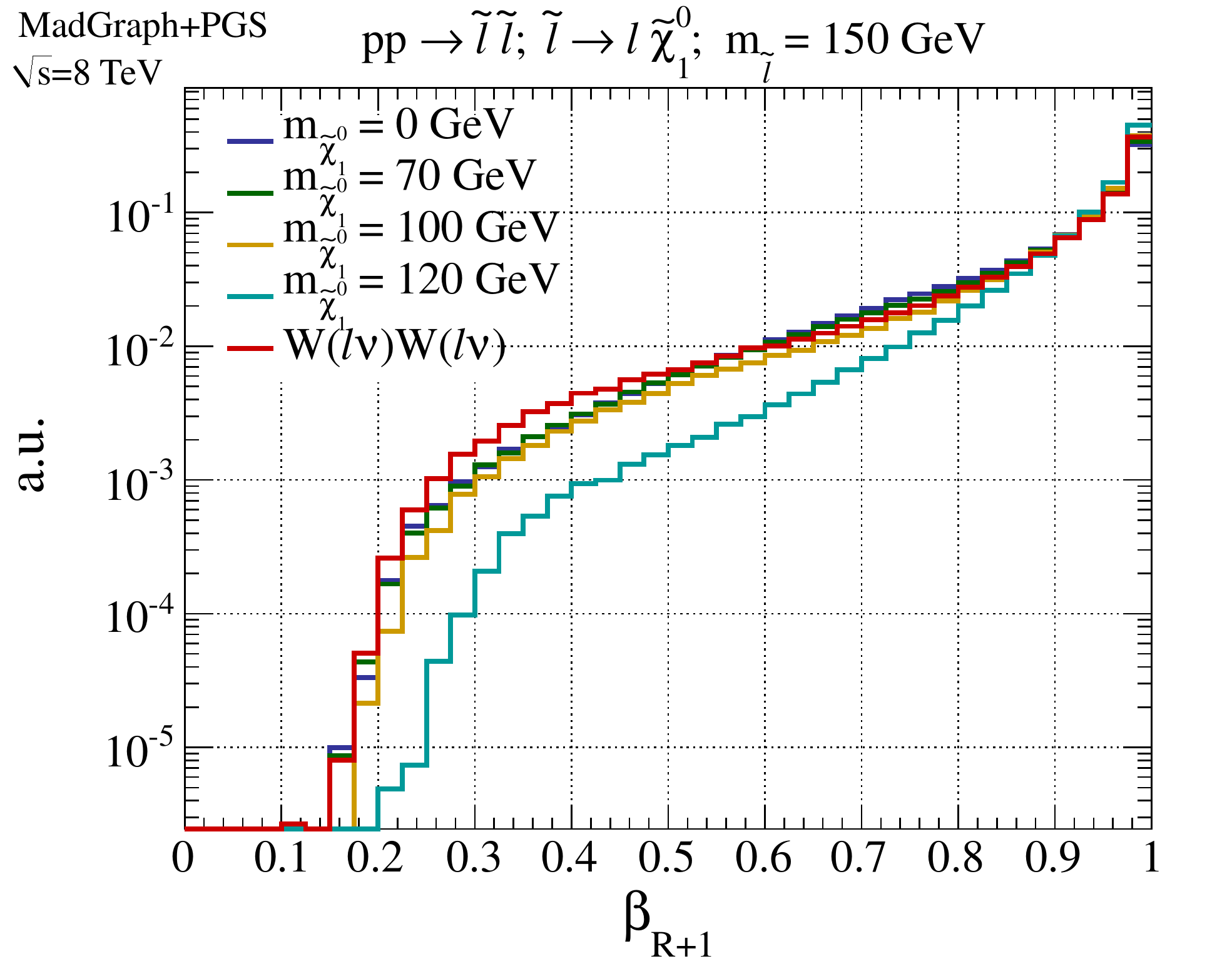}
\includegraphics[width=0.4\columnwidth]{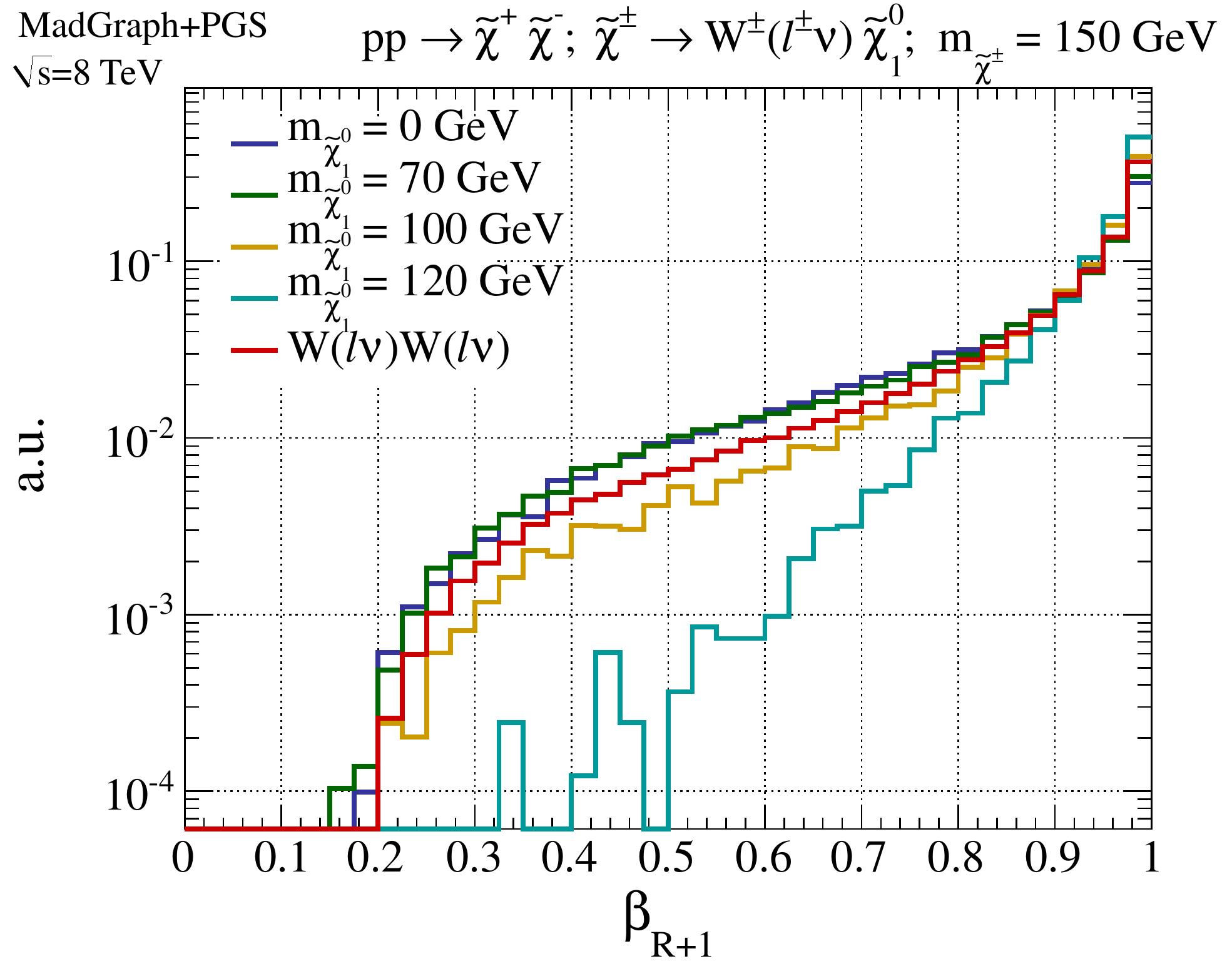}
\includegraphics[width=0.4\columnwidth]{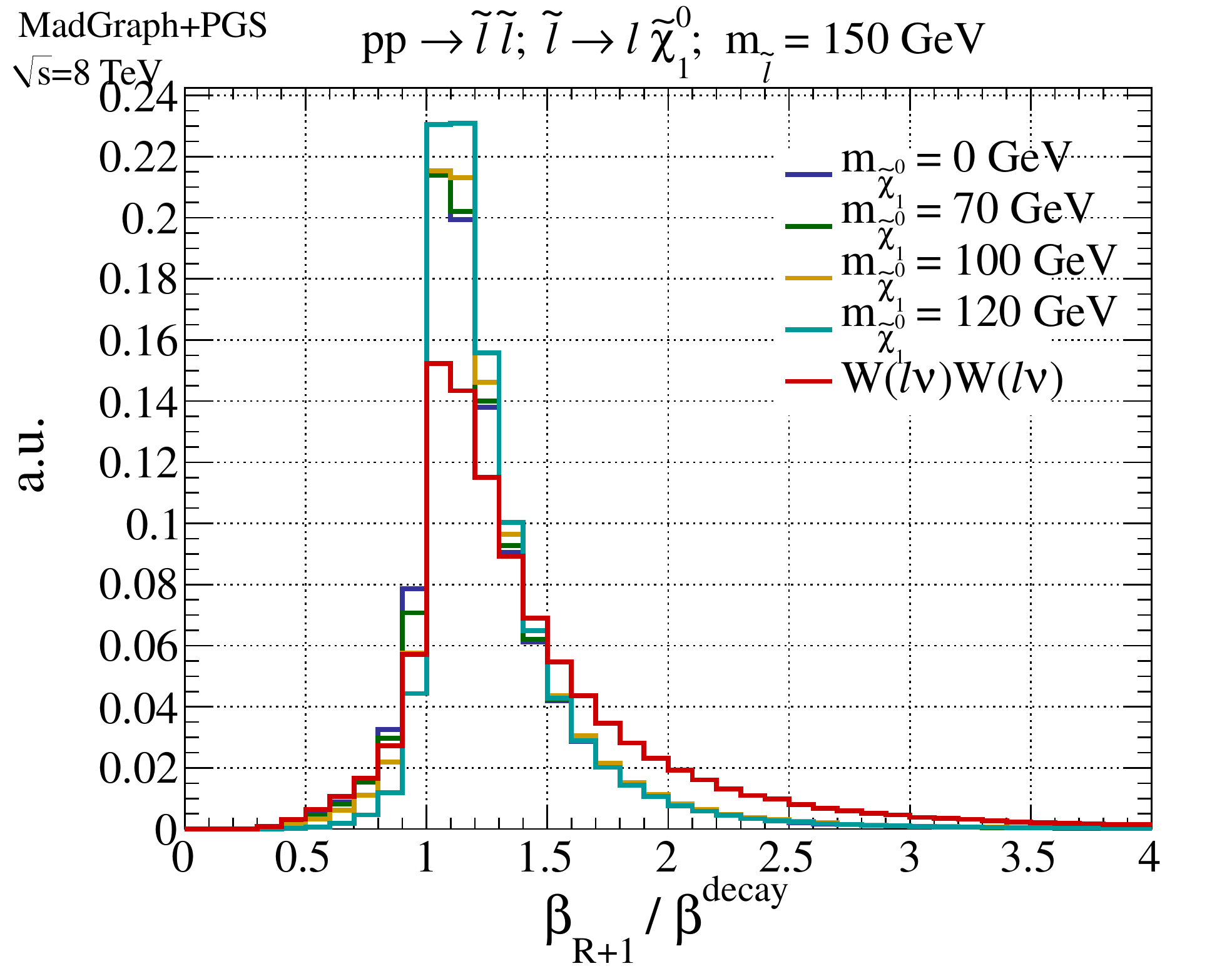}
\includegraphics[width=0.4\columnwidth]{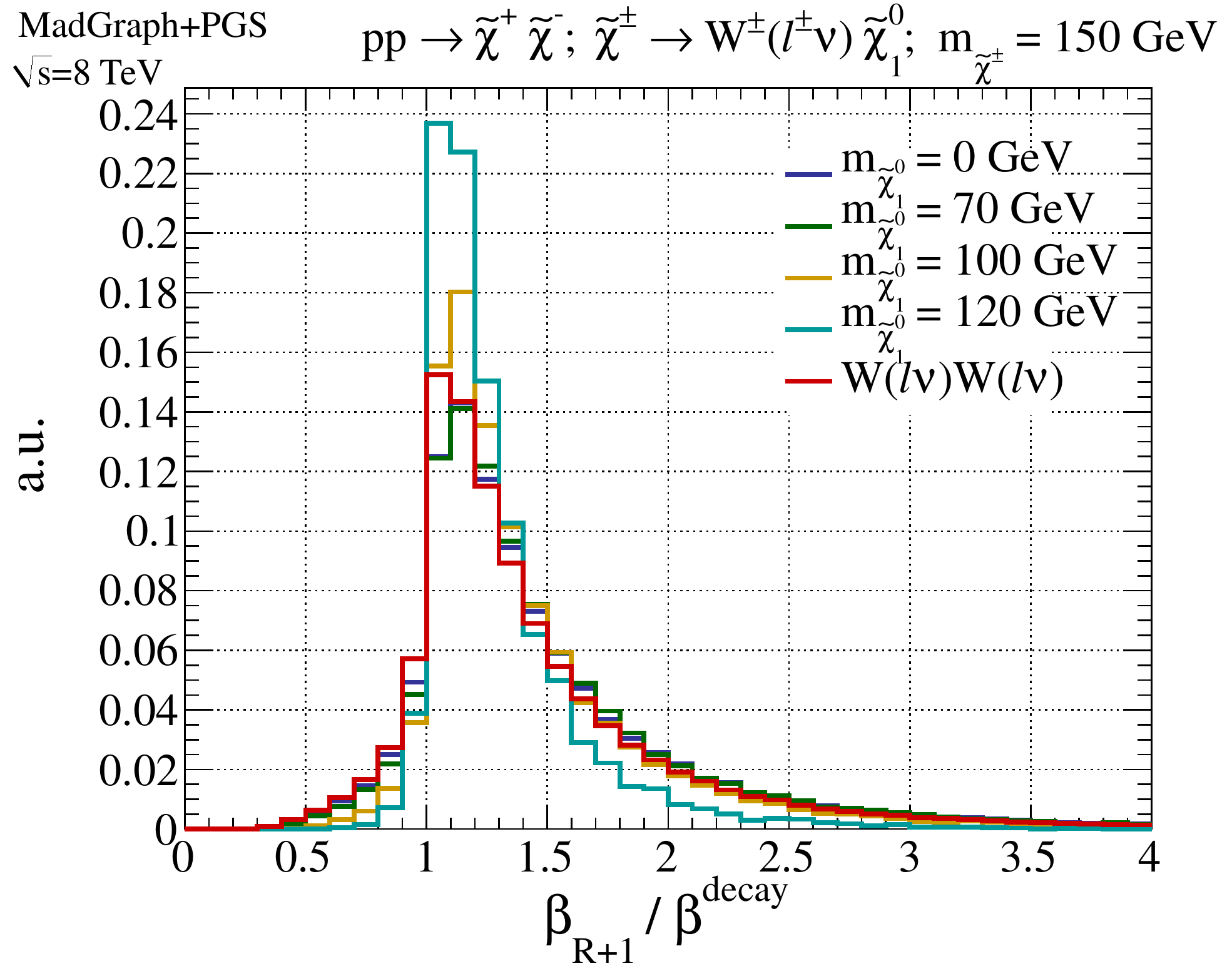}
\caption{Top Row: Distributions of $\beta_{R+1}$ for 150~GeV selectrons (left) or charginos (right) decaying into neutralinos and electrons, for a range of neutralino masses. Also shown is the distribution of the $W^-W^+$ background. Bottom Row: Distributions of normalized $\beta_{R+1}/\beta^{\, \rm decay}$ (right) for 150~GeV selectrons (left) or charginos (right) decaying into neutralinos, again for a range of neutralino masses. \label{fig:betap1}}
\end{figure}

Clearly, building these razor frames requires many assumptions, approximations, and choices that may appear to be {\it ad hoc}. We take the attitude that this technique is justified if, in the end, we find variables that well-approximate the true  values.  In Figure~\ref{fig:beta}, we plot the distributions of $\beta_R$ for both the primary $W^-W^+$ background and slepton or chargino signal production, in all cases decaying to two charged leptons and missing energy. We also plot the boosts $\beta_R$ normalized to the true transverse boost to the CM frame $\beta^{\, \rm CM}_T$.  The equivalent plots for $\beta_{R+1}$ (including normalization to $\beta^{\, \rm decay}$) are shown in Figure~\ref{fig:betap1}.

As expected for a proton-proton collider, the distributions of signal and background events all tend towards small boosts. For signal events, we see that we are systematically overestimating -- albeit slightly -- the magnitude of the boost $\beta_R$ as compared to the true value $\beta^{\, \rm CM}_T$ . This effect is more pronounced when the splitting between the parent and daughter is small. We also mis-estimate the boost by a larger amount for charginos as compared to sleptons. This makes sense, as in constructing $\sqrt{\hat{s}}_R$, we made the assumption that the visible and invisible invariant masses are equal. This becomes increasingly incorrect as the invisible system's mass increases. The presence of extra invisible particles (neutrinos) in the chargino decays also will systematically skew that measurement. We will shortly take advantage of these systematic differences between the mass of the invisible system in the background and signal events to increase our discrimination power using a new set of variables. 

\begin{figure}[ht]
\includegraphics[width=0.35\columnwidth]{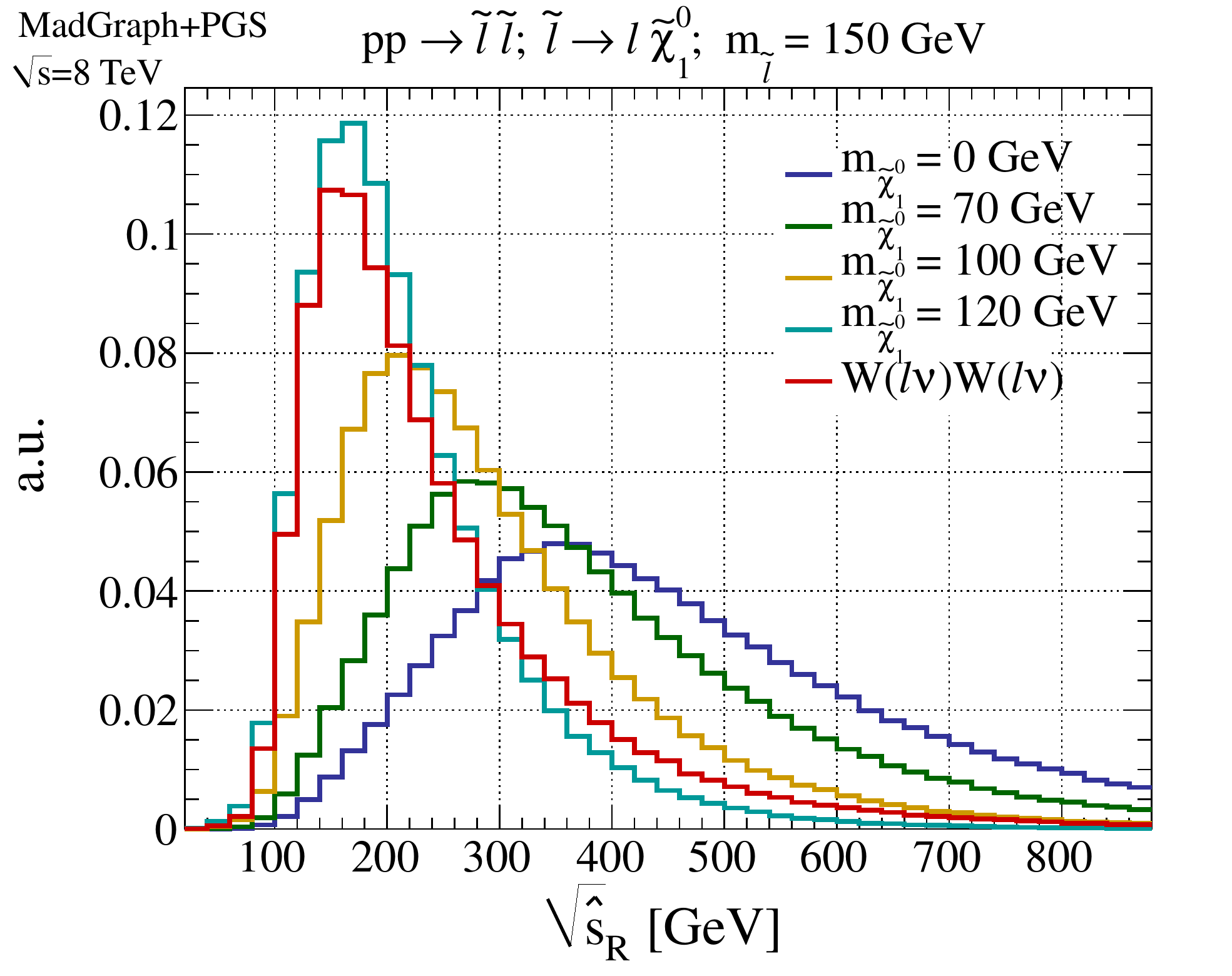}
\includegraphics[width=0.35\columnwidth]{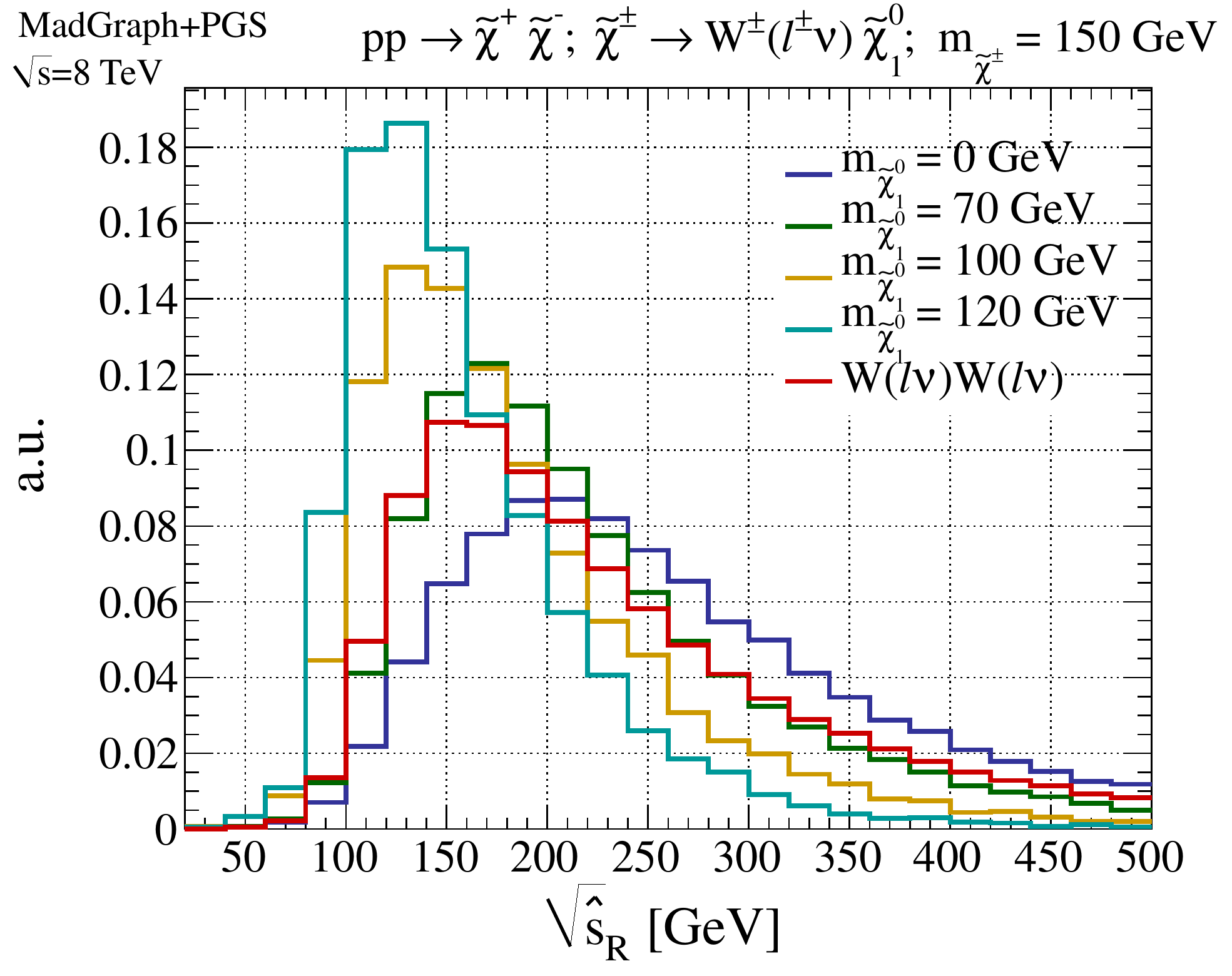}
\includegraphics[width=0.35\columnwidth]{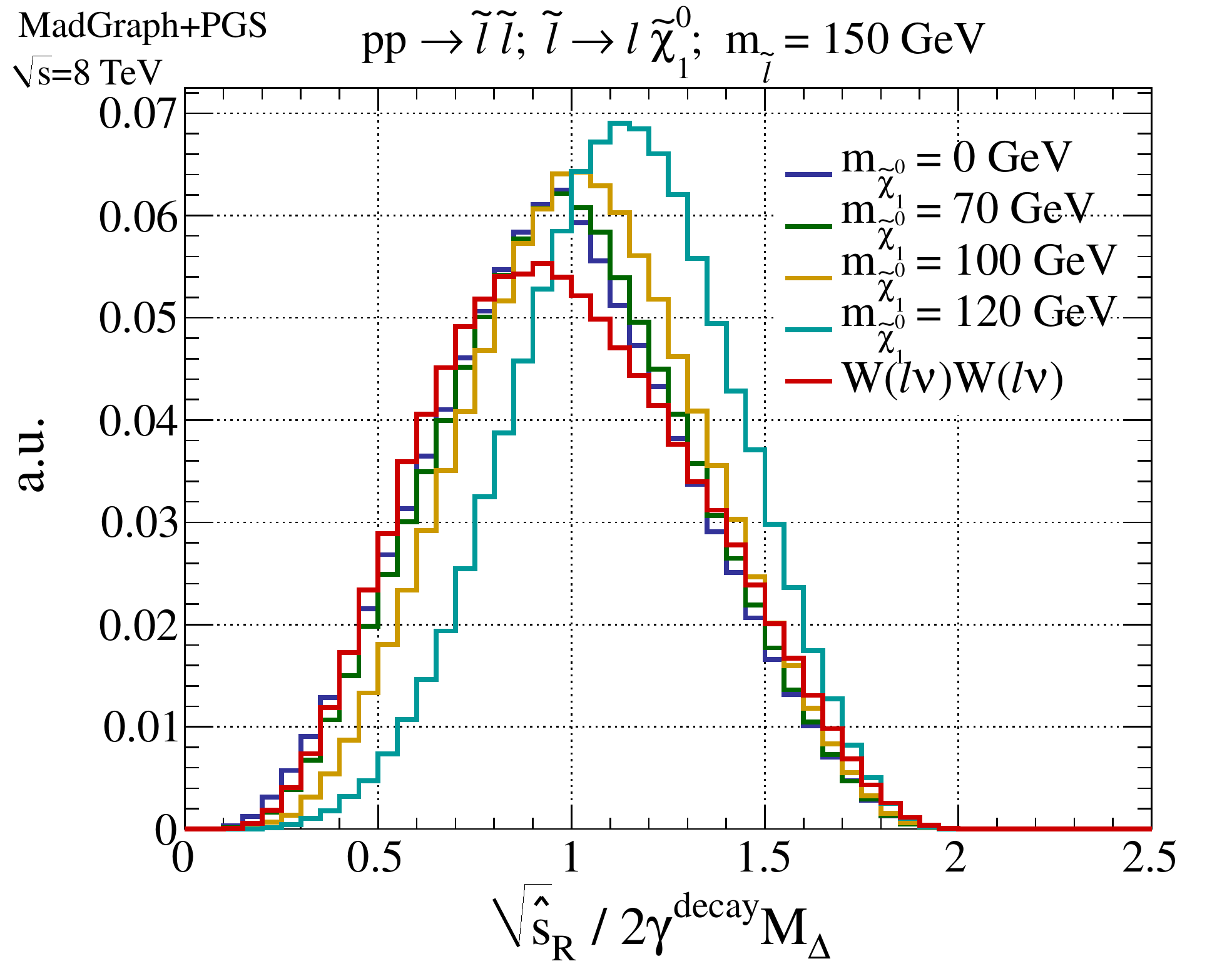}
\includegraphics[width=0.35\columnwidth]{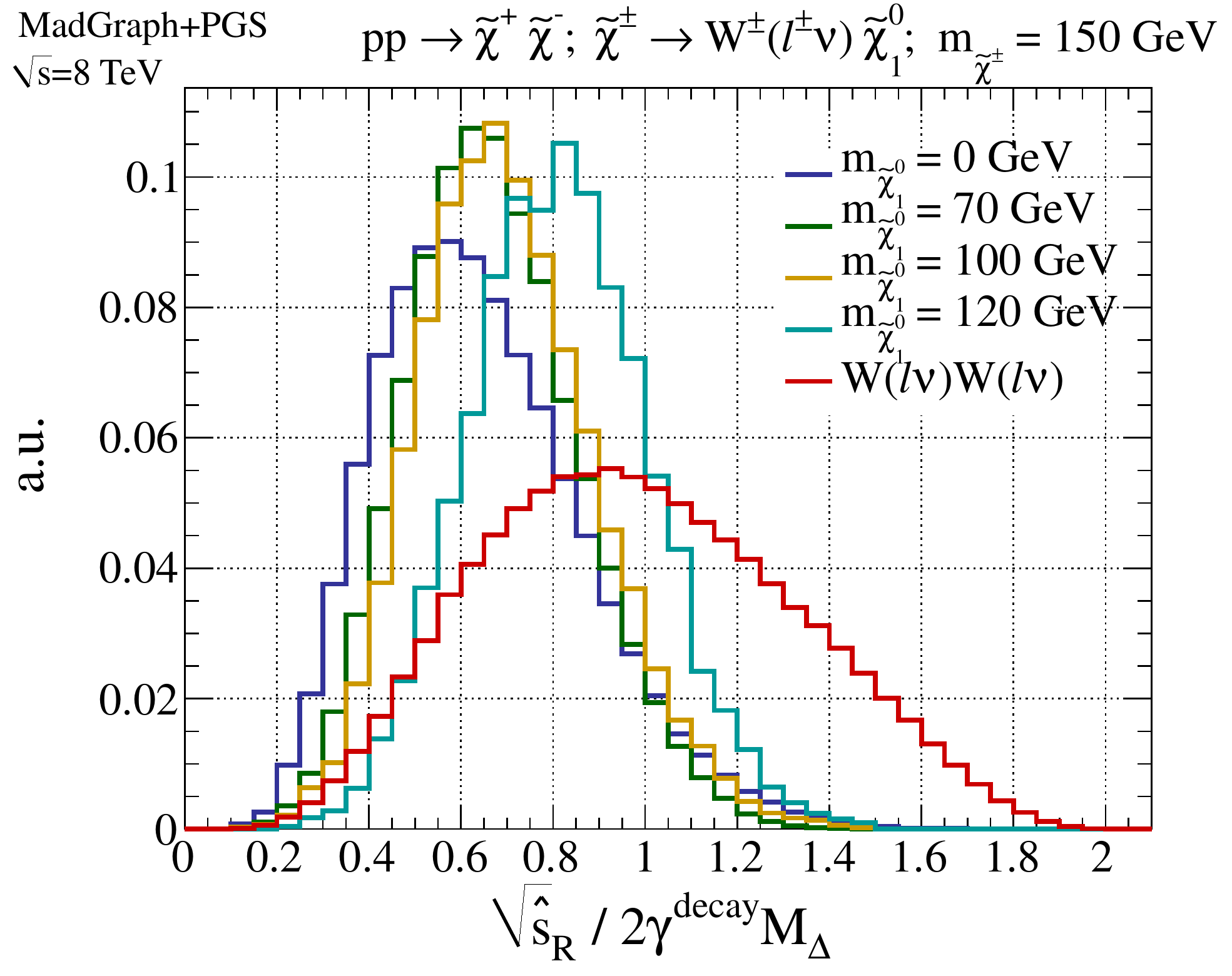}
\caption{Top Row: Distributions of $\sqrt{\hat{s}}_R$ for a 150~GeV slepton (left) or chargino (right) and a range of neutralino masses. Also shown is the distribution of the $W^-W^+$ background.  Bottom row: Distributions of $\sqrt{\hat{s}}_R$ normalized to $2 \gamma^{\rm decay} M_\Delta$ for selectrons (left) and charginos (right), again for a range of neutralino masses. \label{fig:compare_hats}}
\end{figure}

\begin{figure}[ht]
\includegraphics[width=0.35\columnwidth]{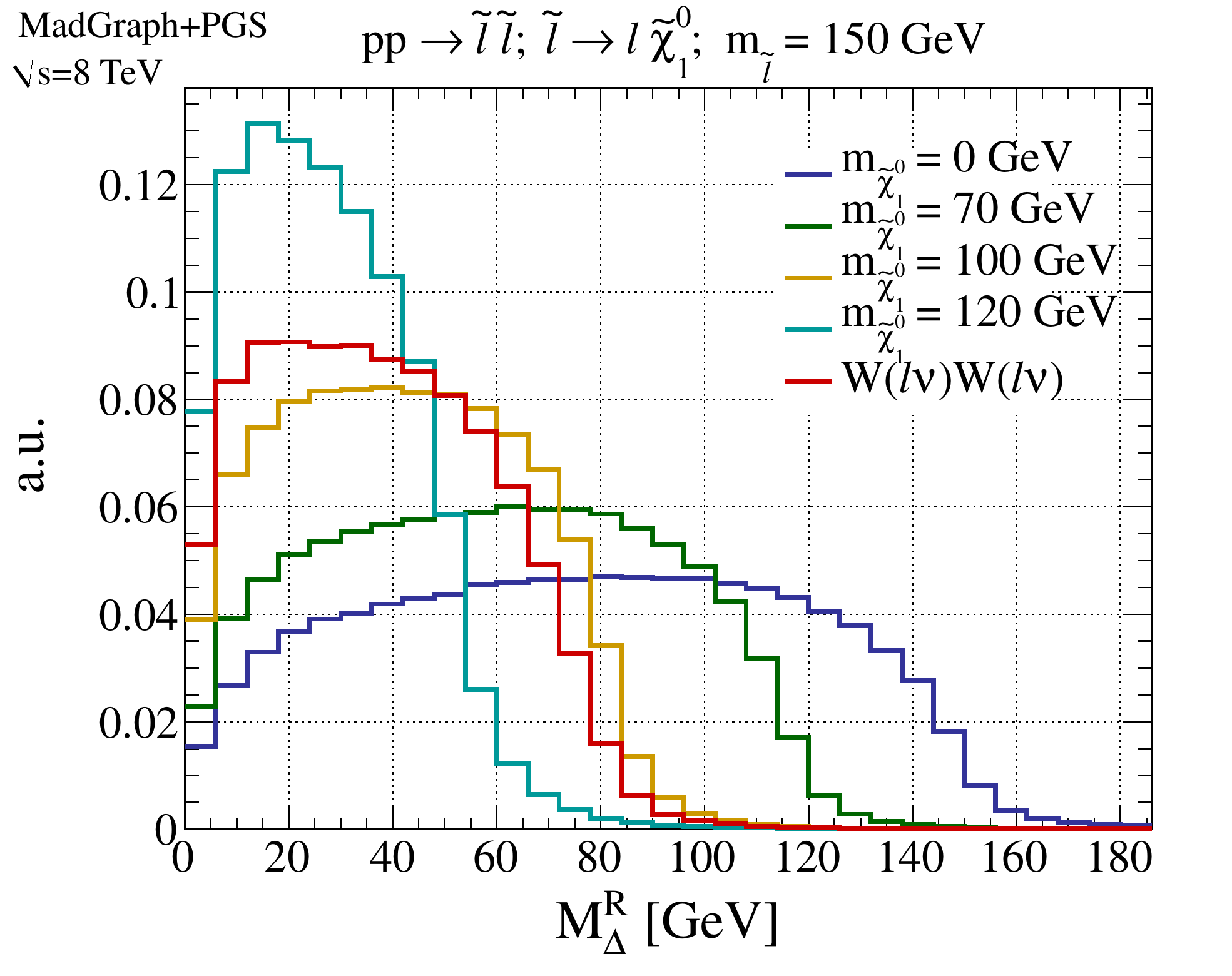}
\includegraphics[width=0.35\columnwidth]{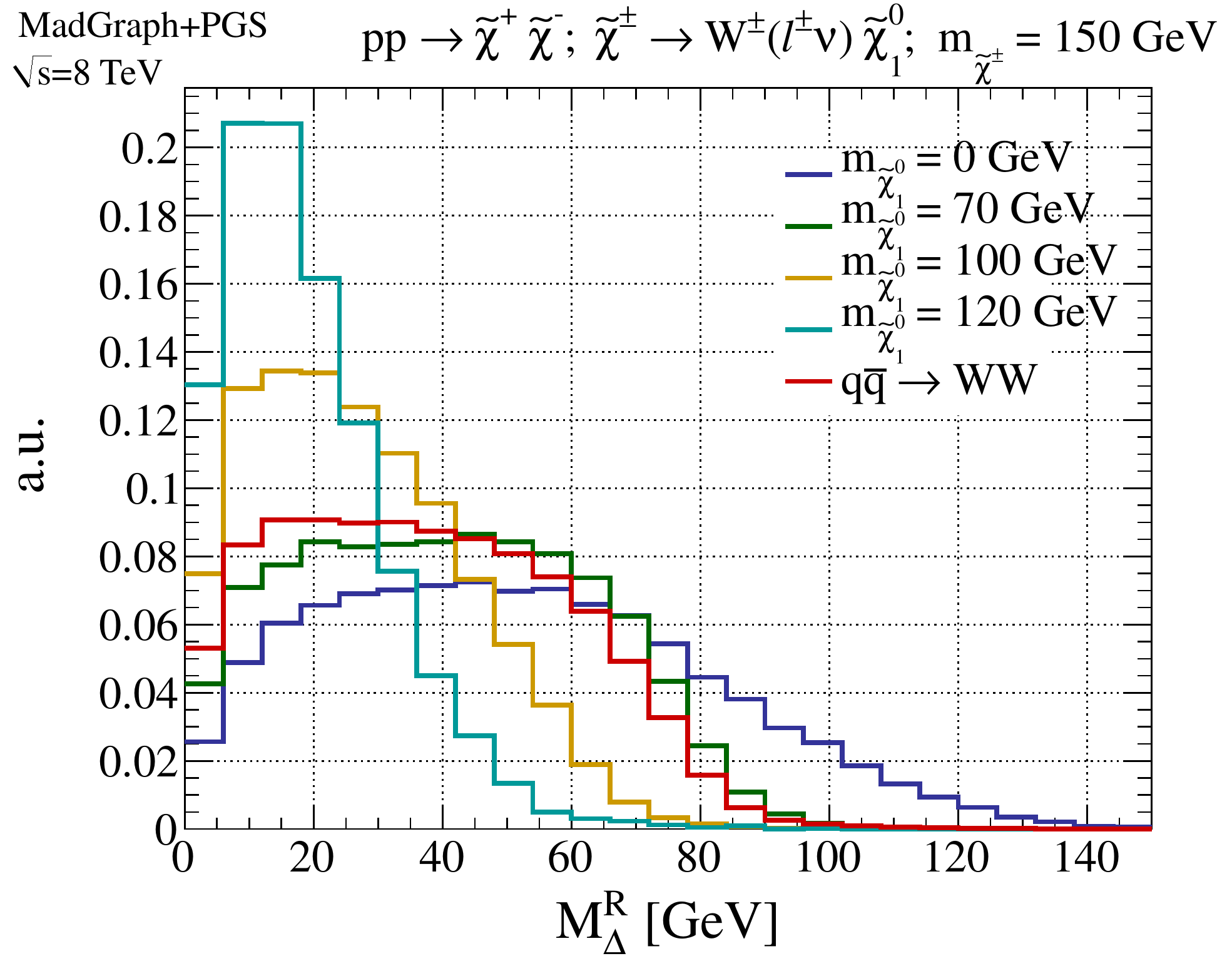}
\includegraphics[width=0.35\columnwidth]{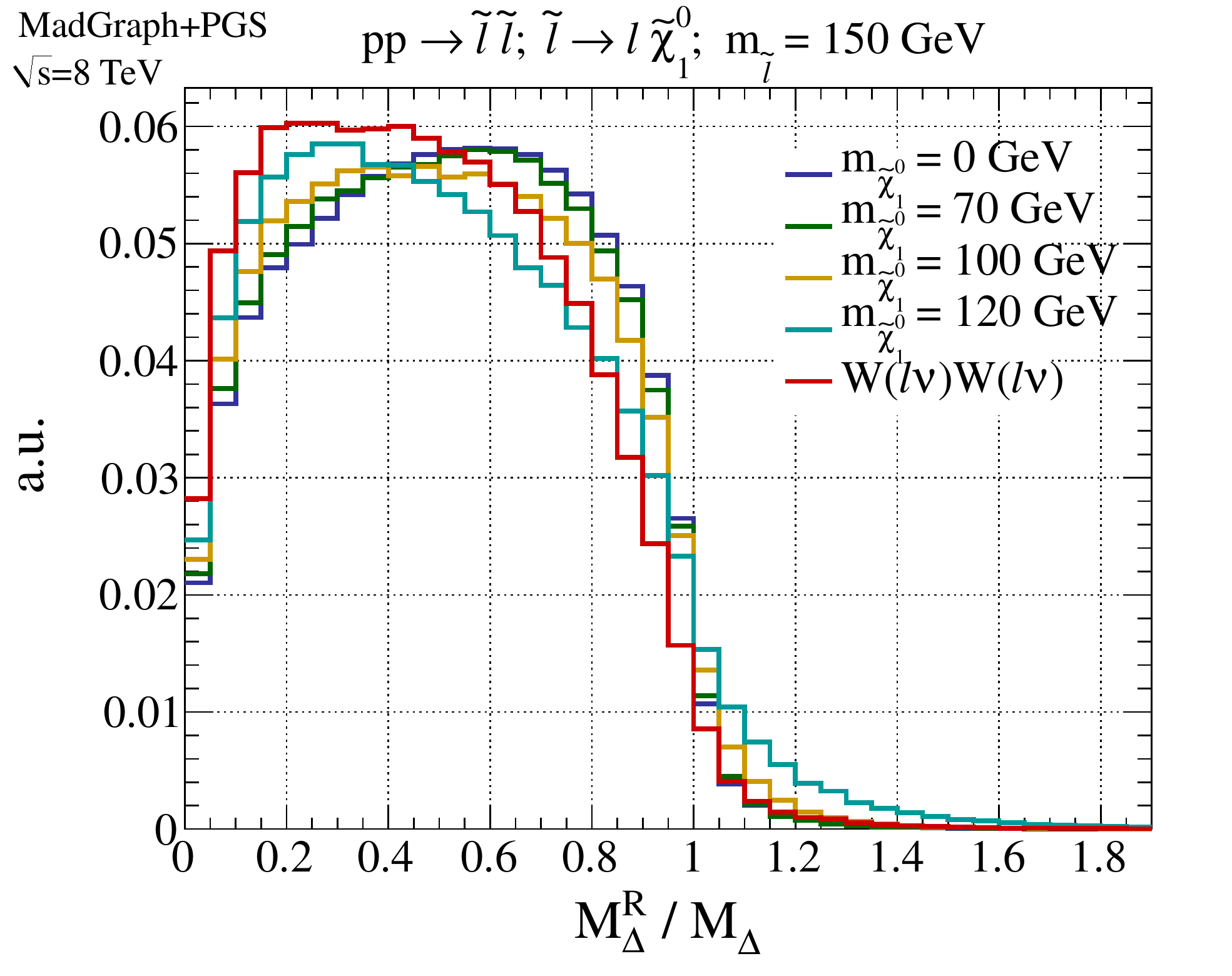}
\includegraphics[width=0.35\columnwidth]{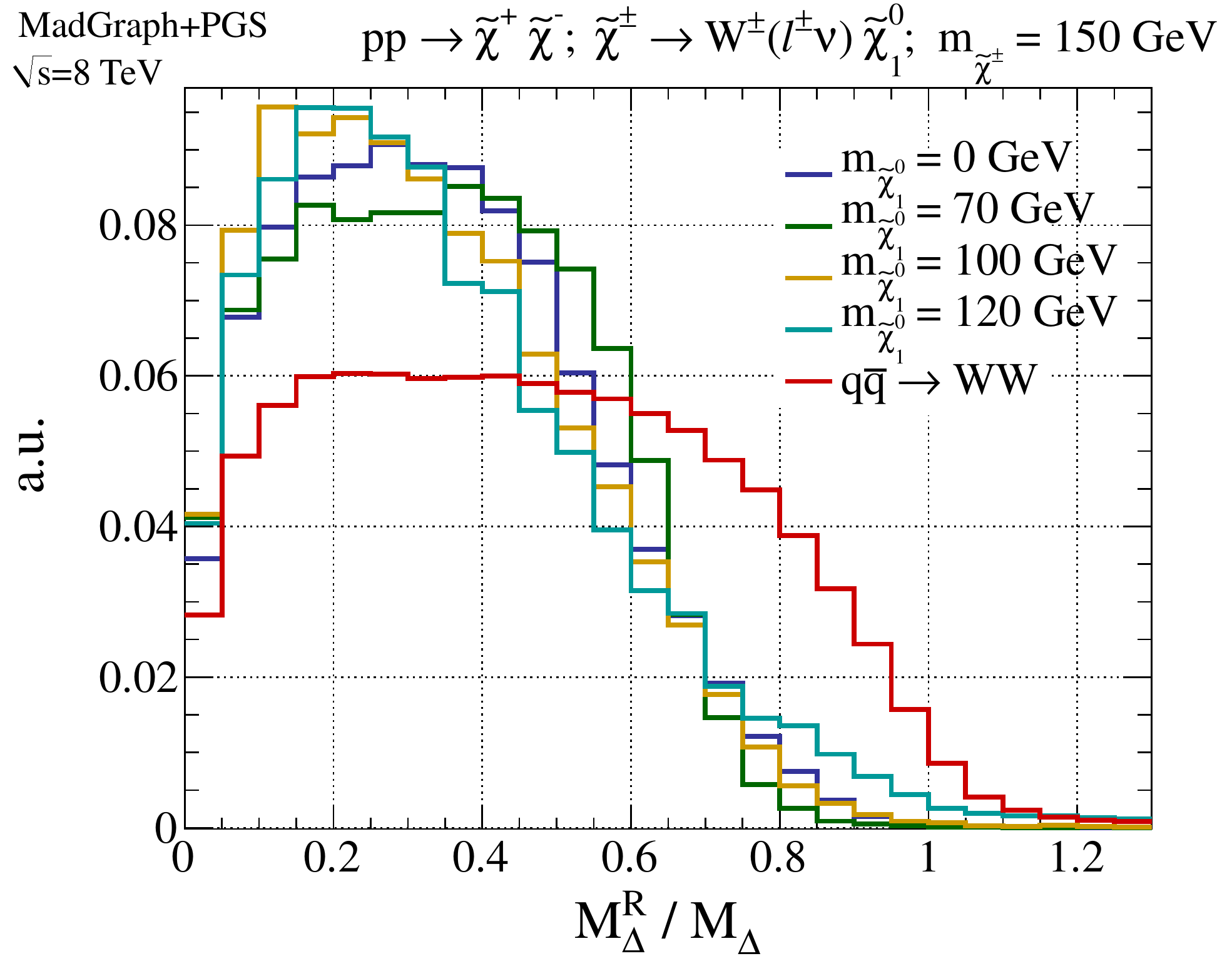}
\caption{Top Row: Distributions of $M_\Delta^R$ for a 150~GeV slepton (left) or chargino (right) and a range of neutralino masses. Also shown is the distribution of the $W^-W^+$ background. For the $W$ background, $M_\Delta = m_W$. Bottom row: Distributions of $M_\Delta^R$ normalized to $M_\Delta$ for selectrons (left) and charginos (right), again for a range of neutralino masses  \label{fig:compare_mdelta}}
\end{figure}

In Figure~\ref{fig:compare_hats} we plot the distributions of $\sqrt{\hat{s}}_R$ for a range of signal points and the $W^-W^+$ background, as well as the ratio of this razor variable to the quantity it is intended to estimate, $2\gamma^{\rm decay}M_\Delta$ (for background, the splitting $M_\Delta$ is the $W$ boson mass $m_W$, as the neutrino is massless for our purposes). As can be seen, the razor approximation is reasonably good, though systematically low for signal points with massive neutralinos, and less accurate for charginos than sleptons, for the reasons discussed previously. In Figure~\ref{fig:compare_mdelta}, we plot the distributions of the variable $M_\Delta^R$, both by itself and normalized to the estimator value of $M_\Delta$. The sharp edge at $M_\Delta^R = M_\Delta$ (seen most clearly in the slepton plot) indicates that this variable is useful in searches for new physics, especially in the regime where $M_\Delta$ is greater than the mass of the $W$ boson.



\begin{figure}[ht]
\includegraphics[width=0.32\columnwidth]{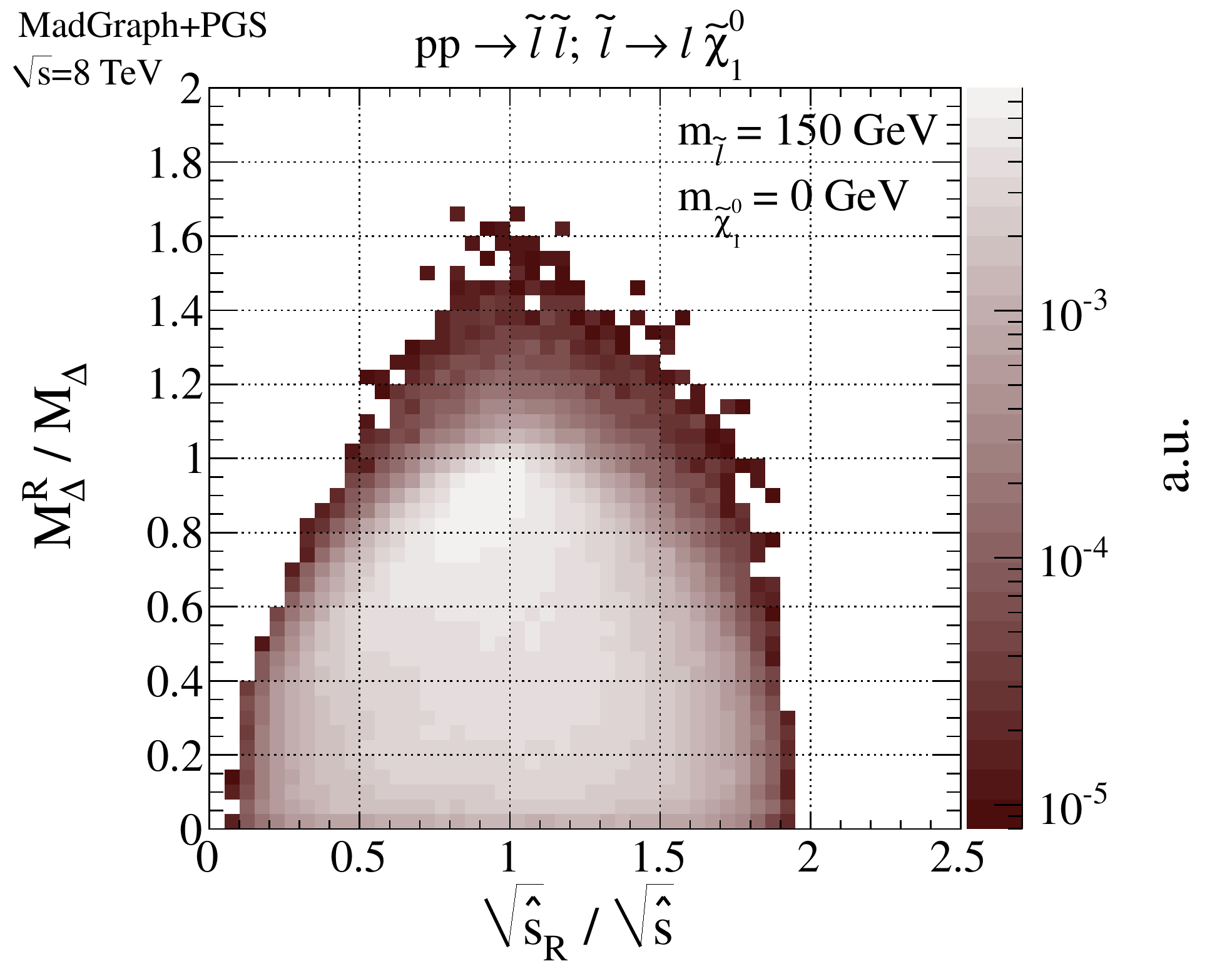}
\includegraphics[width=0.32\columnwidth]{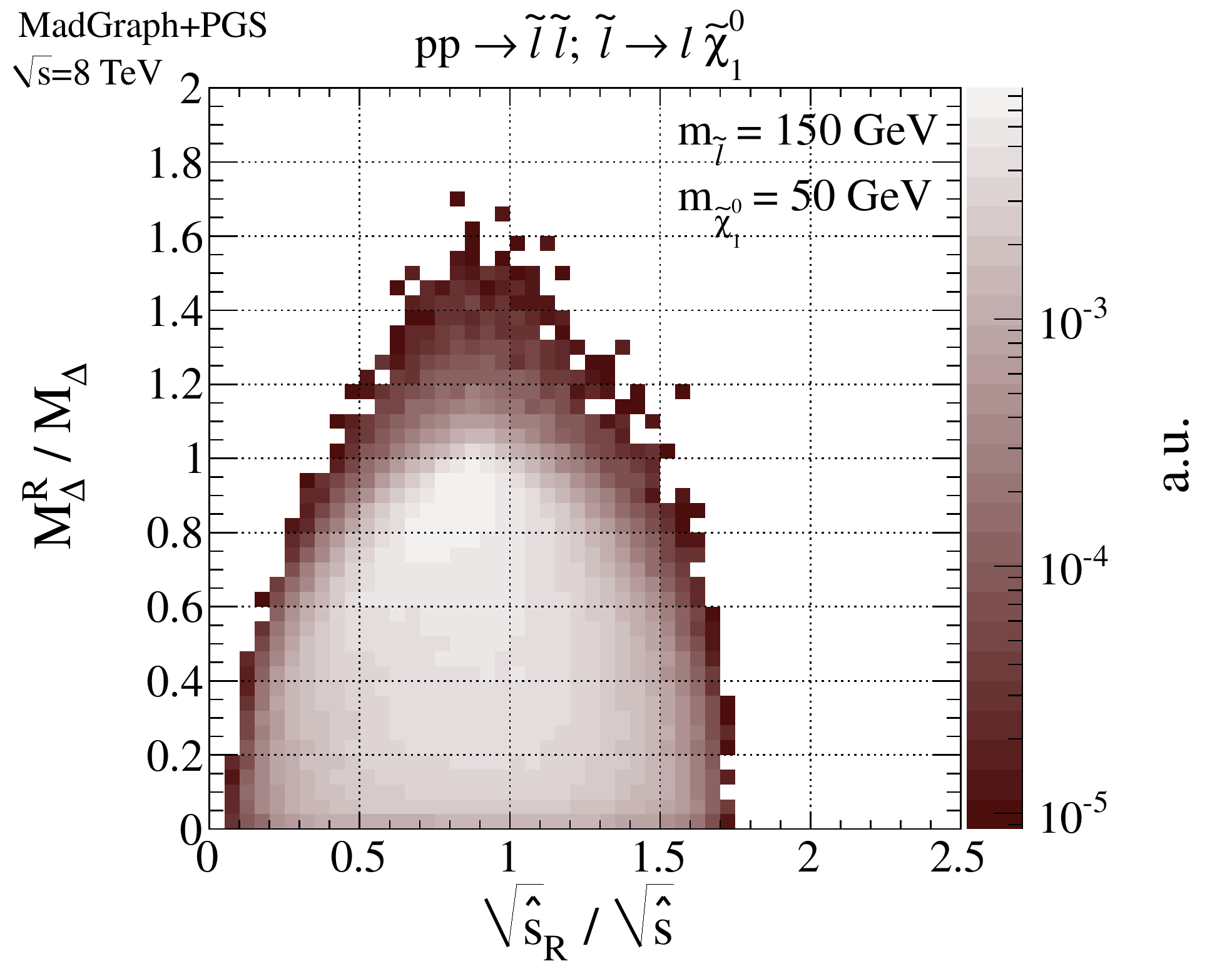}
\includegraphics[width=0.32\columnwidth]{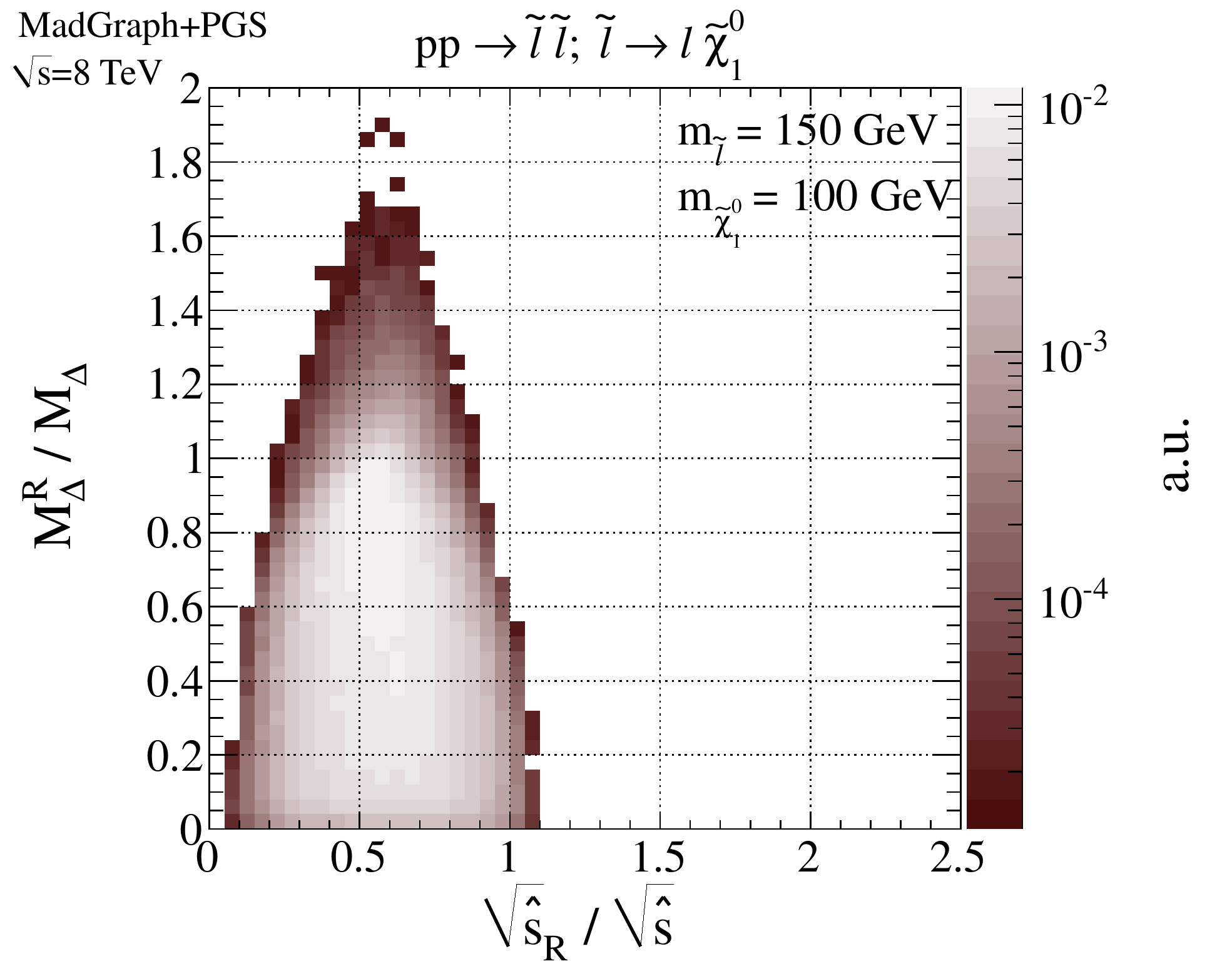}
\includegraphics[width=0.32\columnwidth]{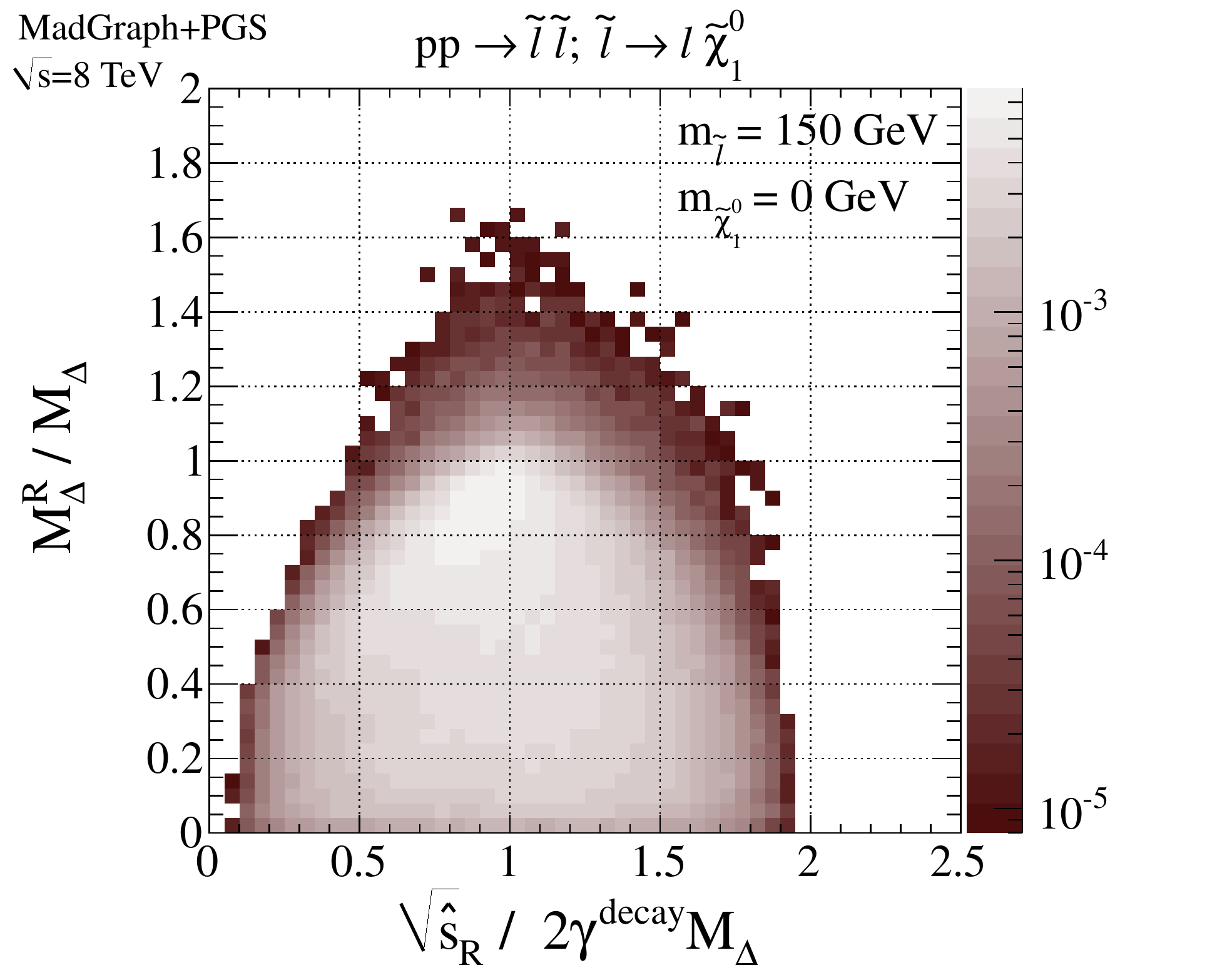}
\includegraphics[width=0.32\columnwidth]{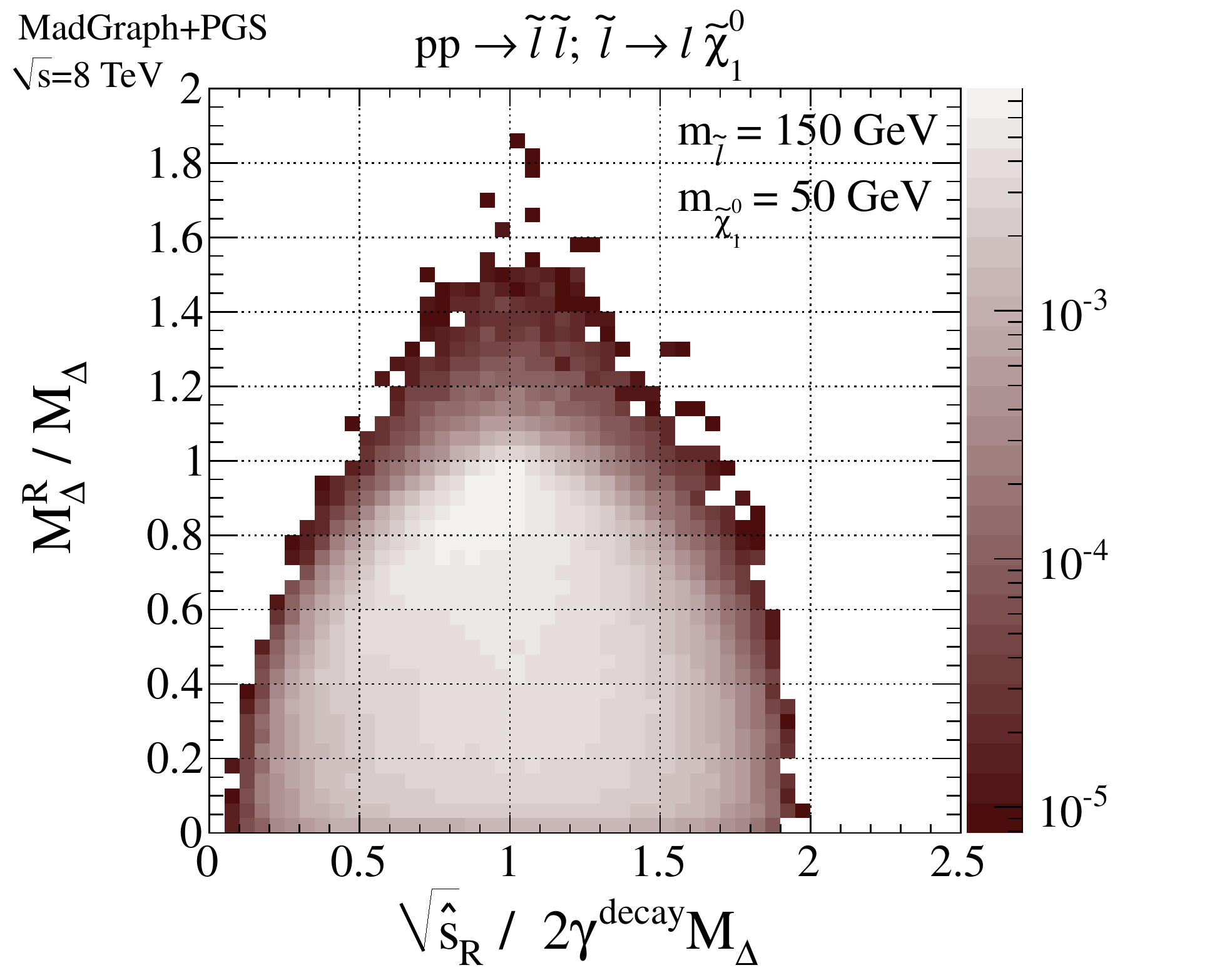}
\includegraphics[width=0.32\columnwidth]{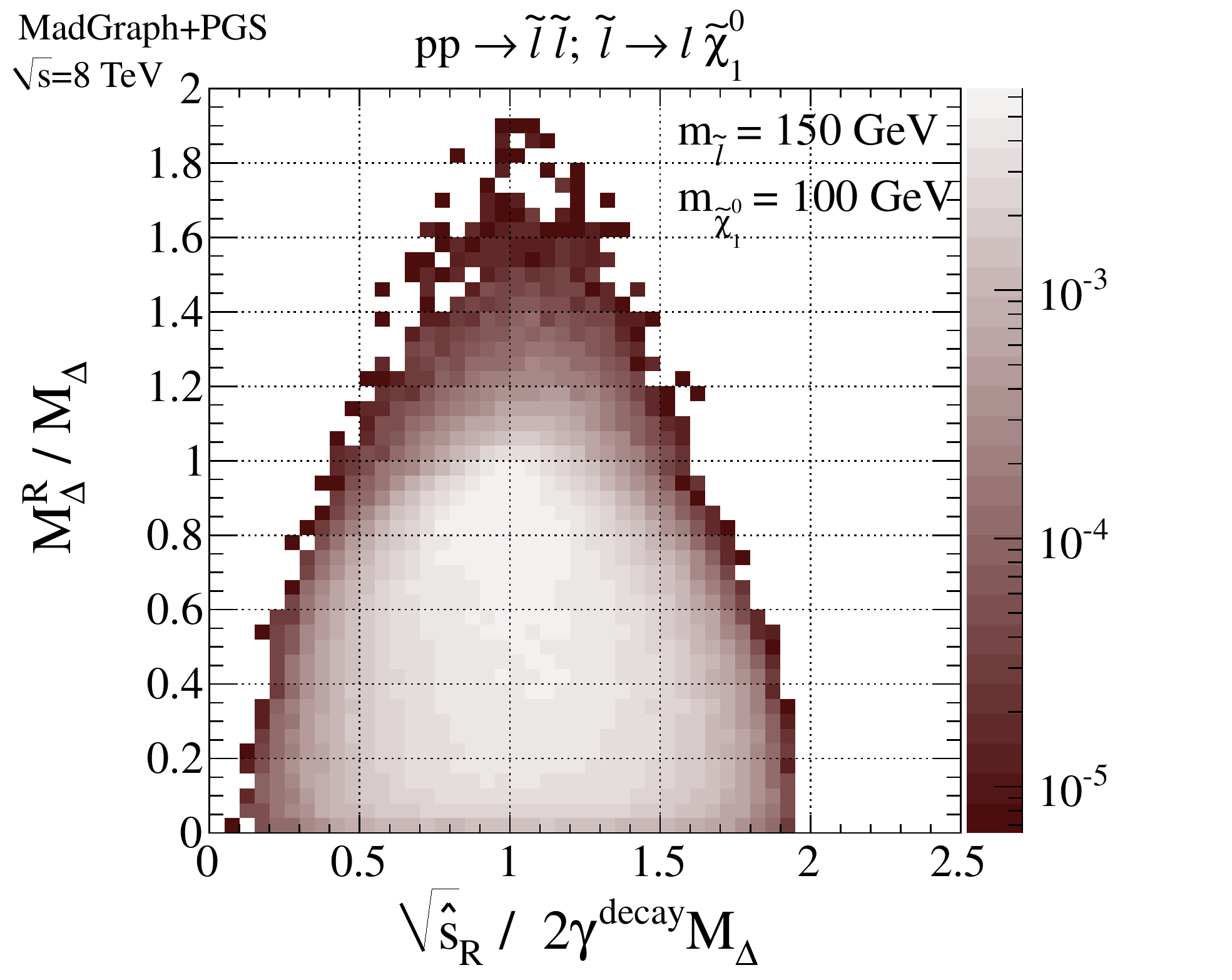}
\caption{Top Row: Distributions of $\sqrt{\hat{s}}_R/\sqrt{\hat{s}}$ vs.~$M_\Delta^R/M_\Delta$ for 150~GeV sleptons and a range of $m_{\tilde{\chi}})$ masses. Bottom Row: Distributions of $\sqrt{\hat{s}}_R/2 \gamma^{\rm decay}M_\Delta$ vs.~$M_\Delta^R/M_\Delta$ for 150~GeV sleptons and a range of $m_{\tilde{\chi}})$ masses.  \label{fig:compare_shat_mdelta}}
\end{figure}

Both $\hat{s}_R$ and $M_\Delta^R$ contain information about the mass splitting $M_\Delta$ for signal events. In Figure~\ref{fig:compare_shat_mdelta}, we plot the two variables (normalized to the physical quantities they estimate). Two things can be seen from these plots. First, the variables $\hat{s}_R$ and $M_\Delta^R$ are not degenerate; though both estimate the same quantity ($M_\Delta$), they contain independent kinematic information in that estimation. Secondly, we see that the scatter of the $\hat{s}_R$ around the true value is minimized near the edge structure of the $M_\Delta^R$ variable. This second piece of information will not be fully utilized in our analyses for computational simplicity, though it may provide a useful handle in the future.

As with the original razor variables $M_R$ and $M_T^R$, one or both of these new razor variables could be used. However, we would ideally like a variables that encapsulated information not about $M_\Delta$, but about the overall mass scale of the new particles in the event. This would help distinguish signal from background, especially in the cases where the mass difference is very small ({\it i.e.}~parent and invisible daughter are nearly degenerate in mass), or when the mass difference approaches the mass of the $W$. 

\begin{figure}[ht]
\includegraphics[width=0.7\columnwidth]{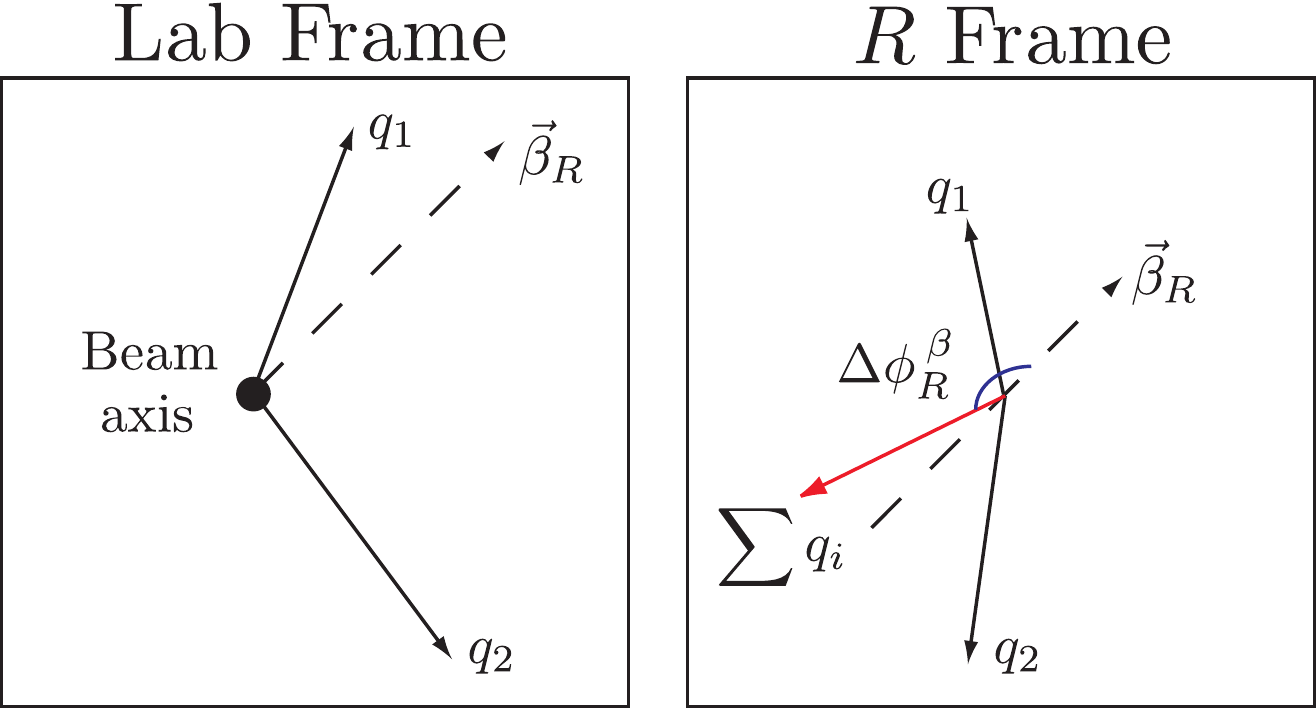}
\caption{Schematic example of the definition of the azimuthal angle $\Delta\phi_R^{\beta}$. The lab frame (seen here down the beam-line) contains two visible objects, $q_1$ and $q_2$. The direction of the boost $\vec{\beta}_R$ (defined in Eq.~\eqref{eq:razorbeta}), in the lab frame is also shown. In the frame $R$, arrived at by performing the boost $\vec{\beta}_R$, the visible momenta $q_1$ and $q_2$ are shown, along with their sum. The azimuthal angle between their sum $q_1+q_2$ and the boost direction $\vec{\beta}_R$ in frame $R$ defines $\Delta \phi_R^{\beta}$. \label{fig:deltaphi}}
\end{figure}

To try to capture more information about the event, we move beyond the mass variables already introduced and look at kinematic angles. In particular, we will be interested in the azimuthal angle between the razor boost $\vec{\beta}_R$ between the lab and $R$ frames and the sum of the visible momenta $\vec{q}_1+\vec{q}_2$, calculated in the razor frame $R$. An illustrative example of the relevant kinematics and angle definition is shown in Figure~\ref{fig:deltaphi}. We call this angle $\Delta \phi_R^{\beta}$, as it is the difference in azimuthal angle between the visible system and the boost $\vec{\beta}_R$, all defined in the razor frame $R$. 

This angle is useful because it inherits information about ratio of masses of the pair produced particles and their invisible daughters, and so can be used in conjunction with a variable such as $M_\Delta^R$ or $\sqrt{\hat{s}}_R$, which have information about the mass difference $M_\Delta$, as previously discussed. The sensitivity of this angular variable to the ratio of masses actually comes from the previously discussed systematic shift of the variable $\sqrt{\hat{s}}_R$ relative to the mass difference $M_\Delta$. As can be seen from Figures~\ref{fig:beta} and \ref{fig:compare_hats}, our estimators of $\beta^{\,\rm CM}$ and $\hat{s}$ ($\beta_R$ and $\hat{s}_R$), do not completely track the center of mass energy of the pair production. $\sqrt{\hat{s}}_R$, for example, is systematically smaller than $\hat{s}$, and $\beta_R$ systematically larger than $\beta^{\, \rm CM}$. This behavior can be easily understood: it is due to the assumption that the energy of the event is evenly split between the visible and invisible systems. For events with invisible particles that are heavy compared to the parent, this assumption will underestimate the energy associated with the missing transverse momentum, and thus $\hat{s}_R$ is an underestimate of $\hat{s}$.

\begin{figure}[ht]
\includegraphics[width=0.45\columnwidth]{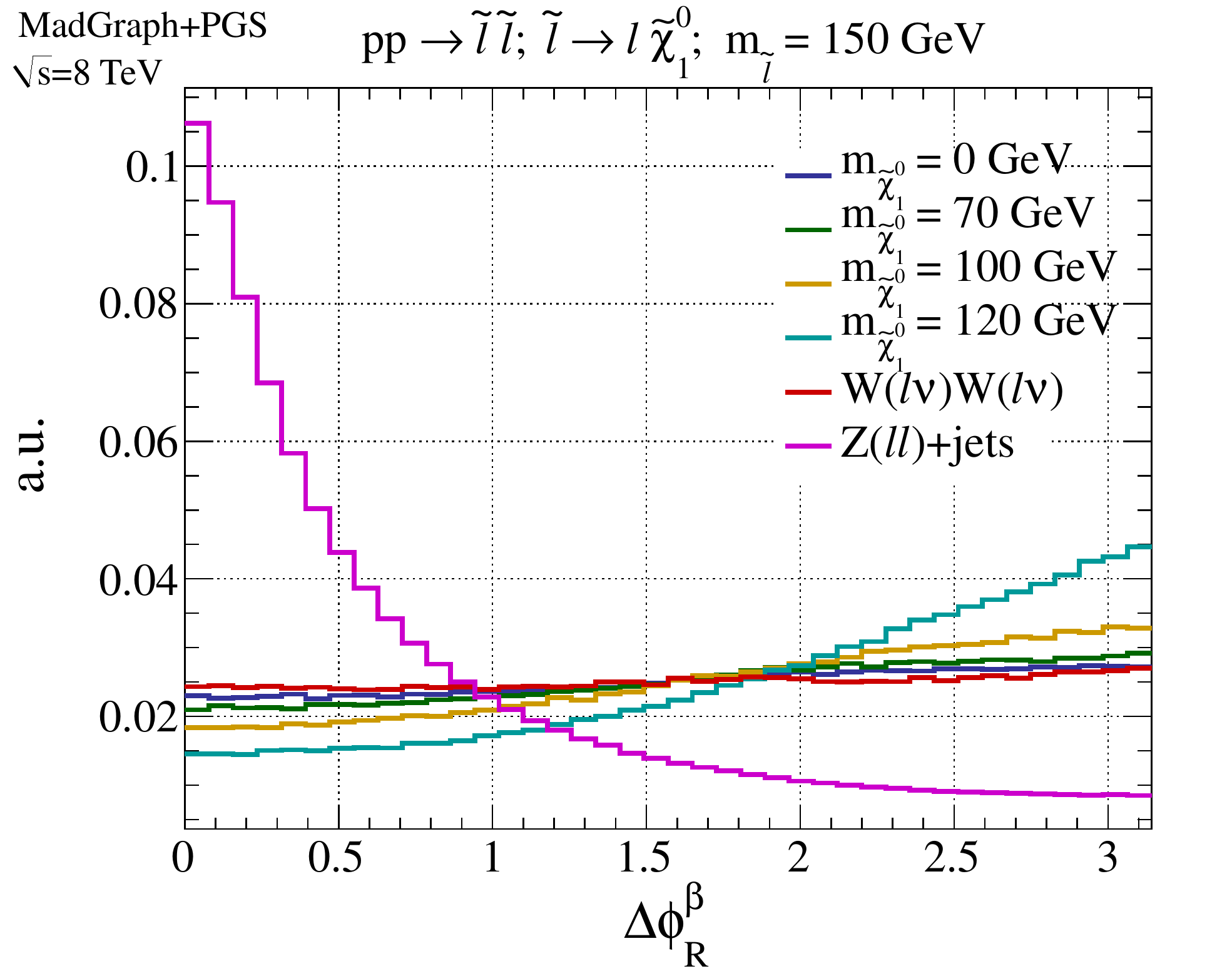}
\includegraphics[width=0.45\columnwidth]{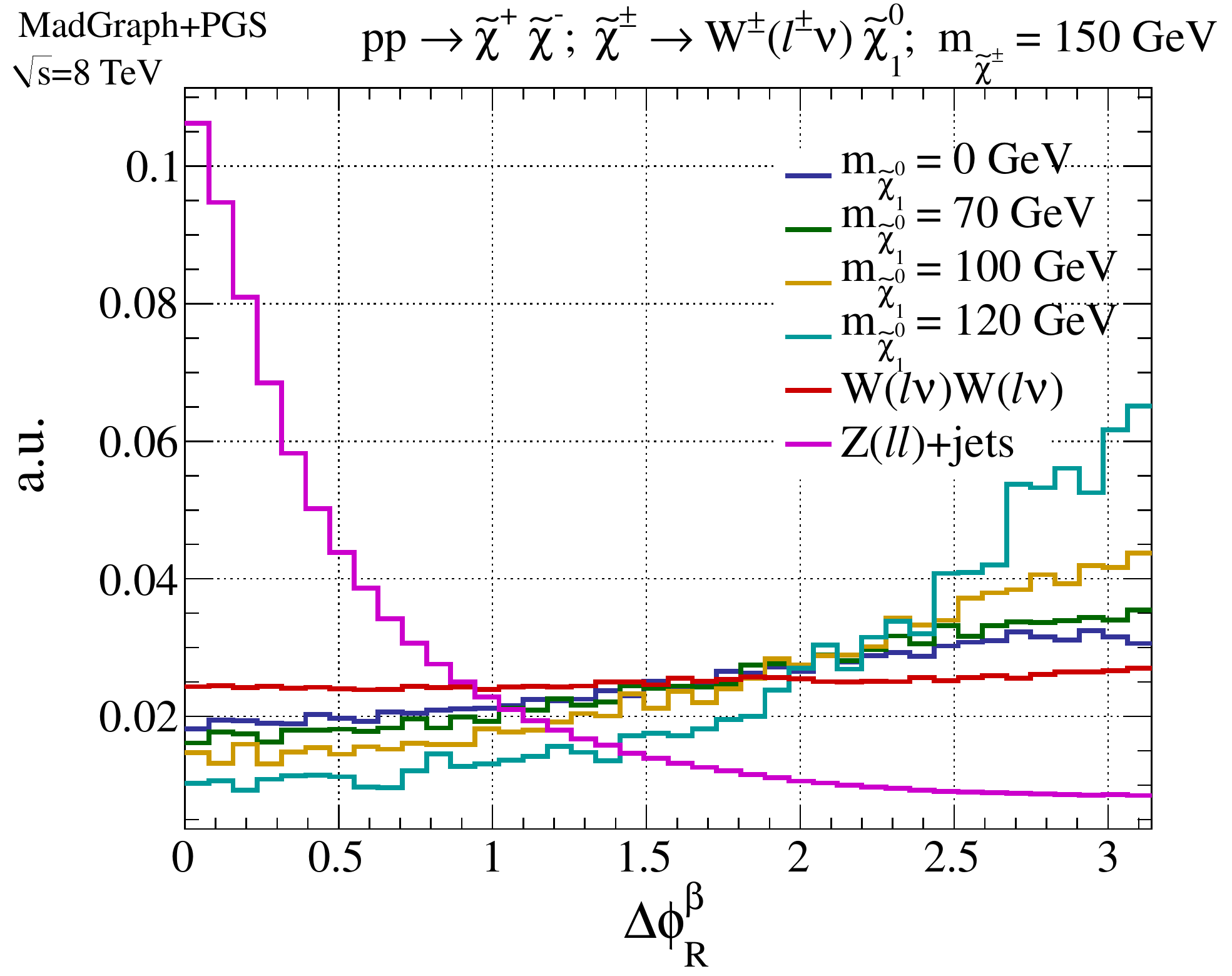}
\caption{Distributions of $\Delta\phi_R^\beta$ for a 150~GeV slepton (left) or chargino (right) and a range of neutralino masses. Also shown are the distributions of the $W^-W^+$ and Drell-Yan $Z$ backgrounds.  \label{fig:deltaRbeta}}
\end{figure}

If $\hat{s}_R < \hat{s}$, then the boost $\vec{\beta}_R$ built using $\hat{s}_R$ will be systematically larger than the correct boost $\vec{\beta}^{\, \rm CM}$. In the CM frame, the distribution of the sum of the visible particles relative to the boost direction should be relatively flat. However, if we are ``over-boosting'' from the lab frame to the approximation of the CM frame, then the sum of the visible momenta will tend to be anti-aligned with the boost direction. That is, for systems where $m_\chi/m_S \ll 1$, we expect that the azimuthal angle between $\beta_R$ and $\sum q_i$ will have a peak near $\Delta\phi_R^{\beta} \sim \pi$. In Figure~\ref{fig:deltaRbeta}, we show the distribution of this angle for a range of neutralino masses (for a fixed slepton or chargino mass). As can be seen, as the ratio $m_\chi/m_S$ approaches one, the peak of the distribution near $\pi$ becomes more pronounced. Note the large drop in statistics for chargino events where the mass of the neutralino approaches that of the parent chargino. With such a mass spectrum, events have difficulty passing the selection criteria, which will be discussed in more detail in the next section. 

Notice also from this figure that sleptons decaying to massless neutralinos are very similar to the $W^+W^- \to \ell^-\ell^+ \nu\nu$ background. This is as expected, as the $WW$ background is a case where the invisible particles (neutrinos) are massless, and so our estimate of $\hat{s}_R$ for this background will overboost to the $R$ frame, just as with the massless signal case. Thus, we do not expect this angle to be of great use in the massless neutralino limit, however, it will be of significant help in distinguishing from background in the near-degenerate limit, where traditional mass variables sensitive to $M_\Delta$ are less effective. We also comment that the Drell-Yan $Z \to \ell \ell$ background, also shown in Figure~\ref{fig:deltaRbeta}, has a strong peak near $\Delta\phi_R^{\beta}\sim 0$. In this case, we are underboosting compared to the correct CM frame, as we are assuming that there is real missing transverse energy in an event that has no invisible particles.

In the $R$-frame, there is one final kinematic variable that we can construct. The variable $\sqrt{\hat{s}}_R$ is our estimate of the total energy available in the pair-production event. In the razor frame $R$, it can be divided up into three components:
\begin{equation}
\frac{\hat{s}_R}{4} = (M_\Delta^R)^2+(q_{1R}+q_{2R})^2+(E_{1R}-E_{2R})^2. \label{eq:shatRexpansion}
\end{equation}
$M_\Delta^R$ and the invariant mass of the visible system $\sqrt{(q_{1R}+q_{2R})^2}$ have already been considered. However, the energy difference of the visible particles, $E_{1R}-E_{2R}$, has not been used. As with $\hat{s}_R$ and $M_\Delta^R$, the overall mass scale of $E_{1R}-E_{2R}$ is sensitive to $M_\Delta$, and is thus degenerate with our other mass variables. We therefore construct a new dimensionless variable
\begin{equation}
|\cos\theta_{R+1}|^2 = \frac{(E_{1R}-E_{2R})^2}{\hat{s}_R/4-(M_\Delta^R)^2} = \frac{\hat{s}_R/4-(M_\Delta^R)^2-(q_{1R}+q_{2R})^2}{\hat{s}_R/4-(M_\Delta^R)^2}. \label{eq:costhetaR1}
\end{equation}
This particular definition (and identification as a cosine of an angle) is because this variable can also be interpreted as the angle between the boost direction $\vec{\beta}_R$ and the direction of $q_1$ or $q_2$ in the frame $R+1$. However, it is more useful to think of this angle as a measure of the energy difference between the two visible particles.

A measure of the energy difference is useful in background rejection, especially in removing $W^-W^+$ events. The reasoning is as follows: for scalar particles, the decay of the parent into the visible and invisible daughters has a flat angular distribution in the parent's rest frame. In the production frame, we do not then expect a large correlation between the energy of the two visible particles. Though in their respective decay frames each has the same energy, the orientation of their momentum relative to the momentum of the parent is uncorrelated, and so $|E_1 - E_2| \propto |\cos\theta_{R+1}|$ will not cluster at zero. The exception is for very large boosts of the parent particle; in this case, the direction of the visible daughter in the decay frame is effective erased by the very large boost. In such cases, both visible particles are colinear with their parent direction and have $E_1 \approx E_2$.

\begin{figure}[ht]
\includegraphics[width=0.45\columnwidth]{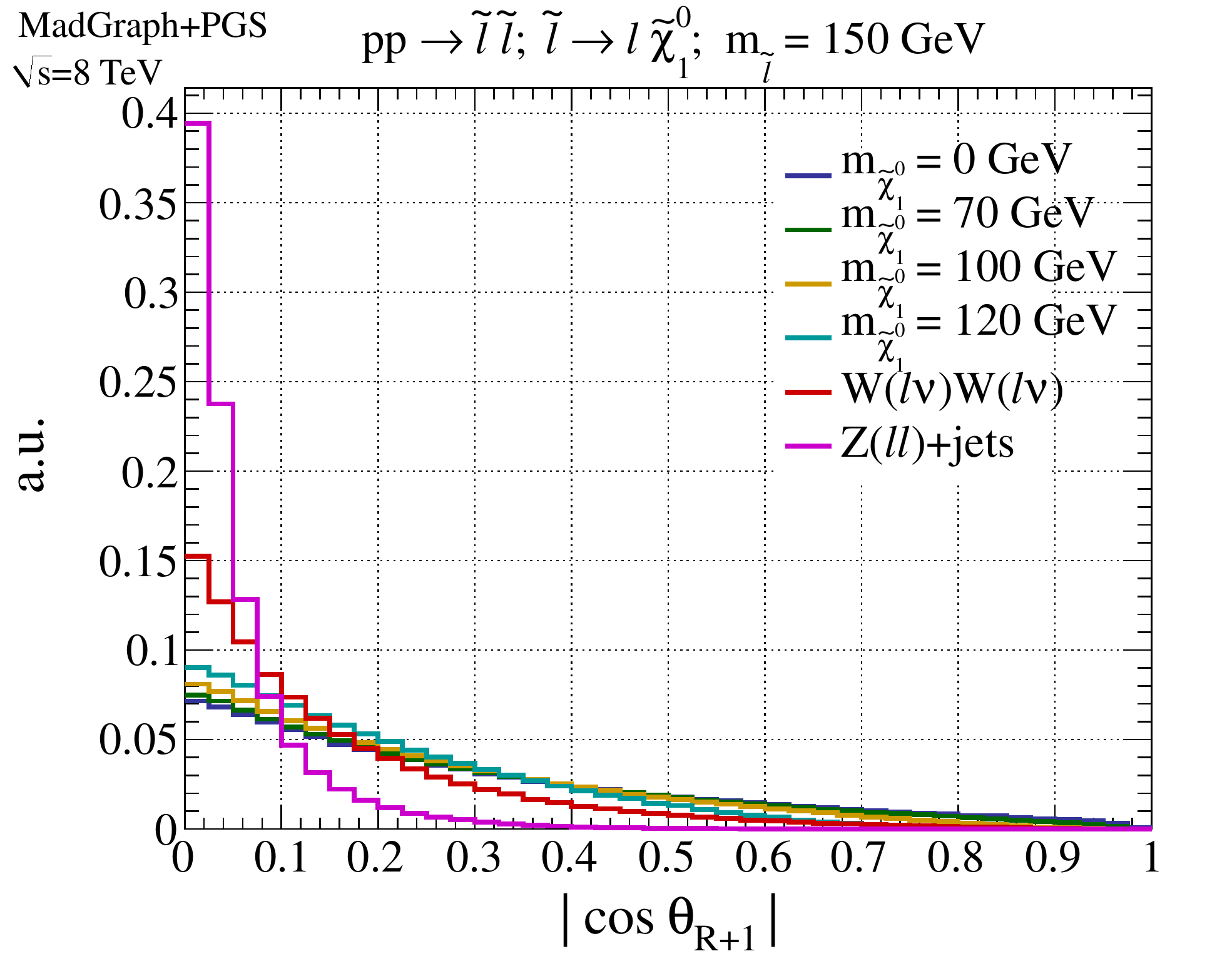}
\includegraphics[width=0.45\columnwidth]{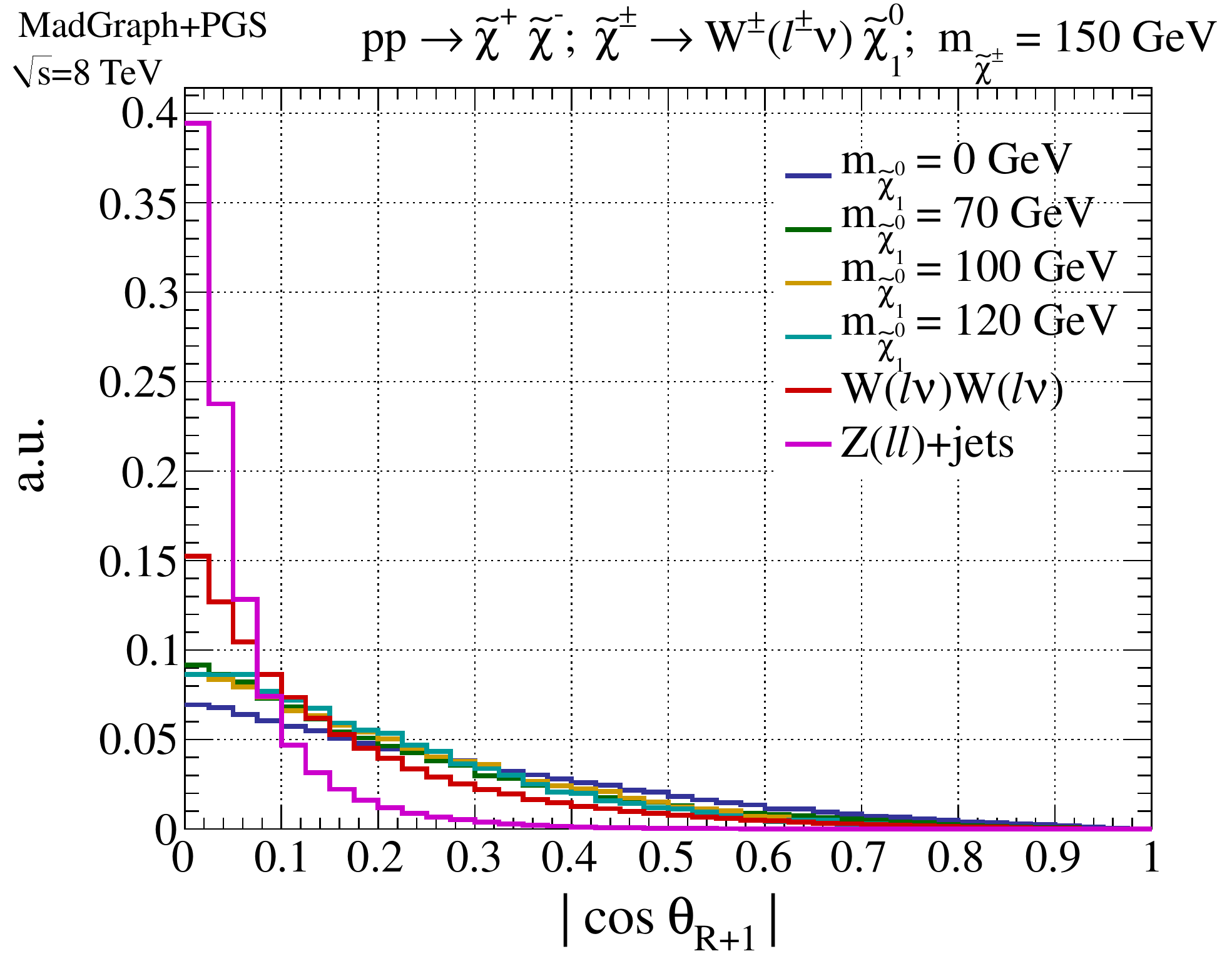}
\caption{Distribution of $|\cos\theta_{R+1}|$ for 150 GeV selectron (left) and chargino (right) pair production, decaying into a range of neutralino masses. Also shown are the $W^-W^+$ and Drell-Yan $Z$ background distributions. \label{fig:costheta}}
\end{figure}

Now consider $W^-W^+$ background. Unlike scalar decay, the vector $W$ boson decaying into fermions has a correlation in the direction of the visible lepton relative to the parent polarization. As the polarizations of the two $W$ bosons in an event are themselves correlated, this means that, after the boost from the decay frames to the production frame (or to our approximation of that frame, the $R$ frame), the two visible leptons will tend to have similar energies: $E_1 \approx E_2$. Therefore, the distribution of $|\cos\theta_{R+1}|$ for this background will be more highly peaked towards zero. The behavior of signal and background in this variable is shown in Figure~\ref{fig:costheta}, for representative signal points.

\begin{figure}[ht]
\includegraphics[width=0.35\columnwidth]{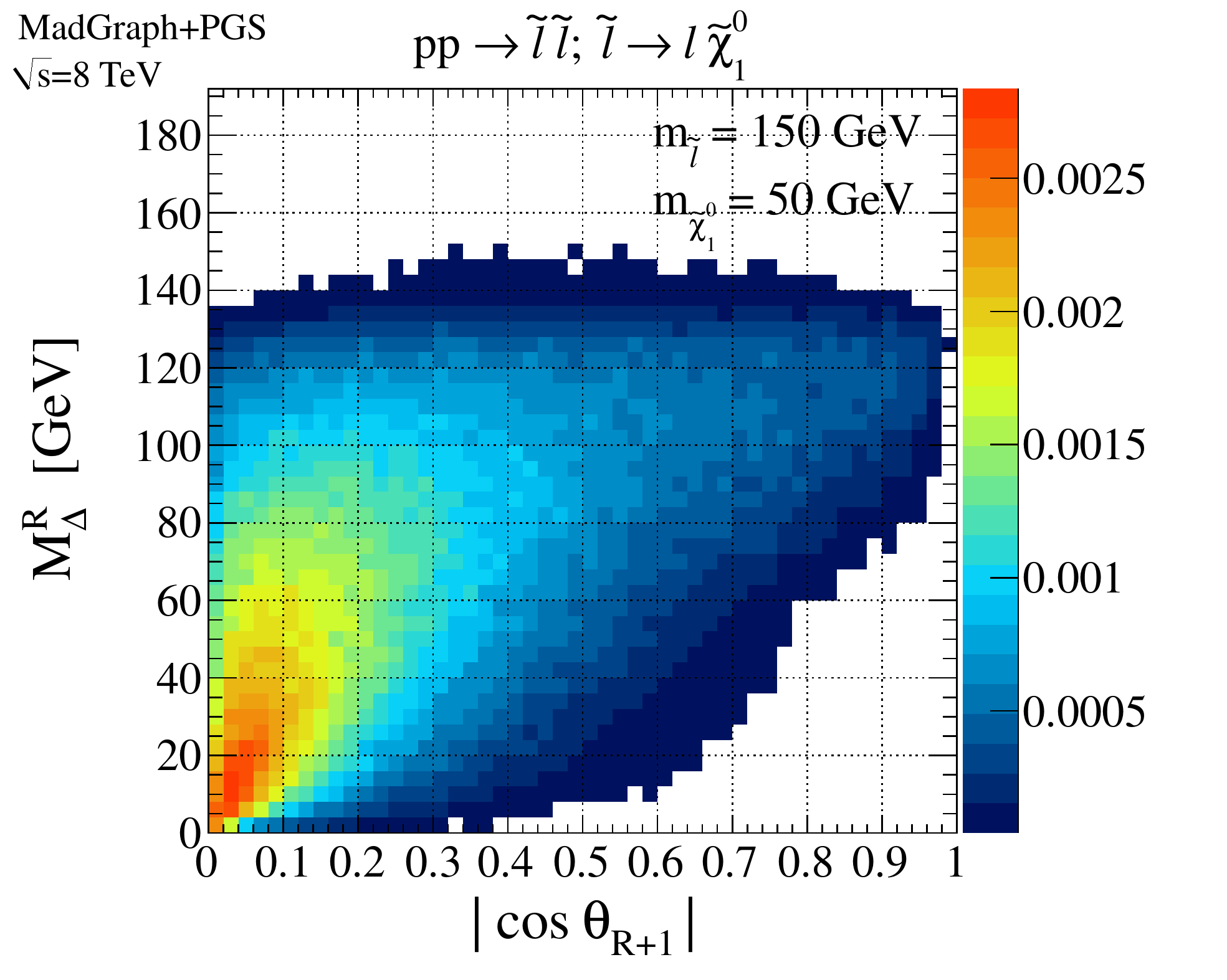}
\includegraphics[width=0.35\columnwidth]{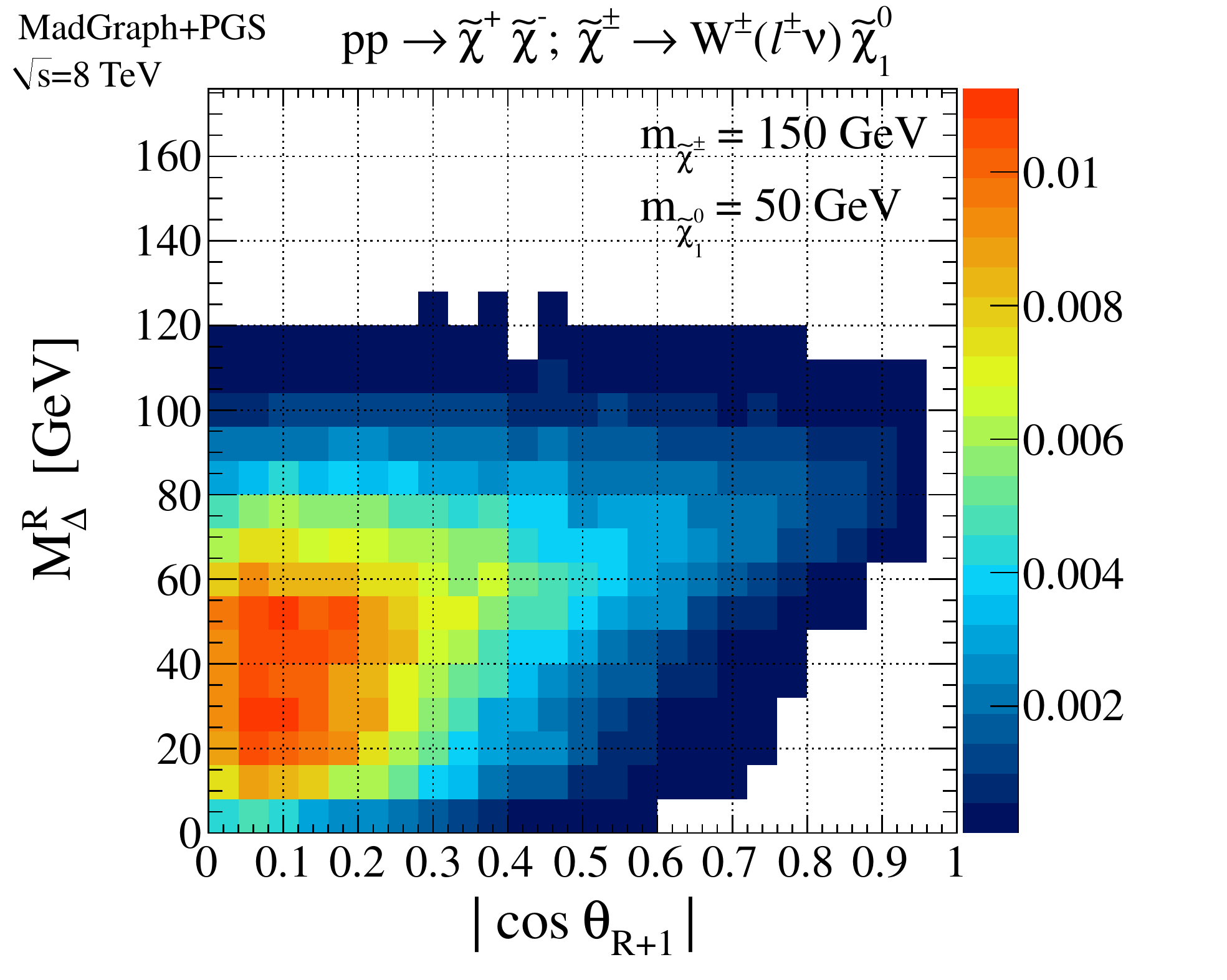}
\includegraphics[width=0.35\columnwidth]{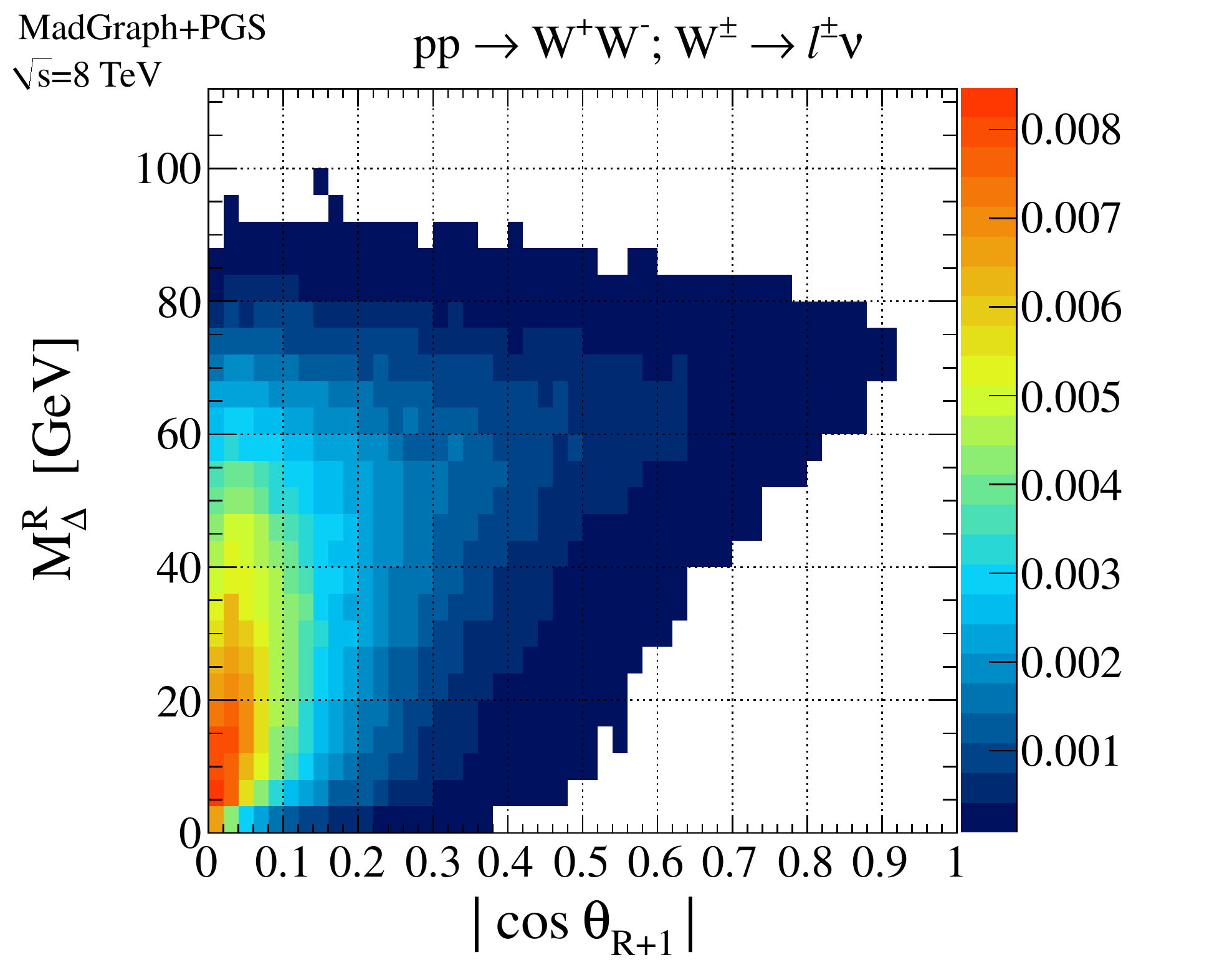}
\includegraphics[width=0.35\columnwidth]{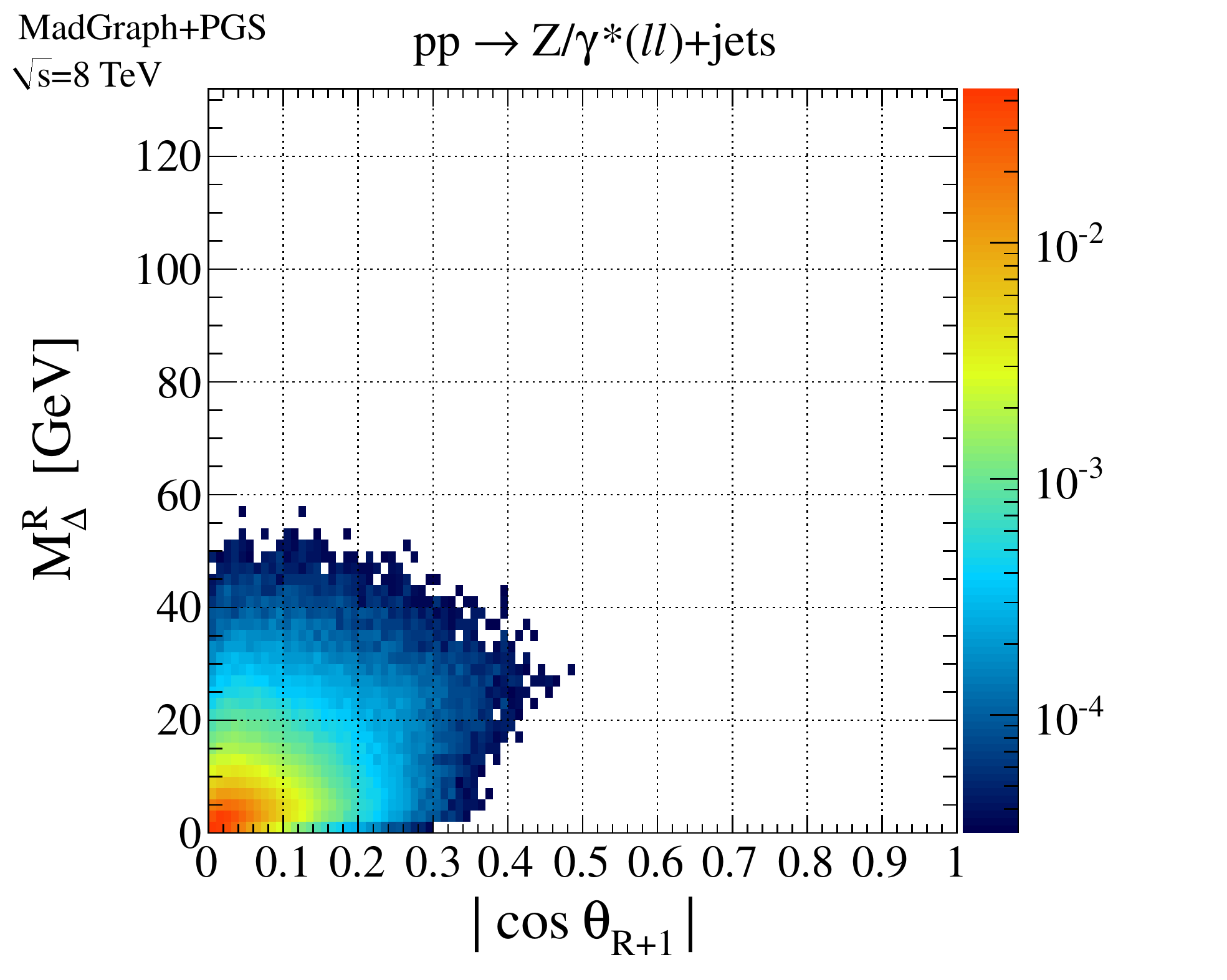}
\caption{Upper Row: Distribution of $|\cos\theta_{R+1}|$ versus $M_\Delta^R$ for 150 GeV selectron (left) or chargino (right) pair production, decaying into 50 GeV neutralinos. Lower Row: $|\cos\theta_{R+1}|$ versus $M_\Delta^R$ for $W^-W^+$ pair production (left) or Drell-Yan $Z$ (right) backgrounds decaying into leptons. \label{fig:costheta_v_MdeltaR}}
\end{figure}

In Figure~\ref{fig:costheta_v_MdeltaR}, we show the distributions of $|\cos\theta_{R+1}|$ with respect to $M_\Delta^R$ for a representative choice of slepton and neutralino masses, and compare with the distribution for the $W^-W^+$ background. As can be seen, when $M_\Delta^R \sim 0$, the signal events cluster near $|\cos\theta_{R+1}| = 0$. This makes sense, as $M_\Delta^R \sim 0$ corresponds to large boosts of the parent particles (as can be seen in Eq.~\eqref{eq:mdeltaR}). As $M_\Delta^R$ approaches $M_\Delta$, we recover the essentially flat distribution of $|\cos\theta_{R+1}|$ we expect from a scalar decay. The $W^-W^+$ background, on the other hand, does not have a flat distribution for $M_\Delta^R \sim m_W$. This, therefore, allows for discrimination of signal and background events even for signal events where $M_\Delta \sim m_W$.

These new angular variables demonstrates the utility of the razor boosts. The mass variables ($\hat{s}_R$ and $M_\Delta^R$) are not completely unique to our work; they have been independently developed in different contexts in the past (see Ref.~\cite{Rainwater:1999sd}). However, by associating these variables with a particular set of boosts, we can approximate the CM of the event. This allows us to build additional variables using this approximation, two of which ($\Delta \phi_R^{\beta}$ and $|\cos\theta_{R+1}|$) turn out to encode further information about the event. Furthermore, as the construction of the $|\cos\theta_{R+1}|$ variable relies on the spin of the new particles being searched for, it has the potential to be used as a measurement of spin if new physics is found. This possibility will be investigated in a future work. 

\begin{figure}[ht]
\includegraphics[width=0.35\columnwidth]{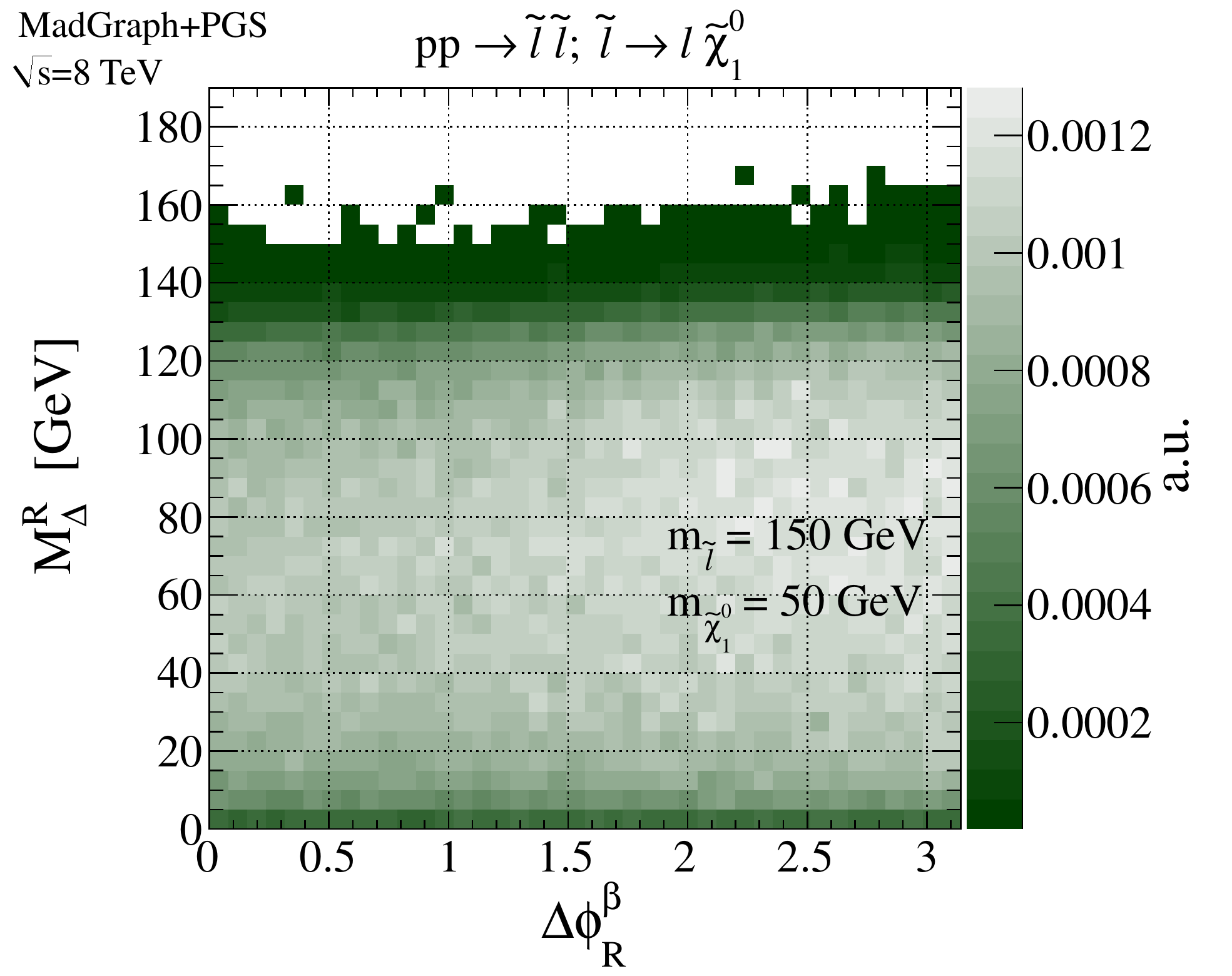}\includegraphics[width=0.35\columnwidth]{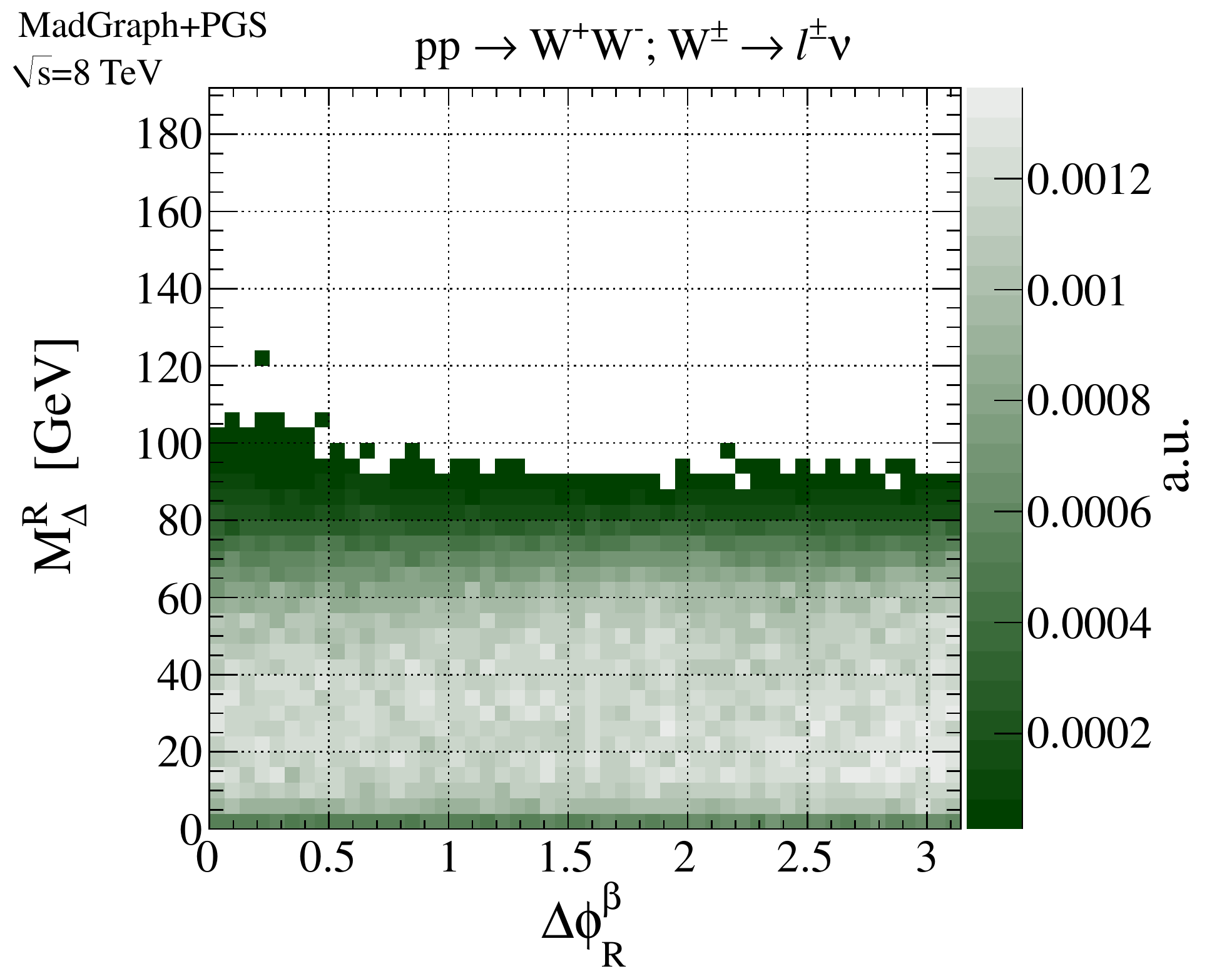}
\includegraphics[width=0.35\columnwidth]{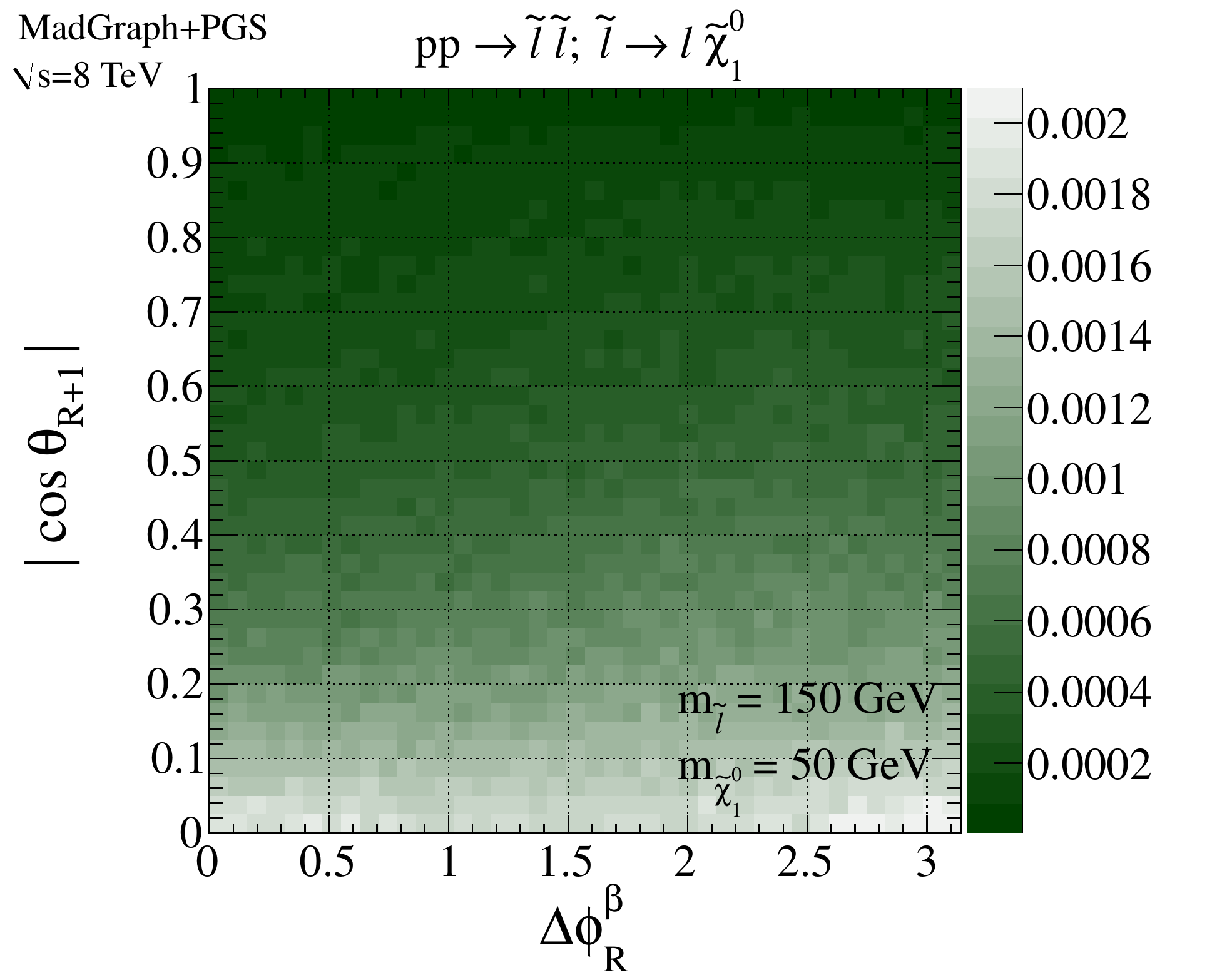}\includegraphics[width=0.35\columnwidth]{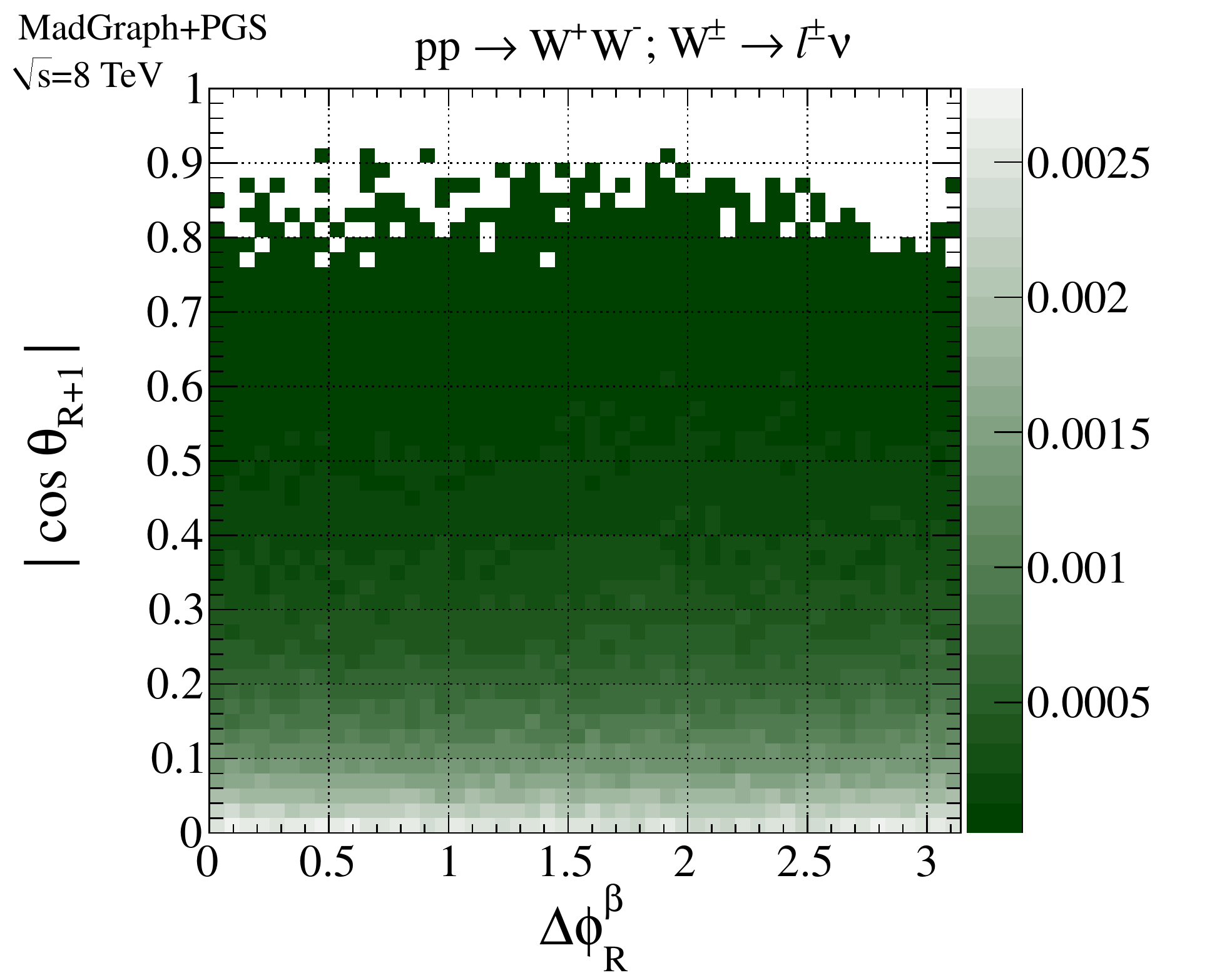}
\caption{Representative distribution of $\Delta\phi_R^\beta$ vs.~$M_\Delta^R$ (top row) and $\Delta\phi_R^\beta$ vs.~$|\cos\theta_{R+1}|$ (bottom row) 
 for 150~GeV selectrons decaying to 50~GeV neutralinos (left) and $W^-W^+$ background (right).  \label{fig:deltaphiMdelta}}
\end{figure}

In this study, we work primarily with the set of four variables $\hat{s}_R$, $M_\Delta^R$, $\Delta \phi_R^{\beta}$, and $|\cos\theta_{R+1}|$. The two mass variables are somewhat degenerate, as both are estimators of the mass splitting between the parent and daughter particles. Though there may be some utility in using all four variables in a single analysis, here we will demonstrate the possible reach of our super-razor search by restricting ourselves to the $M_\Delta^R$, $\Delta \phi_R^{\beta}$, and $|\cos\theta_{R+1}|$ combination only. We will also use $\gamma_{R}$ and $\gamma_{R+1}$ in our improved selection criteria to reduce background contamination  We choose $M_\Delta^R$ over $\hat{s}_R$ for our analysis because, as Figure~\ref{fig:deltaphiMdelta} shows, the variables $M_\Delta^R$ and $|\cos\theta_{R+1}|$ are approximately uncorrelated with $\Delta\phi_R^\beta$ for signal events. This simplifies the shape analysis we will discuss in the next section, as it allows us to decompose a 3D analysis into a 2D $\times$ 1D one.

\clearpage

\section{Event Simulation and Selection \label{sec:simulation}}

\subsection{Sample generation}

\begin{figure}[ht]
\includegraphics[width=0.29\columnwidth]{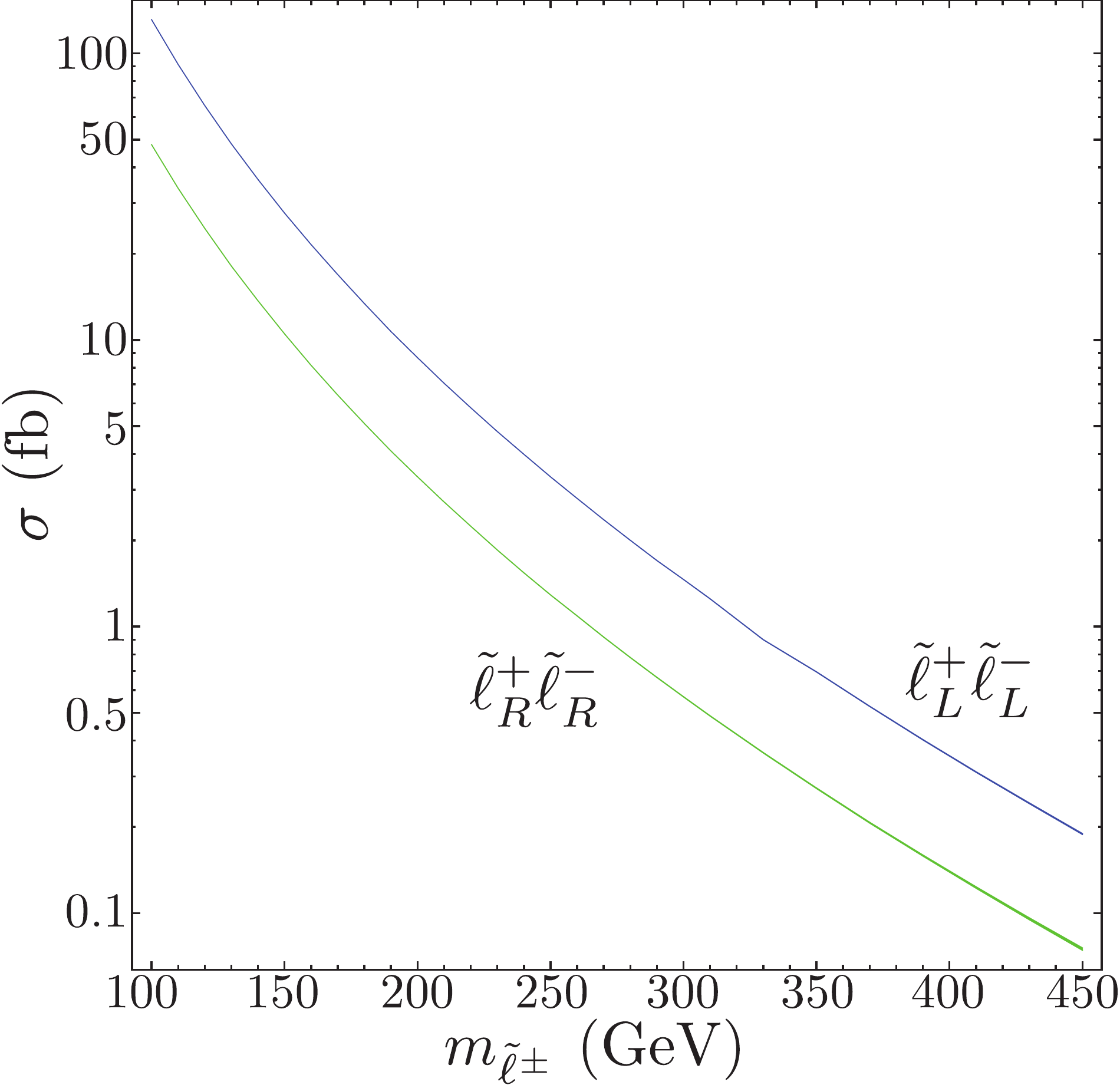}\hspace{5em}\includegraphics[width=0.3\columnwidth]{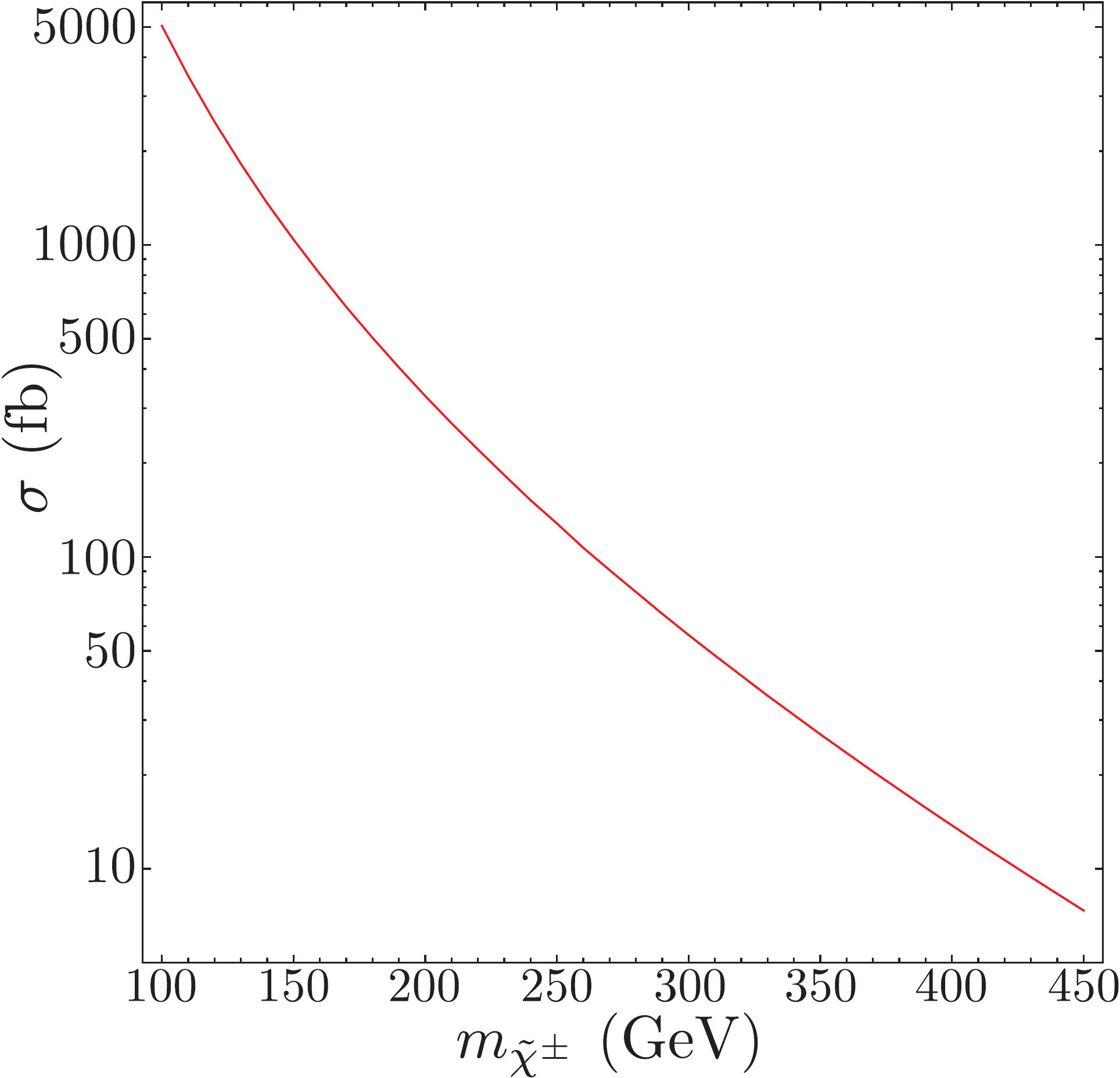} 
\caption{Left: $1^{\rm st}$ or $2^{\rm nd}$ generation left-handed (blue) and right-handed (green) slepton pair production cross sections. Right: Chargino pair production cross section. Cross sections calculated using {\tt Prospino} \cite{Beenakker:1999xh} at NLO for the LHC with $\sqrt{s}=8$~TeV. Theoretical errors indicated by width of lines. \label{fig:sigma}}
\end{figure}

We study the performance of the super-razor variables in the context of searches for new physics appearing in two different scenarios:
\begin{itemize}
\item Pair production of sleptons (selectrons $\tilde{e}^\pm$ or smuons $\tilde{\mu}^\pm$) decaying to electrons or muons, respectively, and neutralinos ($\tilde{\chi}_1^0$) with 100\% branching ratio (BR) for both left- or right-handed sparticles.
\item Pair production of the lightest chargino ($\tilde{\chi}_1^\pm$) decaying into $\tilde{\chi}_1^0$ and a $W$-boson with 100\% BR. We require both $W$-bosons to decay leptonically while accounting for the SM $W$-boson branching ratio.
\end{itemize}
Event samples corresponding to these signal models were generated using {\tt MadGraph5} \cite{Alwall:2011uj} and {\tt Pythia 6.4} \cite{Sjostrand:2006za}, with up to two extra matched jets. 
We consider slepton and chargino masses between the LEP bound ($\sim 100$~GeV) up to 450~GeV, with neutralino masses varying between zero and 20 GeV less than their respective parent sparticle masses. Production cross sections for these signal events were obtained for the LHC with $\sqrt{s} = 8$~TeV at next-to-leading order (NLO) using {\tt Prospino} \cite{Beenakker:1999xh}. All superpartners except for the neutralino and active slepton flavor/chargino were decoupled by setting their mass to 2.5~TeV and the chargino was assumed to be wino-like. In the mass intervals considered, the resulting cross sections range from $\sim 100 - 1$~fb for both flavors of sleptons and $\sim 5000-100$~fb for the chargino, with the cross-sections as a function of sparticle mass illustrated in Figure~\ref{fig:sigma}. 

In order to estimate the sensitivity of the CMS and ATLAS experiments to these putative signals we also generate event samples corresponding to the primary SM backgrounds in the di-lepton final state: di-boson ($W^+W^-$, $W^\pm Z$, and $ZZ$) production, Drell-Yan $(Z/\gamma^*\to\ell\ell)+$jets, and top pair production. Event samples were generated for all channels in {\tt MadGraph5}+{\tt Pythia 6.4},
with up to two extra matched jets and cross-sections calculated from the same generator configuration. 

\subsection{Detector simulation and baseline selection~\label{sec:baseline}}

All of the event samples, for both signal and background processes, are analyzed using the PGS toy detector simulation, from which reconstructed leptons (electrons and muons) are identified and jets are clustered. For all the kinematic distributions and results presented in this work, simulated events are included only if they satisfy baseline selection requirements. 

Each event is required to have exactly two reconstructed leptons with $p_{T} > 20$~GeV and $|\eta| < 2.5$. Events are discarded which have more than two leptons satisfying this requirement. Furthermore, these leptons are required to have opposite charge. The combination of these requirements reduces the yields of selected events corresponding to di-boson backgrounds such as $WZ$ and $ZZ$ where there are either more than two leptons reconstructed or the two leptons arise from the decays of different bosons. Events are assigned to one of three flavor categories corresponding to same flavor (SF) where there are either two reconstructed electrons ($e^-e^+$) or muons ($\mu^-\mu^+$) and opposite flavor (OF), containing
$e\mu$ events. An additional requirement of $m(\ell\ell) > 15$~GeV is applied to events falling in the SF categories in order to reject backgrounds with low-mass di-lepton resonances.

Jets are clustered from simulated calorimeter cells using FastJet \cite{Cacciari:2011ma} and the anti-$k(t)$ algorithm \cite{Cacciari:2008gp}.
Events containing at least one jet with $p_{T} > 25$~GeV and $|\eta| < 2.5$ which is identified as $b$-tagged are discarded from the event sample in order to reduce the contribution from events containing top quarks. The number of reconstructed jets is used to classify events into one of three jet multiplicity categories: 0 jet, $1$ jet and $\ge 2$ jet. This jet counting scheme is based on jets with with $p_{T} > 30$~GeV and $|\eta| < 3$. Furthermore, events are discarded if either of the two reconstructed leptons falls within a cone of $\Delta R \equiv \sqrt{\Delta\eta^{2}+\Delta\phi^2} = 0.4$ around any of the reconstructed jets in the event. Unless otherwise indicated, kinematic distributions include the sum of all three flavor and jet multiplicity categories. 

\subsection{Comparison of different kinematic variables}

\begin{figure}[ht]
\includegraphics[width=0.3\columnwidth]{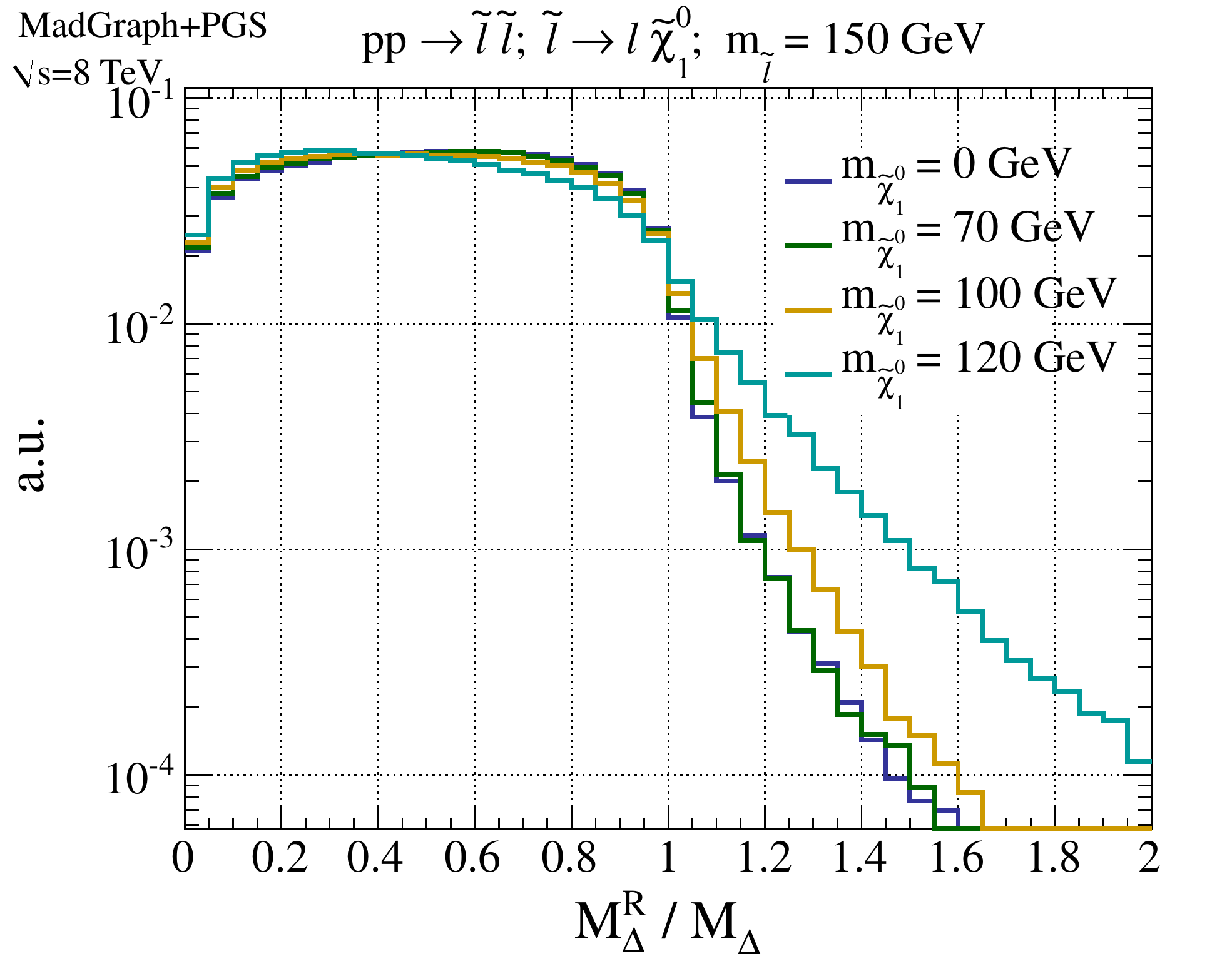}
\includegraphics[width=0.3\columnwidth]{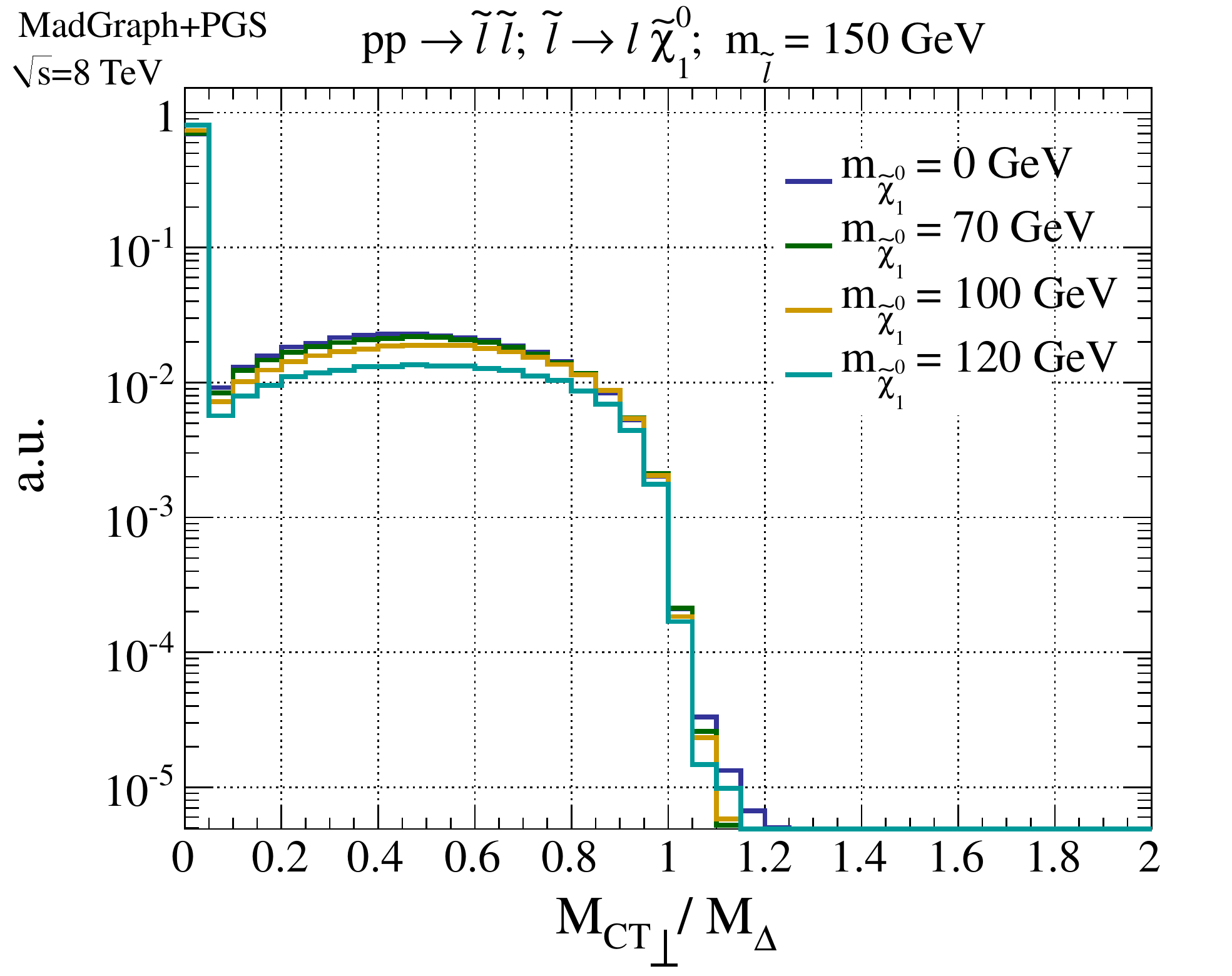} 
\includegraphics[width=0.3\columnwidth]{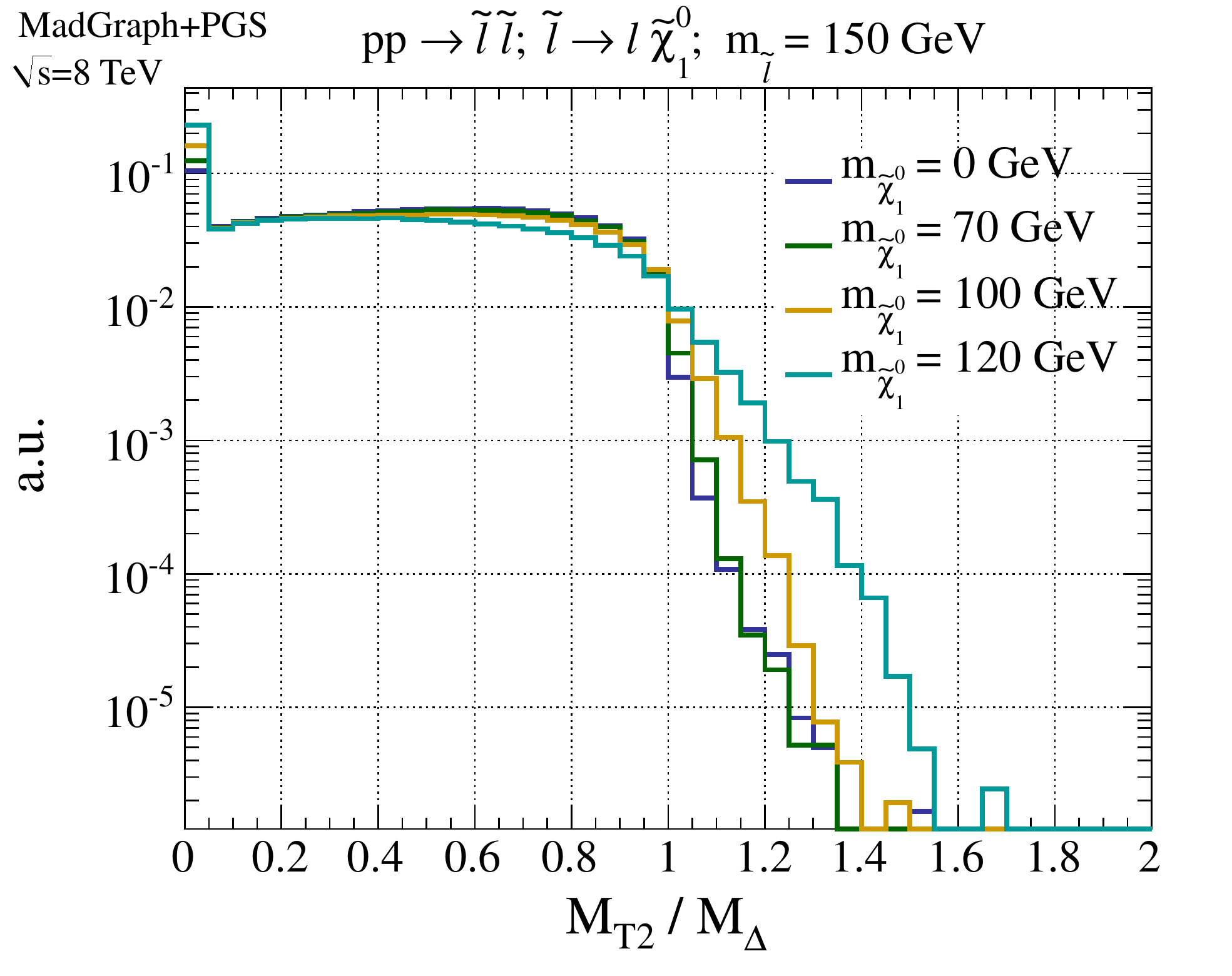} 
\includegraphics[width=0.3\columnwidth]{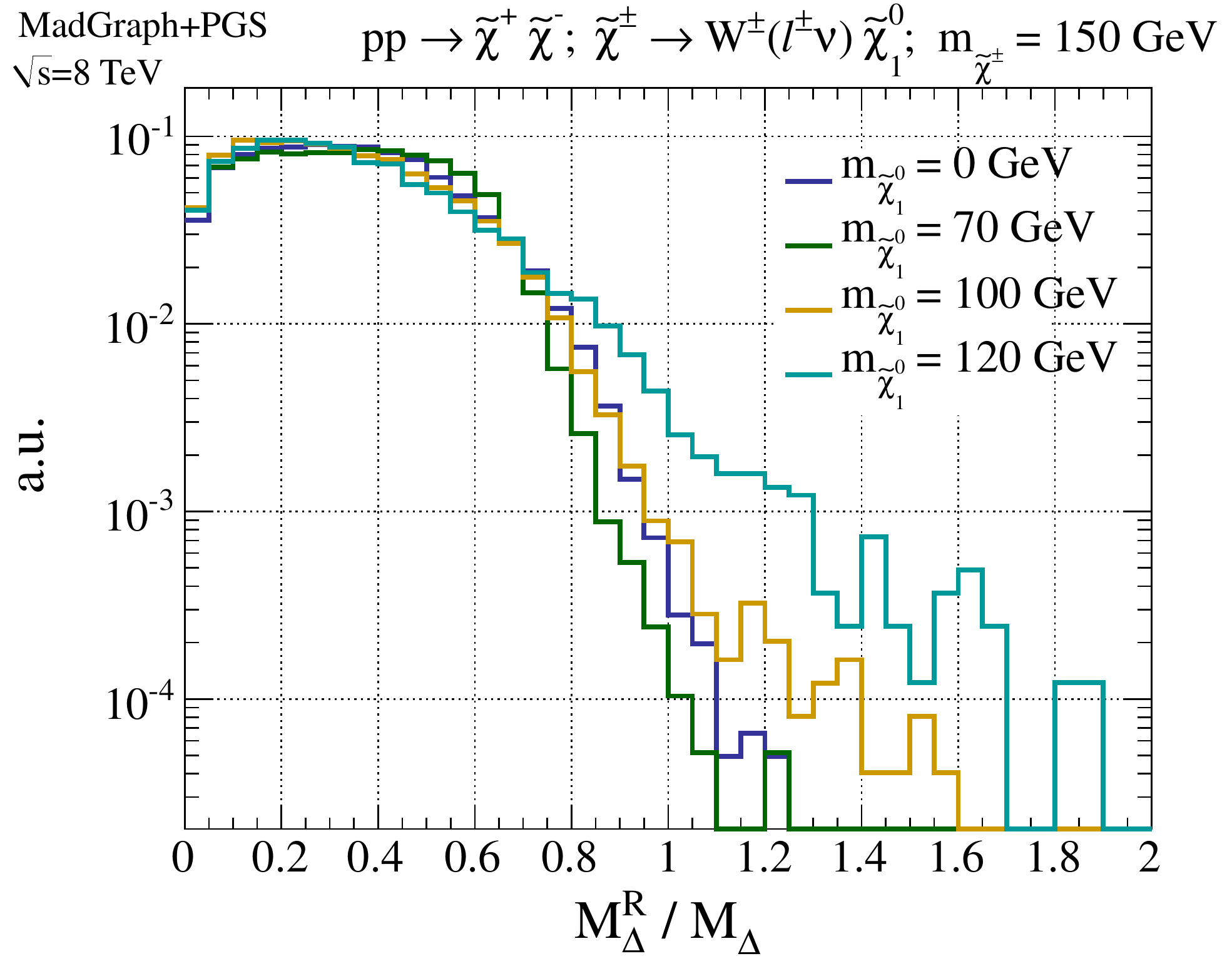}
\includegraphics[width=0.3\columnwidth]{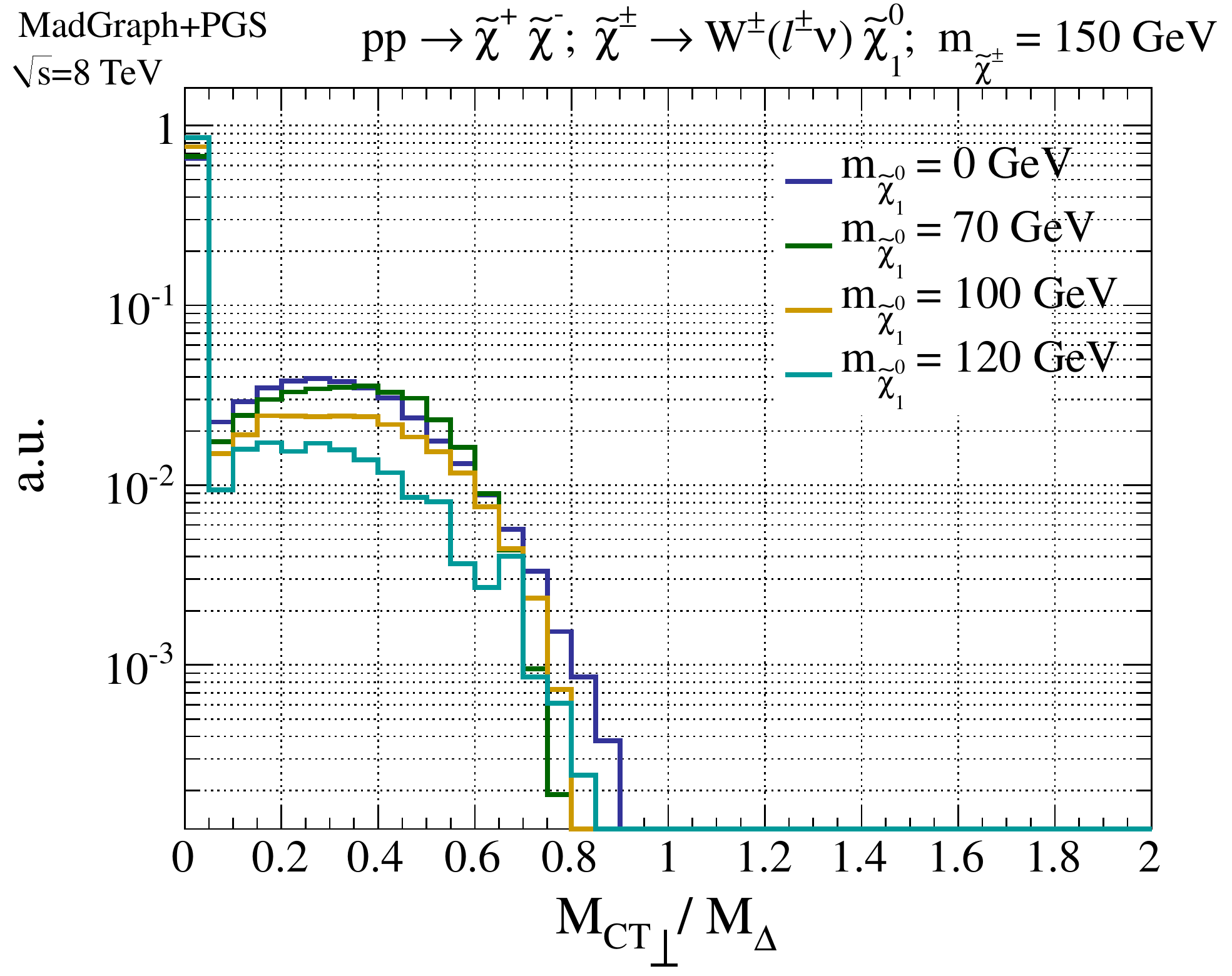} 
\includegraphics[width=0.3\columnwidth]{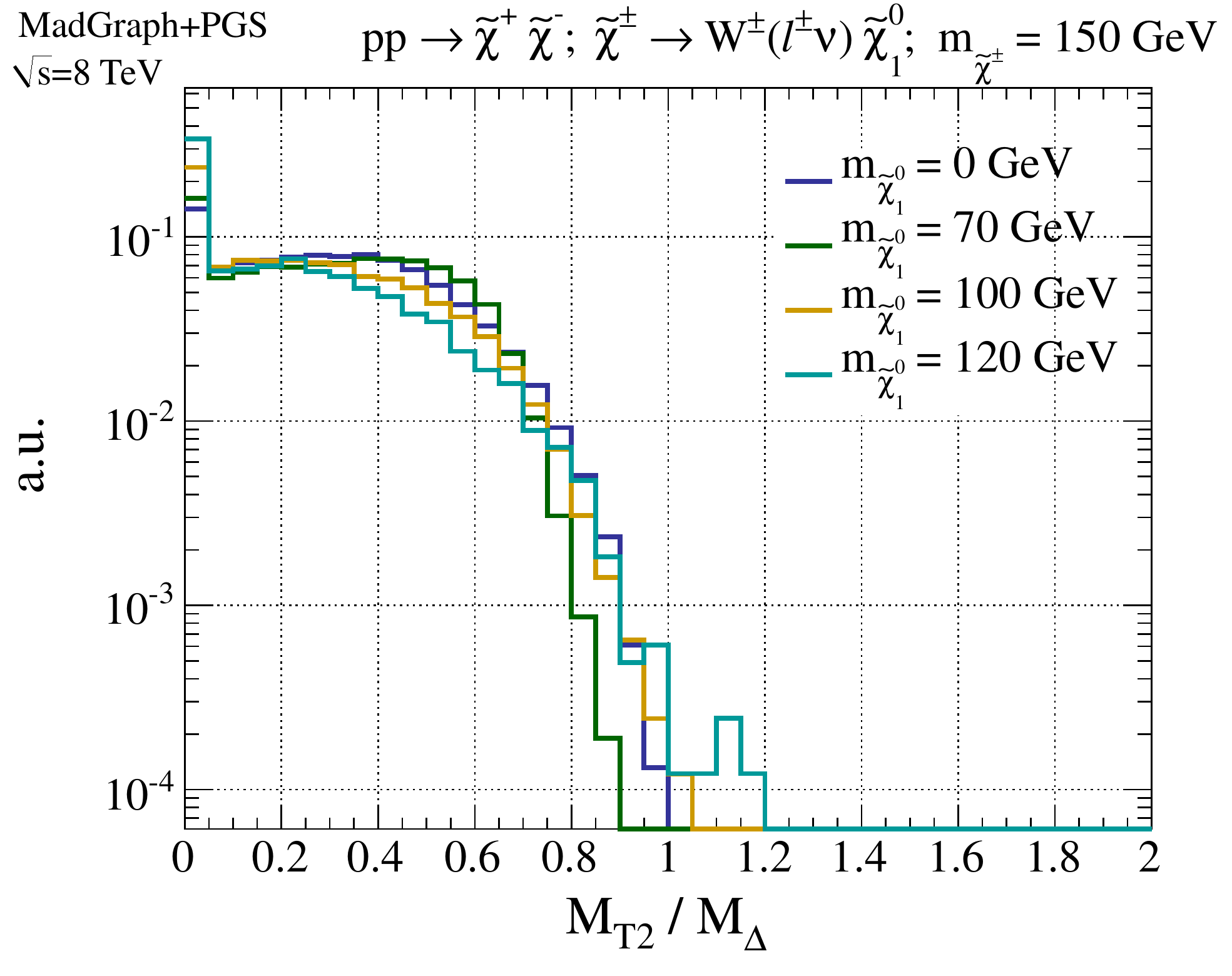} 
\caption{Distributions of $M_{\Delta}$ estimating variables for sleptons (top row) and charginos (bottom row) with mass 150~GeV decaying into neutralinos and leptons, for a range of neutralino masses. Variables include $M_{\Delta}^{R}$ (left), $M_{CT\perp}$ (center) and $M_{T2}$ (right), all normalized to the true value of $M_{\Delta}$ for each sample. \label{fig:compare}}
\end{figure}

We evaluate the potential for the variable $M_{\Delta}^{R}$ to be used in a search for di-slepton and di-chargino production signals by comparing it with similar variables used in CMS and ATLAS searches. The CMS search for slepton production in the di-lepton final state~\cite{CMS-PAS-SUS-13-006} utilizes the variable $M_{CT\perp}$ \cite{Matchev:2009ad,Tovey:2008ui} while the analogous ATLAS analysis \cite{ATLAS-CONF-2013-049} includes requirements on the variable $M_{T2}$~\cite{Lester:1999tx,Barr:2003rg} in definitions of signal regions sensitive to the presence of di-leptons following from slepton decays. The distributions of each of these kinematic variables, $M_{\Delta}^{R}$, $M_{CT\perp}$, and $M_{T2}$, are shown in Figure~\ref{fig:compare} for slepton and chargino signals with various sparticle mass combinations.

The behavior of each of the three variables is similar. Each is sensitive to the quantity $M_{\Delta}$ for signal events, with a sharp edge or endpoint at the true value. The shape of each distribution is largely insensitive to the absolute value of $M_{\Delta}$, such that distributions are nearly identical when scaled by $M_{\Delta}$ (differences are observed when the parent sparticle and the neutralino approach degeneracy). The similarities between these $M_{\Delta}$ sensitive variables are indicative of the fact that they are highly correlated and represent largely redundant information about events. 

An important property of $M_{CT\perp}$ and $M_{T2}$ is their almost complete insensitivity to the transverse momenta of the di-sparticle CM frame ($p_{T}^\text{CM}$) in these events. Regardless of the velocity of the sparticles in the laboratory frame, the position of the $M_{\Delta}$ endpoint in these distributions remains largely unchanged. This property is convenient for interpretation of the putative signal distributions and essential in the construction of these searches, since it also guarantees the invariance of the same kinematic feature for backgrounds like $WW$ and $t\bar{t}$, even for large $p_{T}^\text{CM}$.  For $M_{T2}$, under-constrained kinematic degrees of freedom are assigned through minimization which removes the $p_{T}^\text{CM}$ dependence. Meanwhile, $M_{CT\perp}$ considers only the lepton kinematics along the transverse axis perpendicular to $\vec{p}_{T}^\text{\, CM}$, largely ignoring variations which are sensitive to its magnitude. 

For $M_{\Delta}^{R}$ the same behavior is achieved by explicitly correcting for non-zero $p_{T}^\text{CM}$, transforming the di-lepton system from the laboratory frame to an approximation of the CM frame. By using only Lorentz invariant information in the determination of this transformation, the definition of the resulting reference frame is stable under variations of $p_{T}^\text{CM}$, as are kinematic variables (such as $M_{\Delta}^{R}$) evaluated in it. From Figure~\ref{fig:compare} is clear that the endpoint behavior of these $M_{\Delta}$ estimators has only mild sensitivity to the choice of strategy for removing $p_{T}^\text{CM}$ dependence. 

\begin{figure}[ht]
\includegraphics[width=0.3\columnwidth]{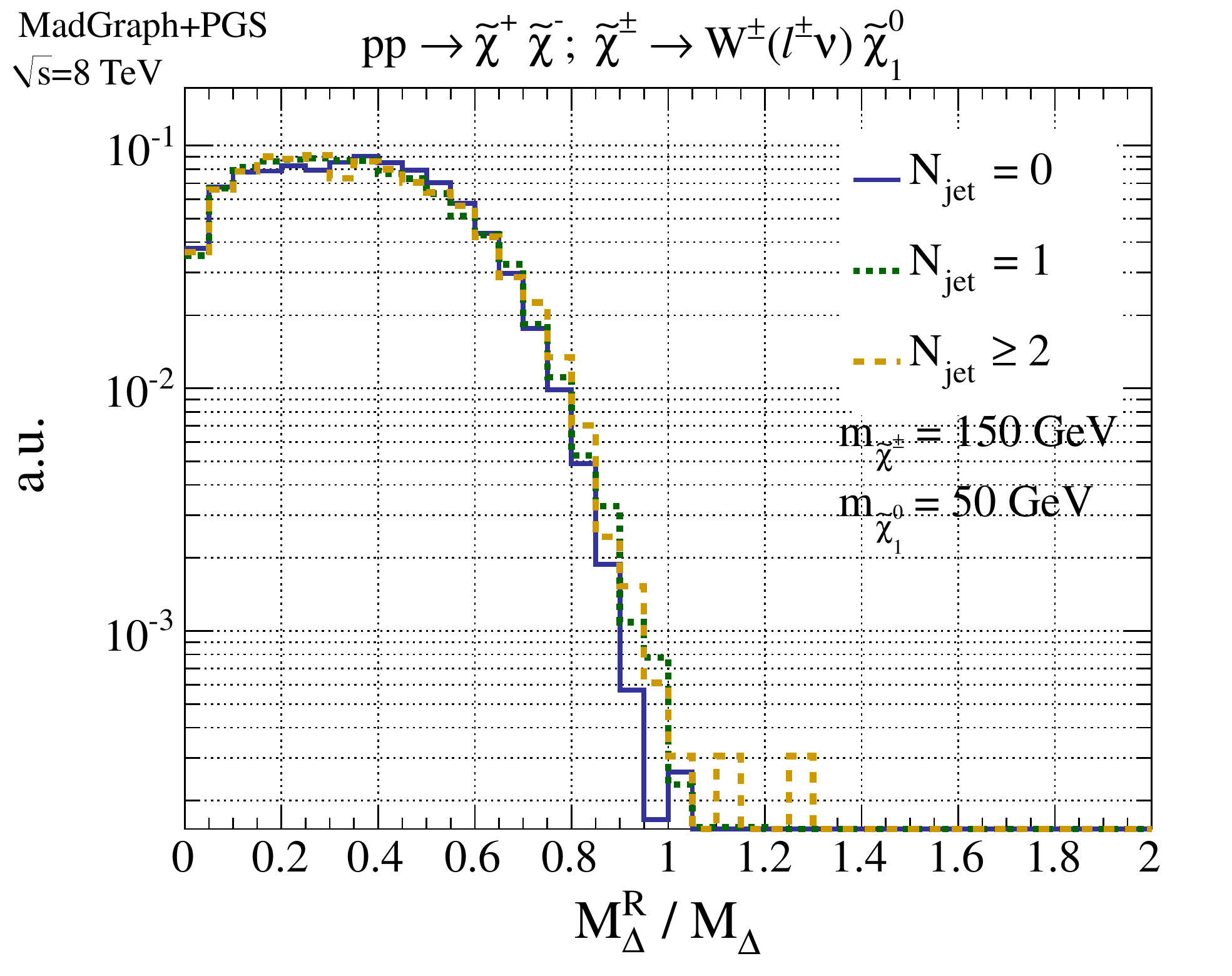}
\includegraphics[width=0.3\columnwidth]{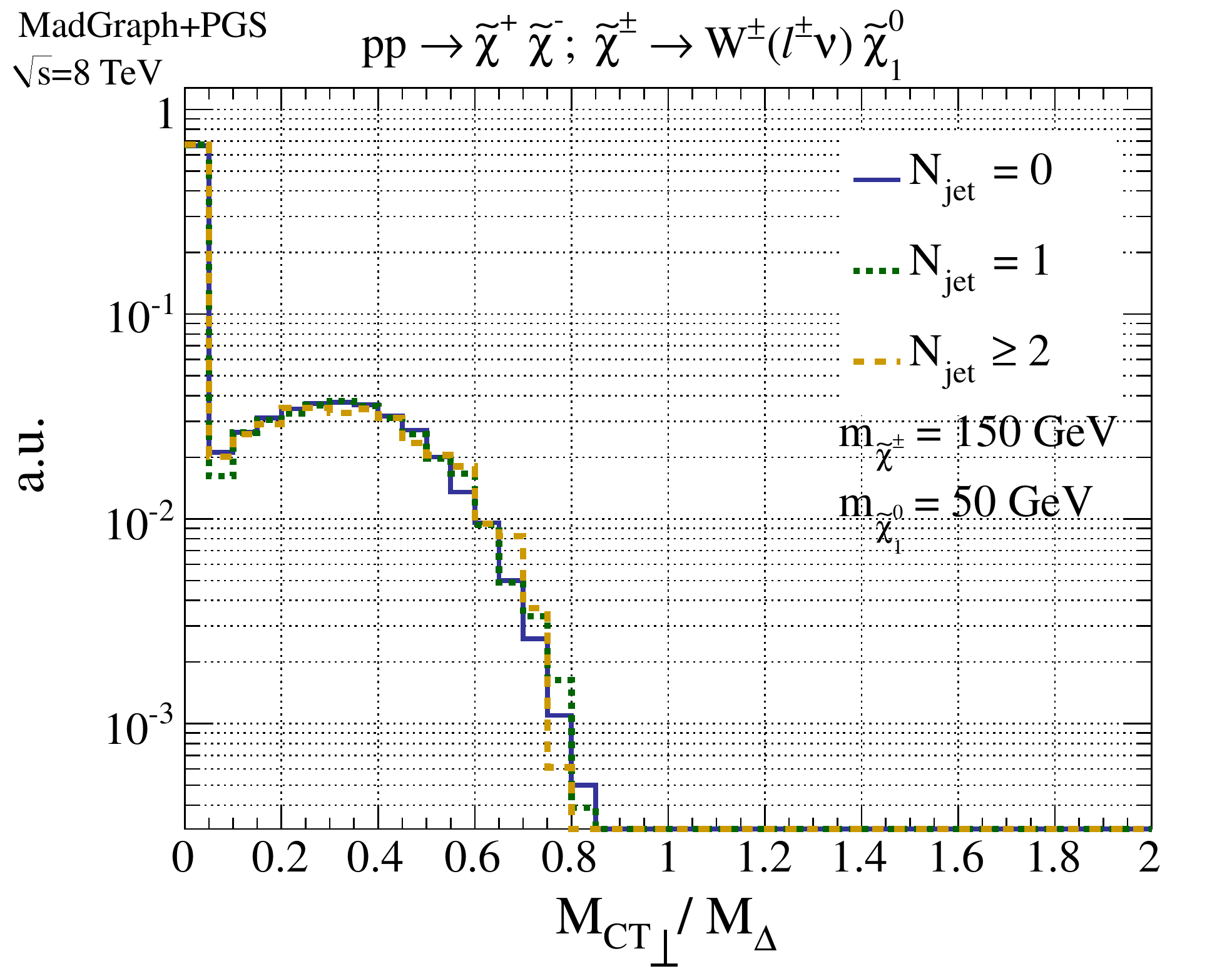} 
\includegraphics[width=0.3\columnwidth]{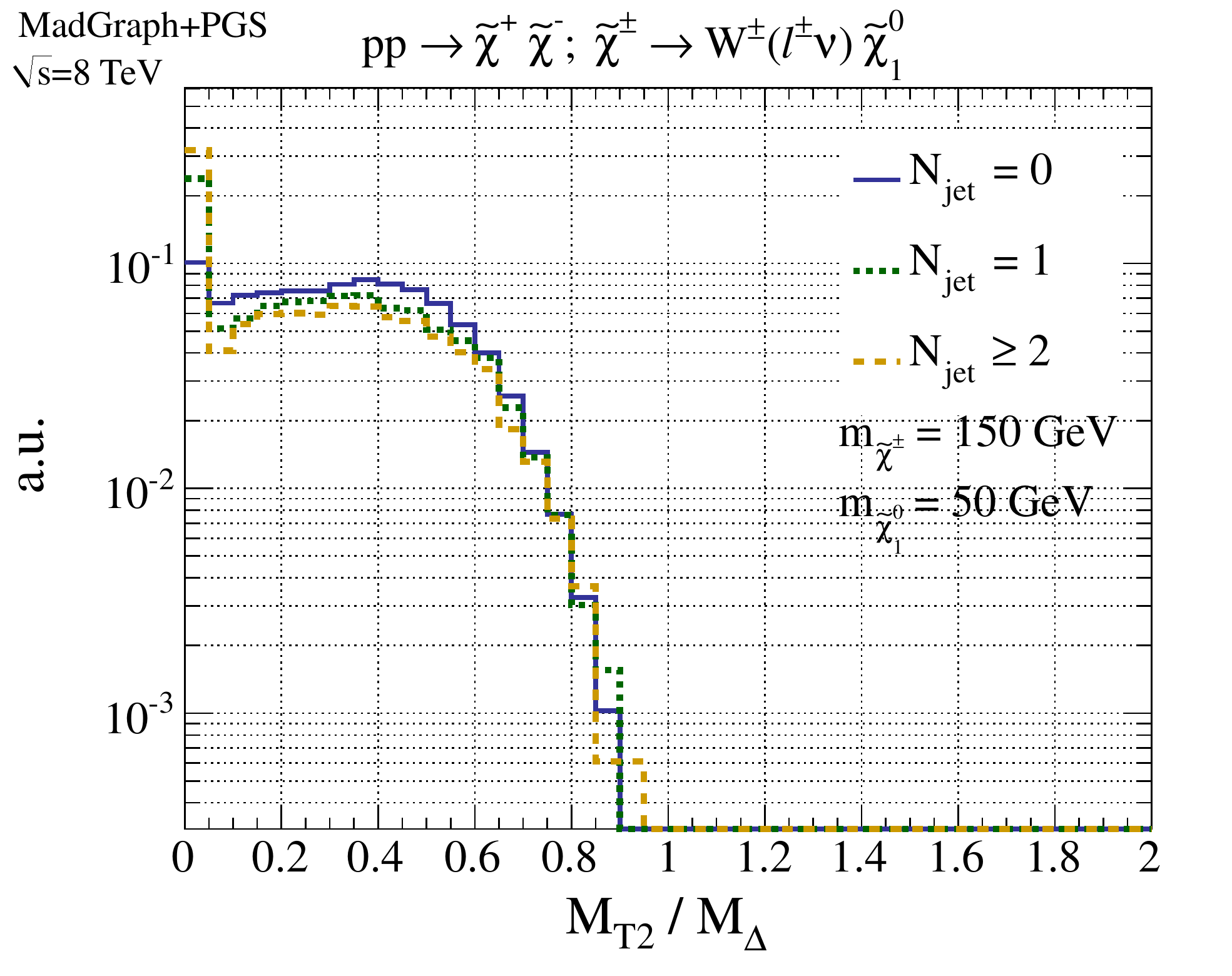} 
\caption{Distributions of $M_{\Delta}$ estimating variables for charginos with mass 150~GeV decaying into 50~GeV neutralinos and leptonically decaying $W$ bosons, as a function of reconstructed jet multiplicity. Variables include $M_{\Delta}^{R}$ (left), $M_{CT\perp}$ (center) and $M_{T2}$ (right), all normalized to the true value of $M_{\Delta}$ for each sample. \label{fig:compare_njet}}
\end{figure}

However, this choice does effect which events can be used to gain sensitivity to $M_{\Delta}$, as can be seen in the presence or absence of an accumulation of events with a value of zero for each of these discriminants. By only considering information along one transverse axis in the event, $M_{CT\perp}$ is exactly zero around 50\% of the time, corresponding to cases where the two leptons are moving in opposite directions along that axis. $M_{T2}$ exhibits similar behavior, though with fewer events having $M_{T2} = 0$. This fraction of events are observed to vary with different signal mass combinations, and also with the magnitude of $p_{T}^\text{CM}$ (see Figure~\ref{fig:compare}). This latter dependence can be seen by comparing the $M_{T2}$ distribution between different jet multiplicity categories, as shown in Figure~\ref{fig:compare_njet}, for each of the three $M_{\Delta}$ estimators, whereby larger jet multiplicity is generally correlated with larger $p_{T}^\text{CM}$. We observe that $M_{\Delta}^{R}$ does not exhibit this behavior.

The accumulation of events at zero in any of these kinematic variables has subtle effects on analysis that use them. Since the behavior is similar for both signal and background events, the ratio of the expected yields is largely insensitive to this effect. What does change is the size of the effective dataset that contains information about $M_{\Delta}$. The 50\% of events with $M_{CT\perp} = 0$ means that the integrated luminosity used in a search is effectively halved, while the increasing number of $M_{T2} = 0$ events at larger $p_{T}^\text{CM}$ results in a similar effect for higher jet multiplicities and/or boosted topologies. The quantitative implications of these dependencies on both selection efficiency and ultimately expected sensitivity of searches are discussed in the following sections.

\subsection{CMS and ATLAS-like event selections\label{sec:CMSATLAS}}

At this stage we have only considered kinematic variables in the context of quite inclusive event selections. In practice, searches for new physics in di-lepton final states include multiple kinematic requirements, each designed to suppress particular backgrounds. These additional requirements often involve other discriminating variables, like $E_{T}^\text{miss}$, which can be highly correlated with the $M_{\Delta}$ estimators described in the previous section. Understanding the efficacy of any of these variables depends on the context of where in kinematic phase-space an analysis is searching. In order to evaluate whether using the variable $M_{\Delta}^{R}$ would yield an improvement in the sensitivity of the CMS and ATLAS searches we define CMS- and ATLAS-like event selections through which we attempt to capture the relevant qualitative features of the kinematic requirements enforced in these analyses. 

In addition to the baseline selection requirements described in Section~\ref{sec:baseline} we consider criteria used by CMS and ATLAS (largely designed to reject Drell-Yan backgrounds):\\
\begin{tabular}{c c}
~~~~~~~~~~~~~~~~~~~~~~~~~~~~~~~~{\bf CMS selection}~~~~~~~~~~~~~~~~~~~~~~~~~~~~~~~~ & {\bf ATLAS selection} \\
& \\
$|m(\ell\ell) - m_{Z}| > 15$~GeV (SF channels) & $|m(\ell\ell) - m_{Z}| > 10$~GeV (SF channels) \\
$E_{T}^\text{miss} > 60$~GeV & $E_{T}^\text{miss,rel} > 40$~GeV
\end{tabular}

where
\begin{eqnarray}      
E_{T}^\text{miss,rel} = \left\{
	\begin{array}{l l}
	E_{T}^\text{miss}  & \quad \mathrm{if}~\sin\Delta\phi_{\ell,j} \ge \pi/2~, \\
	E_{T}^\text{miss} \times \sin\Delta\phi_{\ell,j} & \quad \mathrm{if}~\sin\Delta\phi_{\ell,j} < \pi/2~,
	\end{array} \right. 
	\label{eq:function}
\end{eqnarray} 
and $\Delta\phi_{\ell,j}$ is the azimuthal angle between the lepton or jet closest in the transverse plane to $\vec{E}_{T}^\text{miss}$. Topologically, backgrounds like $WW$ and $t\bar{t}$ are similar to the slepton and chargino signals, with two massive $W$ bosons each decaying to a lepton and neutrino. As a result, the searches' ability to distinguish signal events from these backgrounds is dependent mostly on the difference between $m_{W}$ and $M_{\Delta}$ of each signal, and is accomplished primarily through kinematic variables estimating $M_{\Delta}$. The CMS and ATLAS selection requirements listed above are added specifically to reject $Z/\gamma^*$+jets events, using a $Z$ mass window veto and $E_{T}^\text{miss}$ related selections to eliminate events with mis-measured jets or leptons which result in spurious $E_{T}^\text{miss}$. 

\begin{figure}[ht]
\includegraphics[width=0.35\columnwidth]{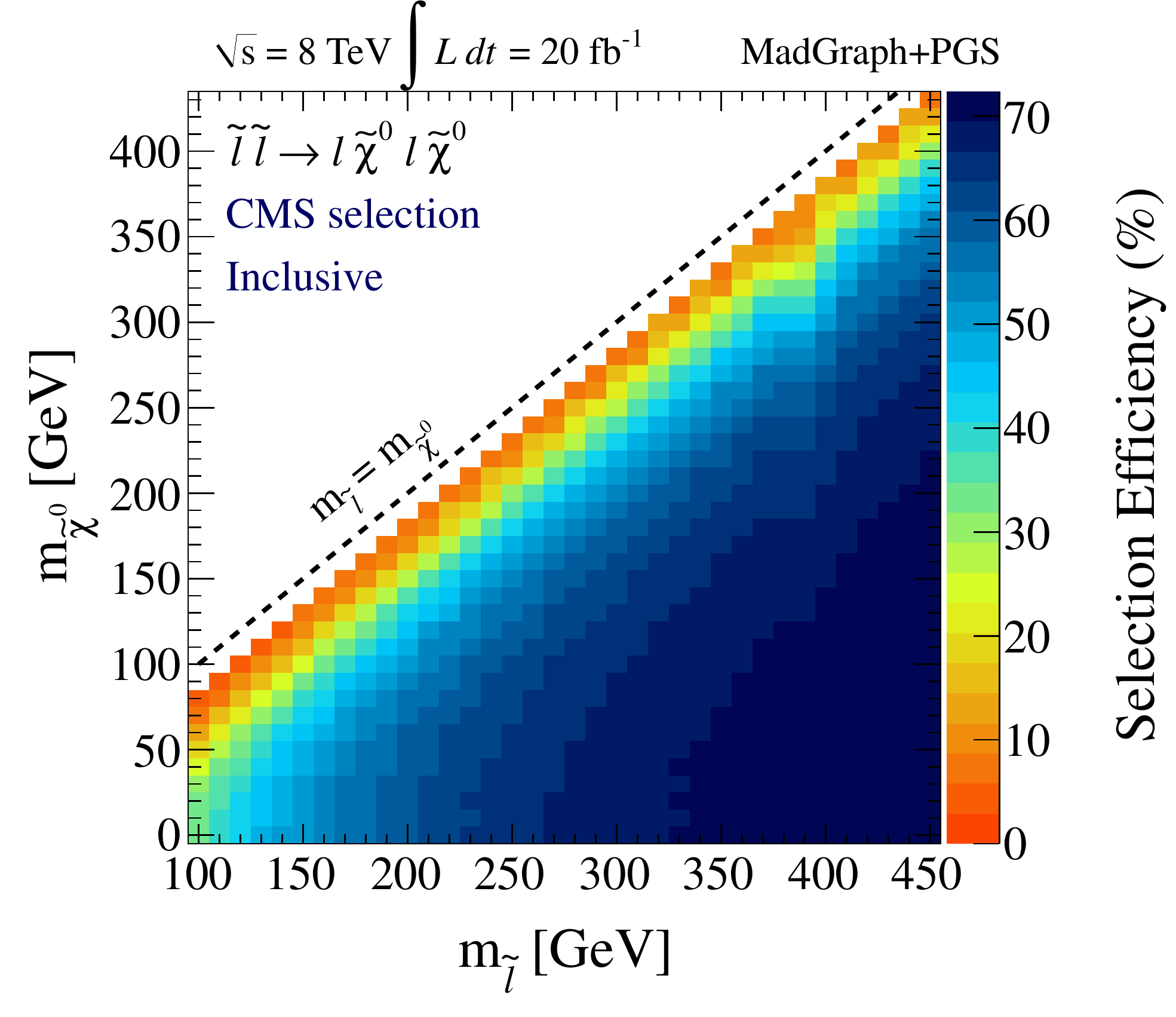}
\includegraphics[width=0.35\columnwidth]{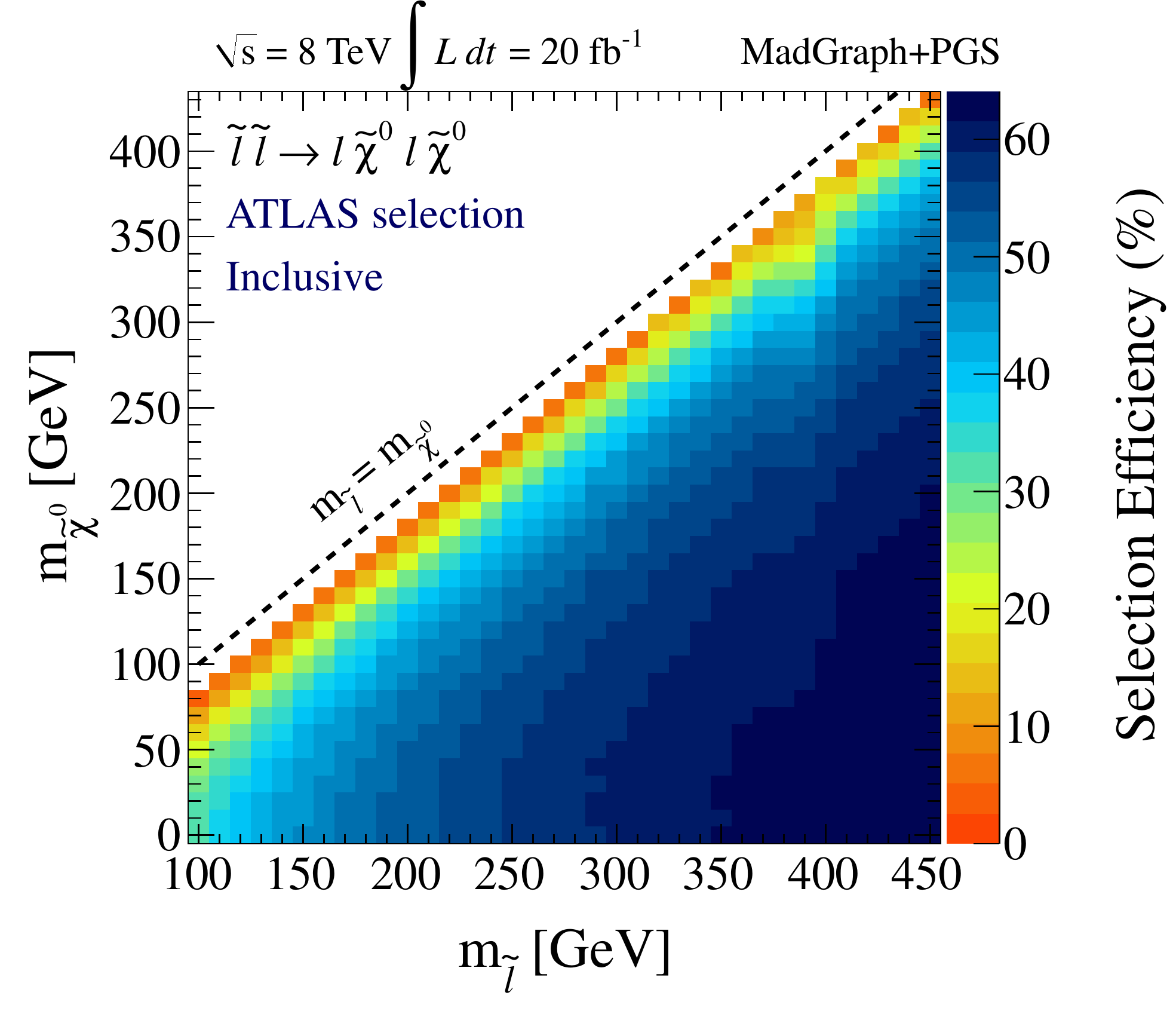}\\
\includegraphics[width=0.35\columnwidth]{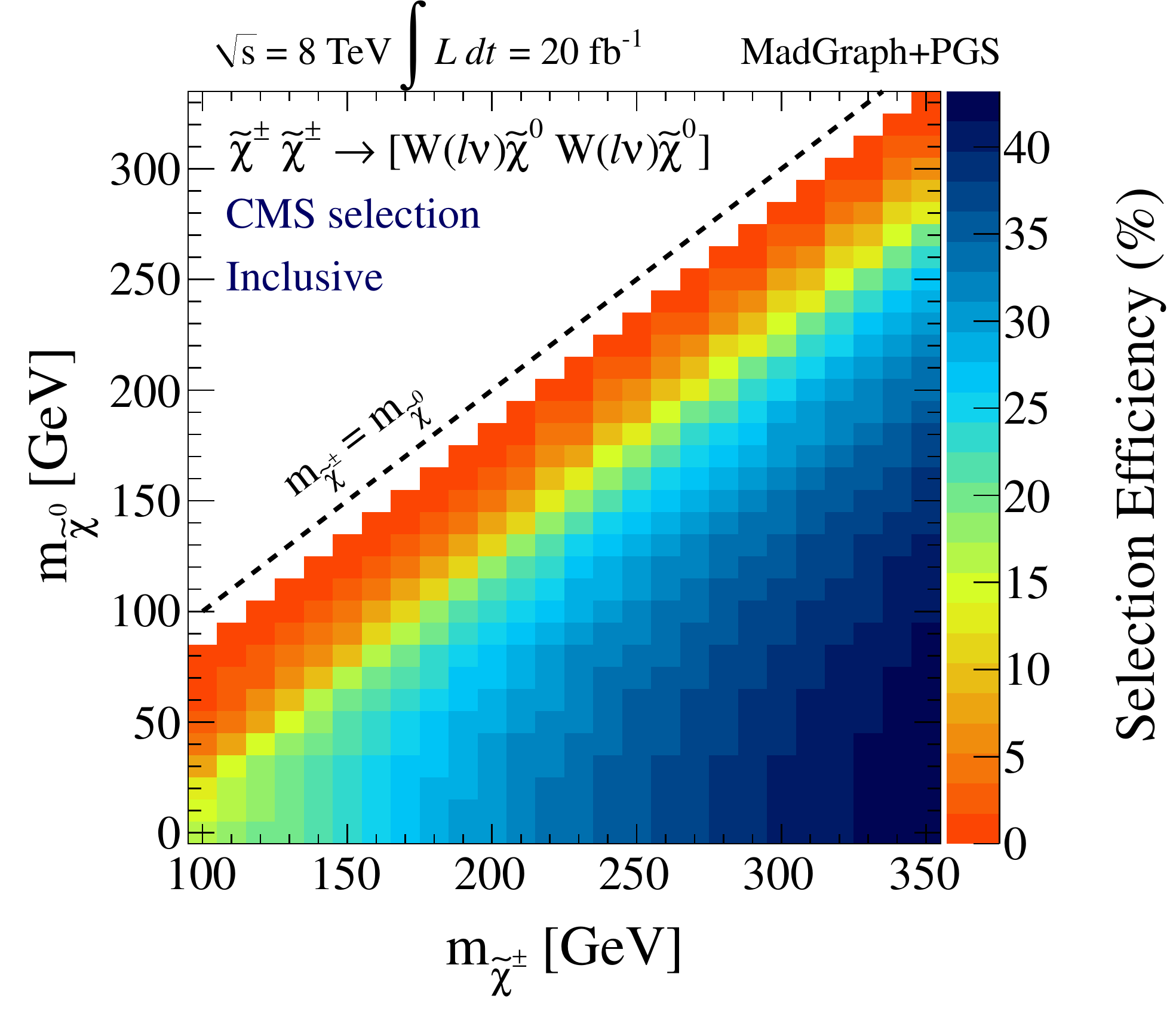}
\includegraphics[width=0.35\columnwidth]{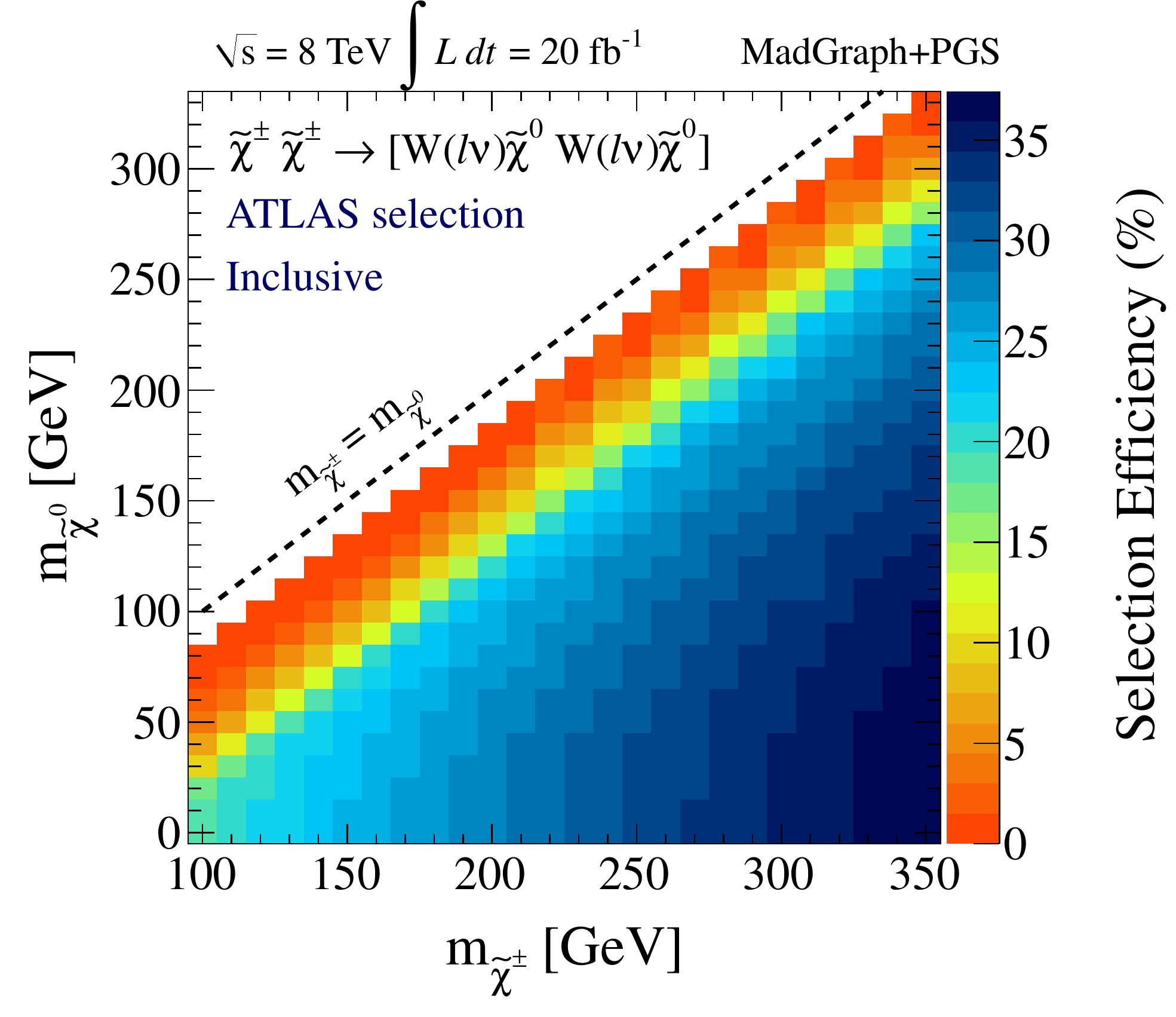}
\caption{Efficiency of the CMS (left) and ATLAS (right) selections for the slepton (top) and chargino (bottom) signal models. Selection efficiencies are calculated as a function of parent sparticle and neutralino mass and include all final state categories. \label{fig:EFF_inclusive}}
\end{figure}

The CMS and ATLAS selection efficiencies for sleptons and charginos, as a function of sparticle masses, are summarized in Figure~\ref{fig:EFF_inclusive}. Without explicit requirements placed on $M_{\Delta}$-estimating variables, the selection efficiencies are relatively flat throughout most of the sparticle mass parameter-space, dropping quickly as parent sparticles and neutrinos approach mass degeneracy. This efficiency drop is a result of portions of the event selection involving energy scale, in this case minimum lepton $p_{T}$ requirements and $E_{T}^\text{miss}$/$E_{T}^\text{miss,rel}$ cuts. For di-slepton production, the leptons are produced in two-body decays of the parent sparticle resulting in the lepton and neutralino momentum distributions scaling closely with $M_{\Delta}$ and a large efficiency gradient once energy scale requirements become comparable to the sparticle mass difference. For chargino models this gradient is less dramatic; with leptons produced in subsequent decays of $W$ bosons (rather than in two-body decays of sparticles) the momentum distributions are broader, with weaker $M_{\Delta}$ correspondence. Furthermore, neutralinos are not the only weakly interacting particles in the final state and the total momentum of the entire system of weakly interacting particles is smaller as more energy is contained in the mass rather than the momentum. The result is that $E_{T}^\text{miss}$ requirements can be especially inefficient for these signals relative to sleptons, with large drops in efficiency extending further from the mass degeneracy diagonal.

\begin{figure}[ht]
\includegraphics[width=0.35\columnwidth]{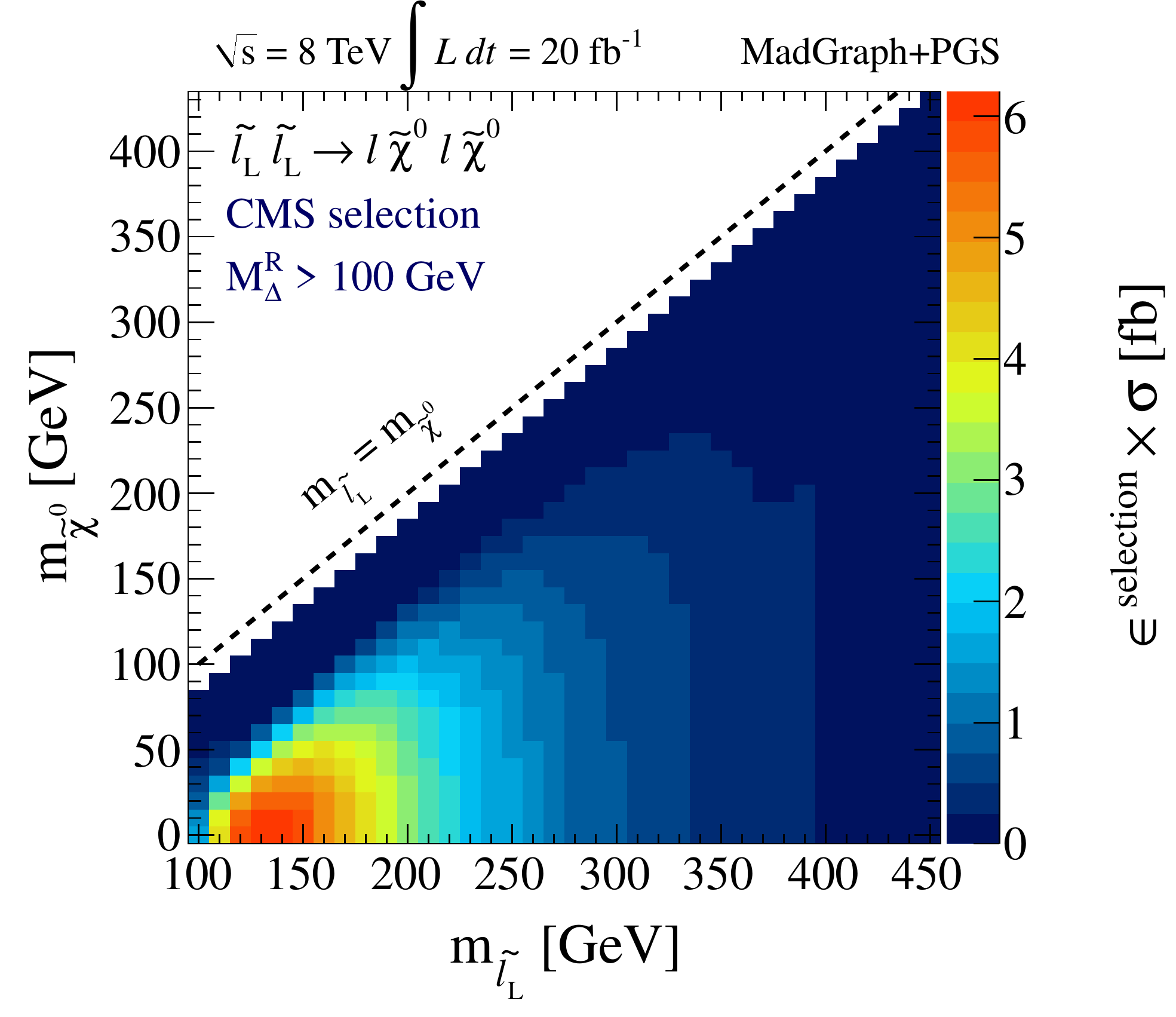}
\includegraphics[width=0.35\columnwidth]{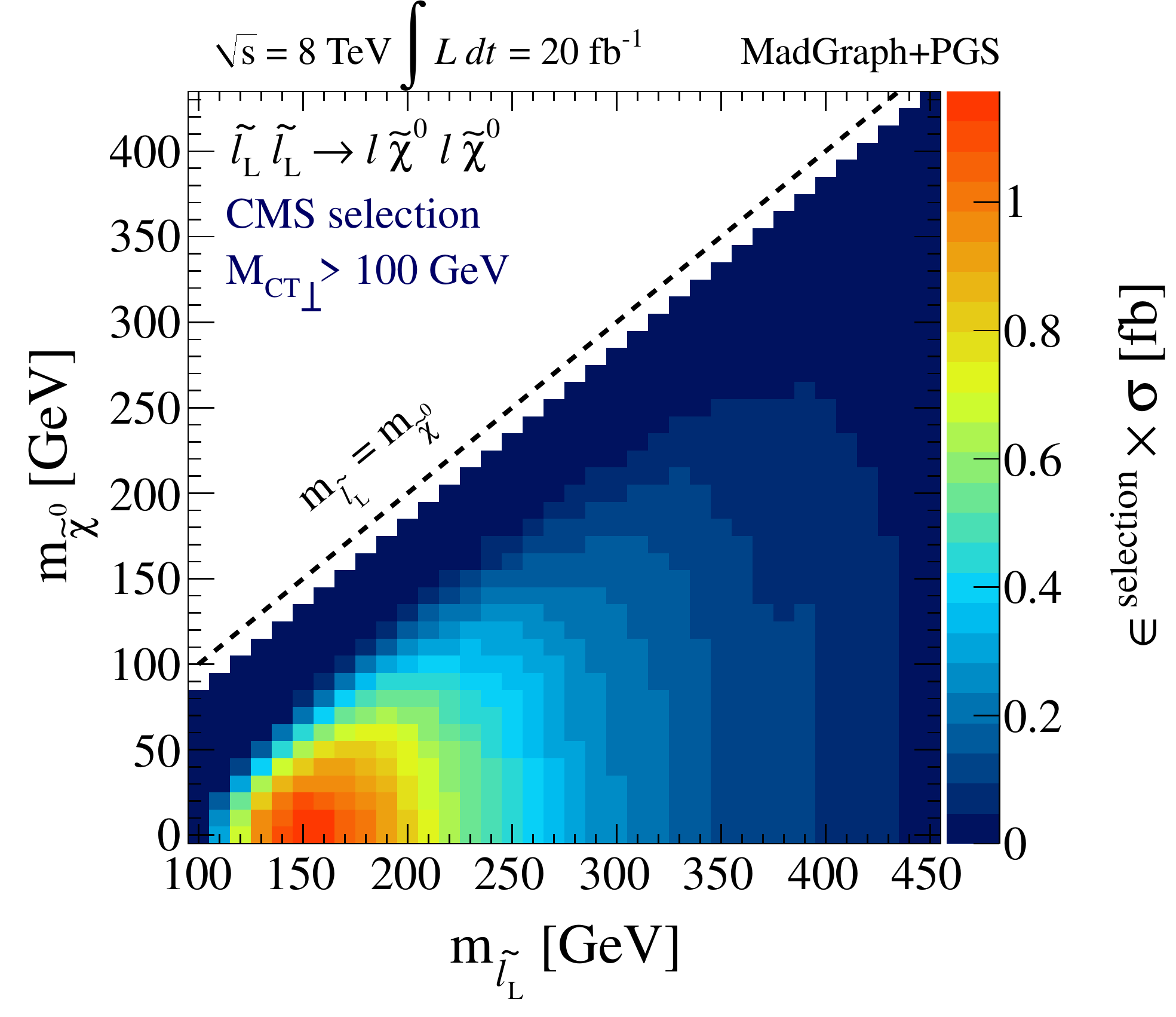}\\
\includegraphics[width=0.35\columnwidth]{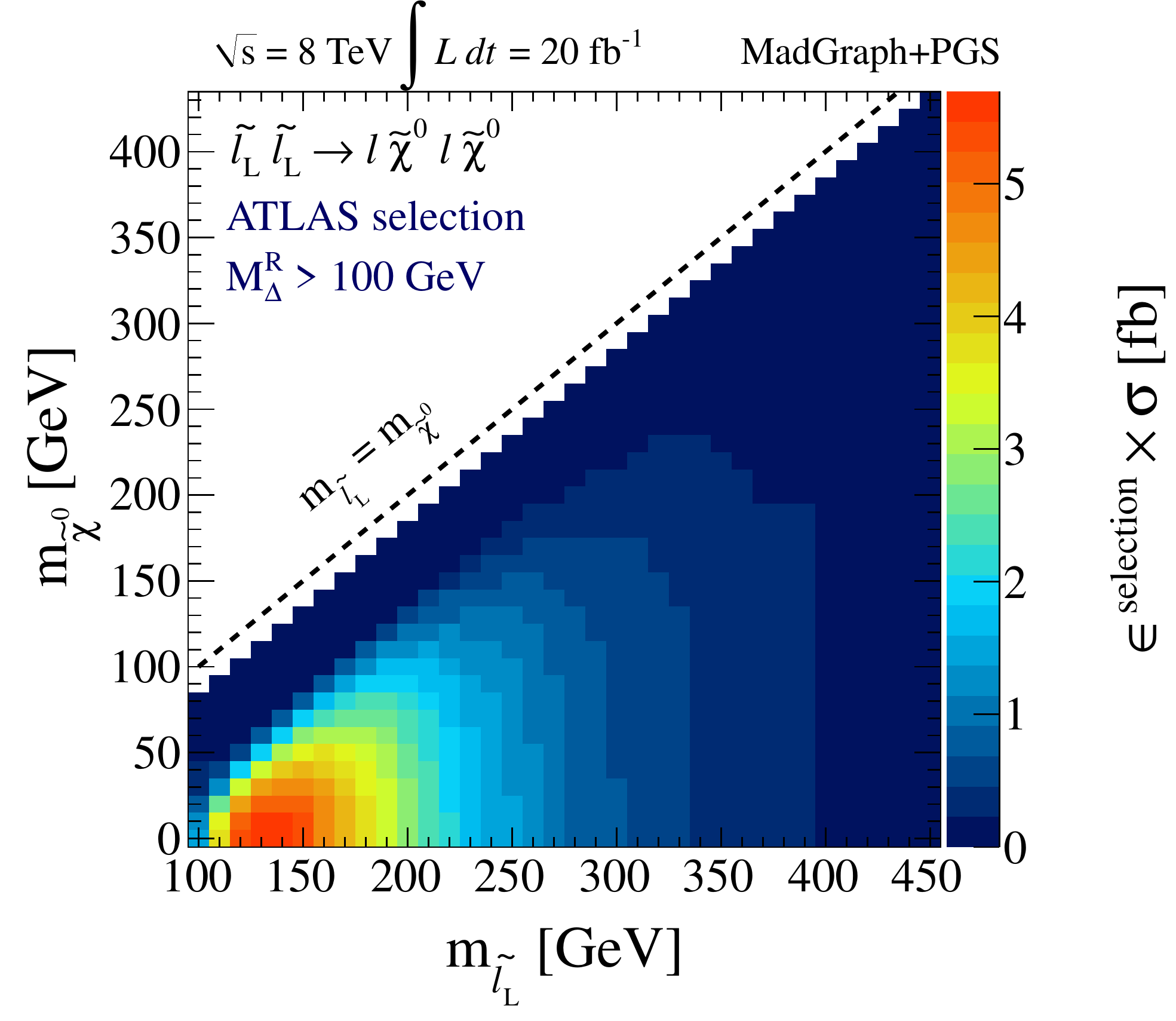}
\includegraphics[width=0.35\columnwidth]{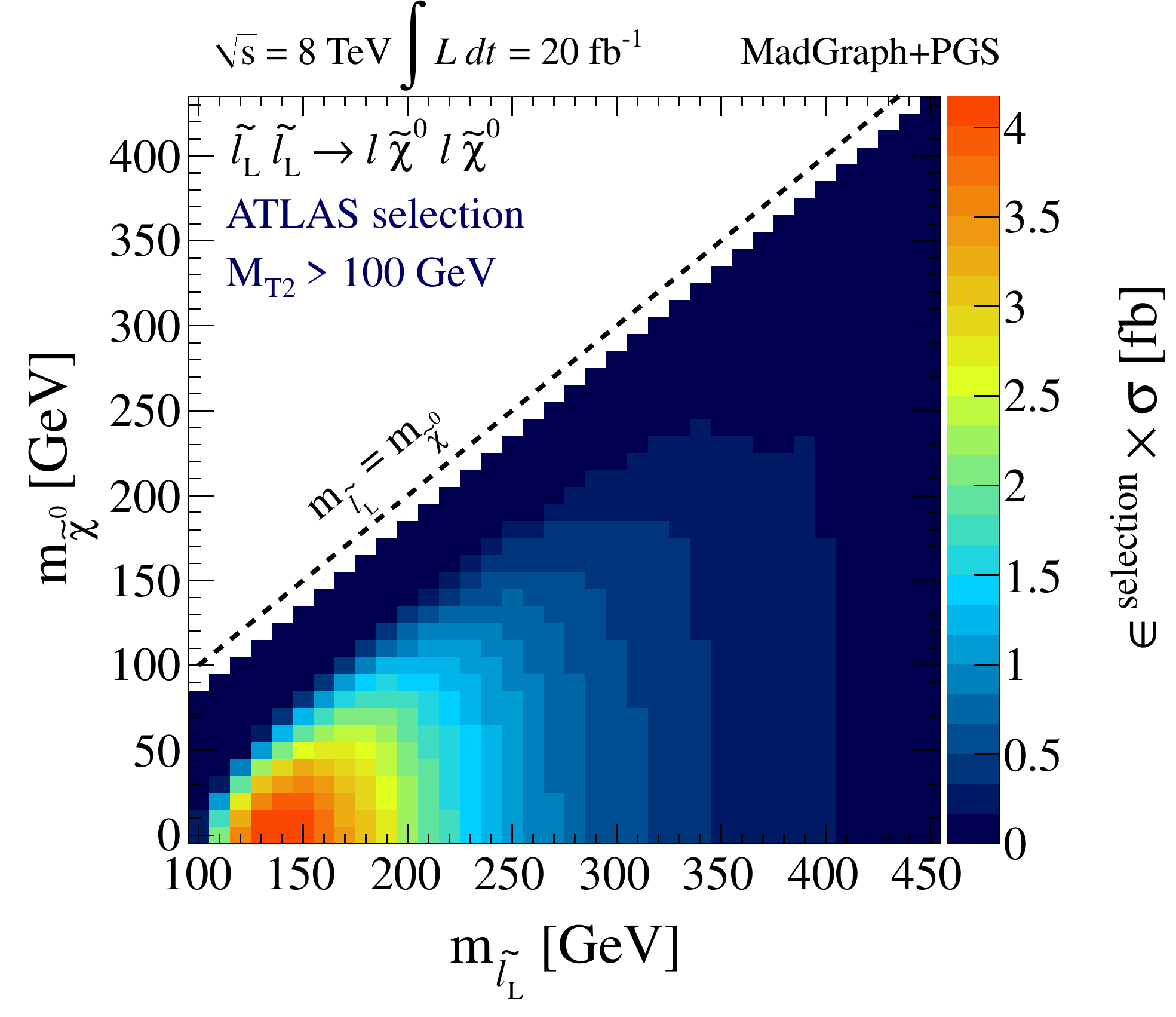}
\caption{Efficiency times cross section for slepton signal samples, as a function of neutralino mass, for the CMS (top) and ATLAS (bottom) selections with additional requirement that the mass sensitive variable ($M_{\Delta}$ - left, $M_{CT\perp}$ - top right, $M_{T2}$ - bottom right) is in excess of 100~GeV. \label{fig:XSEC_M100_slepton}}
\end{figure}

\begin{figure}[ht]
\includegraphics[width=0.35\columnwidth]{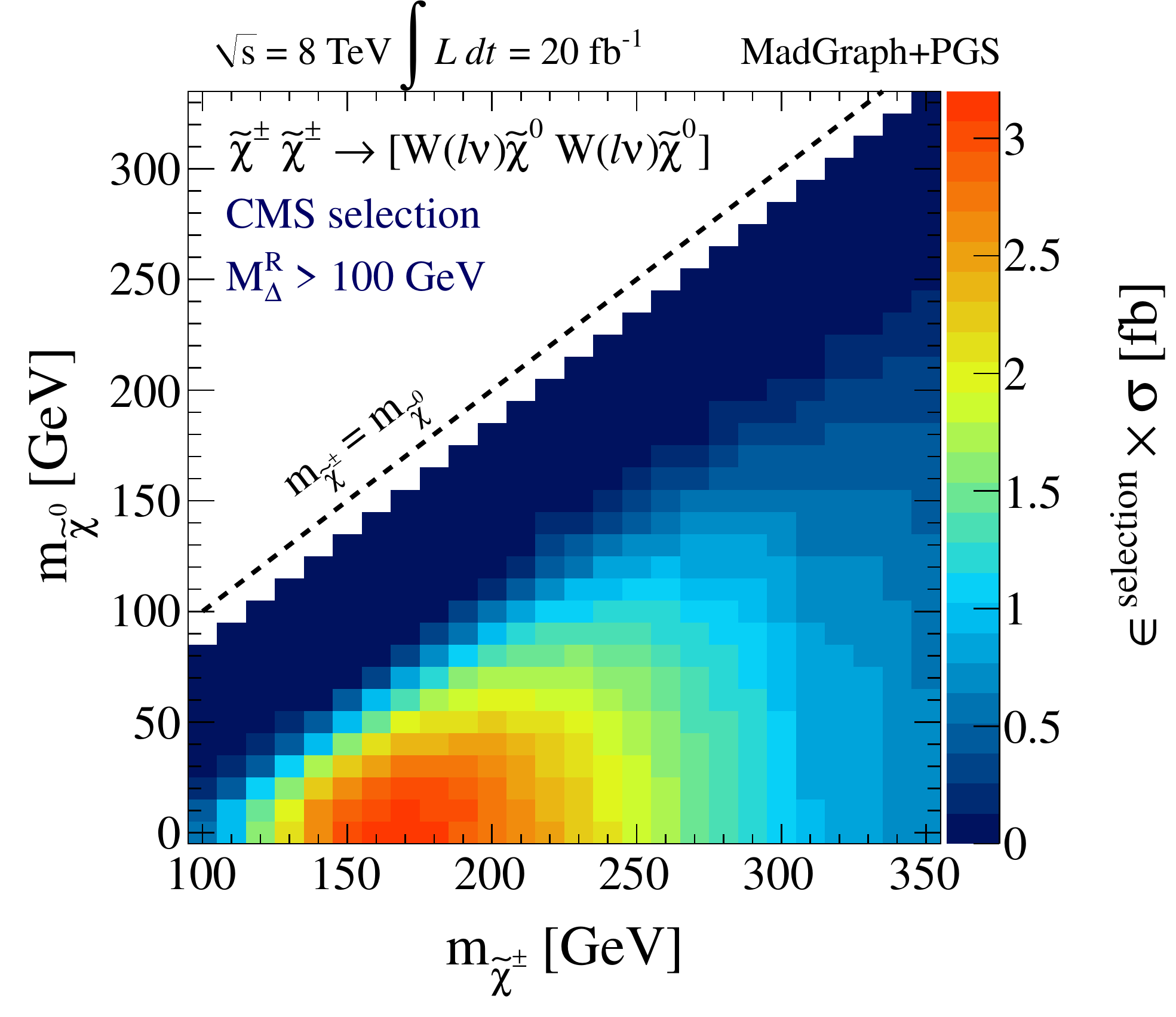}
\includegraphics[width=0.35\columnwidth]{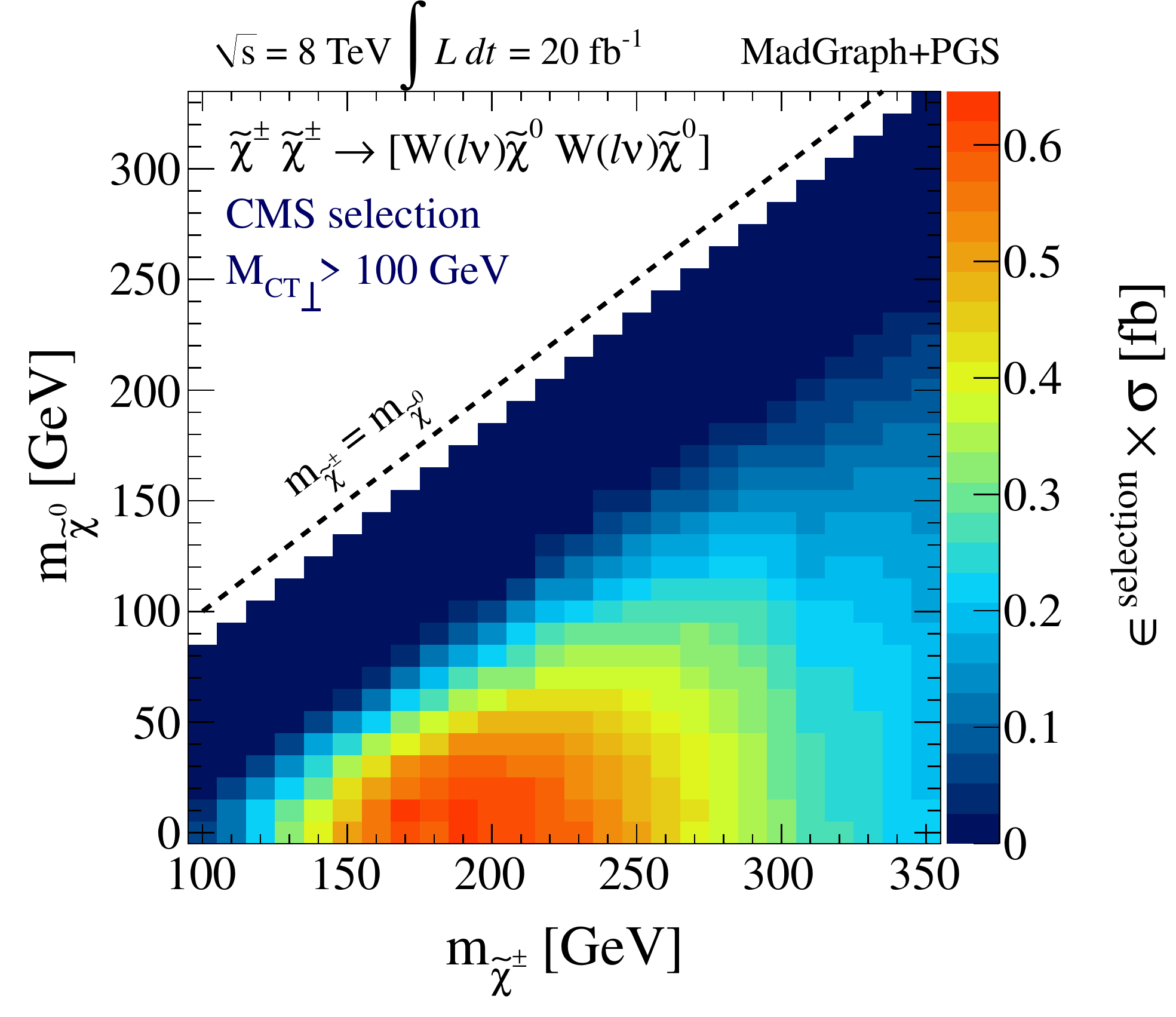}\\
\includegraphics[width=0.35\columnwidth]{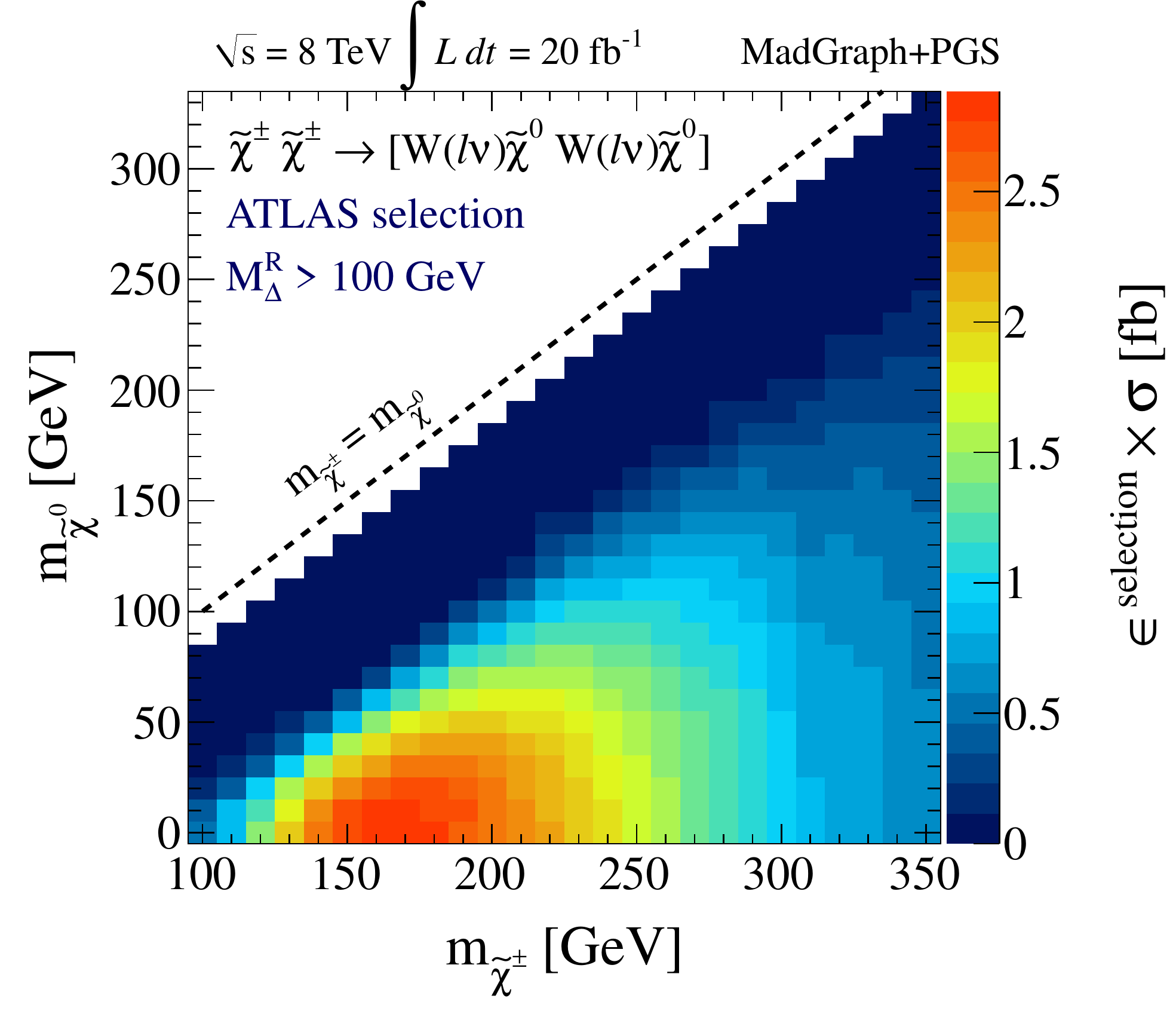}
\includegraphics[width=0.35\columnwidth]{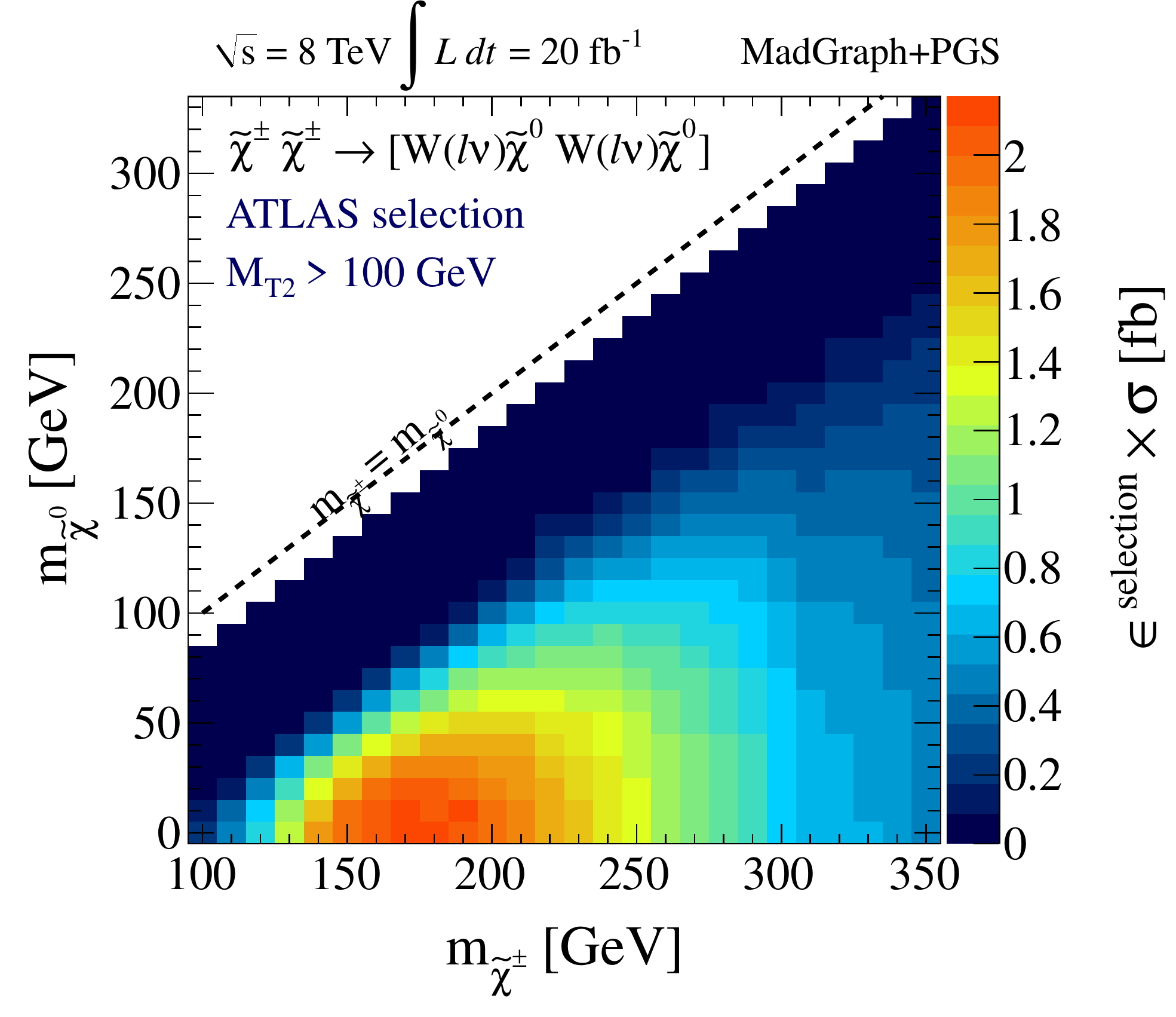}
\caption{Efficiency times cross section for chargino signal samples, as a function of neutralino mass, for the CMS (top) and ATLAS (bottom) selections with additional requirement that the mass sensitive variable ($M_{\Delta}$ - left, $M_{CT\perp}$ - top right, $M_{T2}$ - bottom right) is in excess of 100~GeV. \label{fig:XSEC_M100_chargino}}
\end{figure}

In practice, the sensitivity of search analyses in this final state will not scale exactly with this inclusive efficiency, due to large differences in background yields as a function of $M_{\Delta}$-sensitive variables. Each of the largest backgrounds after the CMS and ATLAS selections ($WW$ and $t\bar{t}$), have $M_{\Delta}$-sensitive variable distributions that have inherited information of the scale $m_{W}$. Therefore the sensitivity scales strongly with $M_{\Delta}$, with significant experimental reach appearing only once $M_{\Delta}$ is in excess of the $W$ mass. The effective cross-sections for signal models after the additional requirement that the $M_{\Delta}$-sensitive variable used in each analysis is in excess of 100 GeV are shown in Figures~\ref{fig:XSEC_M100_slepton} and~\ref{fig:XSEC_M100_chargino} for sleptons and charginos, respectively. The expected sensitivity of the hypothetical searches described in the following sections closely follows these yields. Efficiencies and cross-sections for the SM backgrounds considered in these analyses are summarized in Table~\ref{tab:BKG}.

\subsection{Super-Razor selection without an $E_{T}^\text{miss}$ cut  }

Kinematic variables sensitive to $M_{\Delta}$ can be powerful discriminants between slepton and chargino signals and SM backgrounds when $M_{\Delta}$ is much larger than the $W$ mass, while heavy sparticle production with relatively compressed spectra can more easily
remain hidden under large SM backgrounds. The angular variables introduced in Section~\ref{sec:variables}, $\Delta \phi_{R}^{\beta}$ and $|\cos\theta_{R+1}|$, are designed to address this deficiency. They are sensitive to quantities in events other than $M_{\Delta}$: the ratio of daughter to parent mass and the spin correlations of decaying particles in the event. Thus they can be used to further discriminate between signal and background. 

Each of the super-razor variables, $M_{\Delta}^{R}$, $\sqrt{\hat{s}}_{R}$, $\vec{\beta}_{R}$, $\vec{\beta}_{R+1}$, $\Delta \phi_{R}^{\beta}$, and $|\cos\theta_{R+1}|$, represents a different piece of information about an event, and the collection can be thought of as a kinematic basis. Here we explore a new kinematic selection based on this basis, attempting to increase sensitivity to models with smaller values of  $M_{\Delta}$. In particular, we consider how one can remove explicit requirements on $E_{T}^\text{miss}$. Included primarily to reject Drell-Yan background, such a requirement is inefficient for signal events at low $M_{\Delta}$. Rather than attempting to determine an optimized set of cuts on the super-razor variables, we demonstrate how a selection criteria can be designed through simple choices for each variable based on the backgrounds we are attempting to reject.

We first consider the triplet of variables $M_{\Delta}^{R}$, $\sqrt{\hat{s}}_{R}$, and $\gamma_{R+1}$, which for di-slepton production are meant to estimate $m_{\tilde{\ell}}$, $\sqrt{\hat{s}}$, and $\gamma^\text{decay}$, respectively. For both the true and reconstructed quantities the three variables represent only two unique pieces of information, since they are related by $\sqrt{\hat{s}}_{R} = 2 \gamma_{R+1}M_{\Delta}^{R}$ and $\sqrt{\hat{s}} = 2 \gamma^\text{decay}m_{\tilde{\ell}}$. Which two variables to consider depends on which signal and backgrounds are being investigated. For example, if searching for $H \to W(\ell\nu)W(\ell\nu)$ the variable $\sqrt{\hat{s}}_{R}$ will be resonant at the Higgs mass. If the Higgs is too light to accommodate two on-shell $W$'s then the variables  $M_{\Delta}^{R}$ and $\gamma_{R+1}$ are more difficult to interpret, representing a combination of two $W$ bosons with different masses. For the case of interest in this paper, non-resonant slepton pair production, off-shell sleptons are kinematically suppressed, and so  $M_{\Delta}^{R}$, $\sqrt{\hat{s}}_{R}$ and $\gamma_{R+1}$ are all meaningful. However, they are not equally useful for discriminating between signal and background.
 
For di-slepton pair production the quantity that $M_{\Delta}^{R}$ is attempting to measure is effectively constant event by event, and is useful for discriminating against backgrounds.  On the other hand, $\gamma^\text{decay}$ varies between events, characteristic of non-resonant production, as does  $WW$ and $t\bar{t}$ backgrounds. As a result, $\sqrt{\hat{s}}_{R}$ provides information largely redundant with $M_{\Delta}^{R}$ while $\gamma_{R+1}$ does not strongly discriminate against these large backgrounds. 

\begin{figure}[ht]
\includegraphics[width=0.3\columnwidth]{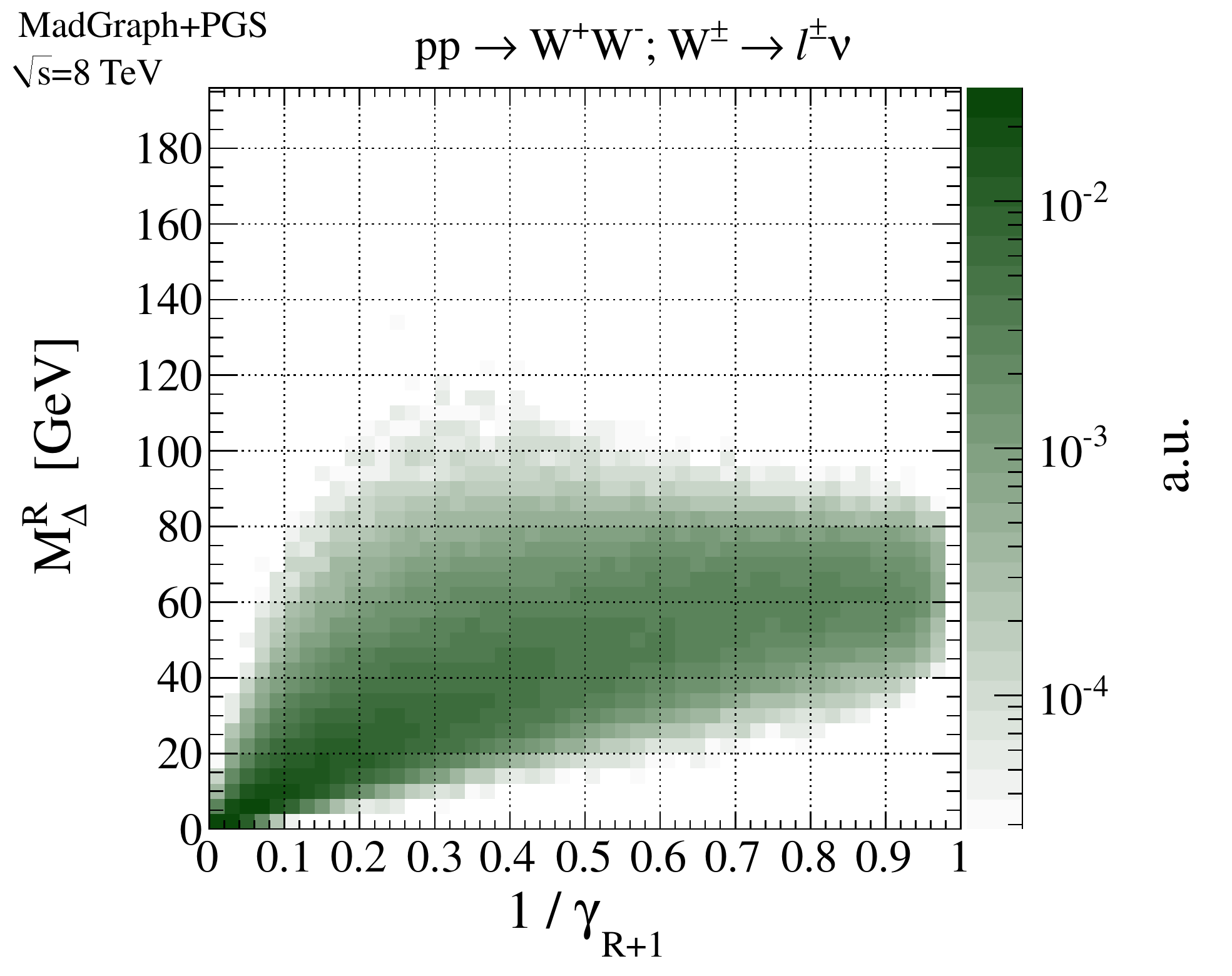}
\includegraphics[width=0.3\columnwidth]{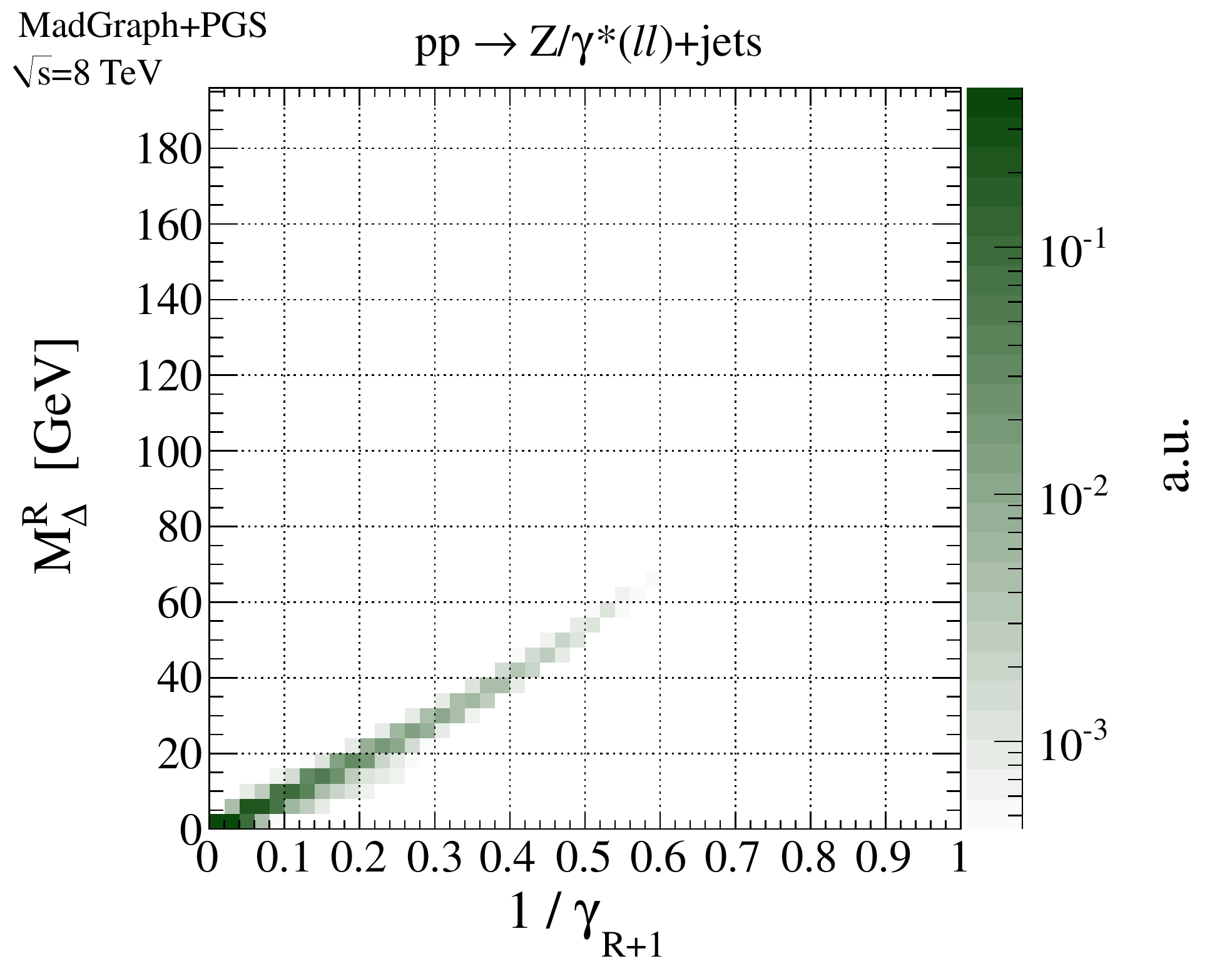} \\
\includegraphics[width=0.3\columnwidth]{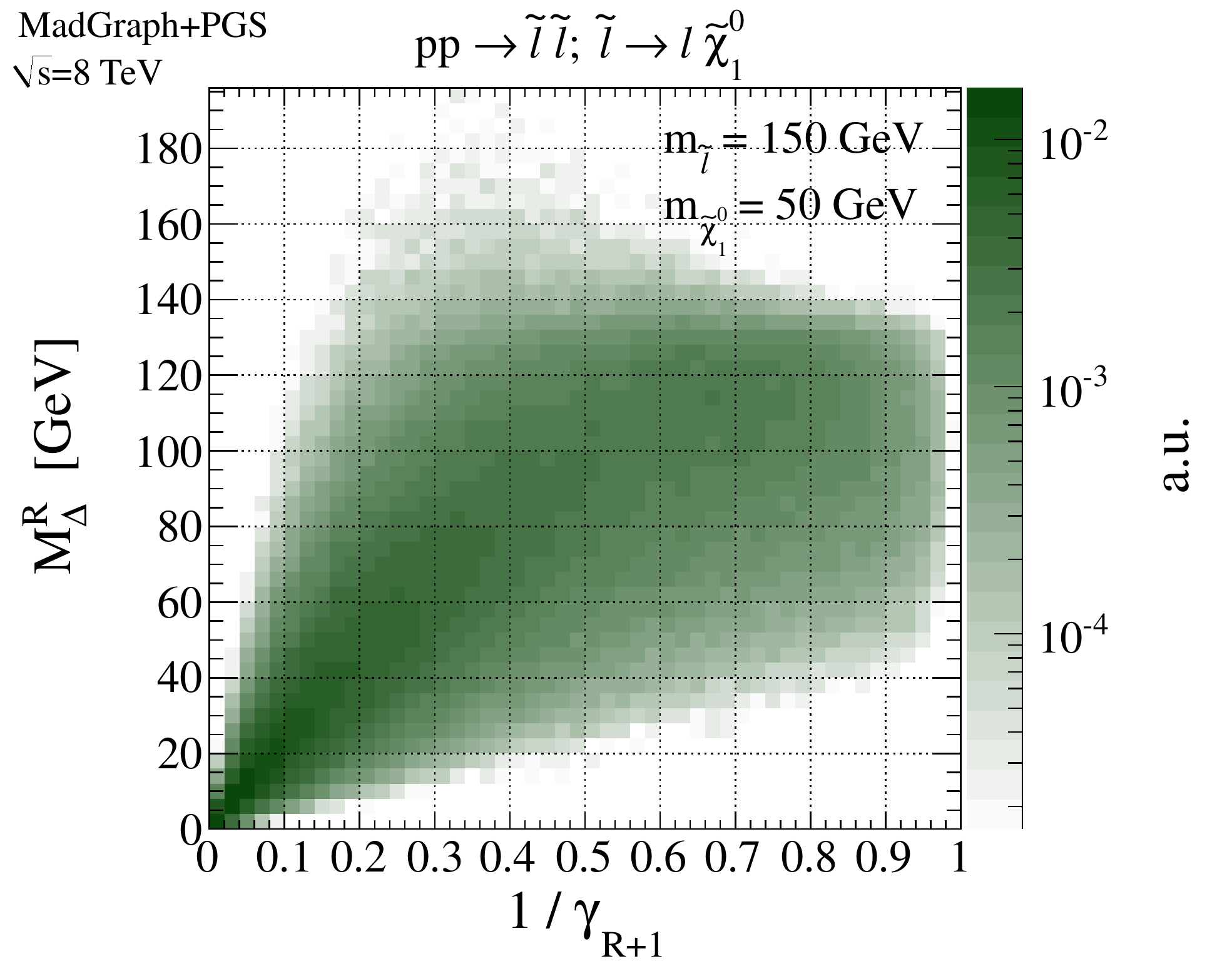}
\includegraphics[width=0.3\columnwidth]{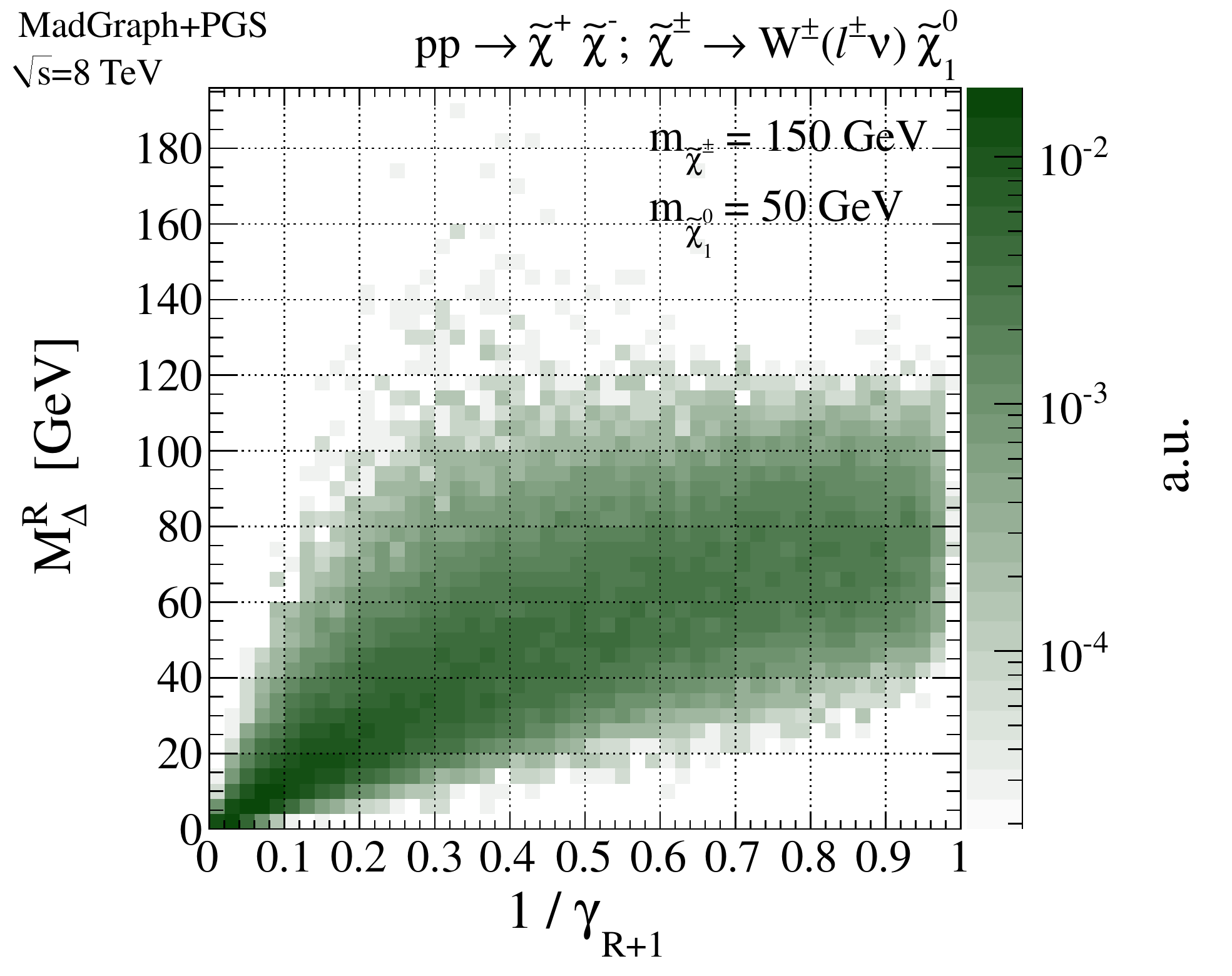}
\caption{Distributions of $M_{\Delta}^{R}$ vs. $1/\gamma_{R+1}$ for simulated events samples with baseline selection applied. Top left: $WW$. Top right: $Z/\gamma*$+jets. Bottom left: di-slepton production with $m_{\tilde{\ell}} = 150$~GeV and  $m_{\tilde{\chi}^0_1} = 50$~GeV. Bottom right: di-chargino production with $m_{\tilde{\chi}^{\pm}_{1}} = 150$~GeV and  $m_{\tilde{\chi}^0_1} = 50$~GeV. \label{fig:Mdelta_v_gamma}}
\end{figure}

The distributions of $M_{\Delta}^{R}$ as a function of $1/\gamma_{R+1}$, for simulated signal and background events, are shown in Figure~\ref{fig:Mdelta_v_gamma}. We observe that, for large $1/\gamma_{R+1}$, the two variables are largely uncorrelated for signal events. For small values of this variable, the signal $M_{\Delta}^{R}$ distribution collapses to zero, corresponding to the case where $\gamma^\text{decay}$ is large and the decay leptons are largely back-to-back in the CM frame. For backgrounds, smaller values of $1/\gamma_{R+1}$ correspond to larger $M_{\Delta}^{R}$ on average. This is especially true for $Z/\gamma^{*}$+jets events; fixed di-lepton invariant mass corresponds to positively correlated lines in the $M_{\Delta}^{R}$ versus $1/\gamma_{R+1}$ plane, as seen in Figure~\ref{fig:Mdelta_v_gamma}. We observe that requiring $1/\gamma_{R+1}$ to be above some small value can remove many background events, especially $Z/\gamma^{*}$+jets, while only removing signal events that fall in an unremarkable region of phase-space (away from their $M_{\Delta}^{R}$ edge).

Resonant di-lepton production from $Z/\gamma^{*}$+jets is a particularly pernicious background. Associated jets produced in the event can boost the di-leptons to topologies that mimic those of di-slepton signals. Spurious $E_{T}^\text{miss}$, often from mis-measurements of jets, prevents this effect from being correctly accounted for. While an absolute requirement on $E_{T}^\text{miss}$ can be used to remove many of these events, this also sets a lower bound $M_{\Delta}$ to which an analysis will be sensitive. Furthermore, $Z/\gamma^{*}$+jets with large $E_{T}^\text{miss}$ that survive such a cut will tend to have large $M_{\Delta}^{R}$, appearing in the region of phase-space we had hoped to query for signal events. To replace a $E_{T}^\text{miss}$ cut, we consider cuts based on scale-less variables, such that background events looking most like signal are also removed and we retain sensitivity to lower values of $M_{\Delta}^{R}$.

The variable $\vec{\beta}_{R}$ is highly sensitive to the mis-measurements which make $Z/\gamma^{*}$+jets a difficult background to remove with $E_T^\text{miss}$ cuts. This boost approximates the transverse portion of the Lorentz transformation from the lab frame to CM frame. Mis-calculations in the reconstruction of the direction and magnitude of this boost leave di-leptons from $Z/\gamma^{*}$ decays in imbalanced configurations while they would be back-to-back in their true CM frame. The magnitude of $\vec{\beta}_{R}$ is not a strong discriminant (it is related to the ratio of CM system $p_{T}$ and its reconstructed mass) while its direction is used in the calculation of $\Delta \phi_{R}^{\beta}$, as discussed in Section~\ref{sec:variables}. 

\begin{figure}[ht]
\includegraphics[width=0.3\columnwidth]{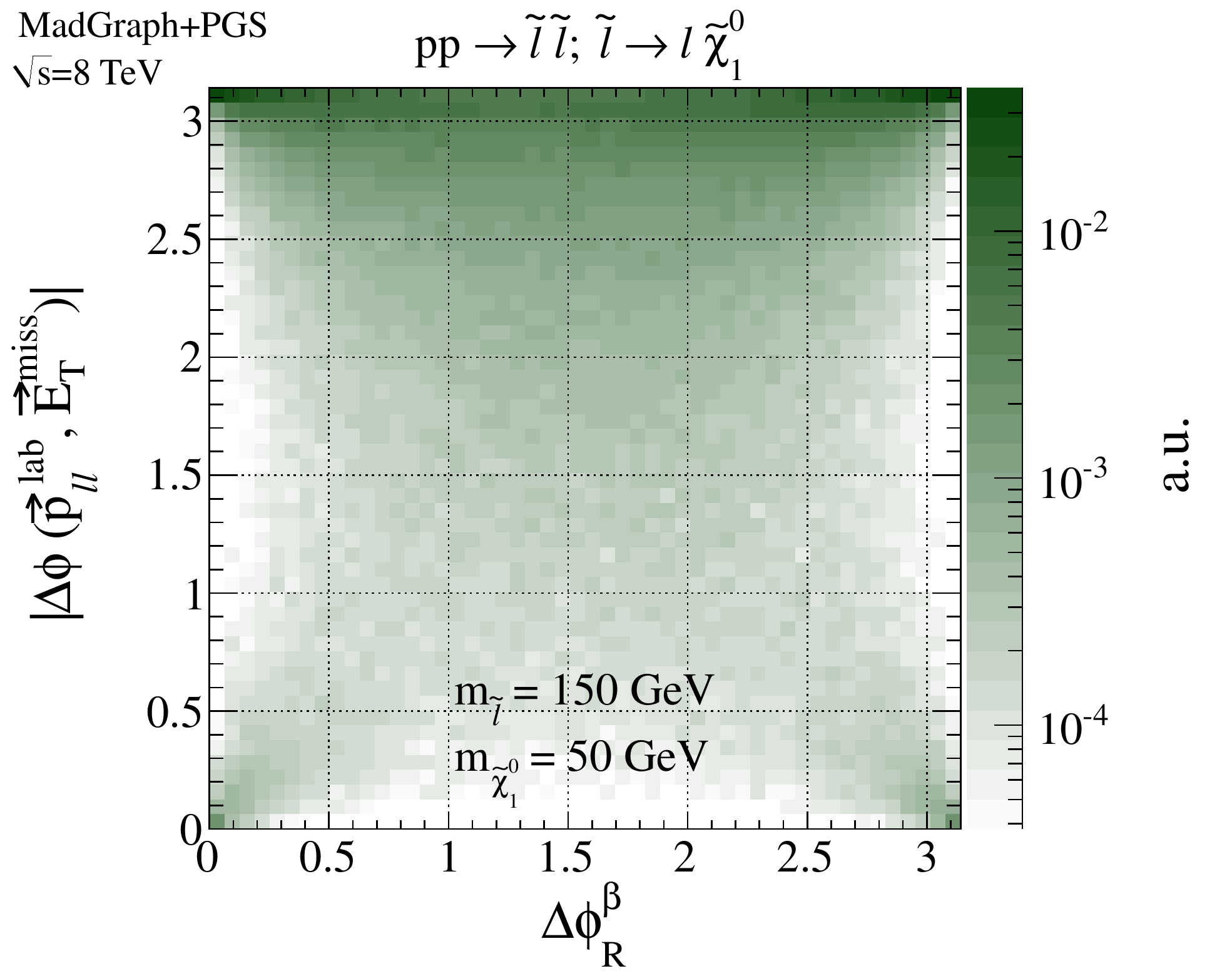}
\includegraphics[width=0.3\columnwidth]{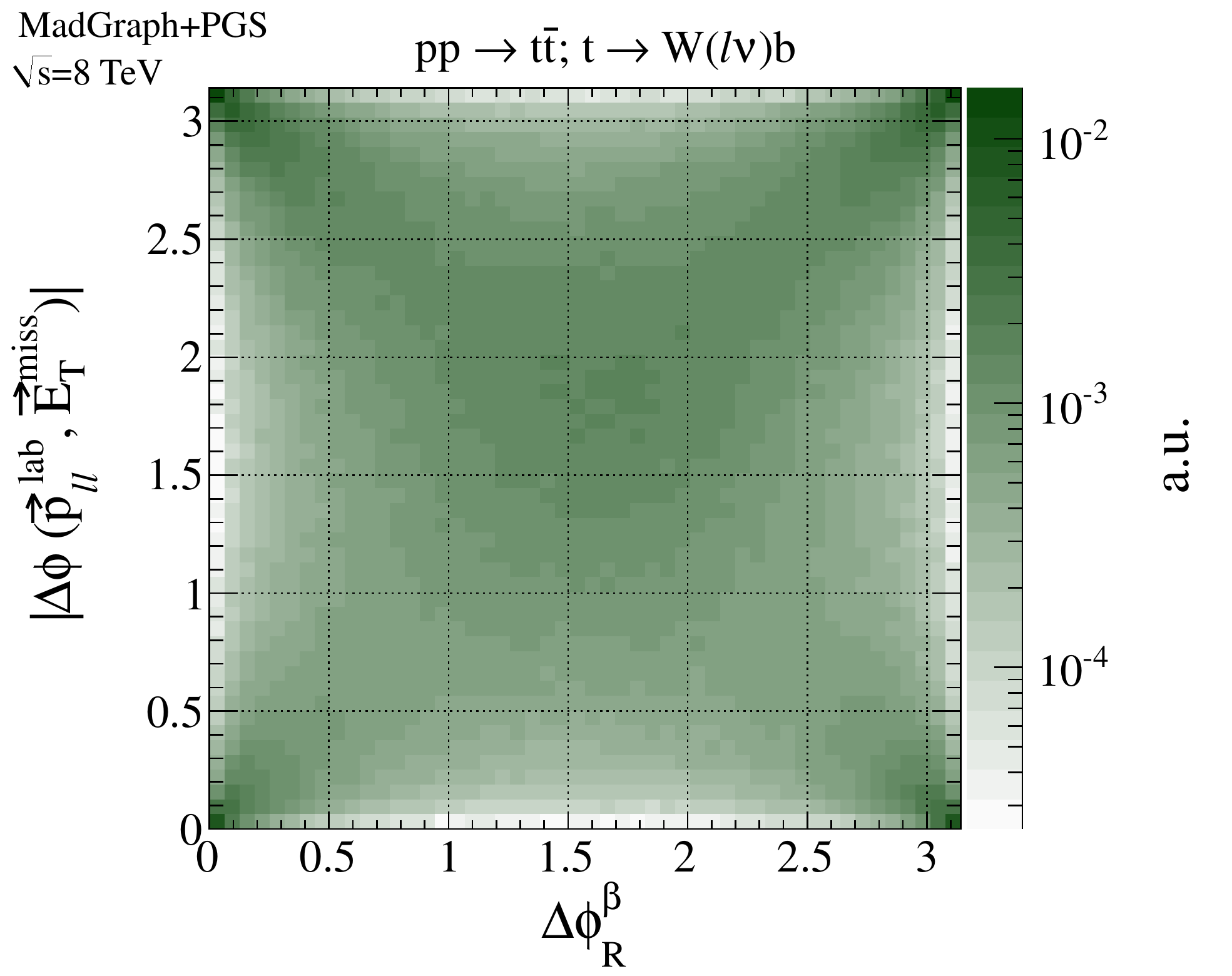}
\includegraphics[width=0.3\columnwidth]{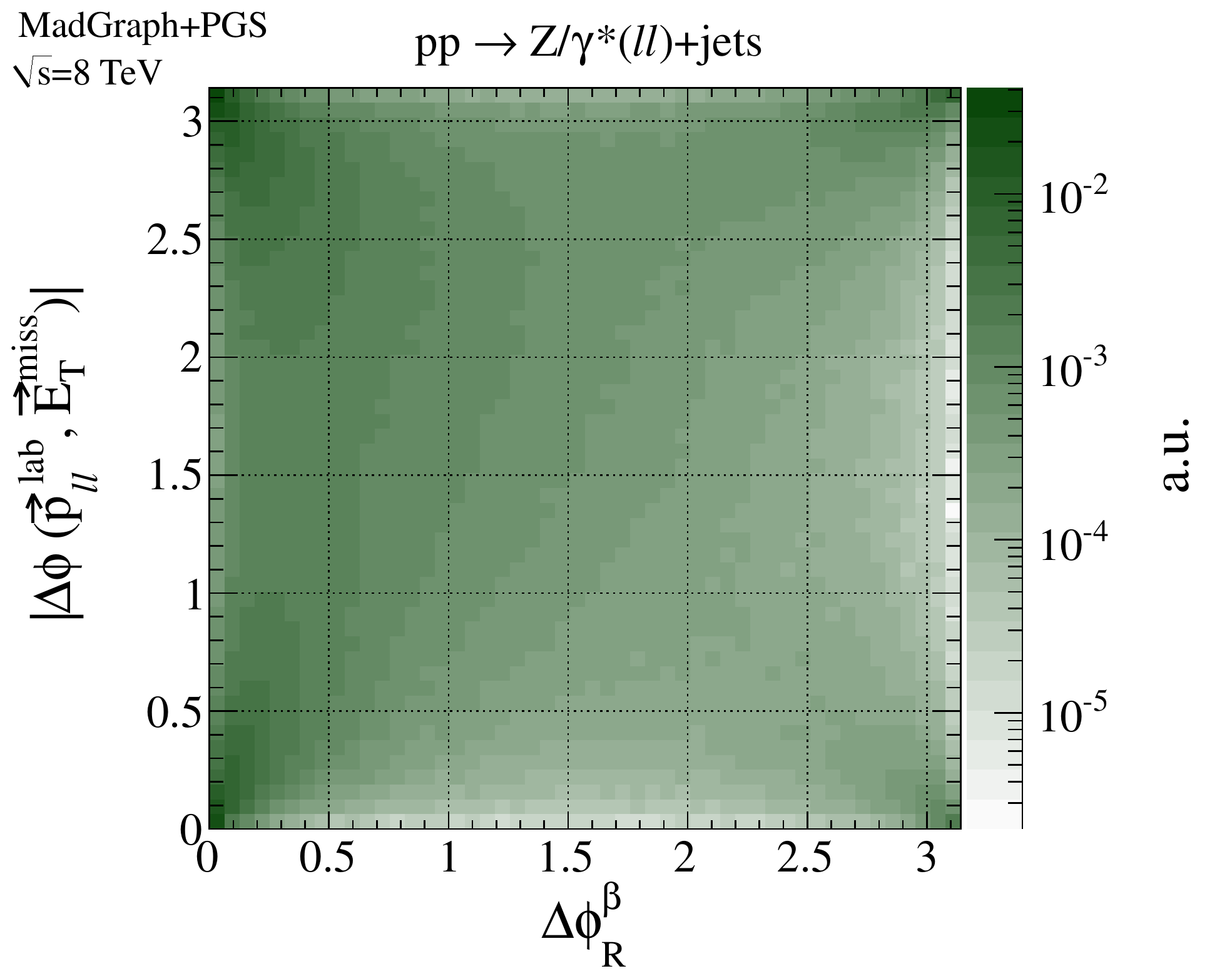}
\includegraphics[width=0.3\columnwidth]{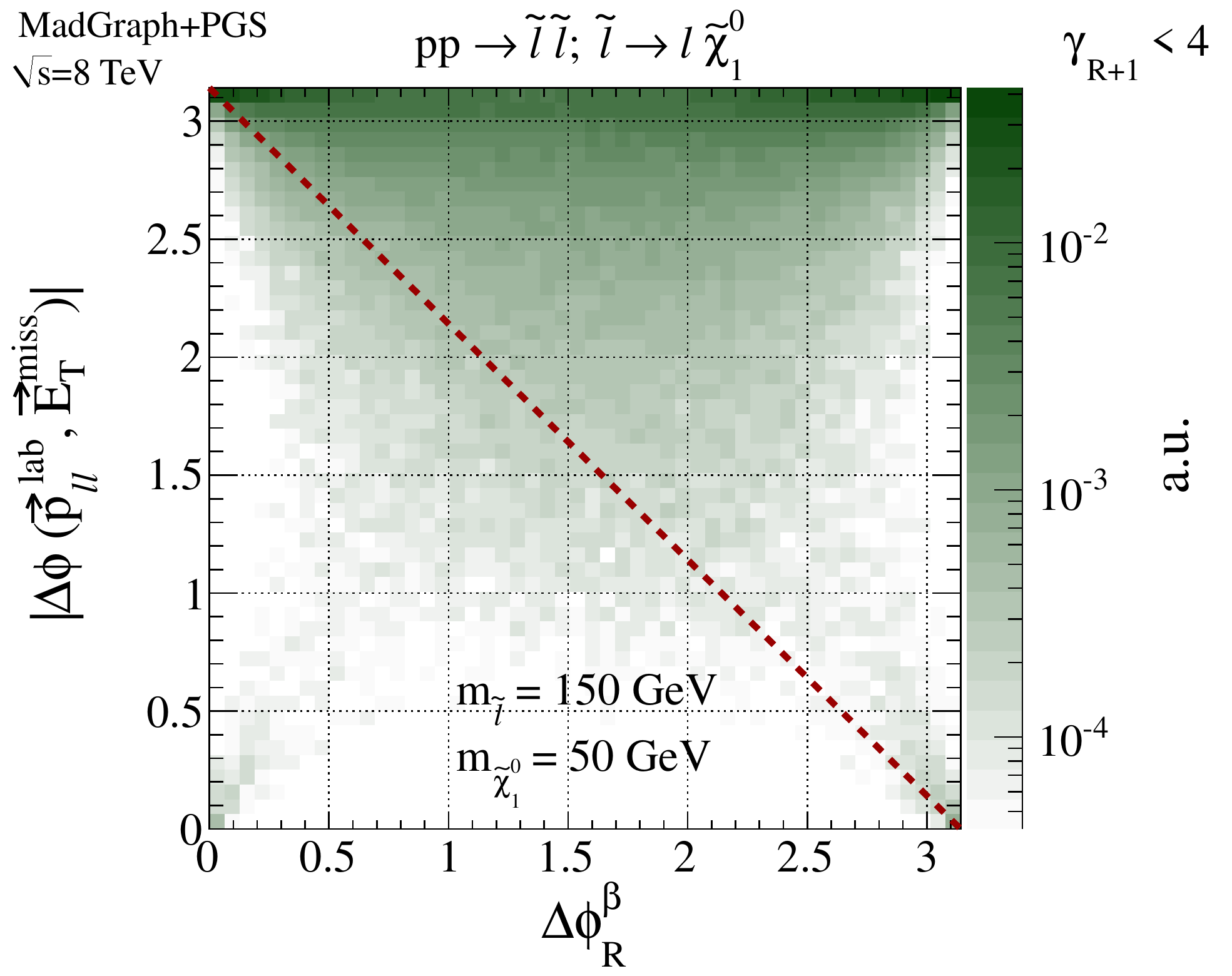}
\includegraphics[width=0.3\columnwidth]{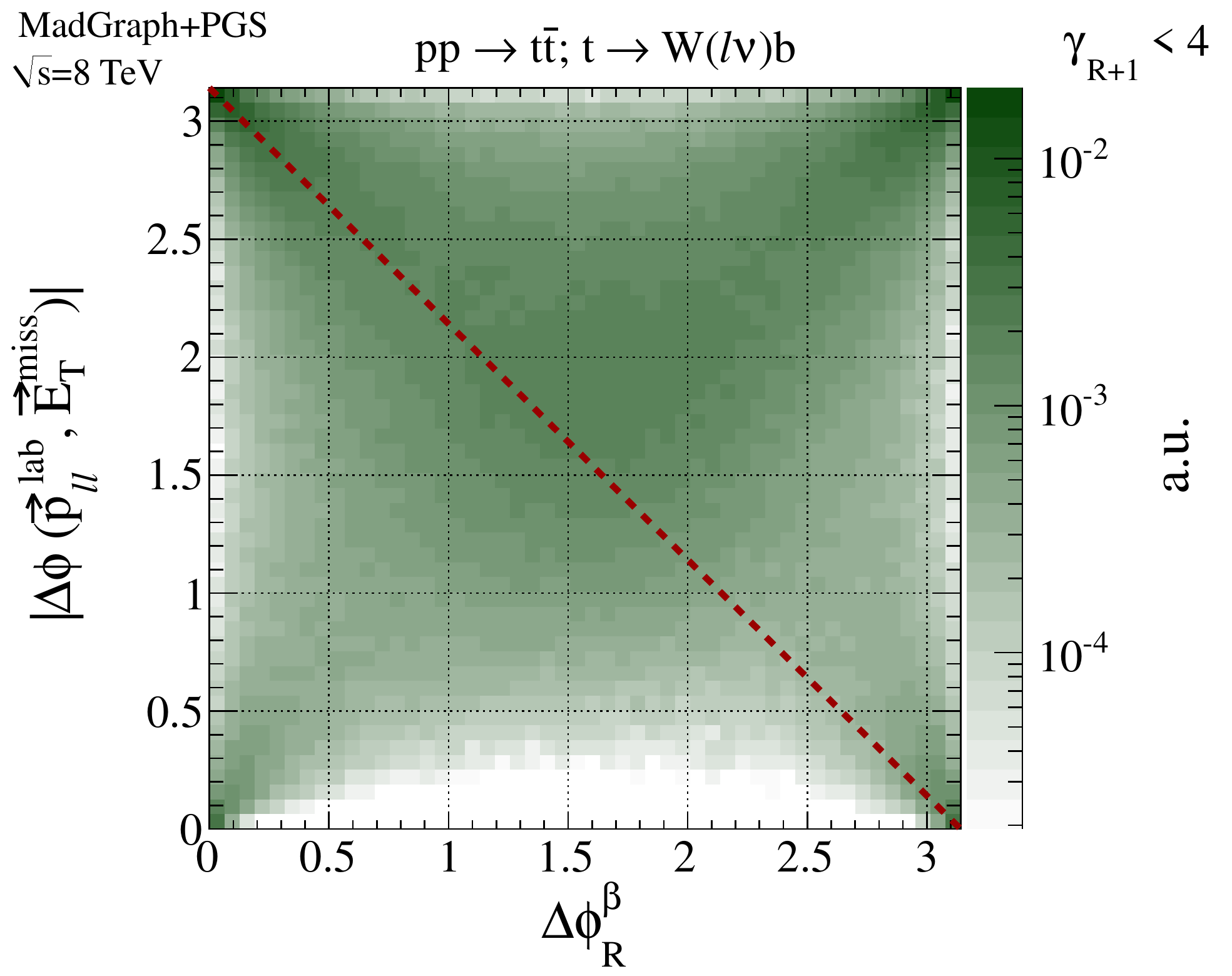}
\includegraphics[width=0.3\columnwidth]{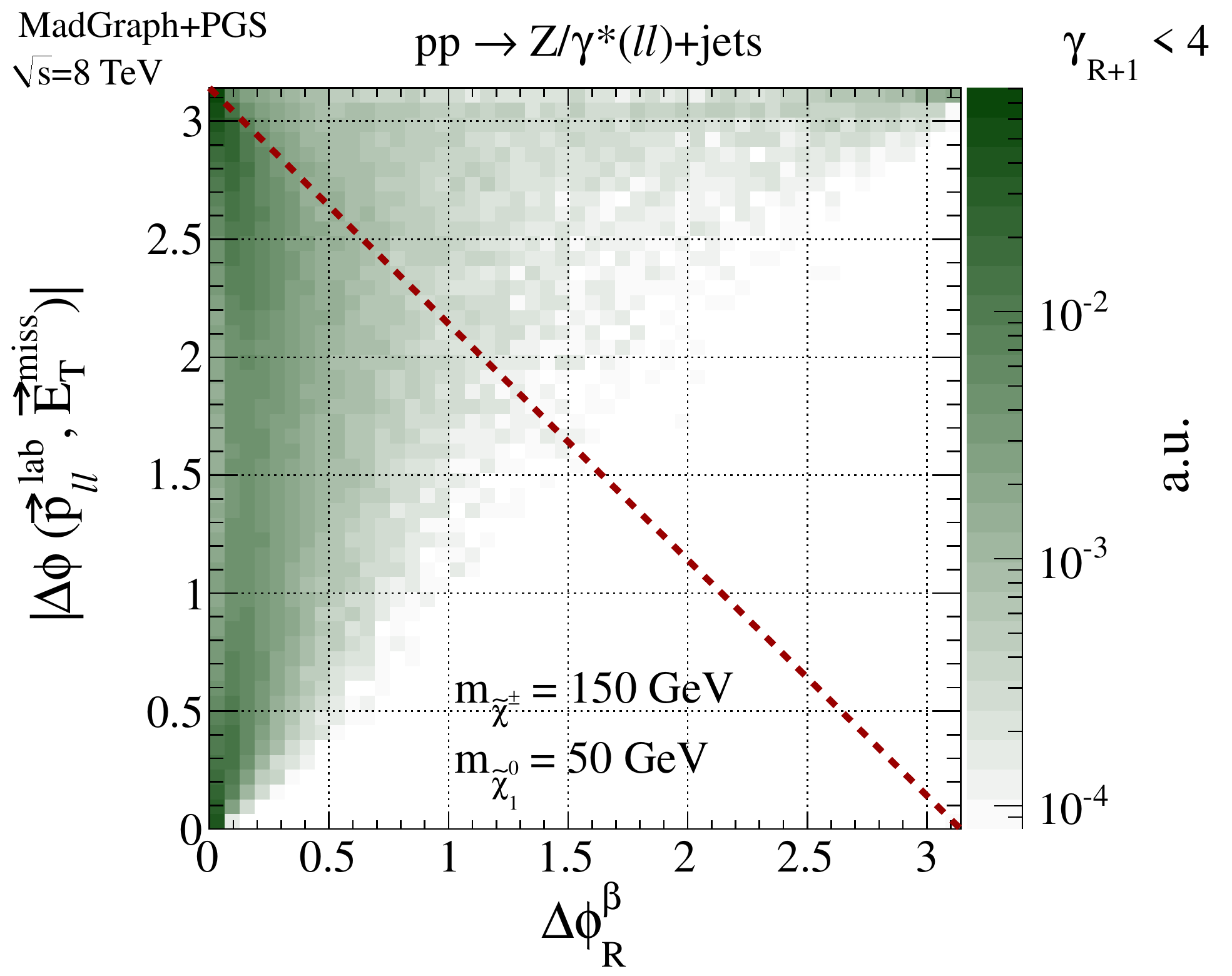}
\caption{Distributions of $|\Delta\phi (\vec{p}_{\ell\ell}^{\, \text{lab}},\vec{E}_{T}^\text{miss})|$ vs. $\Delta \phi_{R}^{\beta}$ for simulated events samples. Top row: inclusive baseline event selection. Bottom row: additional $\gamma_{R+1} > 4$ requirement. Samples correspond to di-slepton production (left), di-leptonic $t\bar{t}$ (center), and $Z/\gamma^{*}$+jets. \label{fig:dphi_v_dphi}}
\end{figure}

Another angle that is useful to consider in diagnosing mis-measured $Z/\gamma^{*}$+jets events is the azimuthal angle between the $E_{T}^\text{miss}$ and di-lepton system in the lab frame, $|\Delta\phi (\vec{p}_{\ell\ell}^{\, \text{lab}},\vec{E}_{T}^\text{miss})|$, in particular for its correlation with $\Delta \phi_{R}^{\beta}$. The distribution of $|\Delta\phi (\vec{p}_{\ell\ell}^{\, \text{lab}},\vec{E}_{T}^\text{miss})|$ is shown as a function of $\Delta \phi_{R}^{\beta}$ in Figure~\ref{fig:dphi_v_dphi}. For signal events the distribution of $|\Delta\phi (\vec{p}_{\ell\ell}^{\, \text{lab}},\vec{E}_{T}^\text{miss})|$ is concentrated at $\pi$; the weakly interacting and di-lepton systems are back-to-back in the CM frame and will remain so in the lab frame without a large CM system transverse momentum. The distribution is more dispersed for $t\bar{t}$ events, as the $W$ bosons are not only recoiling against each other in the CM frame, but also against two $b$-quarks. For $Z/\gamma^{*}$+jets the direction of the $\vec{E}_{T}^\text{miss}$ is largely uncorrelated with the di-lepton system. The strength of the information contained in this two-dimensional plane can be seen when considering only events with $\gamma_{R+1} < 4$, {\it i.e.}~those events which tend towards larger $M_{\Delta}^{R}$ values and are therefore of more significance in an analysis. We observe in the bottom part of Figure~\ref{fig:dphi_v_dphi} that while the di-slepton and $t\bar{t}$ samples retain a similar shape after the $\gamma_{R+1} < 4$ requirement, the remaining $Z/\gamma^{*}$+jets exhibit a very particular correlation between $|\Delta\phi (\vec{p}_{\ell\ell}^{\, \text{lab}},\vec{E}_{T}^\text{miss})|$ and $\Delta\phi_{R}^{\beta}$. The most difficult $Z/\gamma^{*}$+jets events, while still having a relatively flat $|\Delta\phi (\vec{p}_{\ell\ell}^{\, \text{lab}},\vec{E}_{T}^\text{miss})|$ distribution tend to gather at low $\Delta \phi_{R}^{\beta}$. The correlation is such that a cut of $\Delta\phi_{R}^{\beta} + |\Delta\phi (\vec{p}_{\ell\ell}^{\, \text{lab}},\vec{E}_{T}^\text{miss})| > \pi$ removes the majority of $Z/\gamma^{*}$+jets while keeping almost all the significant di-slepton events. This cut is indicated by the dotted red line in the bottom part of Figure~\ref{fig:dphi_v_dphi}, events being rejected if they fall below it.

As discussed in Section~\ref{sec:variables}, the variable $|\cos\theta_{R+1}|$ is also useful for rejecting Drell-Yan background, independently of the $\Delta\phi_{R}^{\beta} + |\Delta\phi (\vec{p}_{\ell\ell}^{\, \text{lab}},\vec{E}_{T}^\text{miss})| > \pi$ requirement. Rather than include requirements on $|\cos\theta_{R+1}|$ in the selection we will use the full distribution as a descriminating variable in an analysis described in the following section. We define the Razor event selection as: \\ \\
{\bf Razor selection:}\\
\begin{tabular}{c c}
~~~~~~~~~~~~~~~~~~~~~~~~~~~~~~~~SF Channels ($ee$,$\mu\mu$)~~~~~~~~~~~~~~~~~~~~~~~~~~~~~~~~ & OF channels ($e\mu$) \\
& \\
$\gamma_{R+1} < 10$ &  \\
$|m(\ell\ell) - m_{Z}| > 10$~GeV  &  None\\
$\Delta\phi_{R}^{\beta} + |\Delta\phi (\vec{p}_{\ell\ell}^{\, \text{lab}},\vec{E}_{T}^\text{miss})| > \pi$ & 
\end{tabular} \\

\begin{figure}[ht]
\includegraphics[width=0.3\columnwidth]{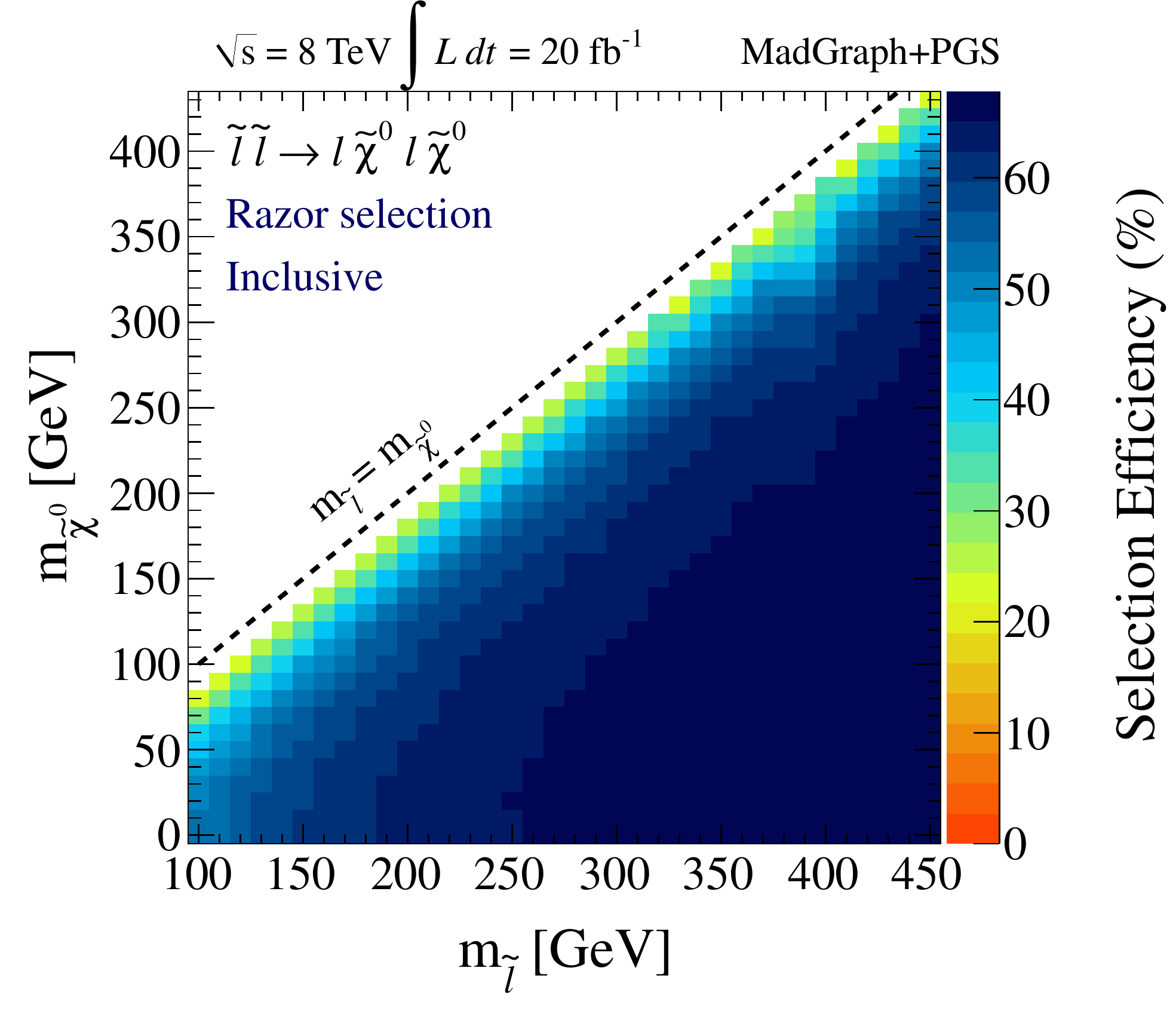}
\includegraphics[width=0.3\columnwidth]{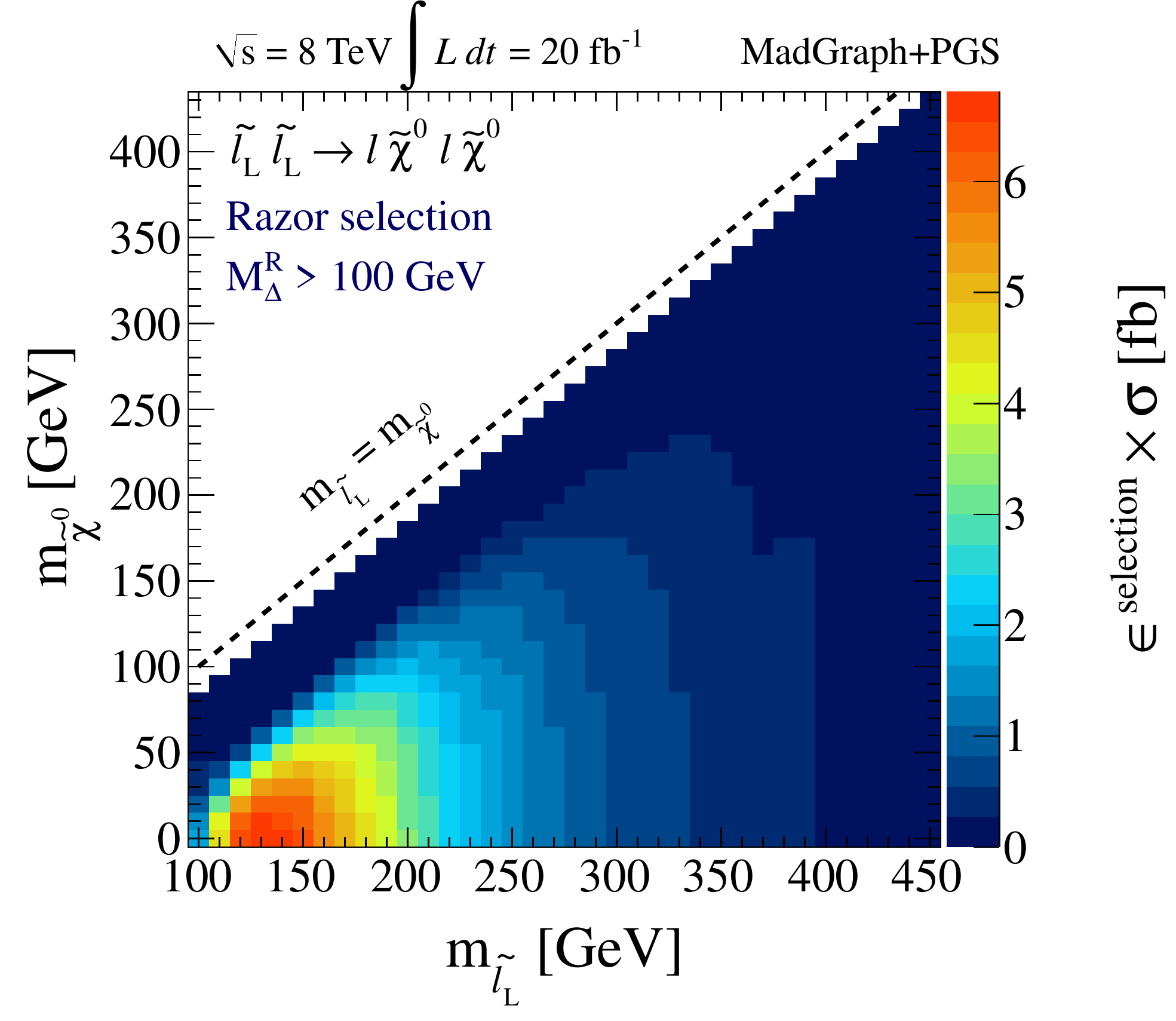}\\
\includegraphics[width=0.3\columnwidth]{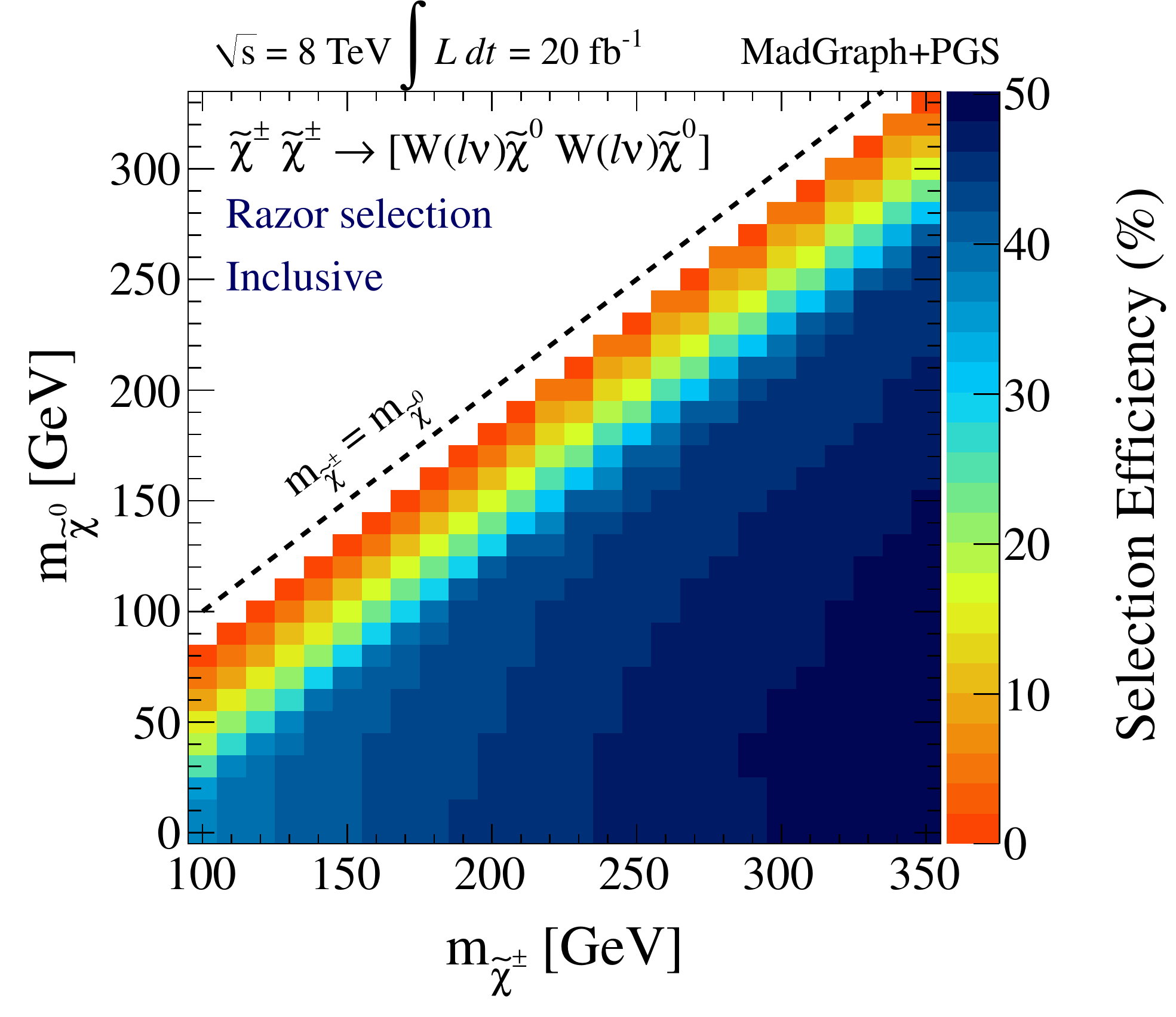}
\includegraphics[width=0.3\columnwidth]{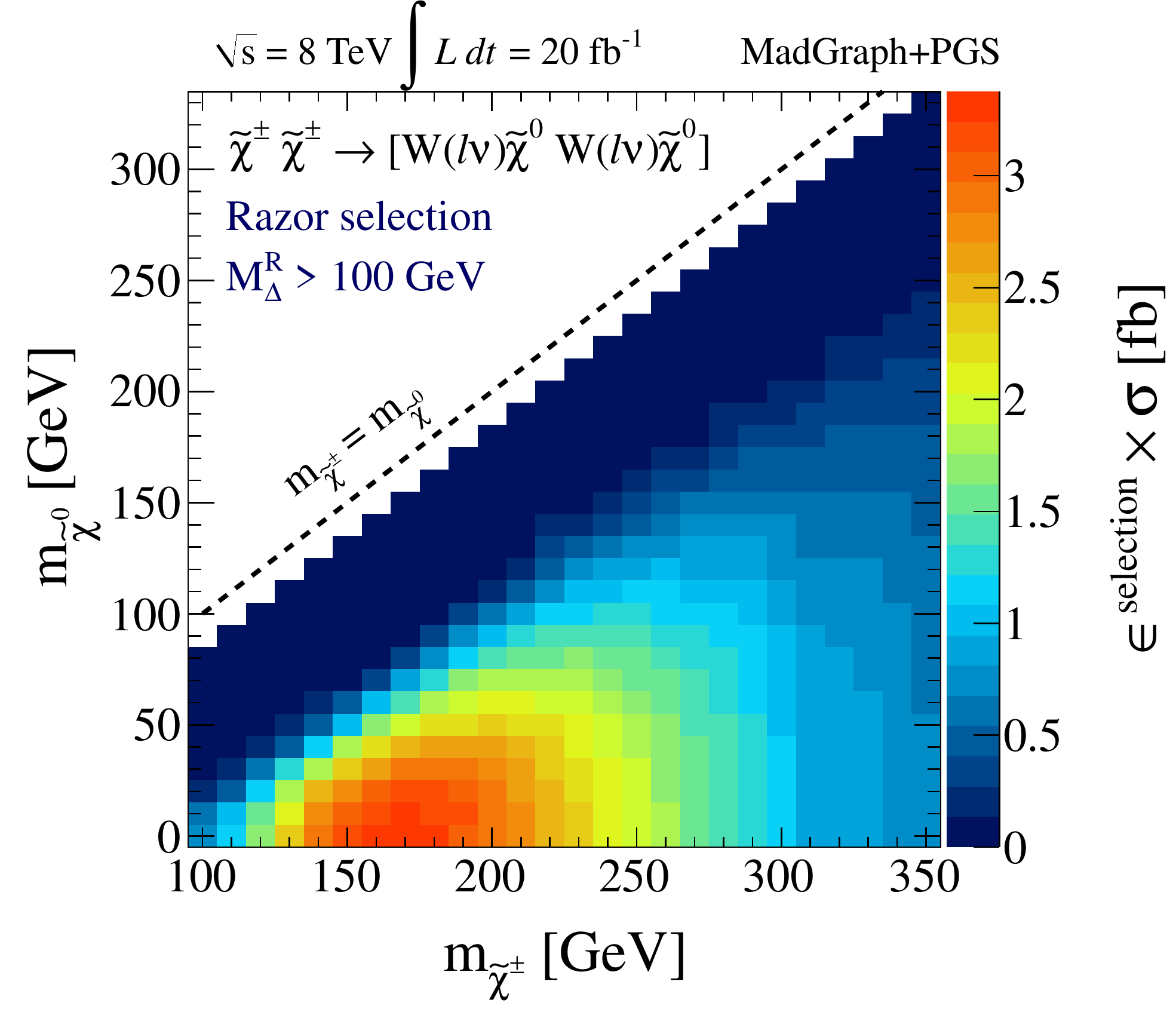}
\caption{Selection efficiencies (left) and efficiency times cross section (right) for left-handed selectrons (upper row) and chargino (lower row) signal samples, as a function of neutralino mass for the Razor selection criteria, described in the text.  \label{fig:EFF_Razor}}
\end{figure}

Used in conjunction with $|\cos\theta_{R+1}|$, this simple event selection sufficiently reduces the $Z/\gamma^{*}$+jets background to a manageable level, without appealing to 
$E_{T}^\text{miss}$ cuts that decrease selection efficiency for signals with lower $M_{\Delta}$. There is likely room for optimization in these cuts, but this combination is sufficient for demonstrating that gains in sensitivity are possible for more compressed spectra, as we will see in the following sections. The efficiencies and expected cross-sections of event yields with the Razor selection applied for the slepton and chargino signal models considered are summarized in Figure~\ref{fig:EFF_Razor}. Analogous values for simulated background processes are provided in Table~\ref{tab:BKG}. We observe that the efficiency for selecting low $M_{\Delta}$ events is improved over the CMS and ATLAS selections, while the number of $Z/\gamma^{*}$+jets at high $M_{\Delta}^{R}$ are reduced.

\begin{table}[ht]
\begin{tabular}{|c|c|c|c|c|c|c|c|} \hline
 \multicolumn{2}{|c|}{$\sigma \times \epsilon$~[fb]}   & \multicolumn{2}{|c|}{CMS selection} & \multicolumn{2}{|c|}{ATLAS selection} & \multicolumn{2}{|c|}{Razor selection} \\ \hline
 \multicolumn{2}{|c|}{}  & \multicolumn{2}{|c|}{Inclusive ($M_{CT\perp} > 100$~GeV)} & \multicolumn{2}{|c|}{Inclusive ($M_{T2} > 100$~GeV)} &\multicolumn{2}{|c|}{Inclusive ($M_{\Delta}^{R} > 100$~GeV)} \\ \hline
Process & Jet mult. & $~~~~~ ee$+$\mu\mu ~~~~~~$ & $e\mu$  & $~~~~~ ee$+$\mu\mu ~~~~~~$ & $e\mu$ & $~~~~~ ee$+$\mu\mu ~~~~~~$ & $e\mu$ \\ \hline
                     & 0 jets            &  230 (0.067) & 280 (0.068) & 430 (0.15) & 500 (0.15) & 910 (0.55) & 1300 (0.54) \\ 
Di-Bosons  & 1 jet              &  120 (0.084) & 150 (0.085) & 120 (0.13) & 142 (0.13) & 170 (0.54) & 380 (0.86) \\ 
                     & $\ge 2$ jets &  55 (0.044) & 70 (0.061) & 37 (0.056) & 44 (0.059) & 47 (0.36) & 130 (0.78) \\ \hline         
                     & 0 jets            & 38 (0.12) & 46 (0.052) & 31 (0.14) & 34 (0.15) & 65 (0.29) & 95 (0.29) \\ 
$t\bar{t}$     & 1 jet              & 140 (0.20) & 180 (0.28) & 110 (0.29) & 120 (0.34) & 180 (0.72) & 340 (0.81) \\ 
                     & $\ge 2$ jets & 290 (0.46) & 360 (0.57) & 170 (0.50) & 200 (0.58) & 310 (1.4) & 650 (1.8) \\ \hline  
                     & 0 jets            &  70 (0.92) & 4.8 (0.037) & 160 (1.5) & 3.3 (0.046) & 1800 (1.6) & 730 ($< 0.001$)  \\ 
$Z/\gamma^*(\ell\ell)$  & 1 jet & 85 (1.3) & 37 (0.010) & 70 (1.5) & 5.2 (0.078) & 110 (0.94) & 110 ($< 0.001$)  \\ 
                     & $\ge 2$ jets & 44 (0.55) & 27 (0.001) & 13 (0.50) & 2.4 ($< 0.001$) & 46 (0.24) & 37 ($< 0.001$) \\ \hline           

\end{tabular}
\caption{Effective cross sections for di-lepton backgrounds at the LHC with $\sqrt{s} = 8$~TeV, after selection requirements (efficiency [$\epsilon$] times cross section [$\sigma$]). Cross sections are listed for each of the event selections (CMS, ATLAS and Razor) as a function of jet multiplicity and lepton flavor, with and without selection requirements on mass sensitive variables. \label{tab:BKG} }
\end{table}

\section{Shape Analysis and Statistical Approach \label{sec:shape}}

In order to test the utility of the super-razor kinematic variables in the context of a search for slepton and chargino production we consider toy experimental analyses. Each of these analyses is a shape analysis, using multiple bins over the range of a kinematic variable of interest, and exploiting differences in changing signal and background expectations over the bins. This approach is used to increase the information being gleaned from these kinematic variables, allowing us to quantify what the maximal performance could look like, irrespective of changing optimized cuts associated with coarser binning. The predictions of these toy shape analyses are potentially optimistic relative to CMS and ATLAS results, due to both increased complexity of the analyses and the shortcomings of the detector simulation utilized here. To account for these differences, large systematic uncertainties are included in the procedure to represent potential experimental uncertainty in the relevant parameters that dictate the shape and yield of signal and background events.

\subsection{Analysis strategy}

For each toy analysis there are one or more kinematic variables identified as the discriminating variable, and the binned distribution of event yields in this variable are the observables in the toy experiment. The expected shape of both signal and background in the variable(s) of interest are required input for this procedure for each process. In our case, these shapes come from simulated event samples of each process. For an actual experimental analysis some can be measured or constrained from control regions. Regardless of their provenance, the uncertainties corresponding to these shapes are as important as the central values as we try to reflect in these toy analyses.

For the CMS and ATLAS analyses, control regions are identified using both object ID and kinematic information in order to isolate particular backgrounds. $Z$ mass windows are used to select $(Z/\gamma^*\to \ell\ell)$+jets backgrounds for normalizing $Z$ mass veto signal regions. Similarly, high jet multiplicity or $b$-tagged jet-enriched selections are used to constrain backgrounds with top quarks. In order to qualitatively capture these control region background constraints we consider multiple lepton flavor ($ee$, $e\mu$, $\mu\mu$) and jet multiplicity ($0$, $1$, $\ge 2$) categories simultaneously in a fit to data, with binned kinematic discriminants for each category. In each fit, high jet multiplicity categories effectively constrain top contributions while di-boson and $Z/\gamma^*$ events at low jet multiplicity are disentangled using relative lepton flavor category yields.

\begin{figure}[ht]
\includegraphics[width=0.3\columnwidth]{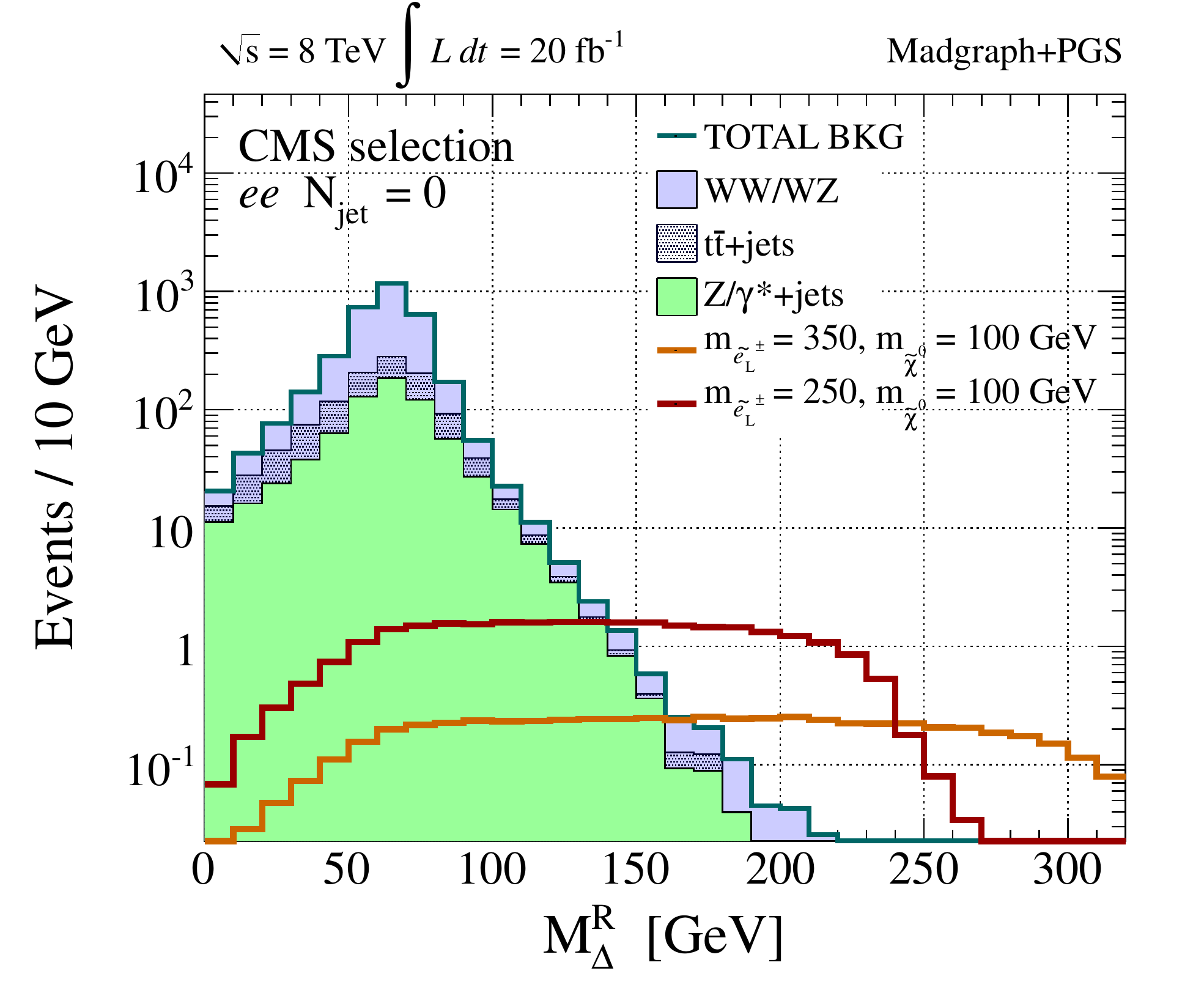}
\includegraphics[width=0.3\columnwidth]{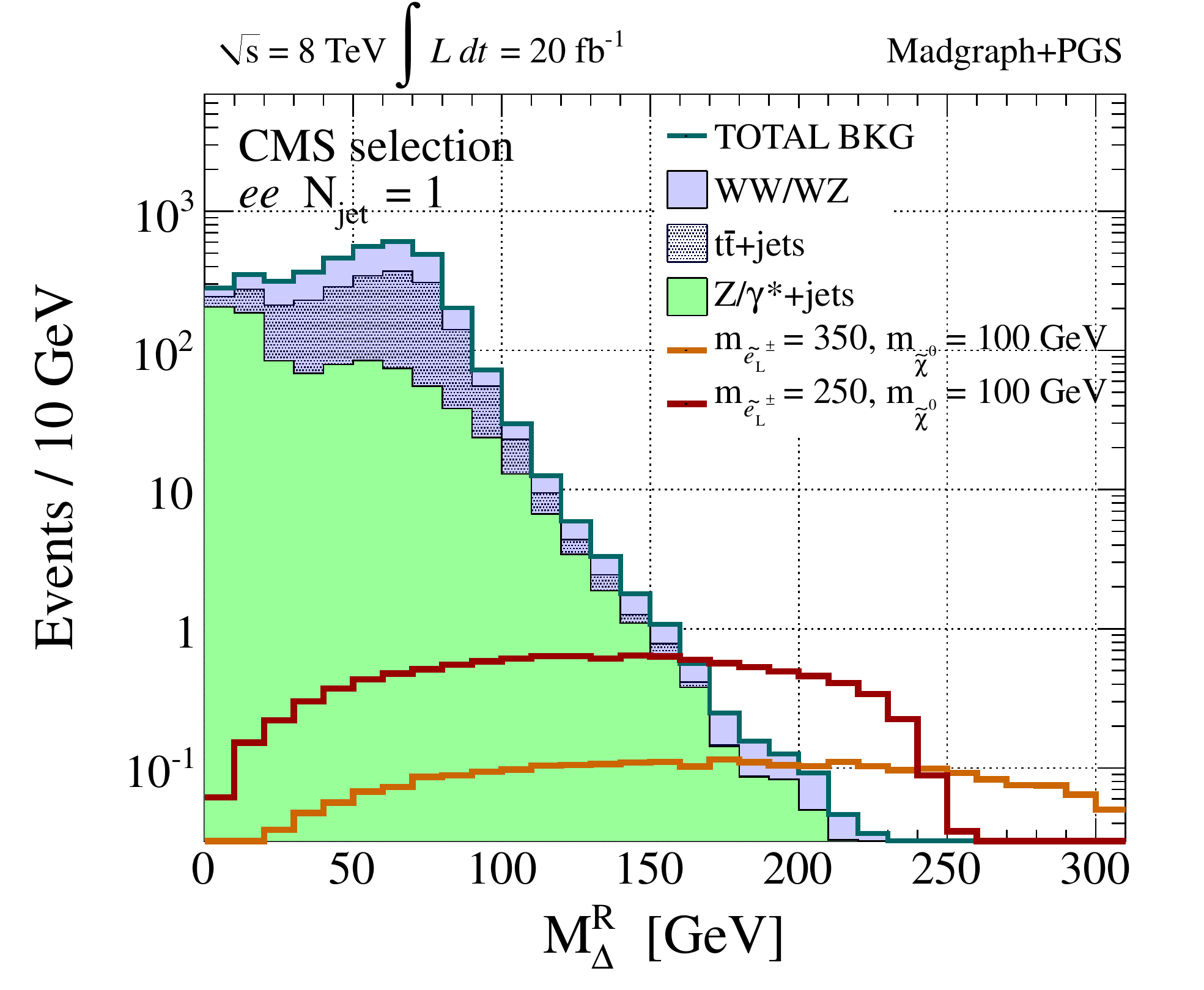}
\includegraphics[width=0.3\columnwidth]{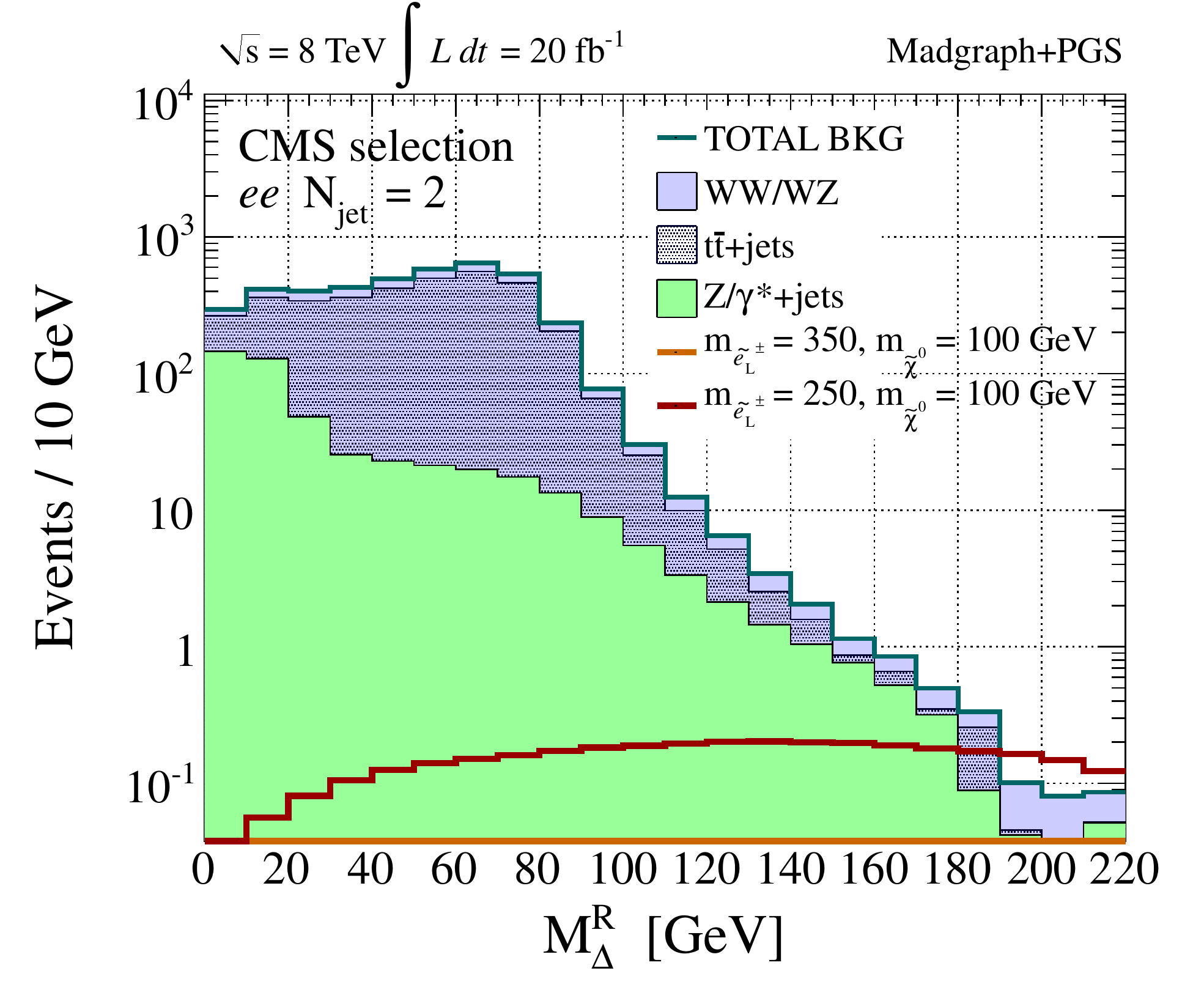}
\includegraphics[width=0.3\columnwidth]{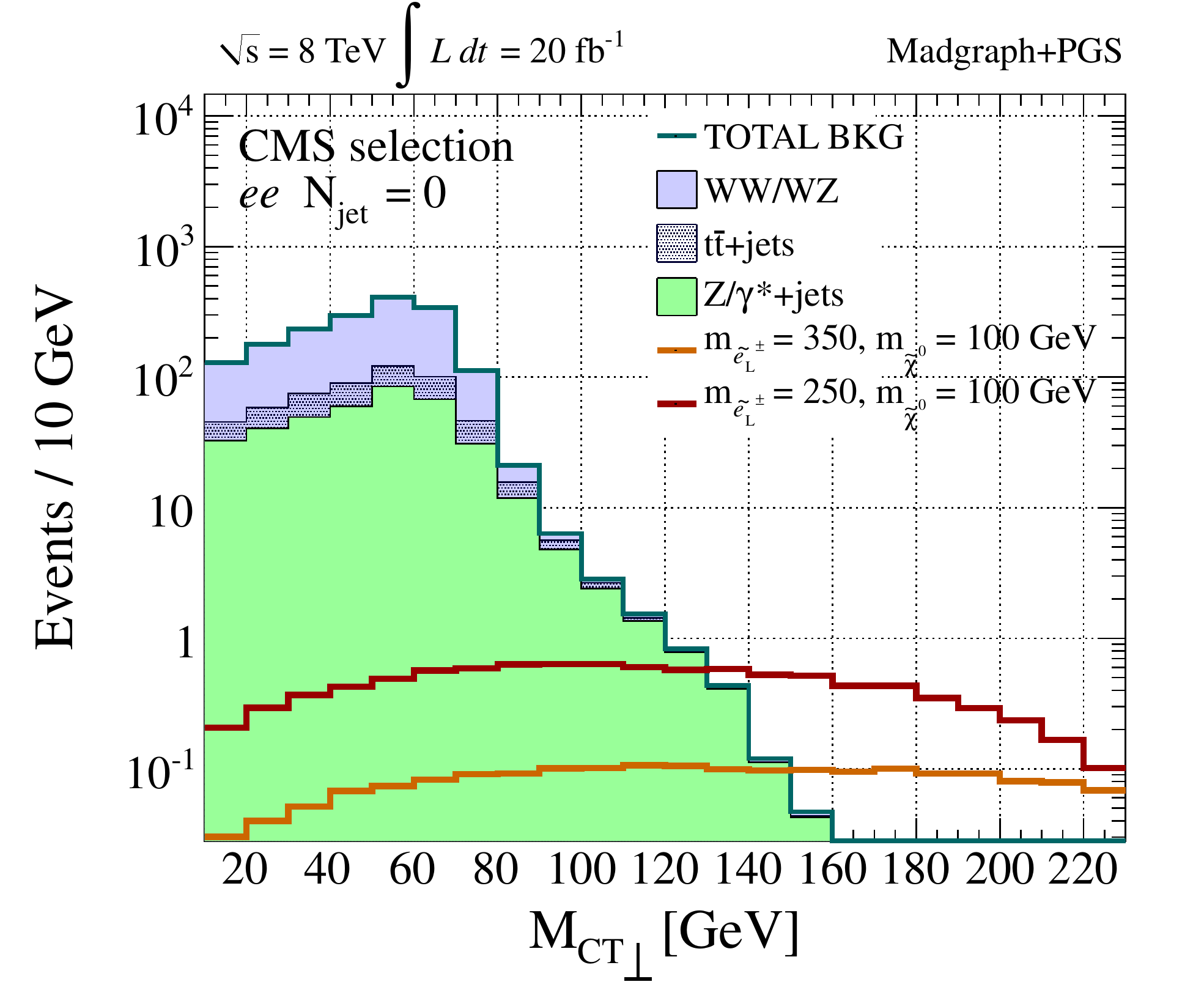}
\includegraphics[width=0.3\columnwidth]{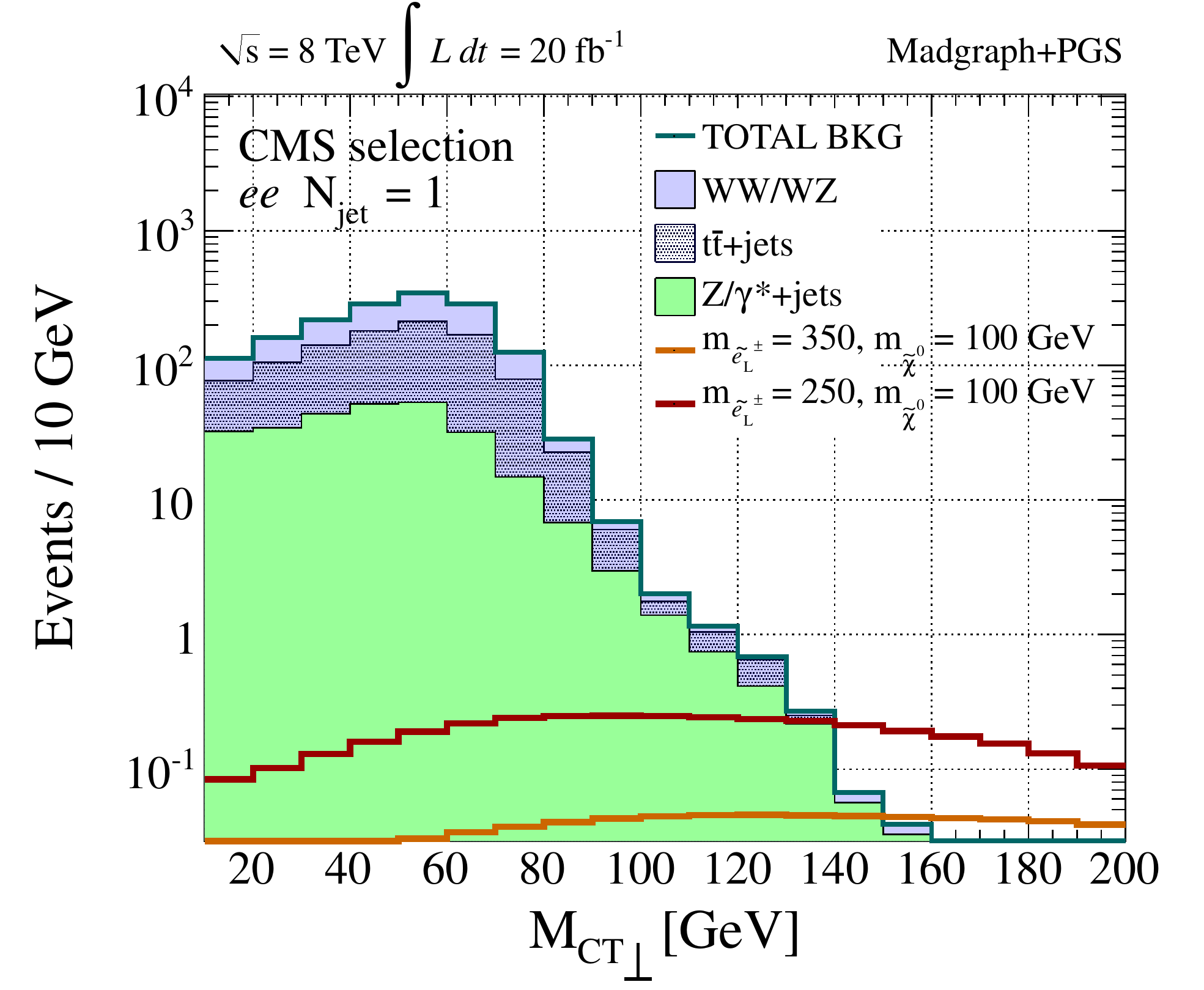}
\includegraphics[width=0.3\columnwidth]{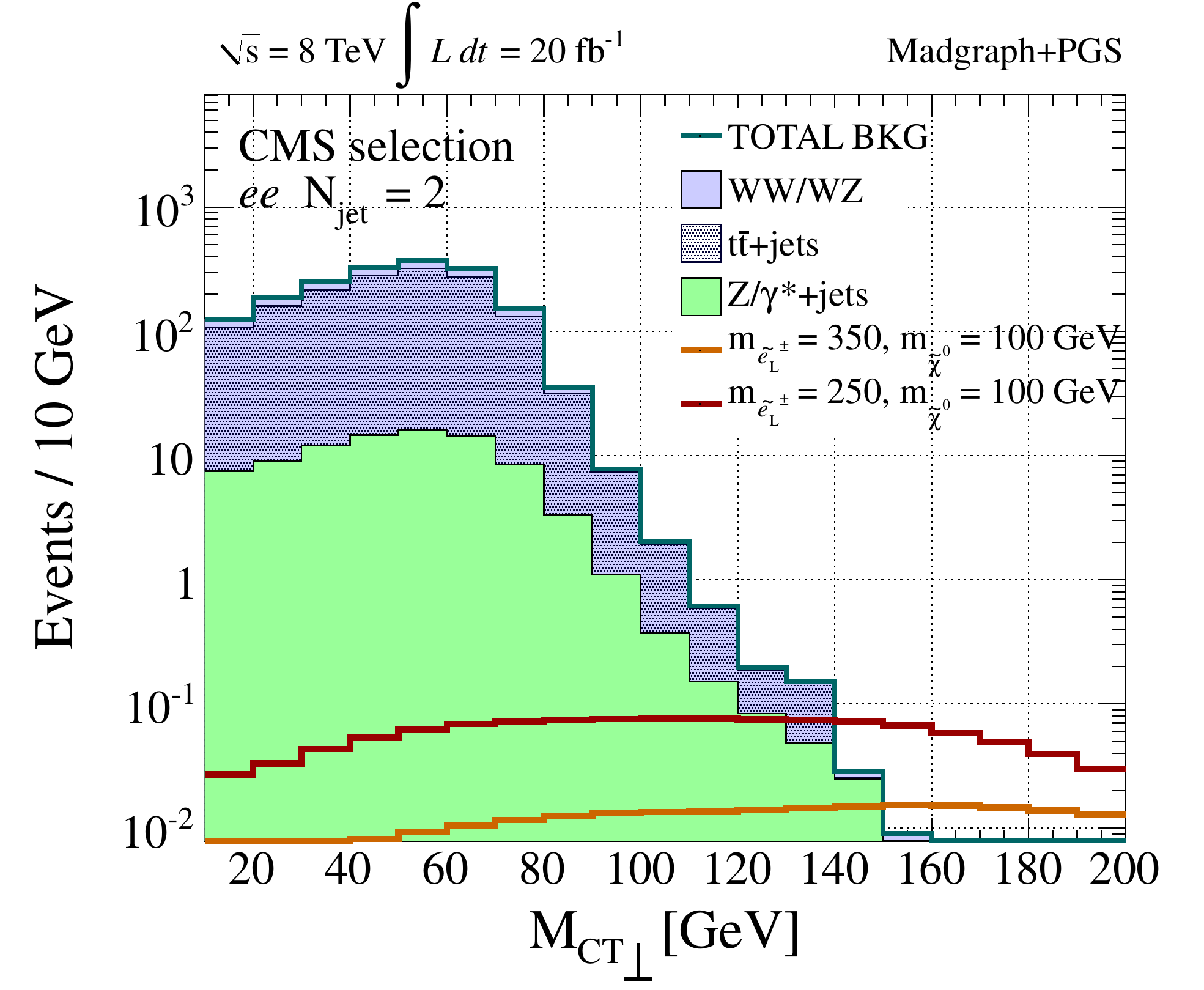}
\caption{Expected background yields in the $ee$ final state passing the CMS selection, normalized to 20 fb$^{-1}$ of data, for different jet multiplicities. Sample left-handed di-selectron signals are included with ($m_{\tilde{\ell}_{L}} = 350$,~$m_{\tilde{\chi}_{1}^0} = 100$) and ($m_{\tilde{\ell}_{L}} = 250$,~$m_{\tilde{\chi}_{1}^0} = 100$)~GeV. Top: $M_{\Delta}^{R}$ distribution. Bottom: $M_{CT\perp}$ distribution. Left: $N_{jet} = 0$. Center: $N_{jet} = 1$. Right: $N_{jet} \ge 2$. \label{fig:EEPDF}}
\end{figure}

\begin{figure}[ht]
\includegraphics[width=0.35\columnwidth]{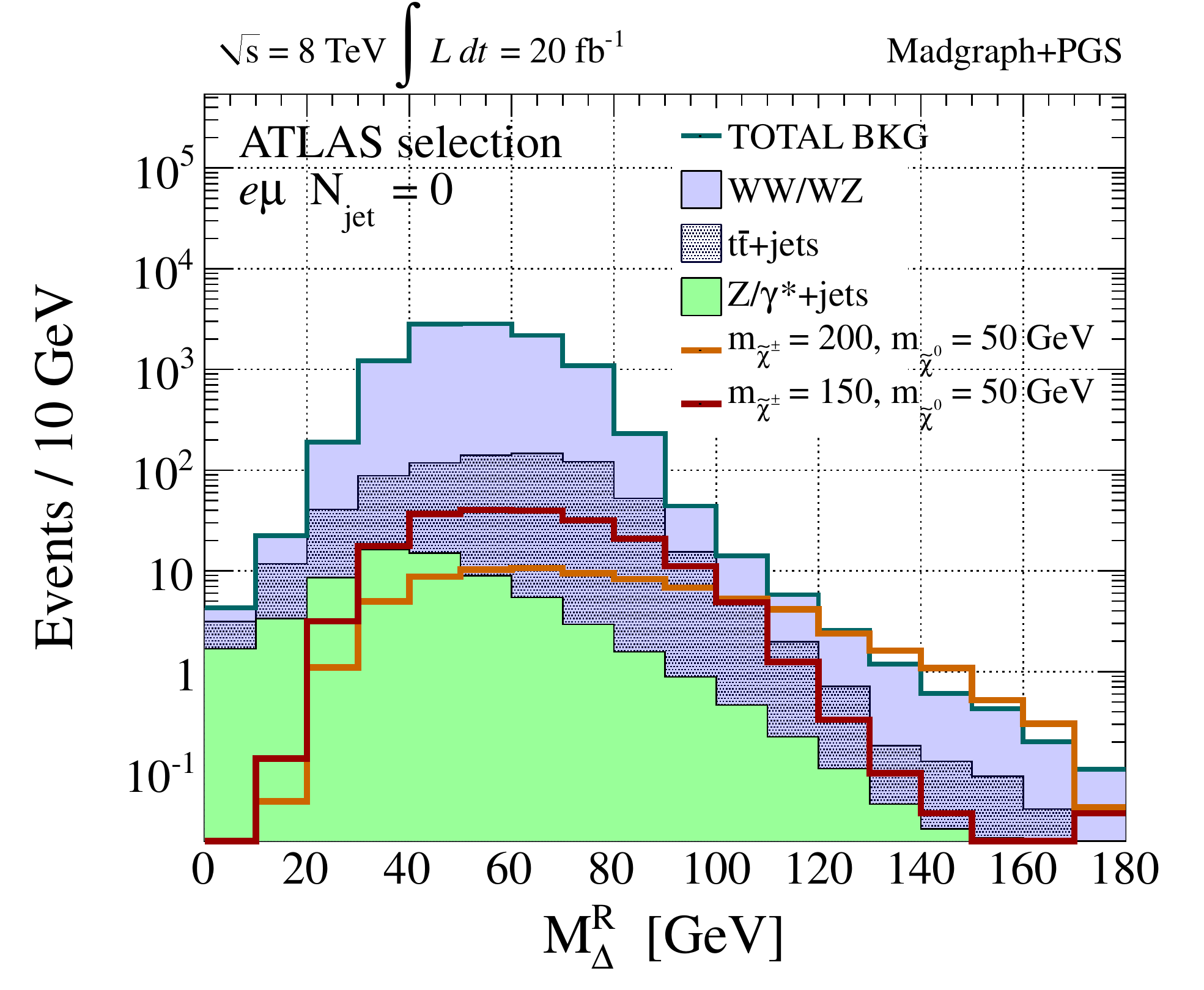}
\includegraphics[width=0.35\columnwidth]{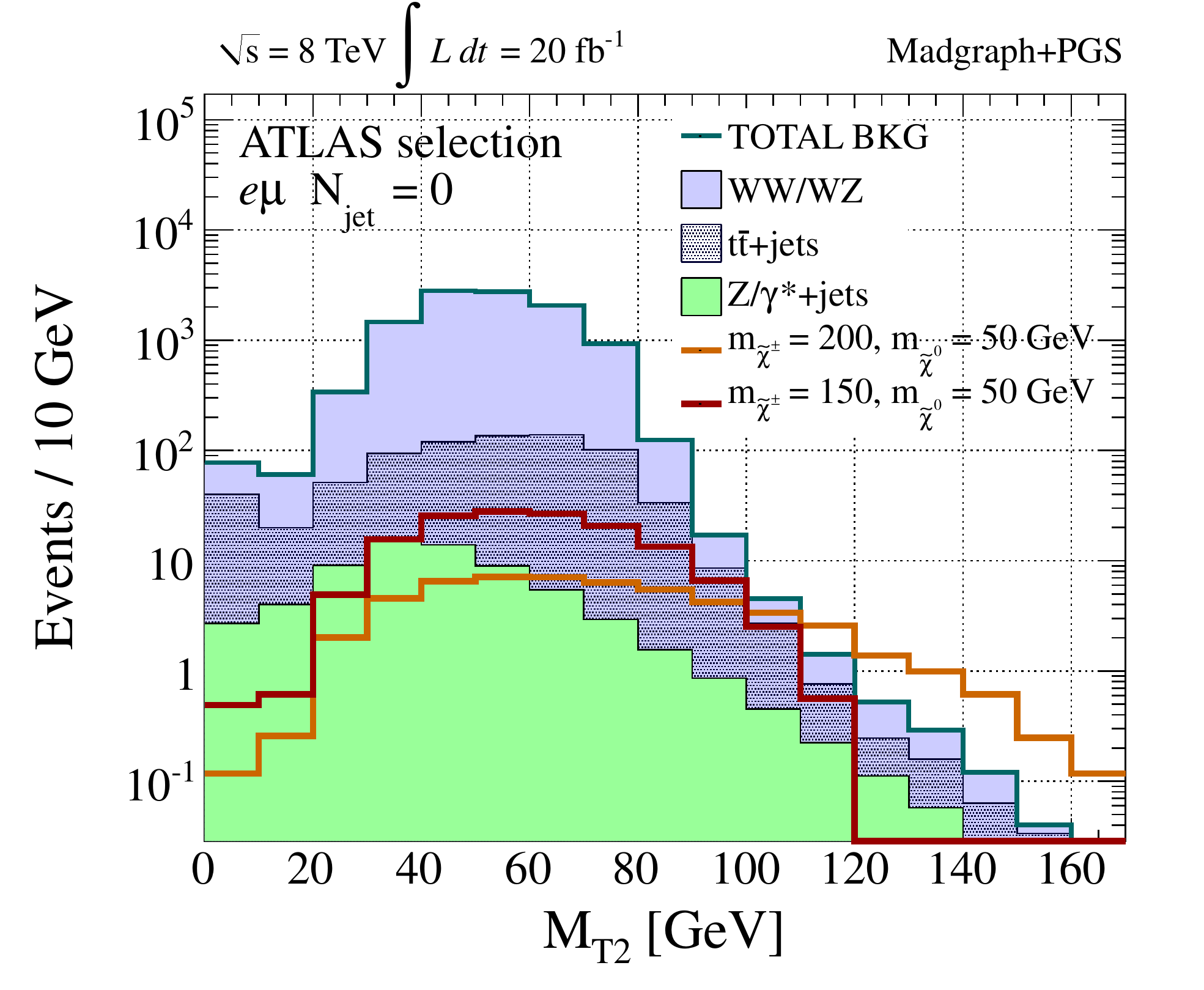}
\includegraphics[width=0.35\columnwidth]{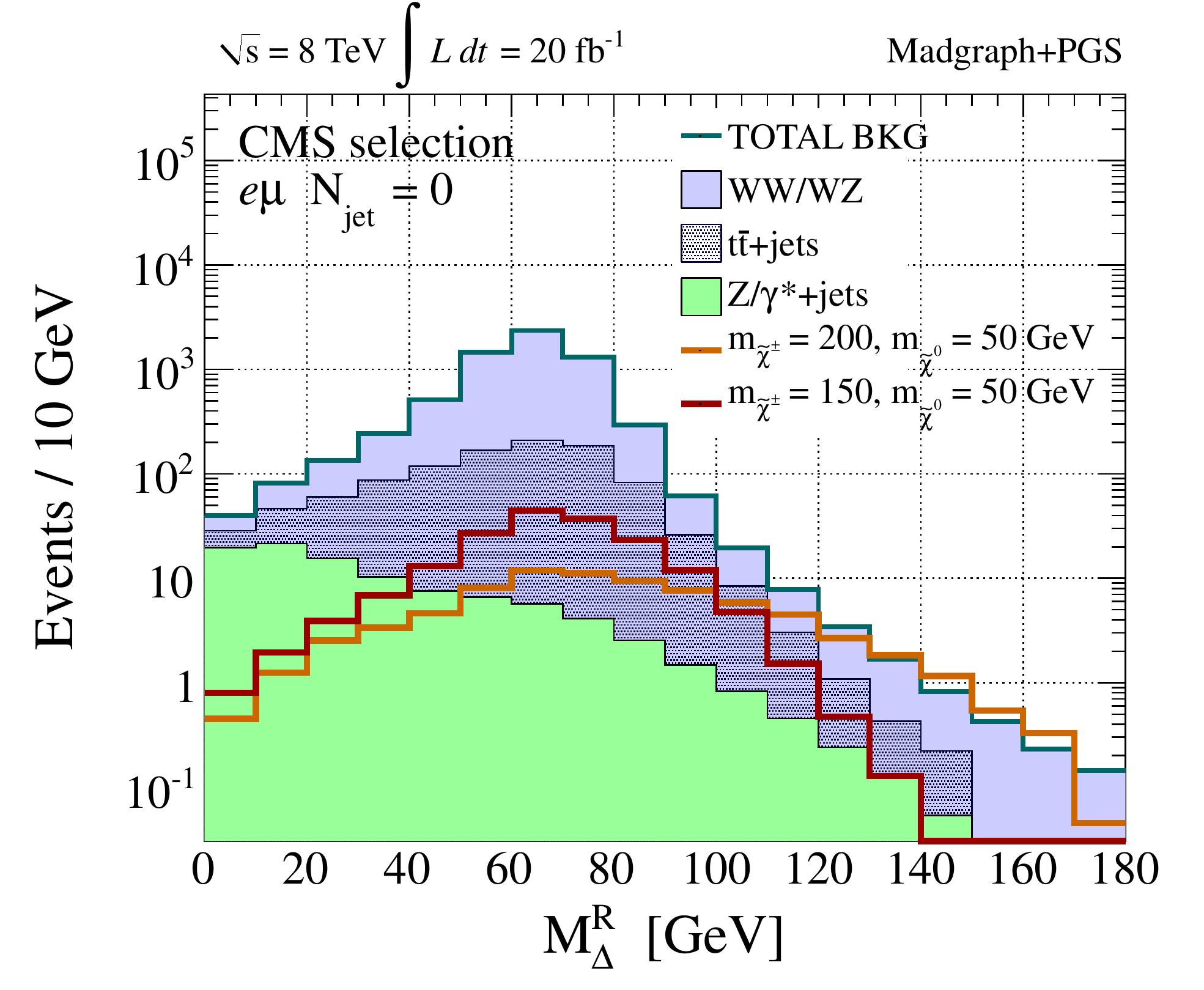}
\includegraphics[width=0.35\columnwidth]{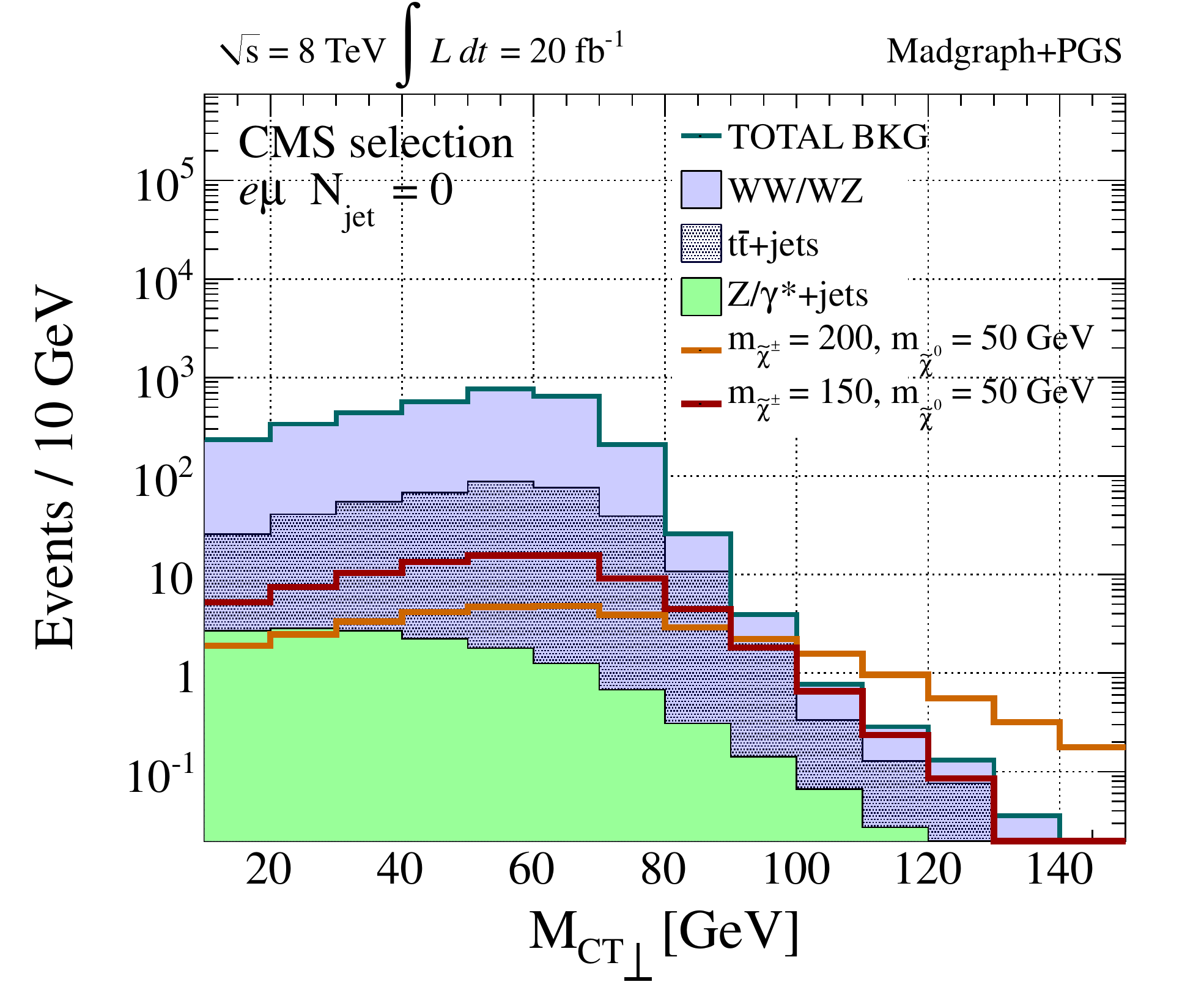}
\caption{Expected background yields in the $e\mu$, $N_{jet} = 0$ final state passing the CMS or ATLAS selections, normalized to 20 fb$^{-1}$ of data. Top left: $M_{\Delta}^{R}$ with the ATLAS selection applied. Sample di-chargino signals are included with ($m_{\tilde{\chi}_{1}^{\pm}}= 200$,~$m_{\tilde{\chi}_{1}^0} = 50$) and ($m_{\tilde{\chi}_{1}^{\pm}} = 150$,~$m_{\tilde{\chi}_{1}^0} = 50$)~GeV.Top right: $M_{T2}$ with ATLAS selection. Bottom left: $M_{\Delta}^{R}$ with CMS selection. Bottom right: $M_{CT\perp}$ with CMS selection. \label{fig:EMUPDF}}
\end{figure}

We first consider one-dimensional analyses, where the kinematic discriminant is chosen to be $M_{\Delta}^{R}$, $M_{CT\perp}$ or $M_{T2}$. The distribution of the variable of interest is binned in 10 GeV steps from zero to 500 GeV. Only events satisfying the baseline selection and the relevant CMS (for $M_{CT\perp}$) or ATLAS (for $M_{T2}$) selection are included. The expected $M_{\Delta}^{R}$ and $M_{T2}$ distributions in the $ee$ final state for sample di-slepton signals and backgrounds are shown in Figure~\ref{fig:EEPDF}. We observe the changing background compositions and diminishing expected signal yield with increasing jet multiplicity. Distributions for each of the $M_{\Delta}$-sensitive variables and selections considered are shown for the $e\mu$, $N_{jet}=0$ final state in Figure~\ref{fig:EMUPDF}.

In addition to one dimensional shape analyses using the variables $M_{\Delta}^{R}$, $M_{CT\perp}$, $M_{T2}$ we also consider a three-dimensional analysis based on $M_{\Delta}^{R}$, $|\cos\theta_{R+1}|$, and $\Delta\phi_{R}^{\beta}$. The two angular variables add complementary information to $M_{\Delta}^{R}$; $\Delta\phi^{\beta}$ introduces sensitivity to the ratio of neutralino and parent sparticle masses while $|\cos\theta_{R+1}|$ helps further resolve the scale $M_{\Delta}$ of a particular sample while also adding discrimination against $WW$ and $t\bar{t}$ using spin correlations (or lack thereof). Both of these angular variables are also useful in rejecting remaining Drell-Yan background events. 

\begin{figure}[ht]
\includegraphics[width=0.35\columnwidth]{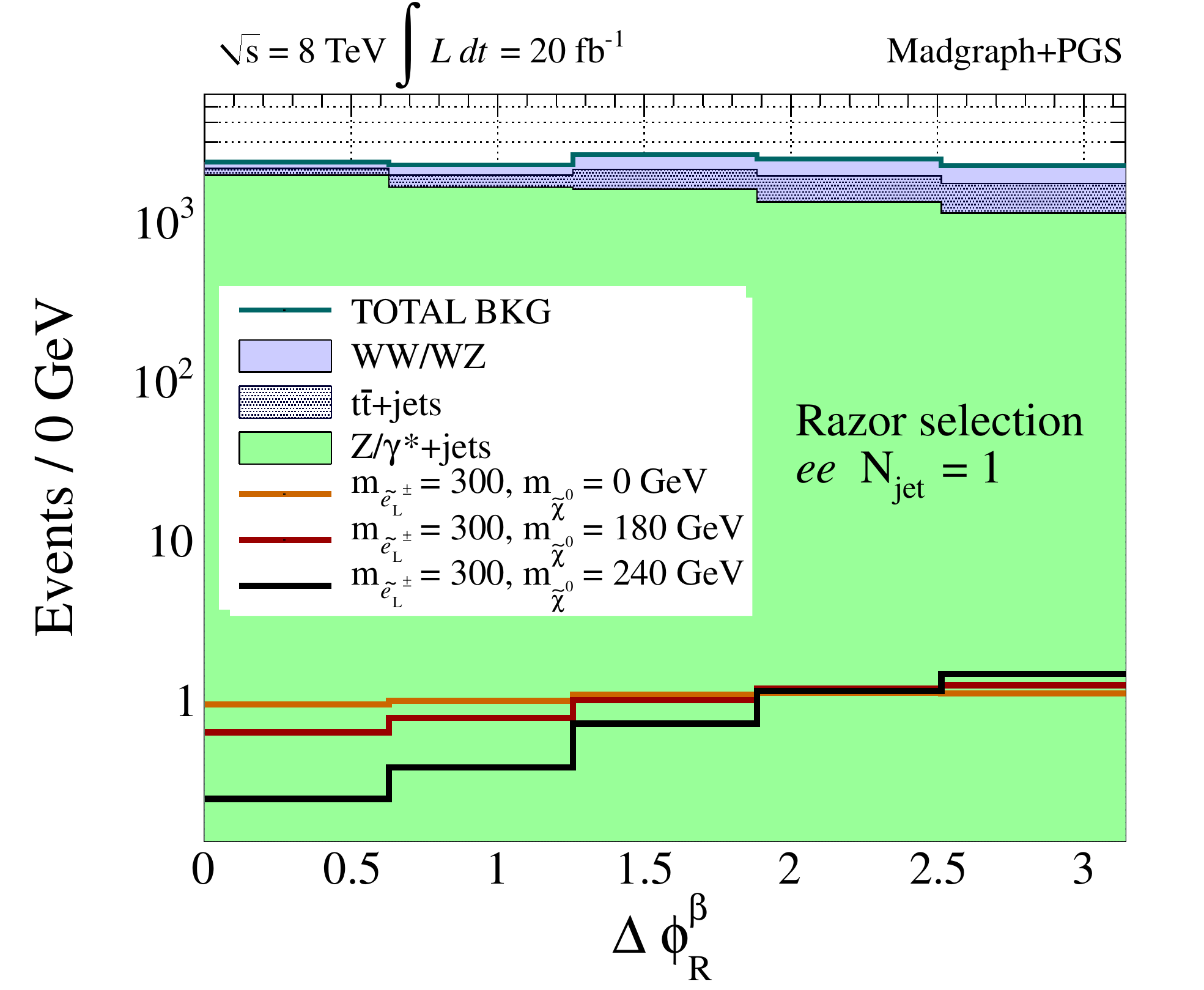}
\includegraphics[width=0.35\columnwidth]{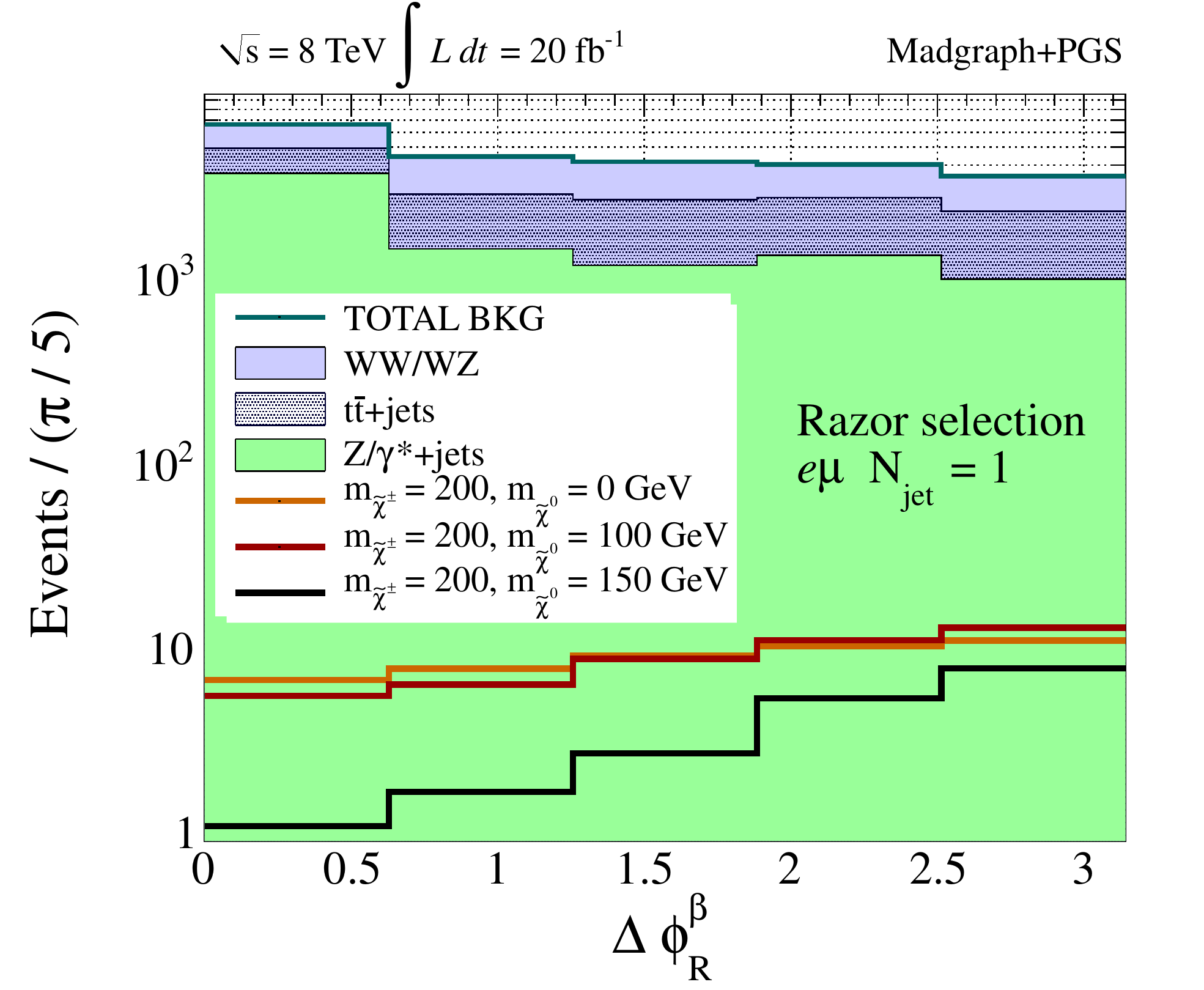}
\caption{Expected background yields in the $N_{jet} = 0$ final state passing the razor selection for $\Delta\phi_{R}^{\beta}$, normalized to 20 fb$^{-1}$ of data. Left: $ee$ final state including sample left-handed di-selectron signals with $m_{\tilde{\ell}} = 300$~GeV and varying neutralino masses. Right: $e\mu$ final state including sample di-chargino signals with $m_{\tilde{\chi}^{\pm}} = 200$~GeV and varying neutralino masses. \label{fig:DPHIPDF}}
\end{figure}

The three dimensional $M_{\Delta}^{R} \times \Delta\phi_{R}^{\beta} \times |\cos\theta_{R+1}|$ analysis uses the razor selection described in the previous section, and represents each of the kinematic discriminants as binned histograms. For both signal and background events we find that the variable $\Delta\phi_{R}^{\beta}$ has only weak correlations with the other two variables. We neglect any residual correlations such that $\Delta\phi_{R}^{\beta}$ distributions are modeled as a one dimensional histogram with five equal-width bins ranging from between zero and $\pi$. Examples of expected event yields for SM backgrounds and representative signal models are shown in Figure~\ref{fig:DPHIPDF}.

\begin{figure}[ht]
\includegraphics[width=0.35\columnwidth]{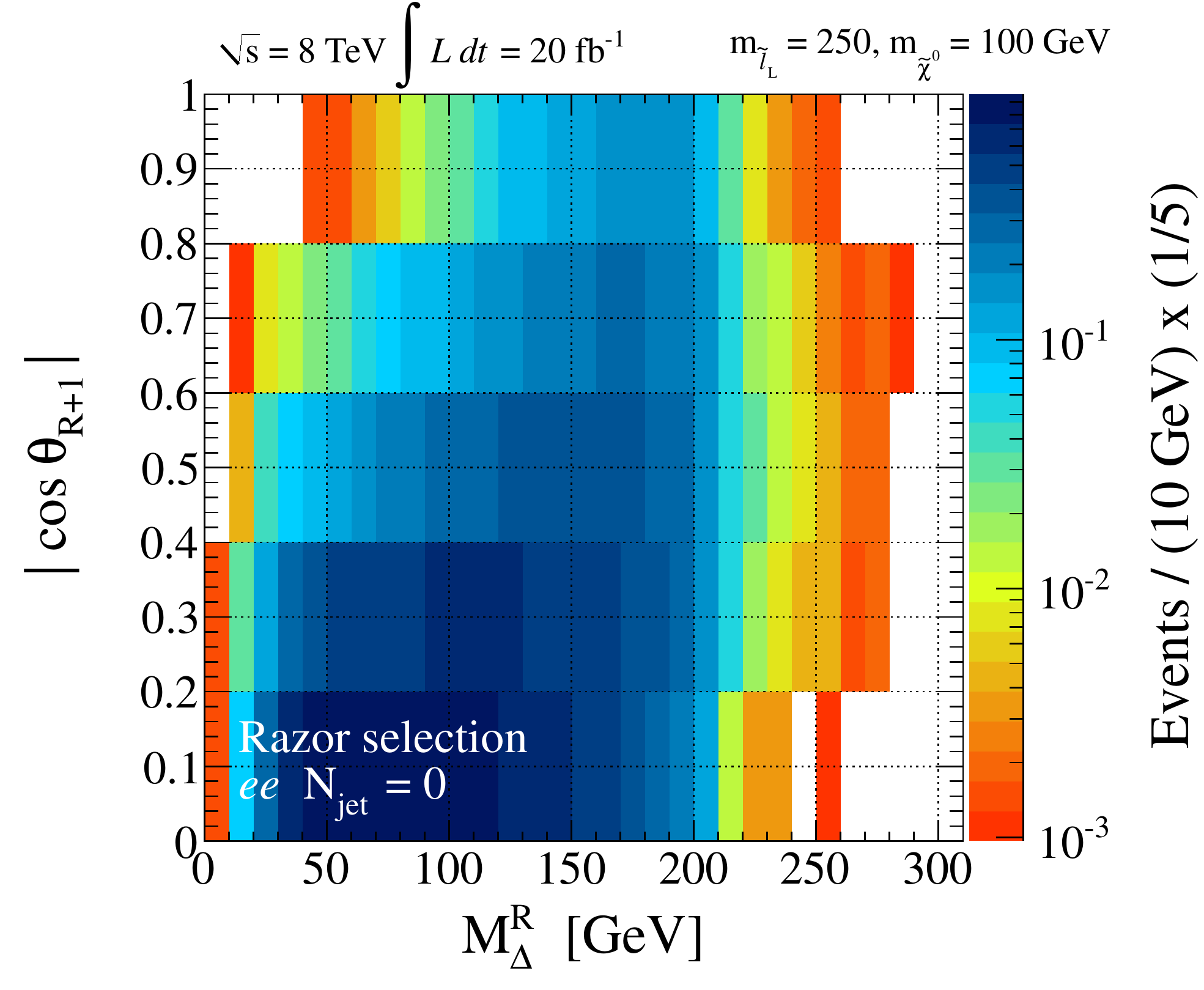}
\includegraphics[width=0.35\columnwidth]{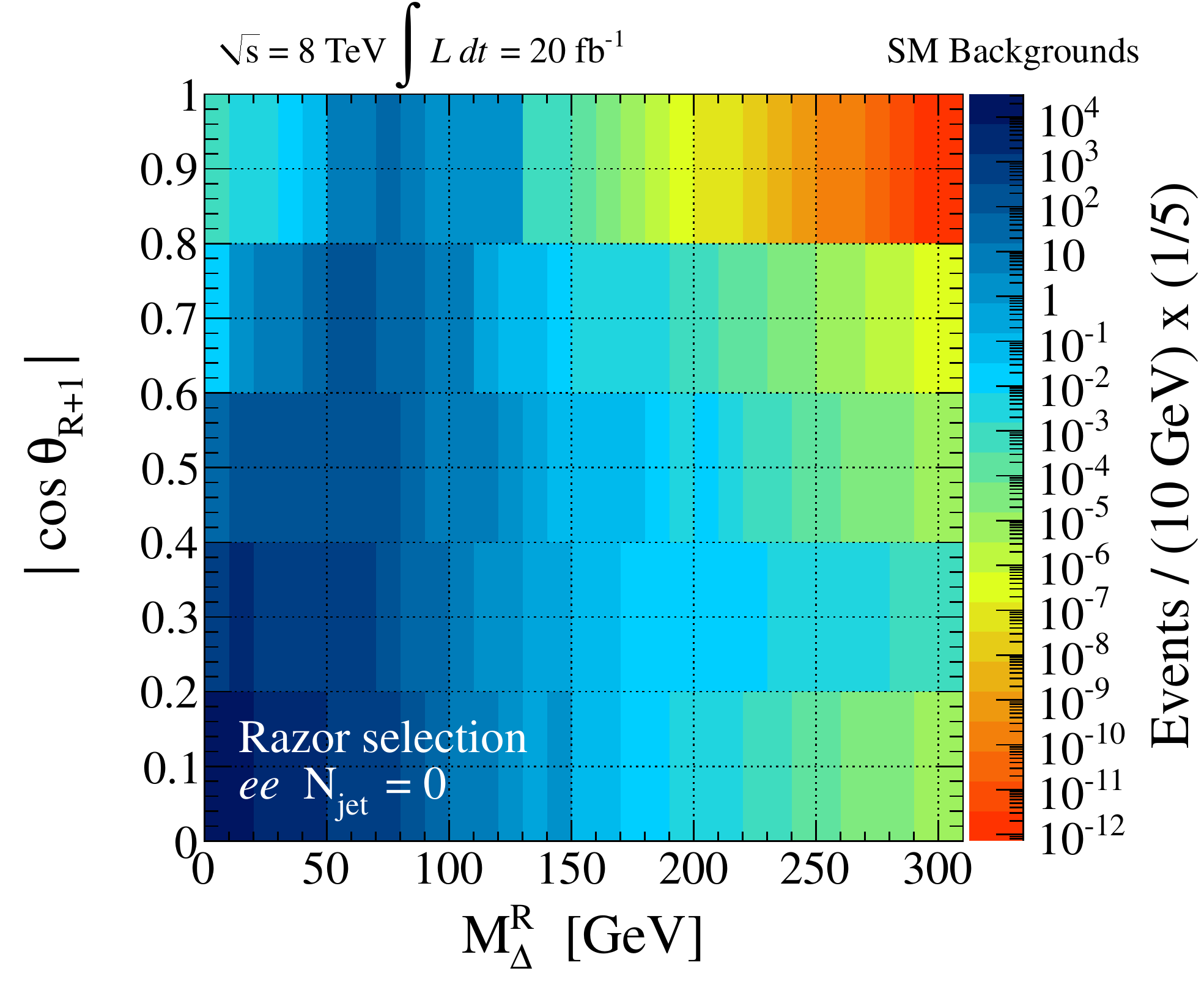}
\includegraphics[width=0.35\columnwidth]{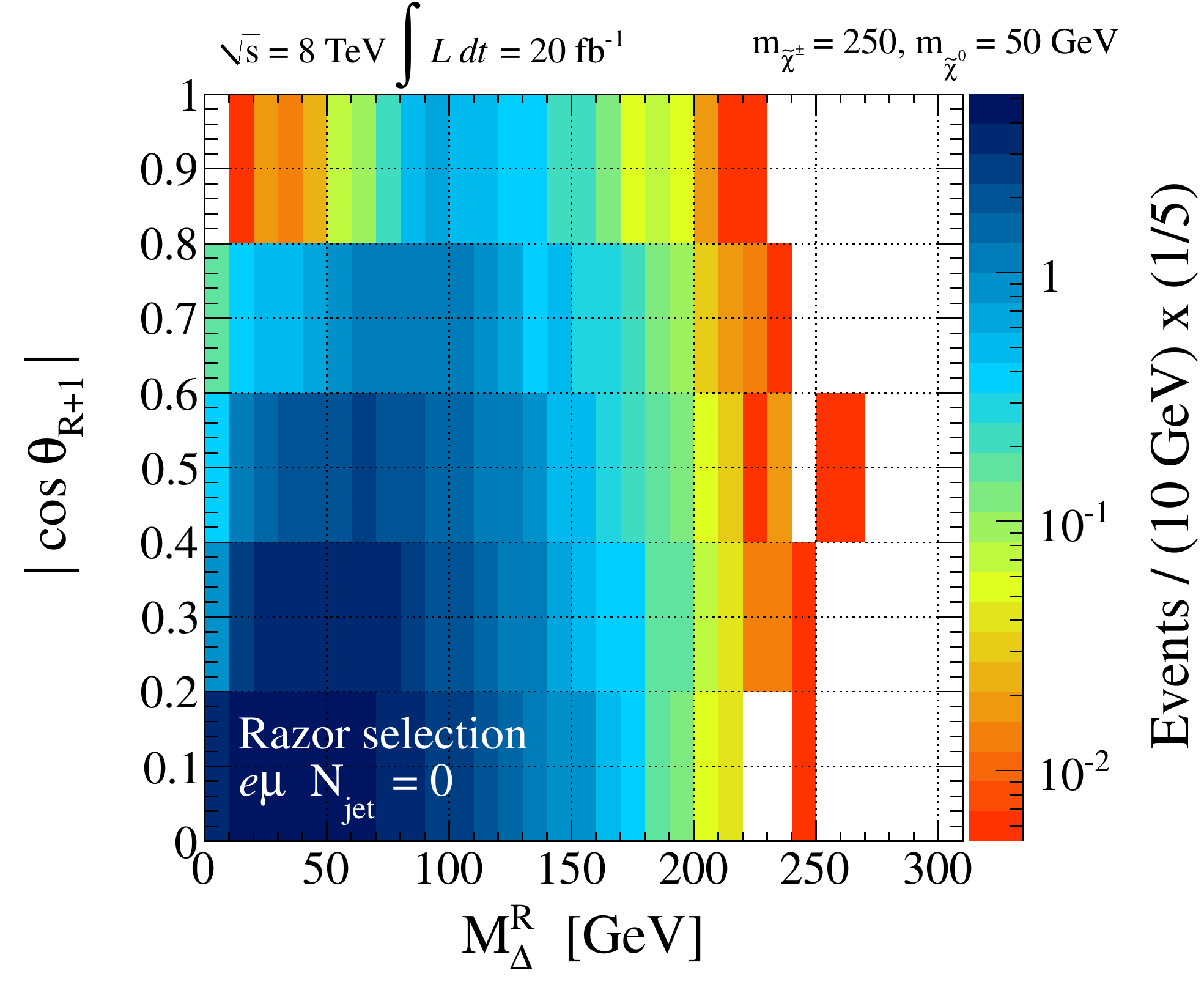}
\includegraphics[width=0.35\columnwidth]{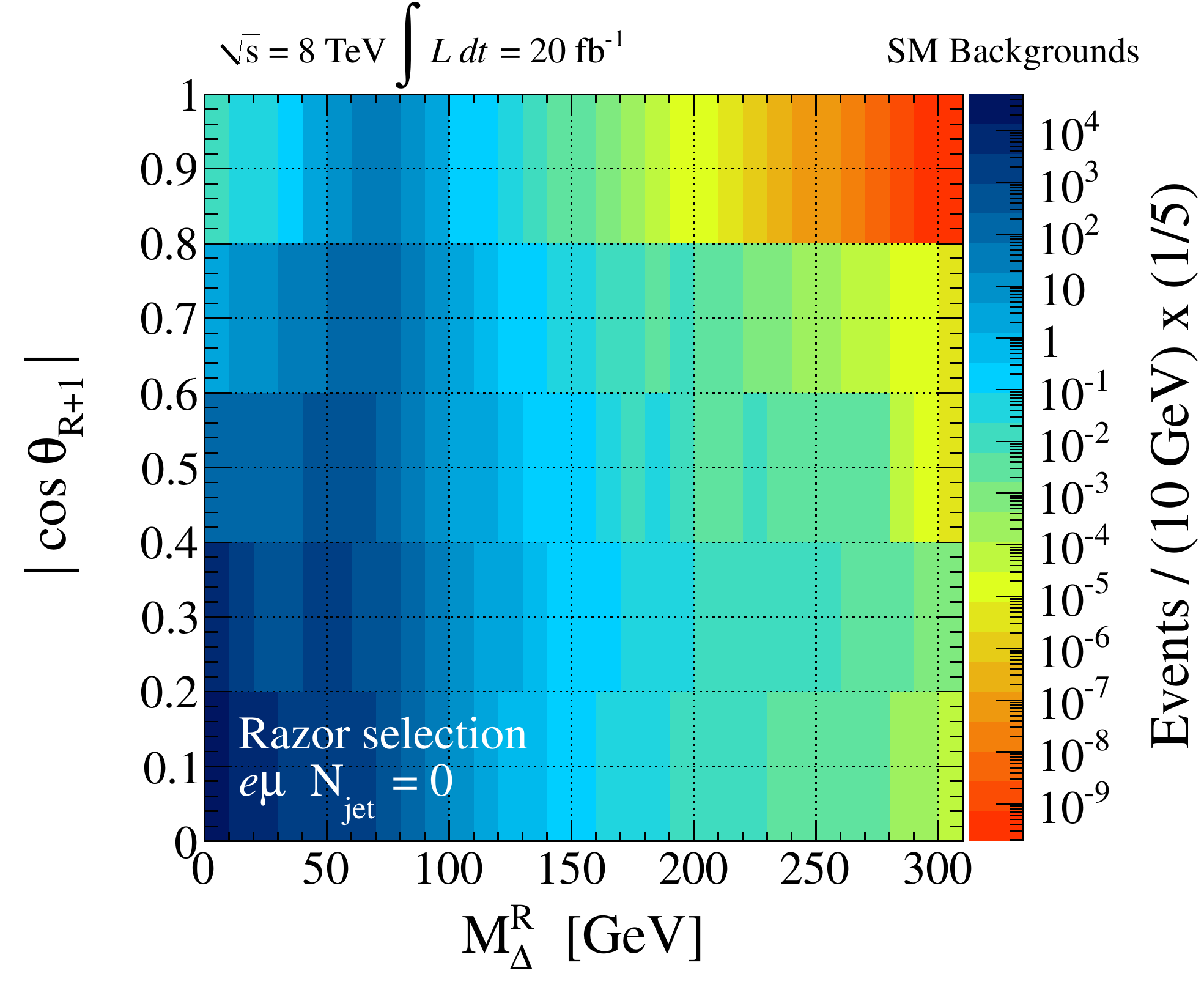}
\caption{Expected background yields in $M_{\Delta}^{R} \times |\cos\theta_{R+1}|$ plane, normalized to 20 fb$^{-1}$ of data. Top: Expected event yields for the $ee$ final state with $N_{jet} = 0$ and the Razor selection applied. Bottom: Analogous figures for the $e\mu$ final state. Left: Kinematic distributions for sample signal models including di-selectron production with ($m_{\tilde{\ell}_L}=250$, $m_{\tilde{\chi}^{0}_{1}}=100$) GeV (top left) and di-chargino production with ($m_{\tilde{\chi}^{\pm}_1}=250$, $m_{\tilde{\chi}^{0}_{1}}=50$)  GeV (bottom left). Right: Total of expected SM background yields. \label{fig:COSTHETAPDF}}
\end{figure}

Strong correlations between $M_{\Delta}^{R}$ and $|\cos\theta_{R+1}|$ mean that these two variables cannot be factorized into one dimensional histograms. Rather, the two variables are modeled as two dimensional histograms with 10 GeV bins ranging from zero to 500 GeV for $M_{\Delta}^{R}$ (as for the one dimensional analysis) and 5 bins between zero and one for $|\cos\theta_{R+1}|$. The expected event yields for the sum of the SM backgrounds and sample signal models in this two dimensional binning are shown in Fig.~\ref{fig:COSTHETAPDF}

\subsection{Fit to toy data and statistical analysis}

For each dataset a fit is performed over all final state categories and bins of the kinematic discriminants simultaneously, measuring the yields of different background contributions. The fit proceeds by maximizing the binned likelihood for the dataset being examined, which can be written as
\begin{equation}
\log {\cal L} = \sum_{i} \log \left( \frac{ b_{i}^{n_{i}}e^{-b_{i}} }{n_{i}!} \right) ~,
\label{eq:simple}
\end{equation}
where $i$ runs over all of the bins and $b_{i}$ and $n_{i}$ are the expected and observed number of events in that bin, respectively. For each toy analysis fit the likelihood is maximized over the yields of each of the backgrounds, subject to constraints between bins so that the full likelihood can be written
\begin{equation}
\log {\cal L}[b_{0},\cdots,b_{N_{p}}] = \sum_{c}\sum_{k} \left[ n_{ck} \log \left(\sum_p b_{p} \hat{b}_{pck}\right) - \sum_p b_{p} \hat{b}_{pck}\right] ~,
\label{eq:full}
\end{equation}
where bins are now indexed by category ($c$) and kinematic discriminant bin ($k$). The total expected number of events for a single process $p$ is $b_{p}$ while $\hat{b}_{pck}$ is the fraction of events from process $p$ expected to fall into bin $ck$, such that the number of expected events in a bin $i$ from Equation~\eqref{eq:simple}, $b_{i}$, has become $\sum_p b_{p} \hat{b}_{pck}$. While the total normalization of each process is independent from the others, the probability distribution function (pdf) of each process, $\hat{b}_{pck}$, provides constraints between different categories and bins of the kinematic discriminant.

Two fits are performed on each dataset, one corresponding to the background-only hypothesis and the other to the signal plus background hypothesis, where the signal corresponds to whichever model is being tested. The background only fit can be represented as
\begin{equation}
\log {\cal L}_{b} = \max_{b_{\mathrm{DB}},b_{t\bar{t}},b_{\mathrm{DY}}} {\cal L} [b_{\mathrm{DB}},b_{t\bar{t}},b_{\mathrm{DY}}]
\label{eq:Lb}
\end{equation}
where $b_{\mathrm{DB}}$, $b_{t\bar{t}}$, and $b_{\mathrm{DY}}$ represent the normalizations for di-boson, $t\bar{t}$ and Drell-Yan backgrounds, respectively. Similarly, the signal plus background fit maximizes the likelihood
\begin{equation}
\log {\cal L}_{s+b} = \max_{b_{\mathrm{DB}},b_{t\bar{t}},b_{\mathrm{DY}}} {\cal L} [b_{s} = \hat{N}_{S},b_{\mathrm{DB}},b_{t\bar{t}},b_{\mathrm{DY}}]
\label{eq:Lspb}
\end{equation}
which differs from $\log {\cal L}_{b}$ in Equation~\eqref{eq:Lb} by the addition of a signal contribution with total yield $b_{s}$. This yield is not floated in the fit; rather, it is fixed to the expected number of signal events for a given model, $\hat{N}_{S}$. The two maximized likelihoods, ${\cal L}_{b}$ and ${\cal L}_{s+b}$, are combined to form the test-statistic used to quantify the separation between the two hypotheses for a given model and dataset, the log-likelihood ratio $\lambda$
\begin{equation}
\lambda = \log \left( {\cal L}_{s+b} / {\cal L}_{b} \right)
\end{equation}

Systematic uncertainties are included in this procedure through marginalization. In this scheme, the kinematic discriminant pdf shapes and normalizations used in the likelihood evaluation remain fixed at their nominal values. During the toy dataset generation process these same shapes and normalizations are systematically varied according to expected uncertainties. We consider several sources and qualitative types of systematic uncertainties. A 10\% uncertainty is applied independently to each SM background process cross-section. The effect of this uncertainty is largely mitigated during the maximization of the likelihoods (where normalizations are floated). For backgrounds with multiple sub-contributions, like di-boson production, the relative sub-process yields are fixed in the likelihood evaluation resulting in an effective shape uncertainty. Each of the expected signal yields is also varied based on a calculation of the theoretical cross-section uncertainty. 

In addition to overall normalization uncertainties there are also a collection of variations which change the shape of background pdfs, both by varying the relative yields in different final state categories and by altering the shapes of the kinematic discriminants themselves. A 2\% uncertainty is assigned for the reconstruction and identification of each lepton, uncorrelated between lepton flavors.  This uncertainty is assumed to be correlated between different processes. Similarly, a 10\% uncertainty is assigned for the reconstruction and identification of each additional jet, effectively varying the relative yields between different jet multiplicity categories, independently for each process. This is meant to account for not only experimental effects relevant to jet counting, such as jet energy scale (JES) and resolution but also theoretical uncertainties in the production of strong emissions. To introduce uncertainty in the shape of kinematic discriminants we propagate the effects of potential JES uncertainties to the $E_{T}^\text{miss}$ and kinematic variable calculation. For each simulated event all of the reconstructed jets, without a $p_{T}$ threshold, are varied in $p_{T}$ either up or down by 10\%. The different between the original and new jet momenta is added vectorially to the $\vec{E}_{T}^\text{miss}$ and the kinematic variables of interest (for both selection requirements and kinematic discriminants) are recalculated using both the up and down variations separately. Each of the datasets corresponding to these variations are used to re-derive pdfs for the kinematic variables of interest such that the pdf shape for a given toy experiment is taken from a linear combination of the up, down and nominal templates. 

\begin{figure}[ht]
\includegraphics[width=0.35\columnwidth]{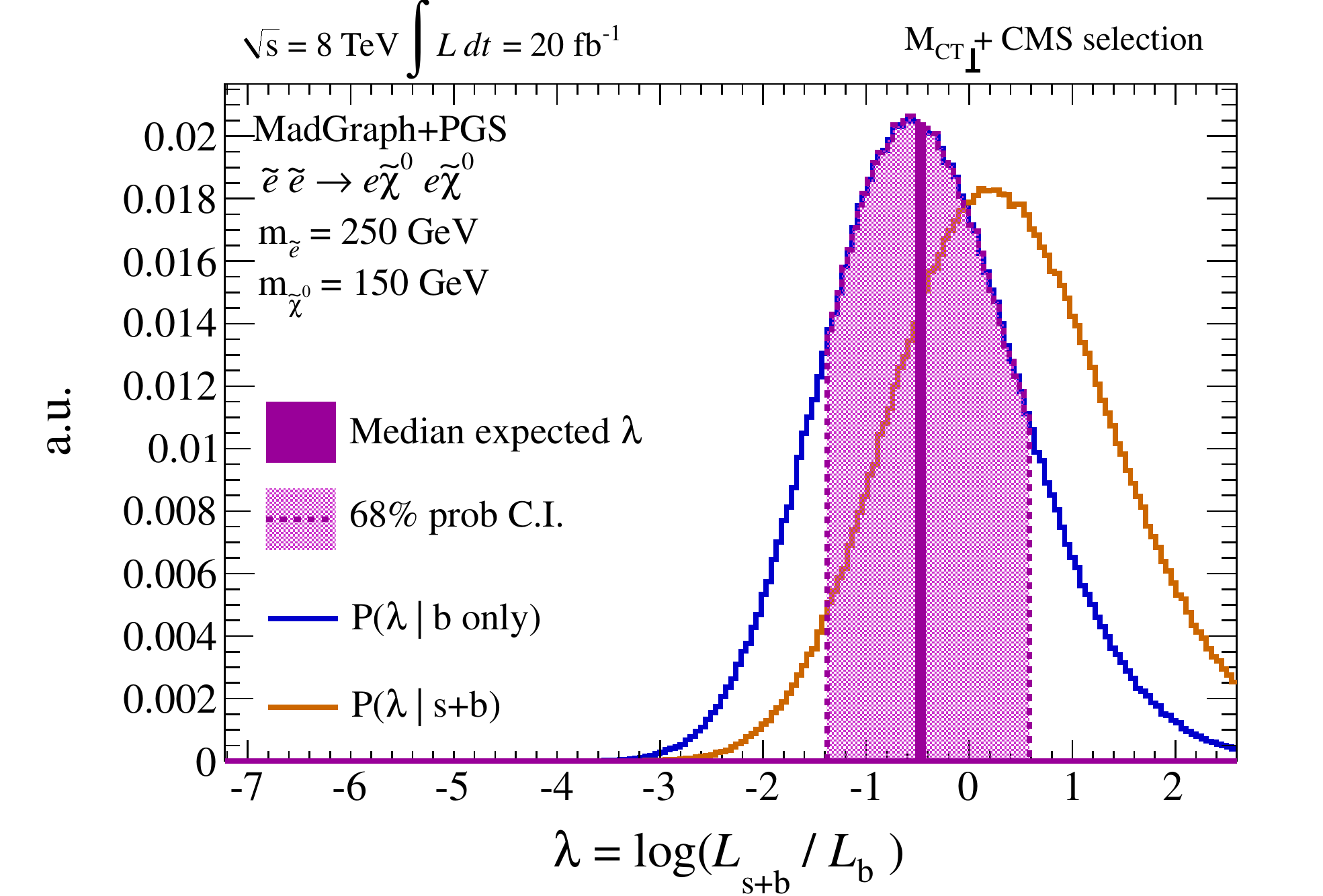}
\includegraphics[width=0.35\columnwidth]{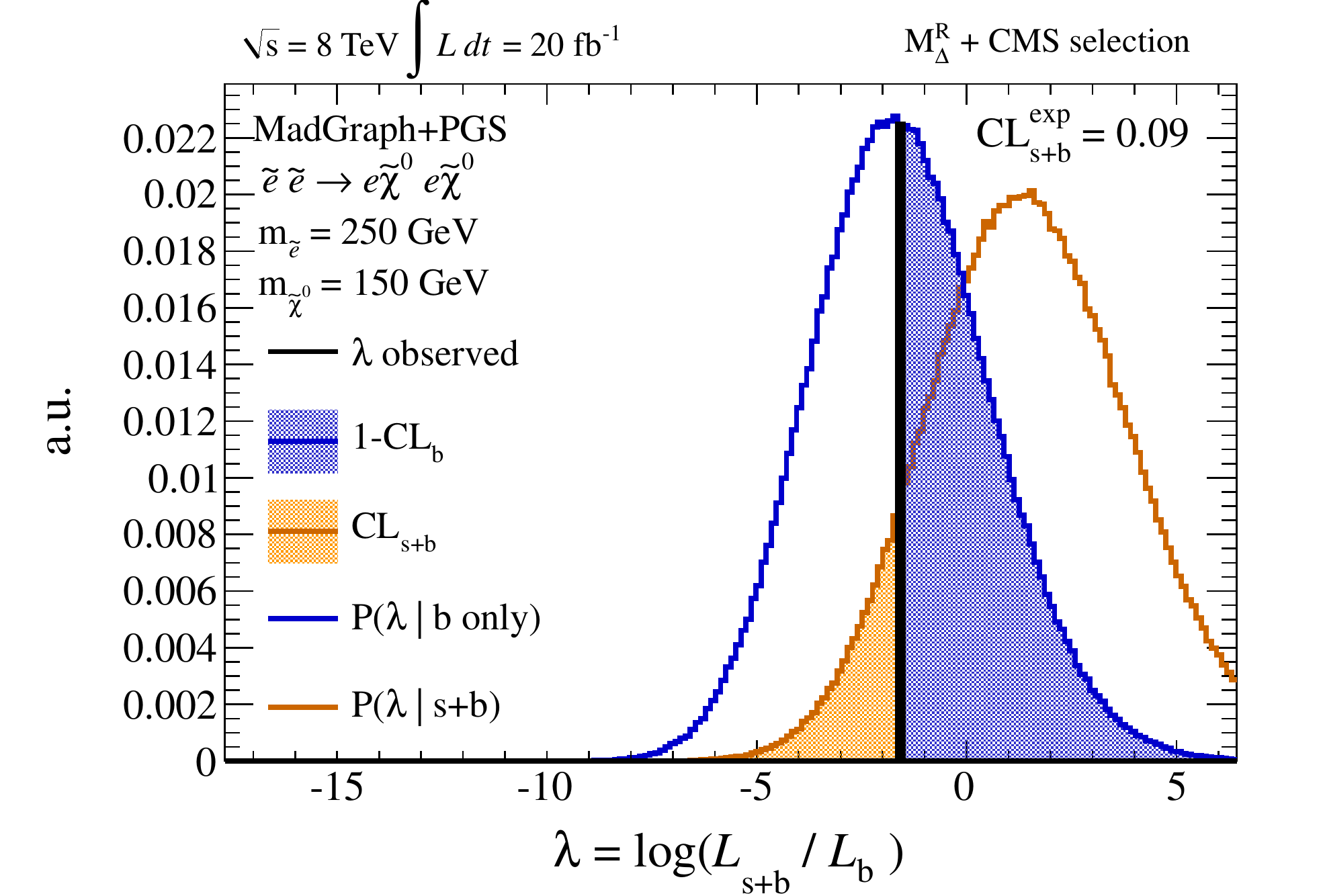}
\caption{Distributions of $\lambda$, assuming both signal and background-only scenarios. Left: Test-statistic distributions for the one-dimensional $M_{CT\perp}$ analysis searching for di-selectron production with ($m_{\tilde{\ell}_L}=250$~GeV, $m_{\tilde{\chi}^{0}_{1}}=150$~GeV). Right: $\lambda$ distributions for the one-dimensional $M_{\Delta}^{R}$ analysis, using the same mass point. \label{fig:bells}}
\end{figure}

For each signal model a series of toy pseudo-experiments are performed. For each pseudo-experiment, the parameters describing each of the systemic uncertainties are varied, yielding a new set of pdfs and normalizations for each process. These are then used to generate two toy data samples for the pseudo-experiment, with one set including the expected contribution of the signal in the data sample and another without the signal. Each dataset in the pseudo-experiment is then fit to each of the hypotheses (signal or no signal), yielding the maximized likelihoods ${\cal L}_{b}$ and ${\cal L}_{s+b}$ and the test-statistic $\lambda$. Repeating this procedure for many toys allows us to estimate the expected distribution of $\lambda$ in the case that there is only background in the data sample, $P(\lambda | b~\mathrm{only})$ and when there is also signal, $P(\lambda | s+b)$. Example distributions of $\lambda$ for pseudo-experiments corresponding to representative signal models and analyses are shown in Figure~\ref{fig:bells}.

The expected sensitivity of a given search is calculated from these test-statistic distributions. In order evaluate the probability of observing a given $\lambda$ value in an experiment ($\lambda^{\mathrm{exp}}$) the expected pdfs of $\lambda$ are used to calculate the quantities CL$_{b}$ an CL$_{s+b}$ as
\begin{eqnarray}
\mathrm{CL}_{b} = \int_{-\infty}^{\lambda^{\mathrm{exp}}} P(\lambda | b~\mathrm{only})~, \nonumber \\
\mathrm{CL}_{s+b} = \int_{-\infty}^{\lambda^{\mathrm{exp}}} P(\lambda | s+b)~.
\end{eqnarray}
CL$_{b}$ is the probability of observing a $\lambda$ at least as background-like as $\lambda^{\mathrm{exp}}$ assuming that there is no signal contribution, while CL$_{s+b}$ is the same probability assuming there is signal injected. In order to quantify the expected sensitivity of an analysis we choose $\lambda^{\mathrm{exp}}$ to be the median expected $\lambda$ assuming it is distributed as $P(\lambda | b~\mathrm{only})$. The resulting CL$_{s+b}$ is then the median expected $p$-value for a given signal hypothesis, with lower values indicating that the model would be excluded at higher significance. These expected $p$-values are converted into a number of $\sigma$, corresponding to a normally distributed set of outcomes, as
\begin{equation}
N~\sigma = \sqrt{2} ~\mathrm{erf}^{-1} (\mathrm{CL}_{s+b})~.
\end{equation}
A particular model is expected to be excluded at 95\% confidence level (C.L.) if the median expected CL$_{s+b}$ is less than 0.05, and $N \sigma \ge 1.96$. The CMS and ATLAS experiments choose to quote results in the context of the CL$_{s}$ convention~\cite{Junk:1999kv,Read:2002hq}, where CL$_{s}$ = CL$_{s+b}$/CL$_b$. For median expectations, CL$_b$ is exactly 1/2, implying that a CL$_{s} \le 0.05$ threshold for excluding a given hypothesis corresponds to a 97.5\% C.L. exclusion, or $N \sigma \ge 2.24$. 

\begin{figure}[ht]
\includegraphics[width=0.35\columnwidth]{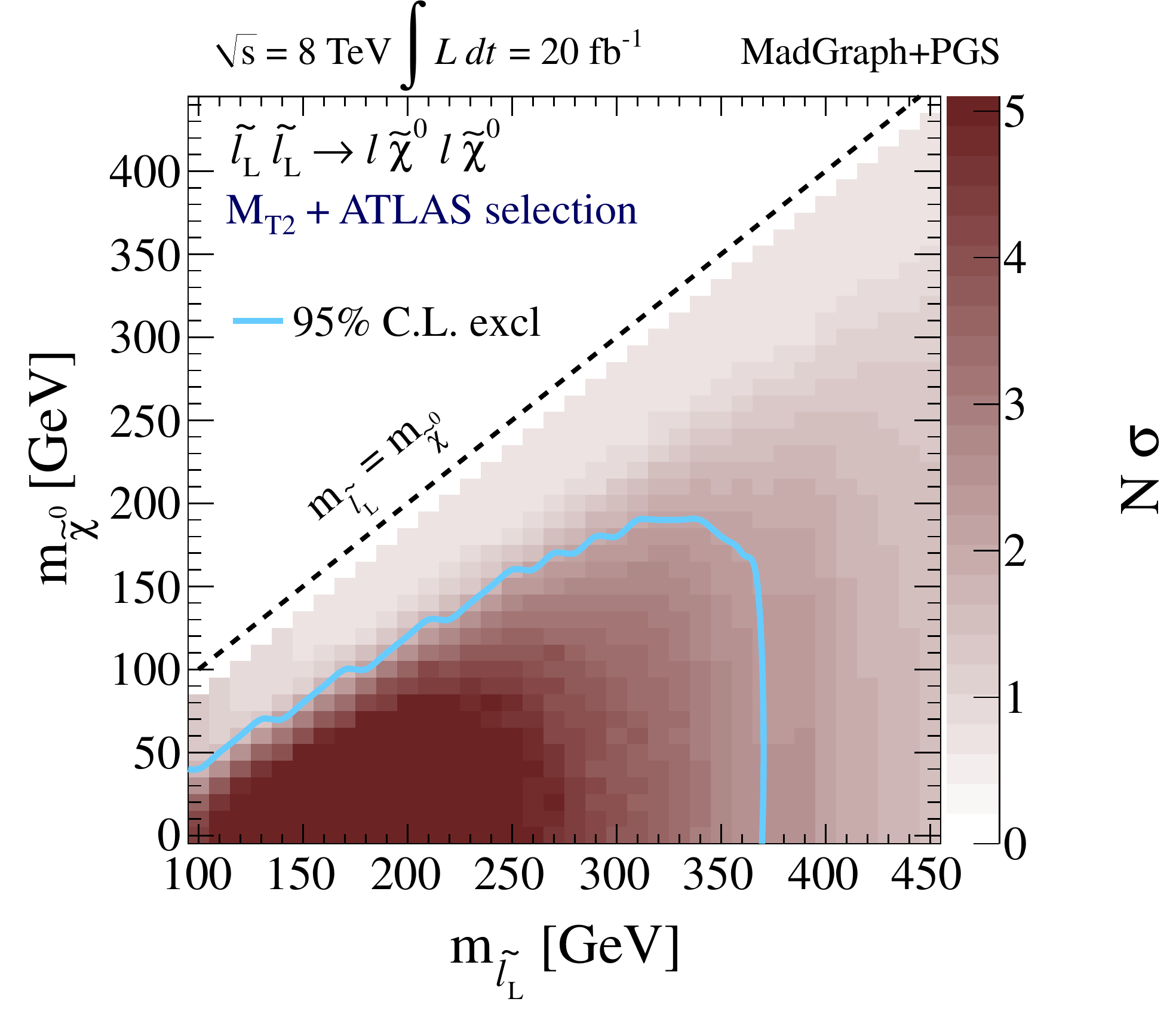}
\includegraphics[width=0.35\columnwidth]{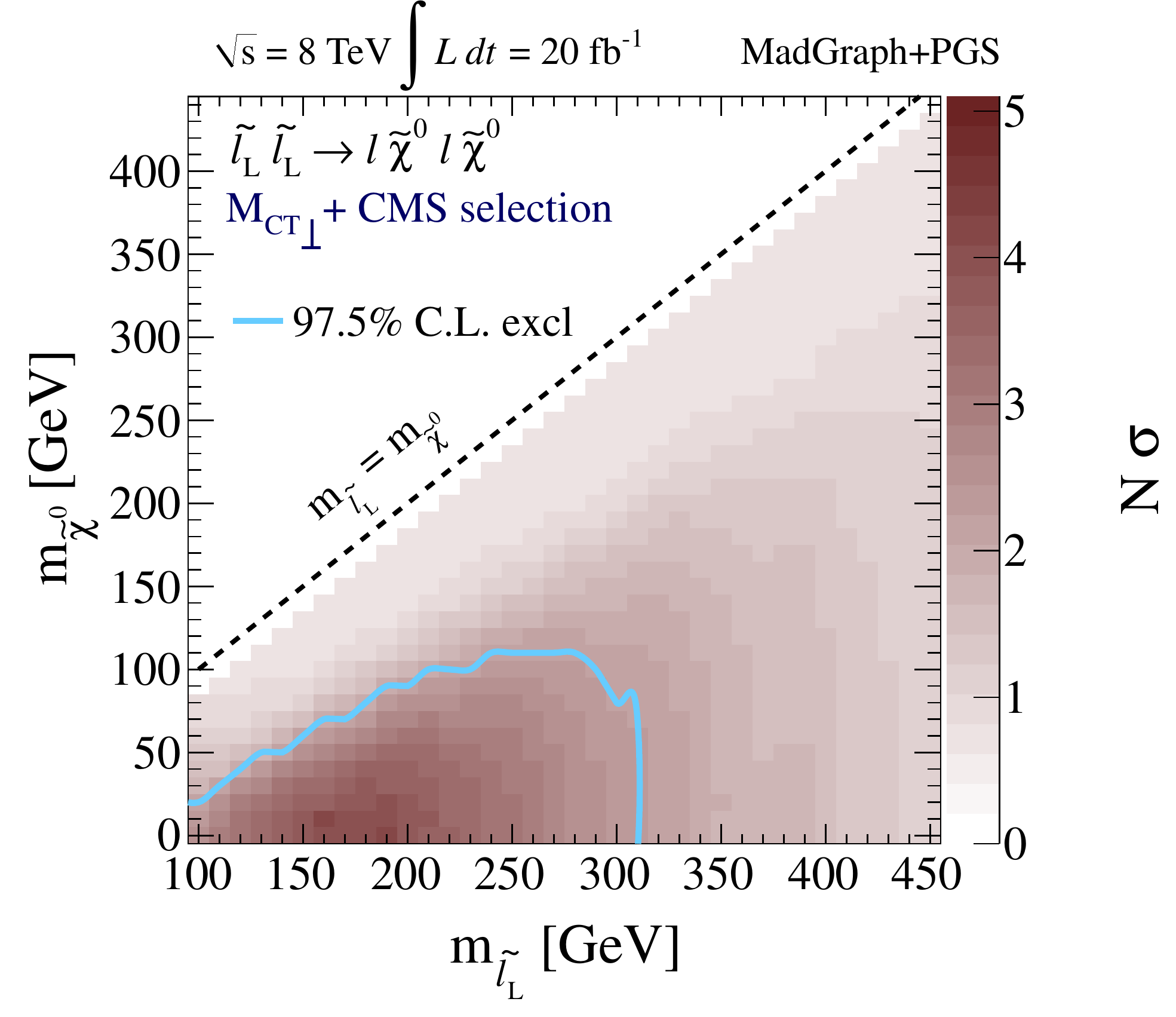}
\caption{ Median expected number of $\sigma$ for excluding the presence of different left-handed di-slepton signals, as a function of slepton and neutralino masses. Left: Expected results using $M_{T2}$ with the ATLAS selection. Right: Expectations when using $M_{CT\perp}$ in conjunction with the CMS selection.\label{fig:CLs} }
\end{figure}

The expected exclusions for di-slepton signals at 97.5\% C.L. for analyses performed with 20 fb$^{-1}$ of integrated luminosity at $\sqrt{s} = 8$~TeV, evaluated using this statistical approach, are shown in Figure~\ref{fig:CLs}. Comparing the excluded models from these toy analyses with those from the actual CMS \cite{CMS-PAS-SUS-13-006} and ATLAS \cite{ATLAS-CONF-2013-049} searches we observe that the results are in reasonable agreement. The expectations from toy experiments tend to be more optimistic than the actual experimental results, which is expected given that kinematic discriminants are being used in a shape analysis and deficiencies in detector simulation likely correspond to underestimated resolution effects, particularly for $E_{T}^\text{miss}$. Regardless, this toy analysis framework allows for a quantitative comparison of different kinematic discriminants in the context of an analysis with realistic experimental effects at least partially accounted for.

\clearpage

\section{Results and discussion \label{sec:conclusion}}

\begin{figure}[ht]
\includegraphics[width=0.4\columnwidth]{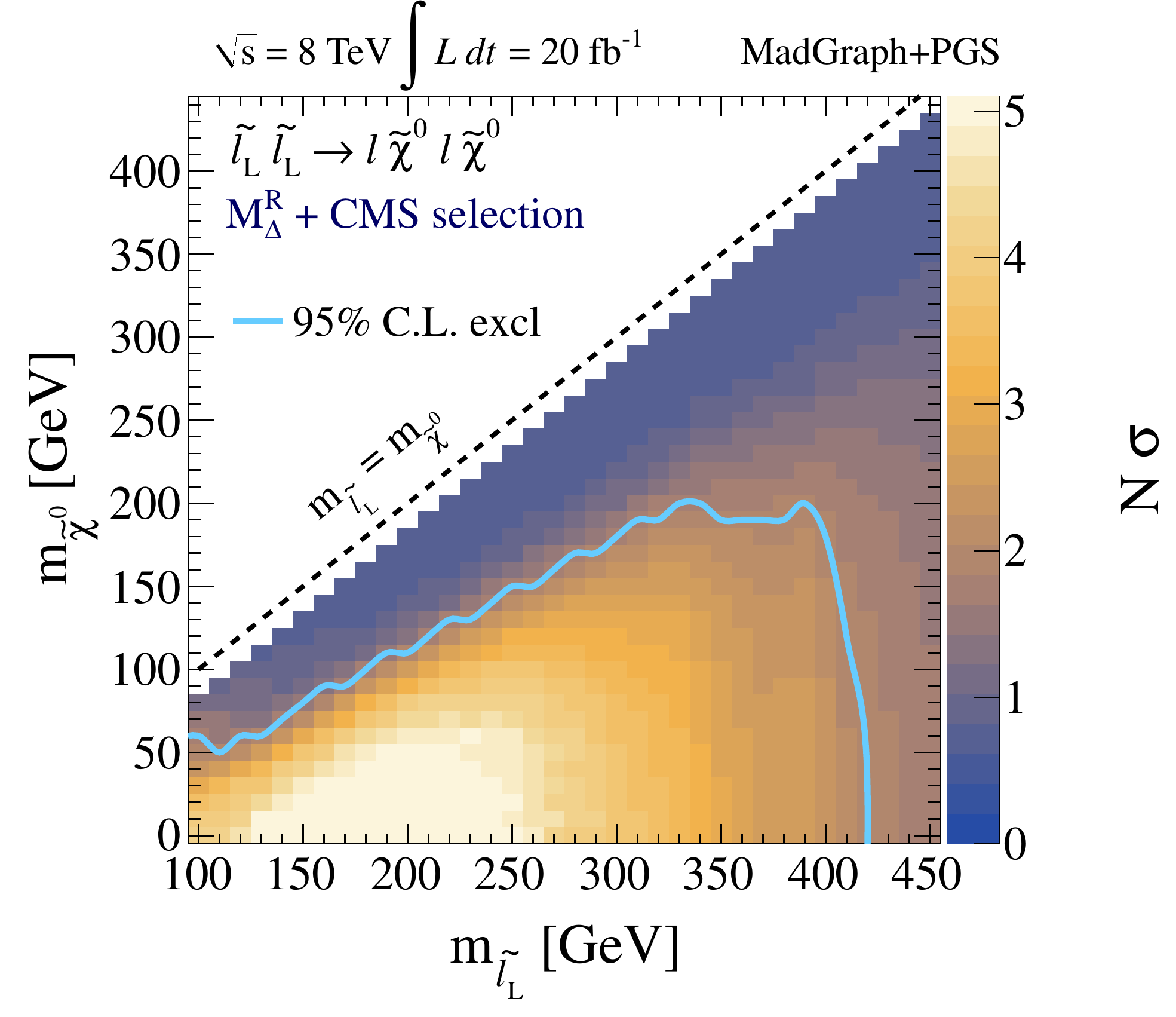}
\includegraphics[width=0.4\columnwidth]{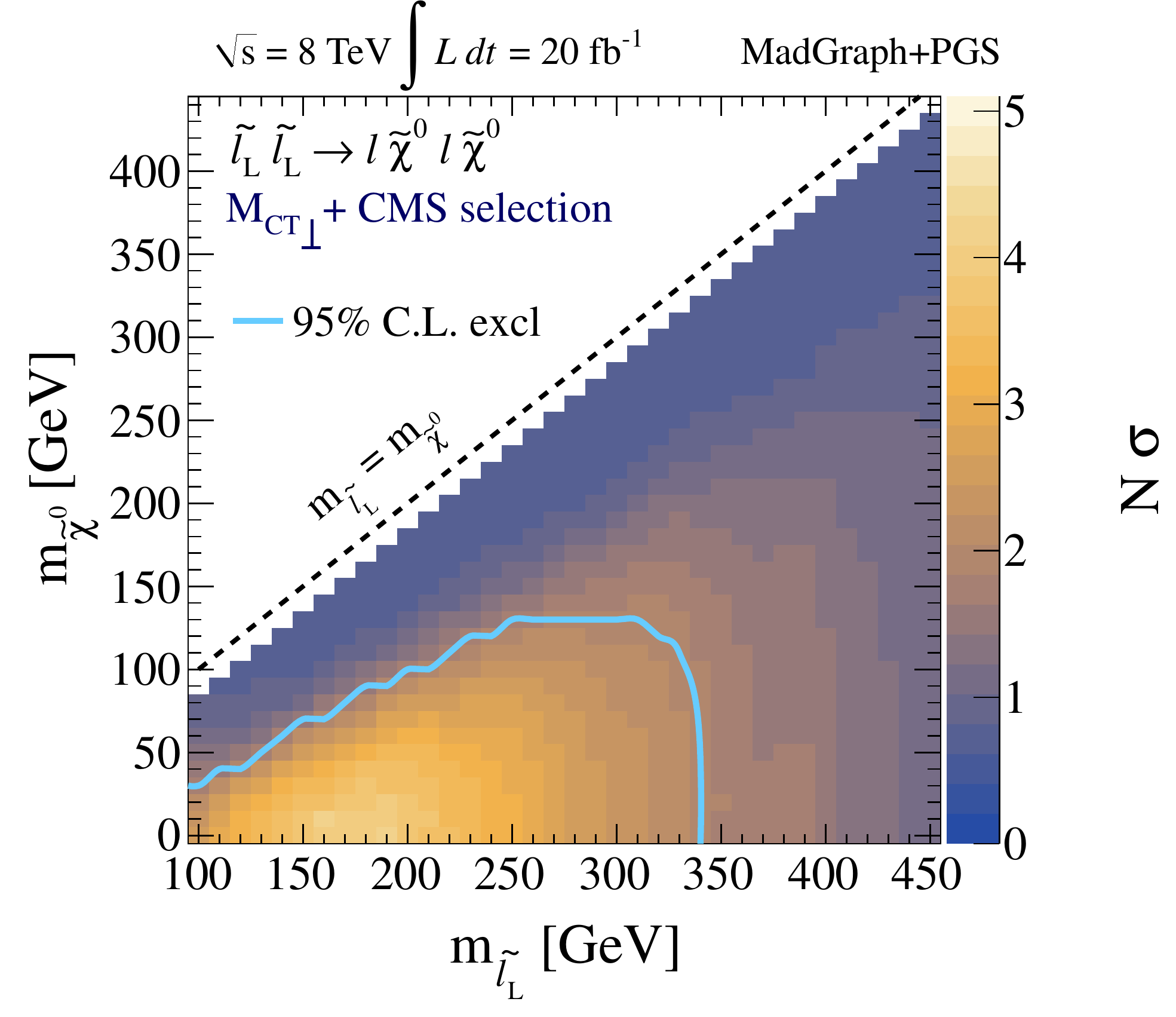}
\includegraphics[width=0.4\columnwidth]{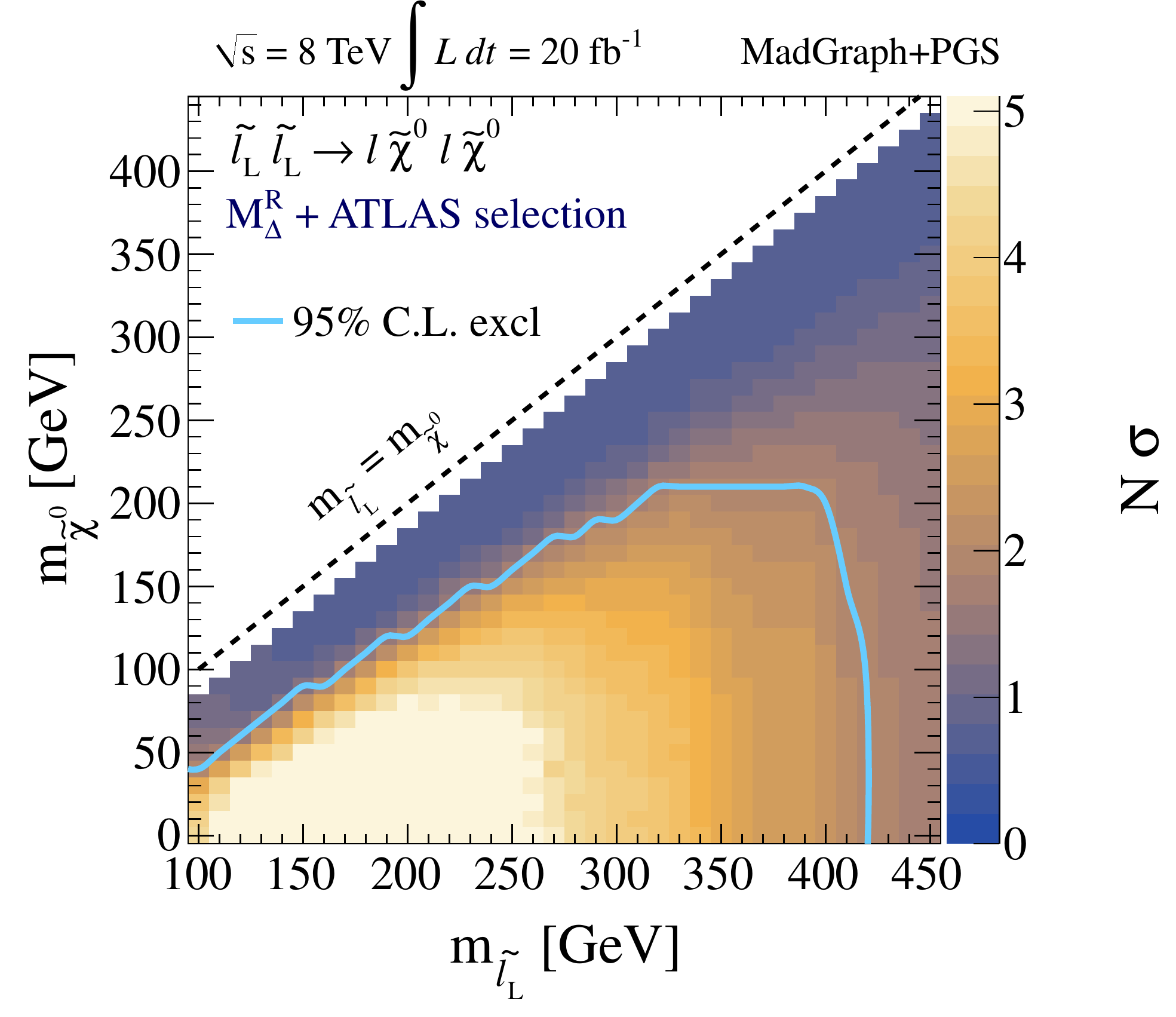}
\includegraphics[width=0.4\columnwidth]{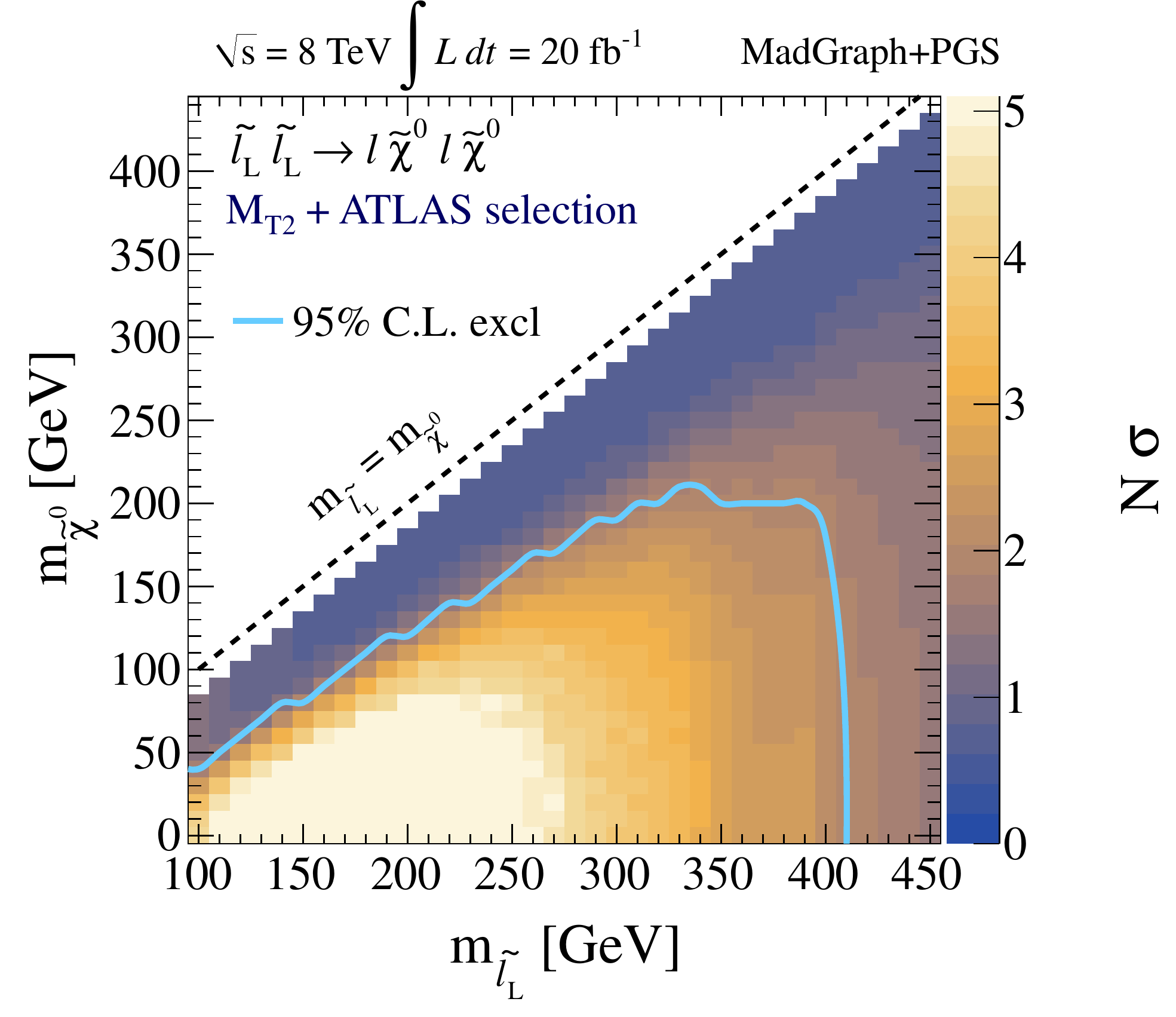}
\caption{Expected exclusion limits (in units of $\sigma$) for left-handed selectrons decaying to leptons and neutralinos using 20~fb$^{-1}$ of 8 TeV data, as a function of both selectron and neutralino masses. Expected limits are shown for our 1D $M_\Delta^R$ analysis using CMS (upper left) and ATLAS (lower left) selection cuts, and directly compared to our expected exclusions using our simulated CMS $M_{CT\perp}$ (upper right) and ATLAS $M_{T2}$ (lower right) analyses. \label{fig:results_slepton_2D_compare}}
\end{figure}

\begin{figure}[ht]
\includegraphics[width=0.4\columnwidth]{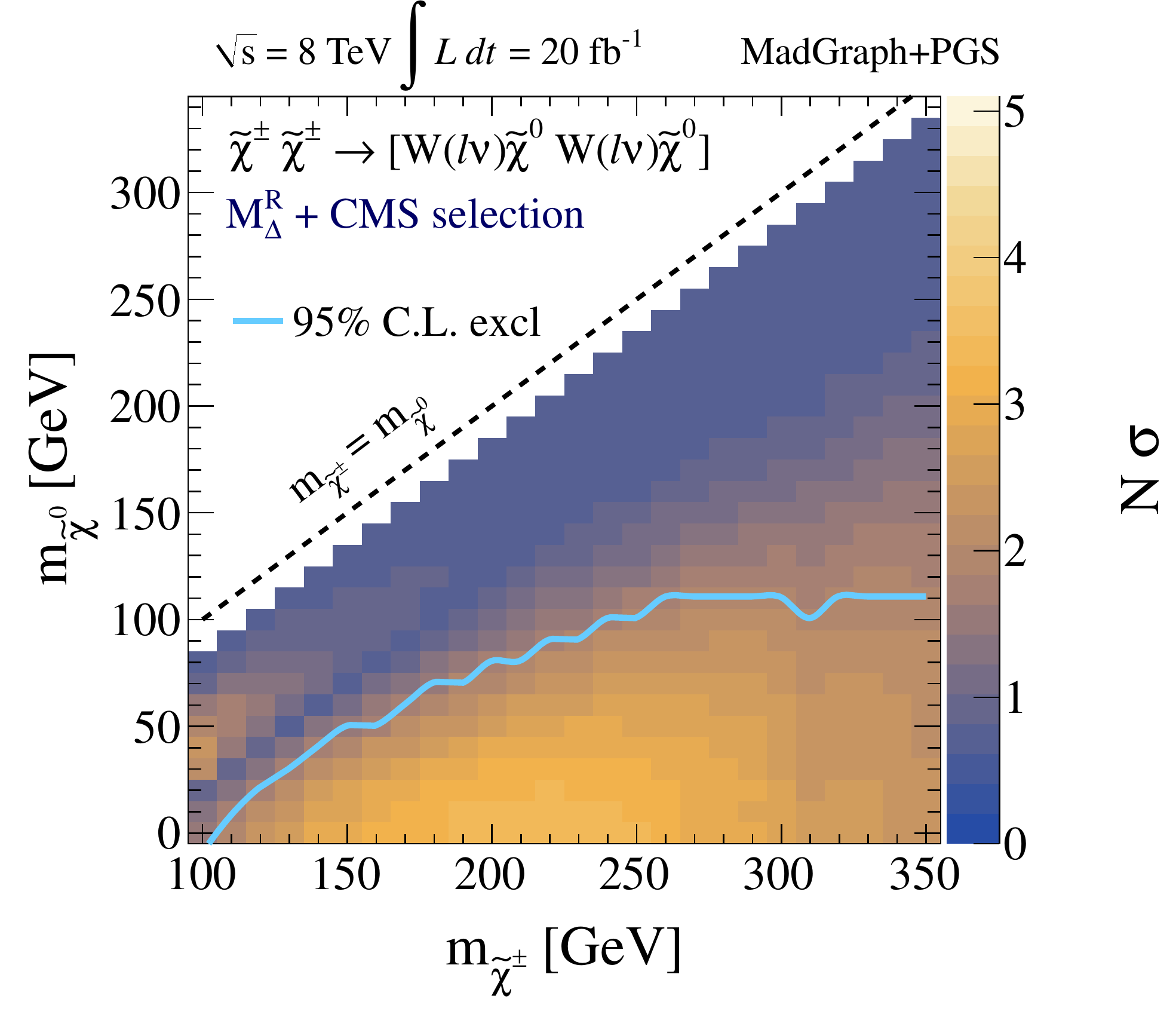}
\includegraphics[width=0.4\columnwidth]{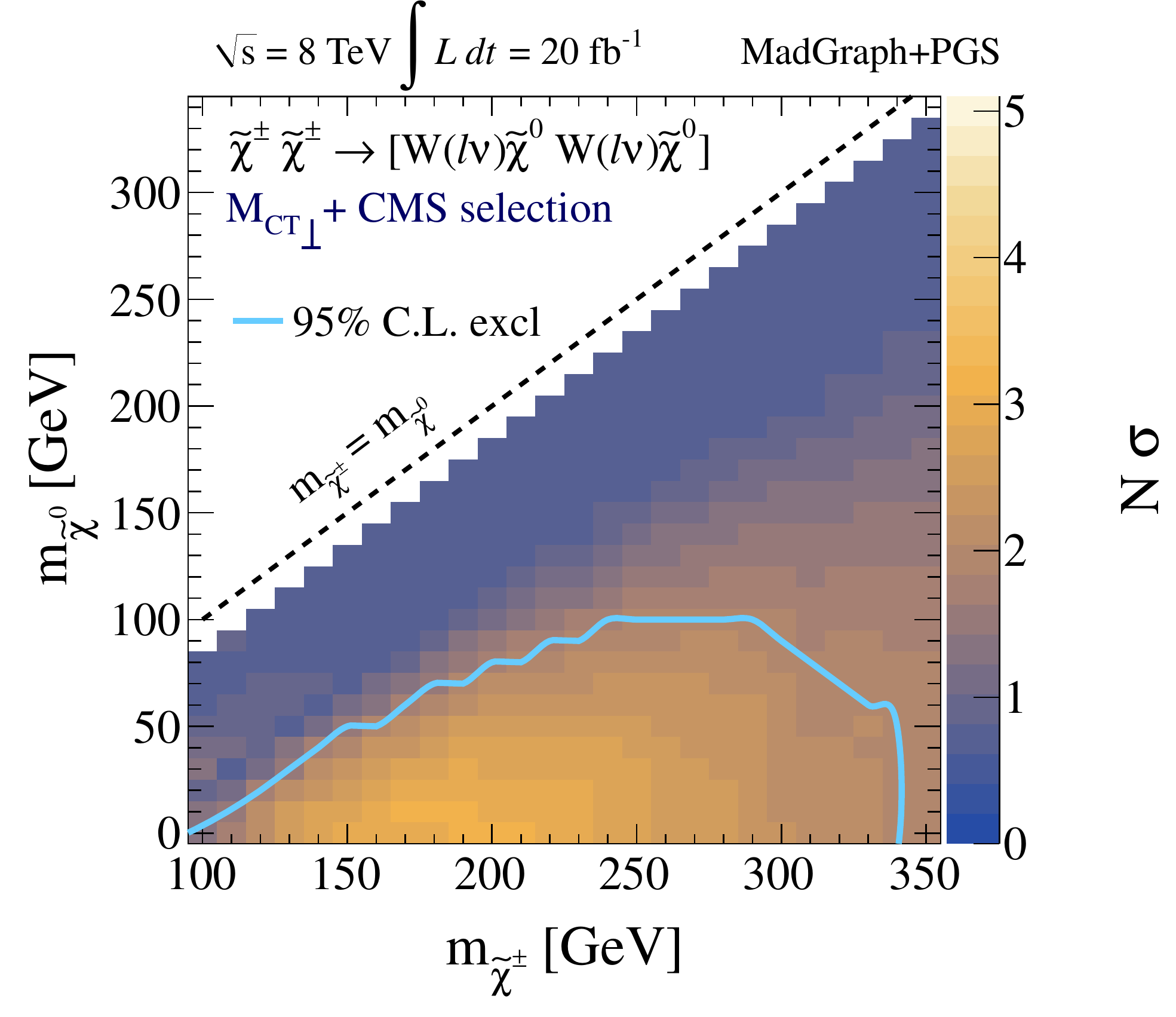}
\includegraphics[width=0.4\columnwidth]{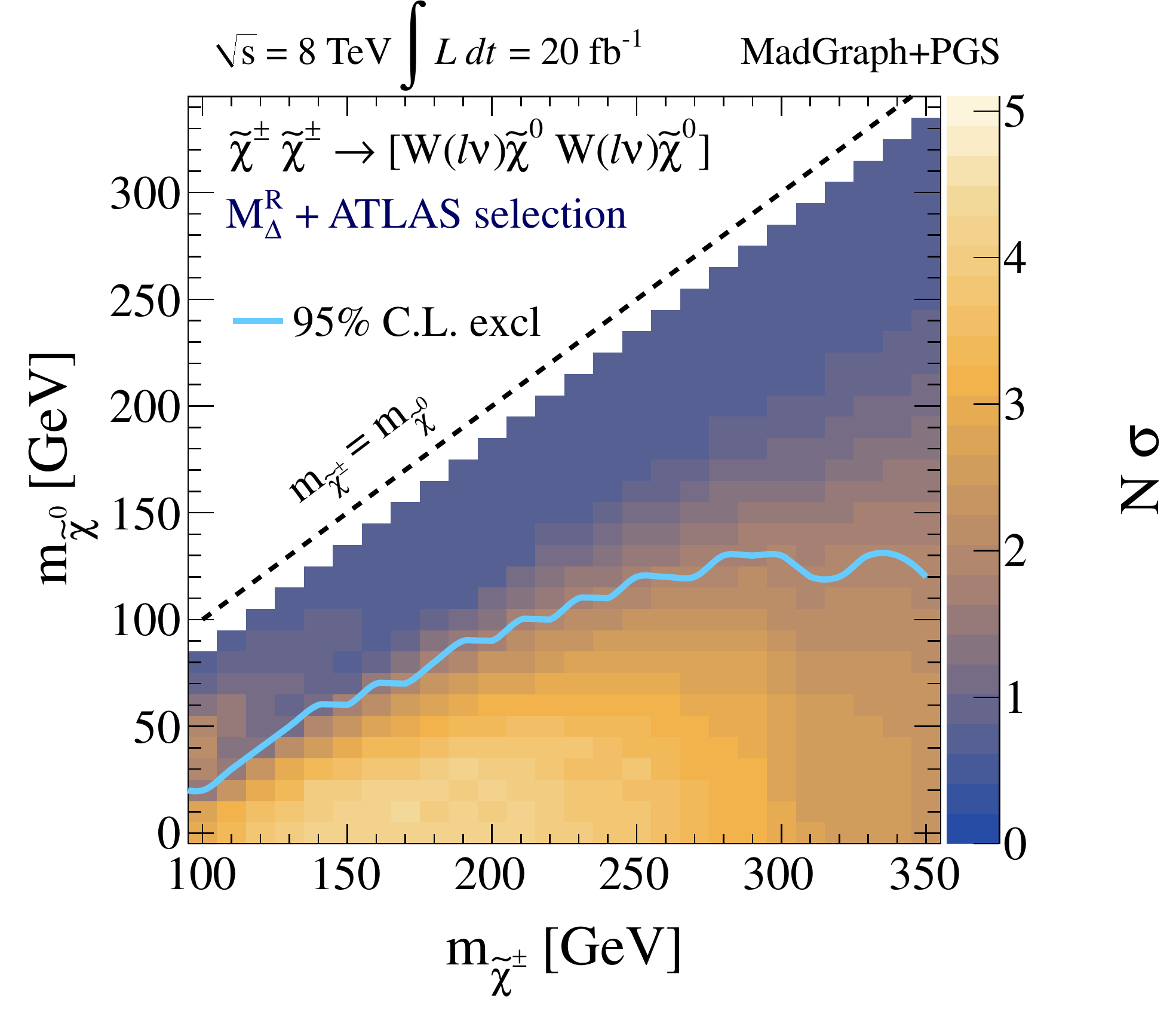}
\includegraphics[width=0.4\columnwidth]{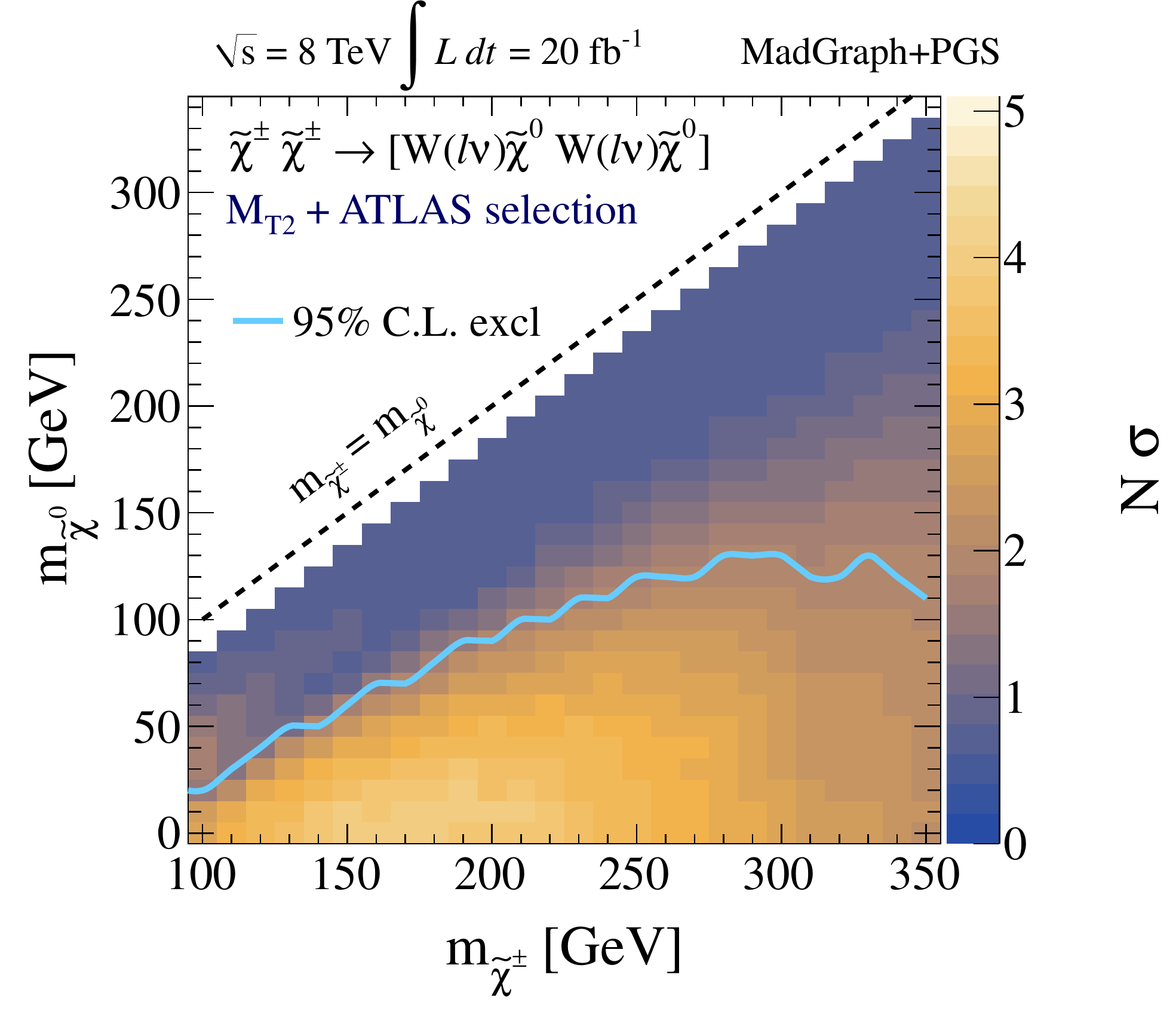}
\caption{Expected exclusion limits (in units of $\sigma$) for charginos decaying to neutralinos and leptonic $W$ bosons using 20~fb$^{-1}$ of 8 TeV data, as a function of both selectron and neutralino masses. Expected limits are shown for our 1D $M_\Delta^R$ analysis using CMS (upper left) and ATLAS (lower left) selection cuts, and directly compared to our expected exclusions using our simulated CMS $M_{CT\perp}$ (upper right) and ATLAS $M_{T2}$ (lower right) analyses.  \label{fig:results_chargino_2D_compare}}
\end{figure}

\begin{figure}[ht]
\includegraphics[width=0.4\columnwidth]{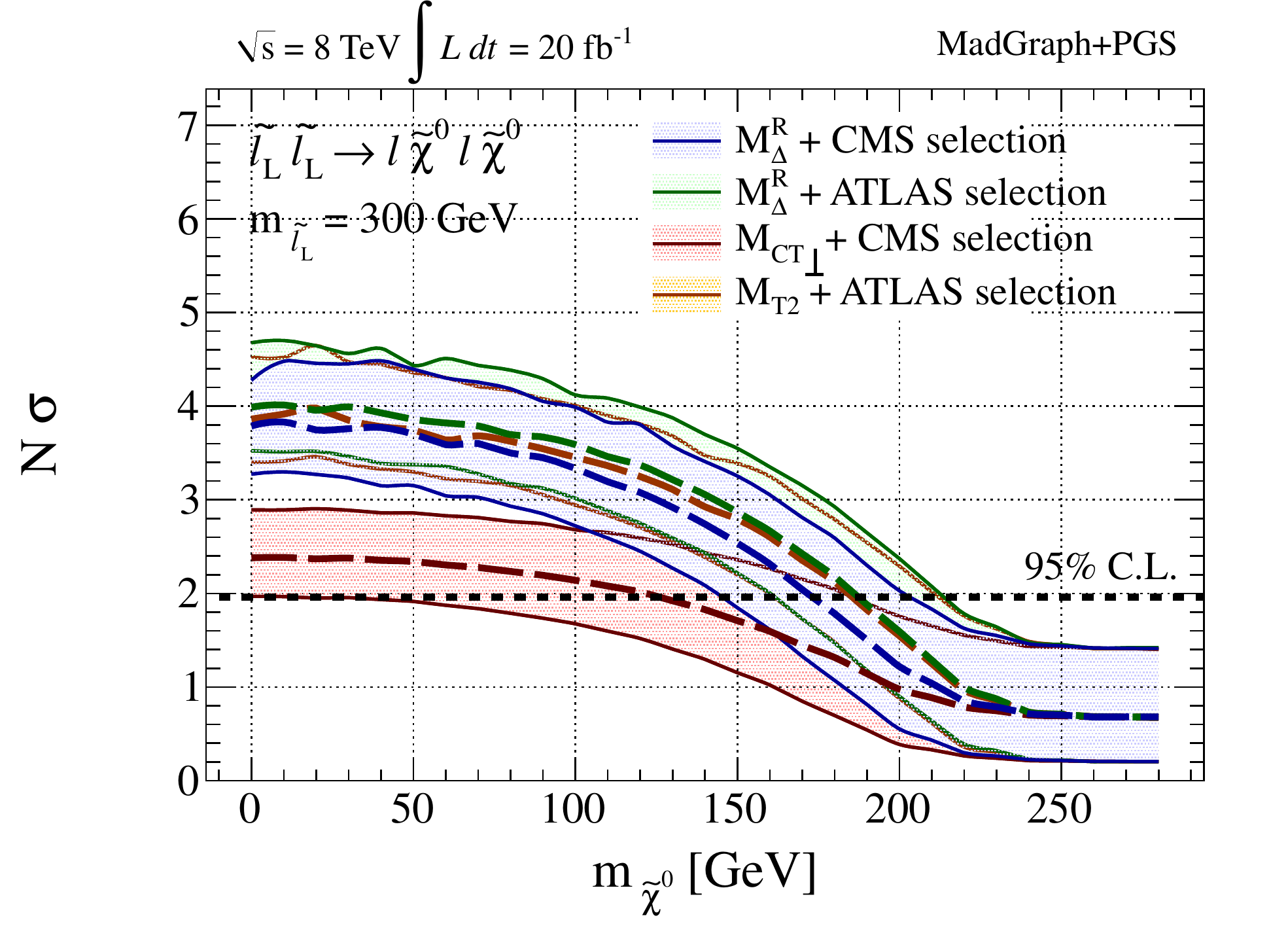}
\includegraphics[width=0.4\columnwidth]{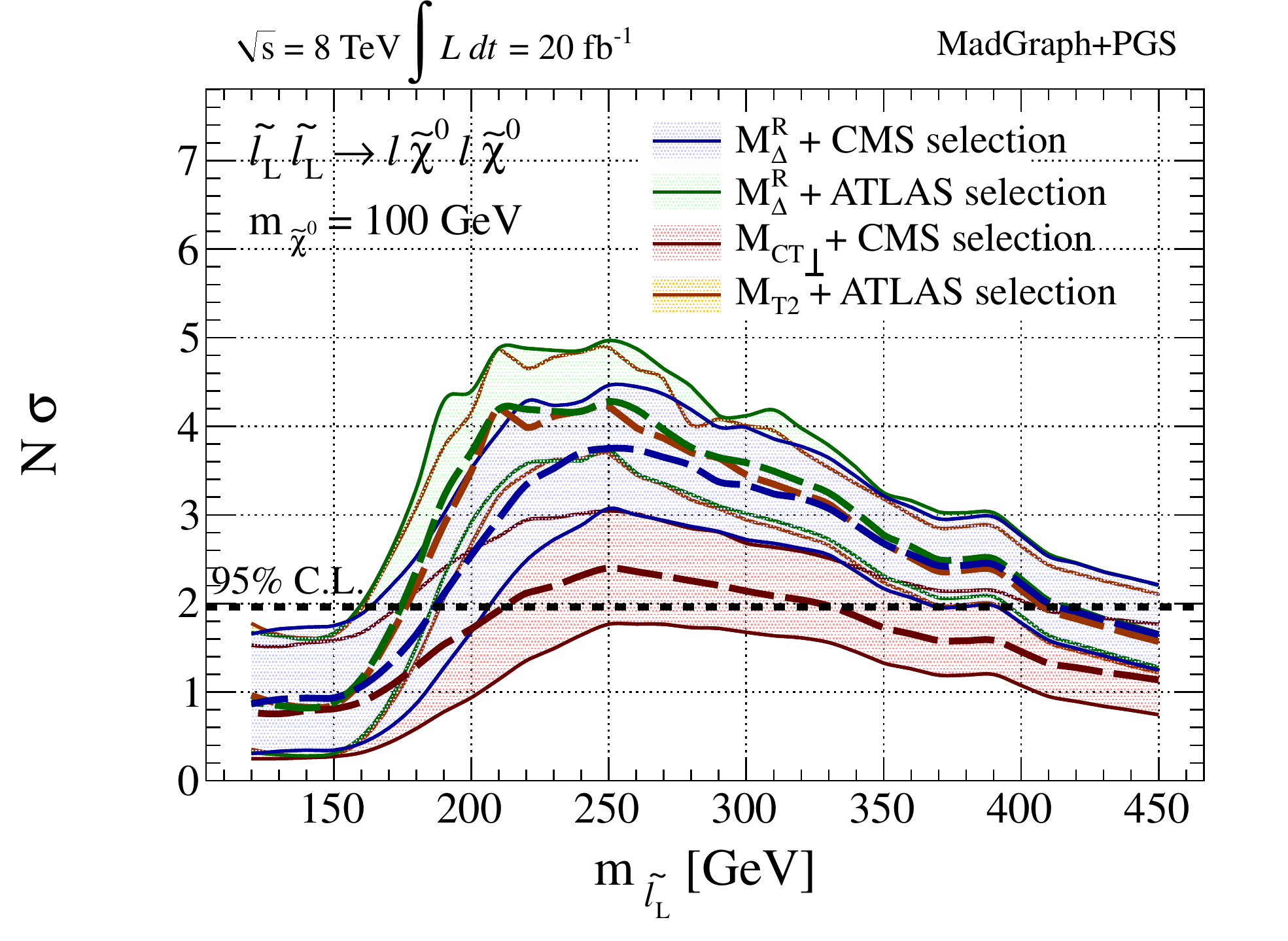}
\caption{Expected exclusion limits (in units of $\sigma$) for left-handed selectrons decaying to leptons and neutralinos using 20~fb$^{-1}$ of 8 TeV data, as a function of neutralino mass with 300 GeV selectrons (left) or as a function of selectron mass with 100 GeV neutralinos (right). Expected limits are shown for our 1D $M_\Delta^R$ analysis using CMS (blue) and ATLAS (green) selection cuts, and directly compared to our expected exclusions using our simulated CMS $M_{CT\perp}$ (red) and ATLAS $M_{T2}$ (orange) analyses. \label{fig:results_1D_compare}}
\end{figure}

Our 1D shape analyses using the mass variables $M_\Delta^R$, $M_{CT\perp}$, and $M_{T2}$ allow
a fair and realistic comparison of their discriminating power.
We begin by plotting the expected exclusion sensitivity for left-handed selectrons or charginos decaying to neutralinos, as a function of selectron/chargino and neutralino masses, assuming 20~fb$^{-1}$ of data from a single experiment at the 8~TeV LHC. Charginos are assumed to decay into $W$ bosons and an invisible neutralino, followed by Standard Model decays of the $W$ bosons into leptons. Results for left-handed smuons would be similar to those for the selectron, but we assume only a single species of slepton for our analysis. In Figures~\ref{fig:results_slepton_2D_compare} and \ref{fig:results_chargino_2D_compare}, we show the expected exclusion reach (at 95\% confidence level) of the ATLAS $M_{T2}$ and CMS $M_{CT\perp}$ analyses compared to the new technique using $M_\Delta^R$. In making the comparisions we use the same sets of ATLAS or CMS cuts as the existing experimental searches, which are not optimized for our analysis. 
Even with this disadvantage the expected exclusion limits using the super-razor variable $M_\Delta^R$ outperform the $M_{CT\perp}$ searches in terms of both absolute slepton or chargino mass and near the degenerate limit (when the mass of the parent is close to the mass of the invisible daughter). We show selected slices of these analyses in Figure~\ref{fig:results_1D_compare}, fixing either the selectron or neutralino mass, and varying the other. This allows a more direct comparison of our new variable $M_\Delta^R$ to the alternative techniques. Again the sensitivity using $M_\Delta^R$ outperforms
that obtained from $M_{CT\perp}$. For these 1D analyses the performance using $M_{T2}$ is only
slightly worse than that obtained with $M_\Delta^R$.

\begin{figure}[ht]
\includegraphics[width=0.4\columnwidth]{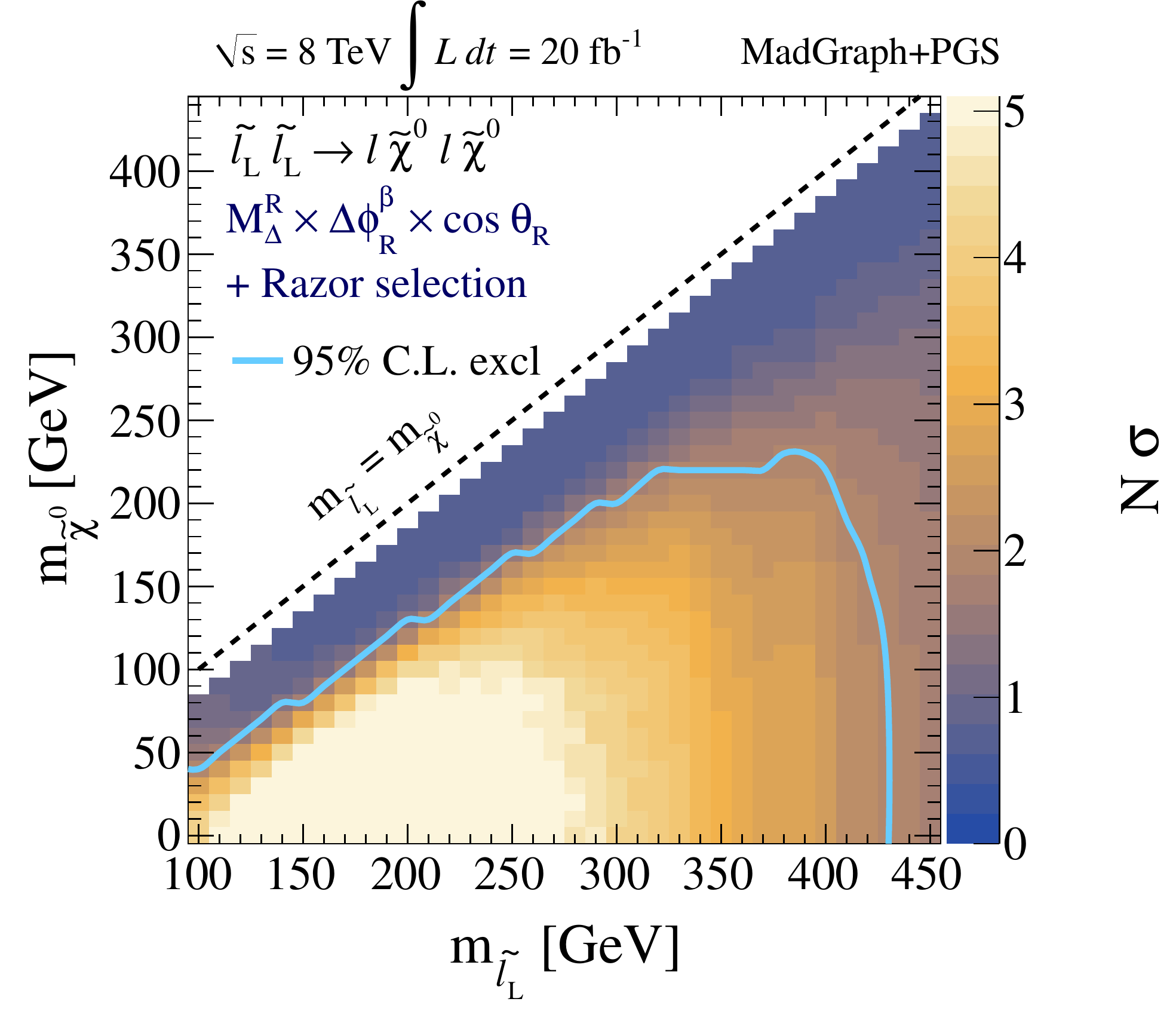}
\includegraphics[width=0.4\columnwidth]{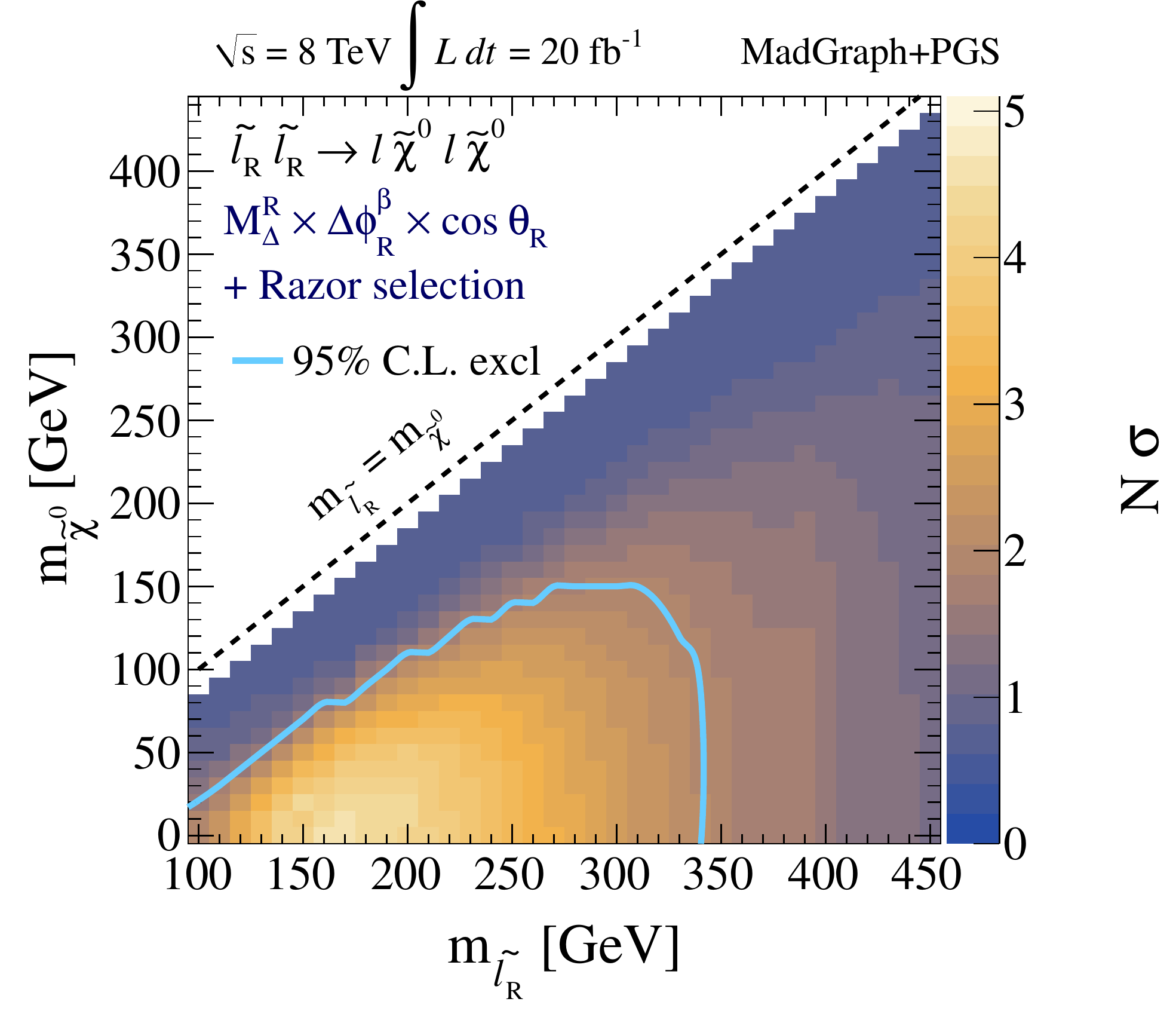}
\includegraphics[width=0.4\columnwidth]{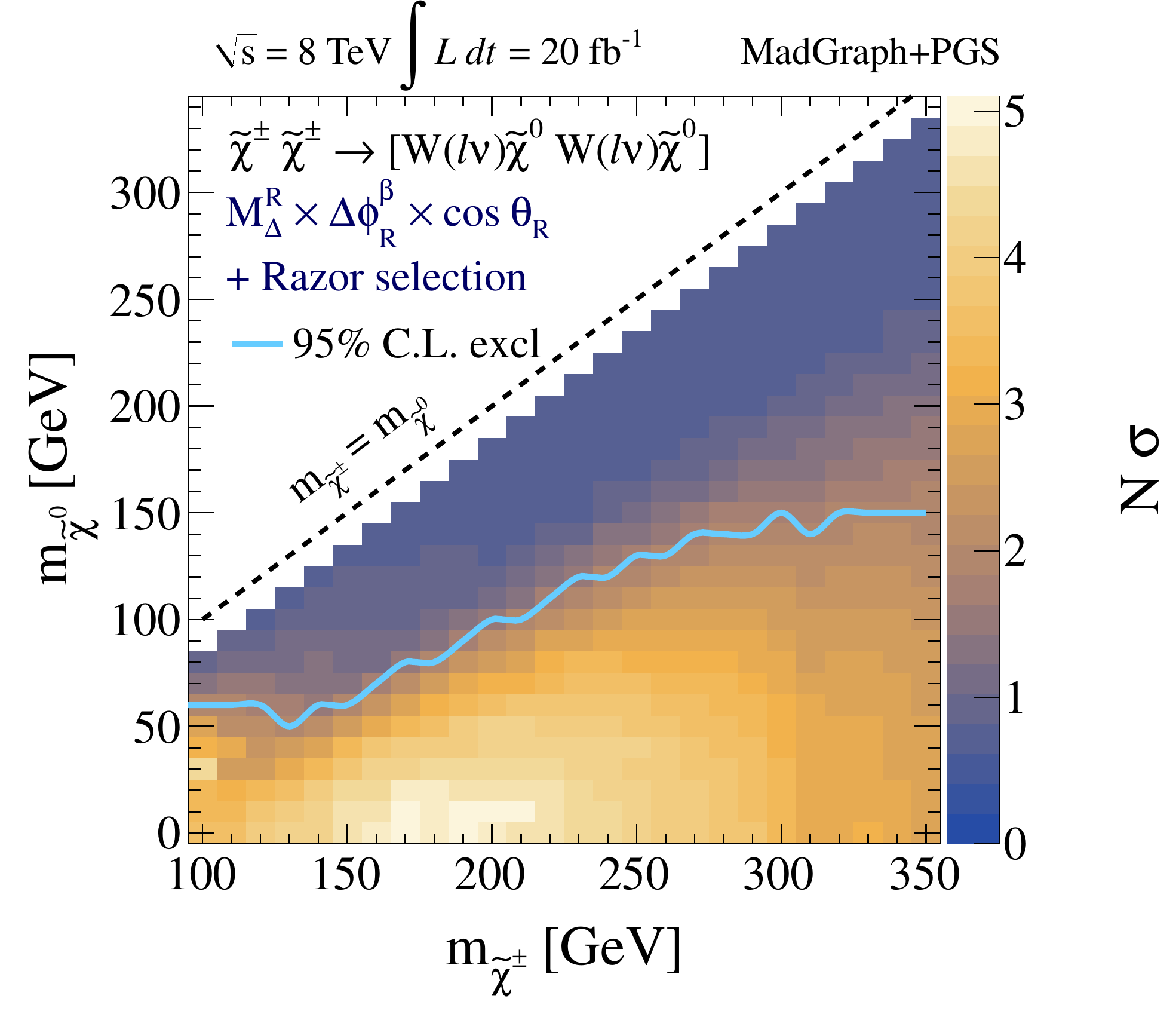}
\caption{Expected exclusion limits (in units of $\sigma$) for left-handed selectrons (upper left) right-handed selectrons (upper right), and charginos decaying to neutralinos and leptonic $W$ bosons (bottom center) using 20~fb$^{-1}$ of 8 TeV data, as a function of both selectron/chargino and neutralino masses. Expected limits are derived using our multi-dimensional $M_\Delta^R$, $\Delta\phi_R^\beta$ and $|\cos\theta_{R+1}|$ analysis super-razor analyses with the razor selection cuts described in the previous section.  \label{fig:results_2D_best}}
\end{figure}

We can understand these 1D results by again consulting the kinematic distributions shown
in Figure \ref{fig:compare} of Section \ref{sec:simulation}. The fact that approximately 50\% of
signal events end up in the zero bin for $M_{CT\perp}$ gives a loss in statistics that is not
compensated by the clean kinematic edge. For $M_{T2}$ the corresponding effect is much smaller,
resulting in performance very similar to that achieved with $M_\Delta^R$.

In Figure~\ref{fig:results_2D_best}, we show the exclusion reach of the full super-razor analyses, using our multi-dimensional shape analysis which employs $M_\Delta^R$, $\Delta\phi_R^\beta$ and $|\cos\theta_{R+1}|$, and the new super-razor selection described in Section \ref{sec:shape} in order to maximize the sensitivity over background. Exclusions are shown for both left- and right-handed selectrons, as well as charginos decaying to $W$ bosons and neutralinos.  The exclusion sensitivities include the
effects from systematic errors on kinematic shapes, and on reconstruction of jets and leptons,
as described in Section \ref{sec:shape}. Again we emphasize that the
super-razor selection has no $E_{T}^\text{miss}$ cut.

\begin{figure}[!ht]
\includegraphics[width=0.40\columnwidth]{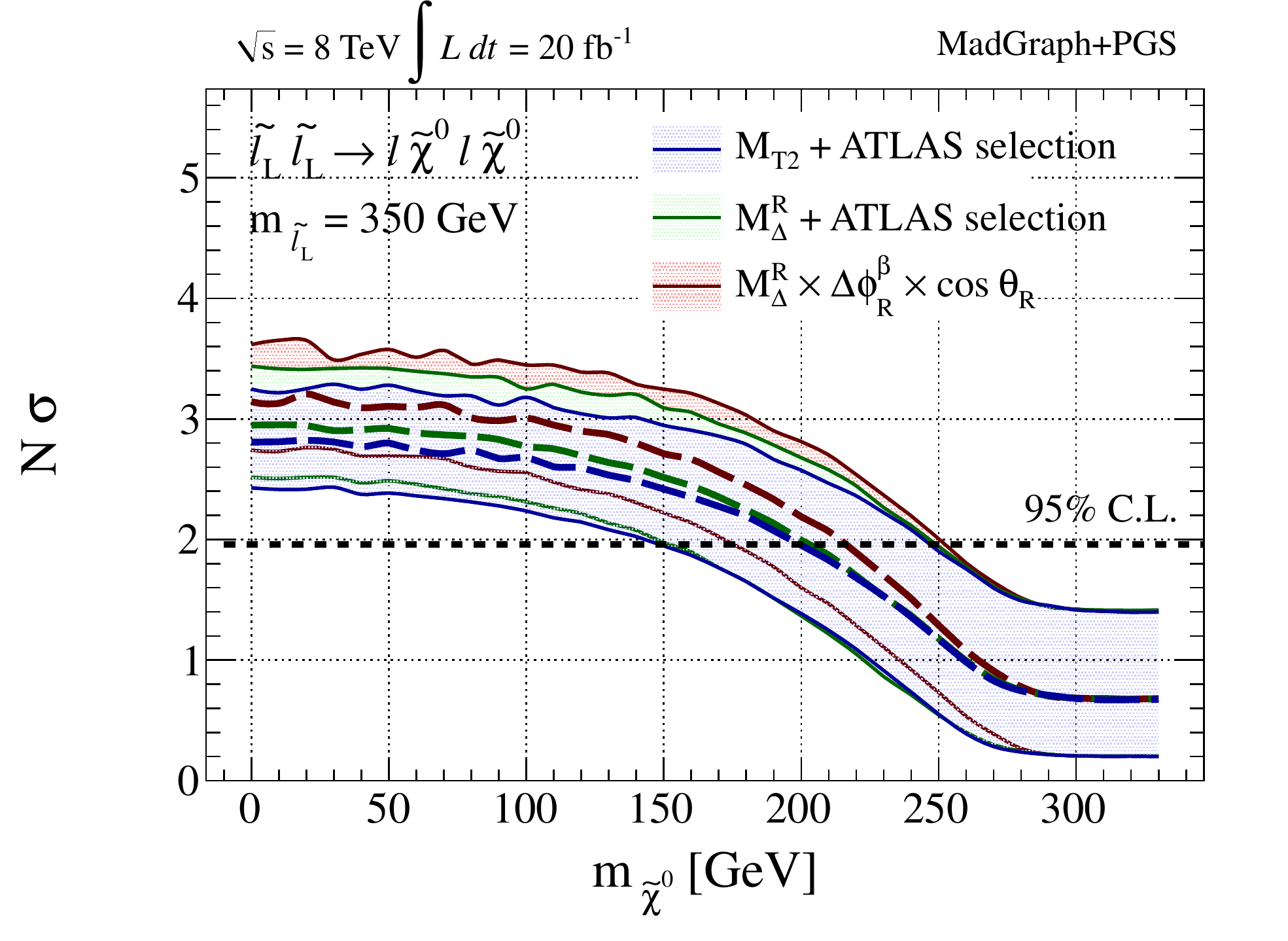}
\includegraphics[width=0.40\columnwidth]{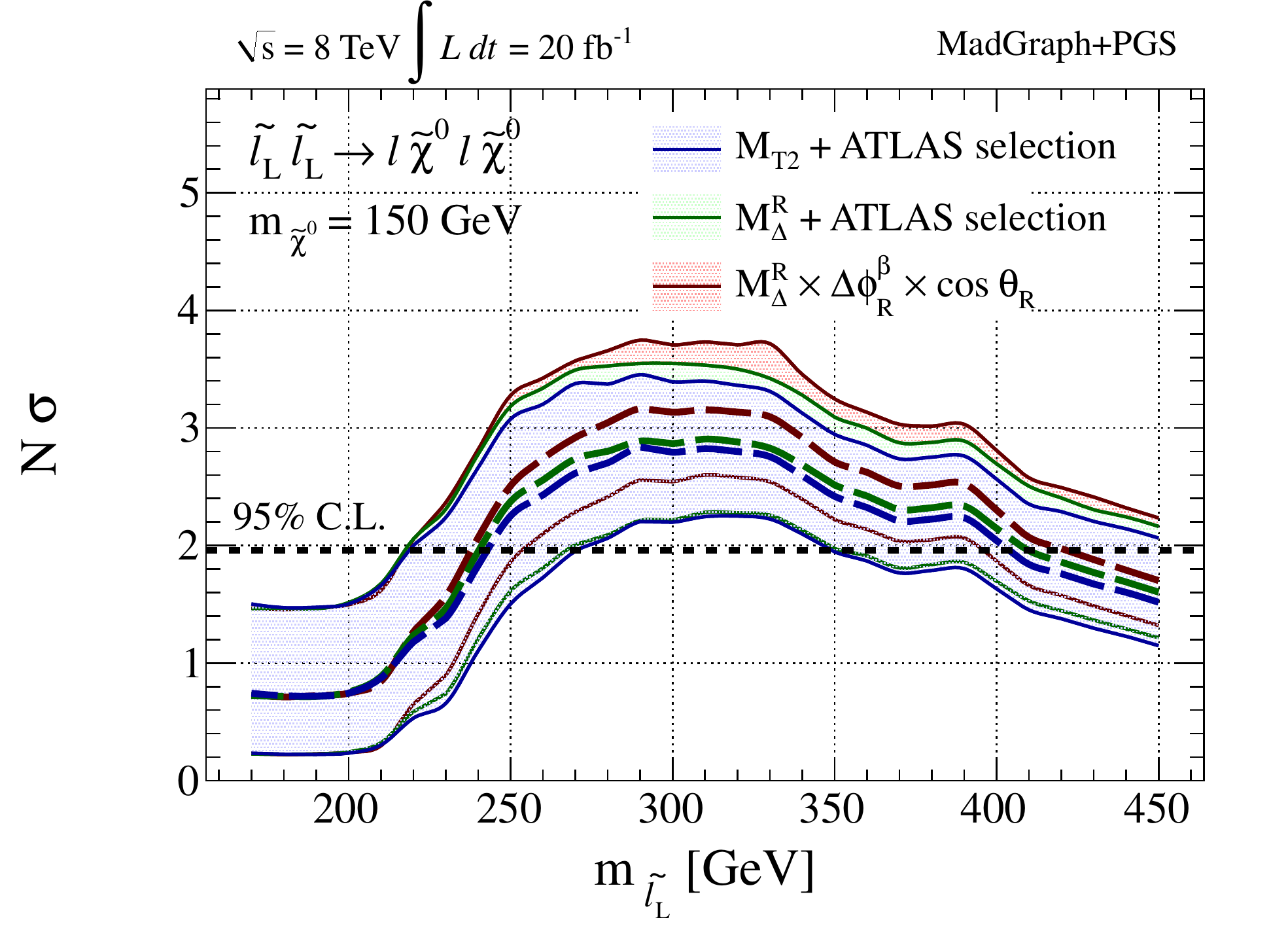}
\includegraphics[width=0.40\columnwidth]{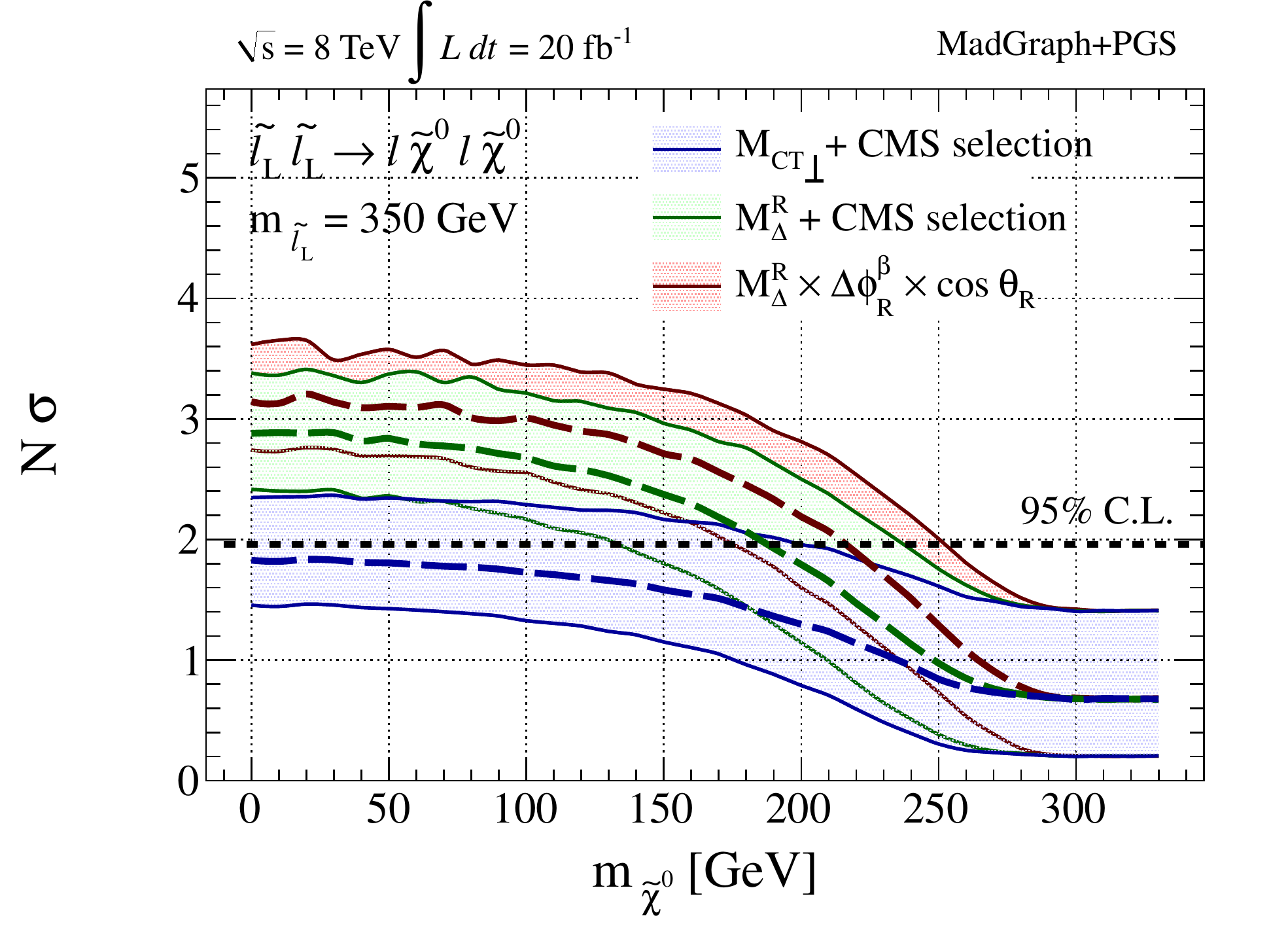}
\includegraphics[width=0.40\columnwidth]{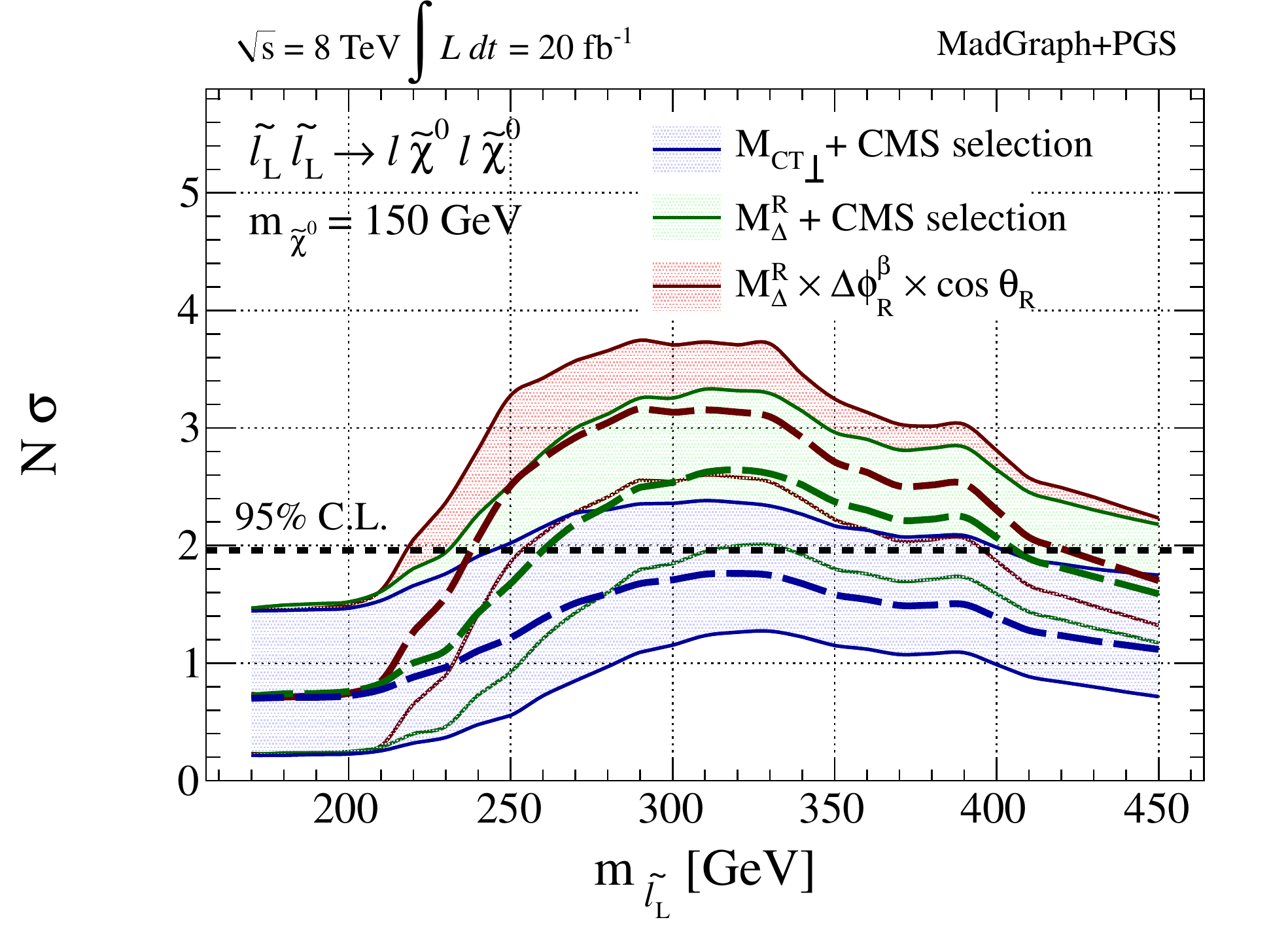}
\caption{Expected exclusion limits (in units of $\sigma$) for left-handed selectrons decaying to leptons and neutralinos using 20~fb$^{-1}$ of 8 TeV data, as a function of neutralino mass with 350 GeV selectrons (upper and lower left) or as a function of selectron mass with 150 GeV neutralinos (upper and lower right). Expected limits are shown for our multi-dimensional razor analysis (red), and compared to either ATLAS (upper plots) or CMS (lower plots) mass variables and selection criteria. \label{fig:results_slepton_1D_compare}}
\end{figure}

\begin{figure}[!hb]
\includegraphics[width=0.40\columnwidth]{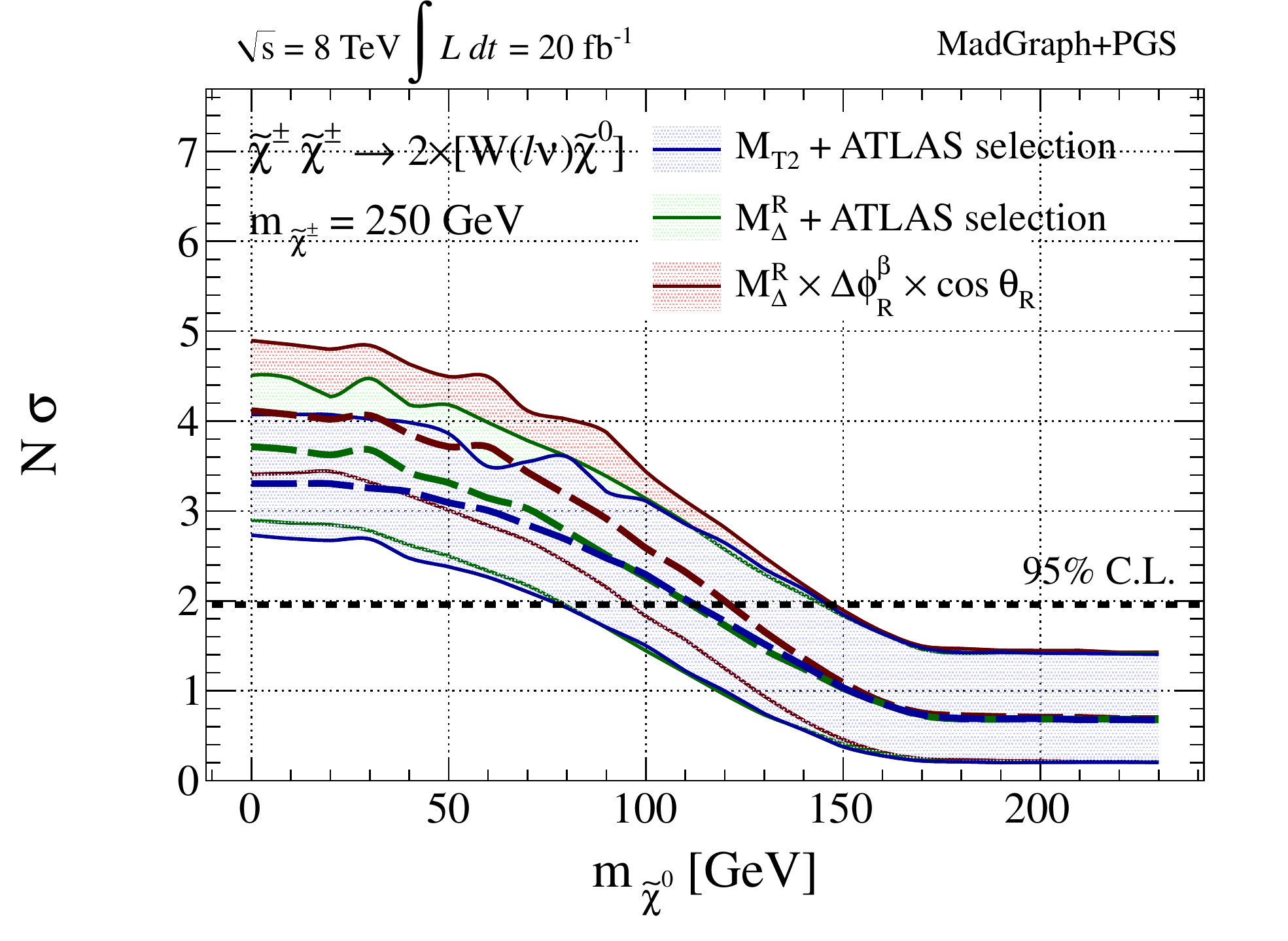}
\includegraphics[width=0.40\columnwidth]{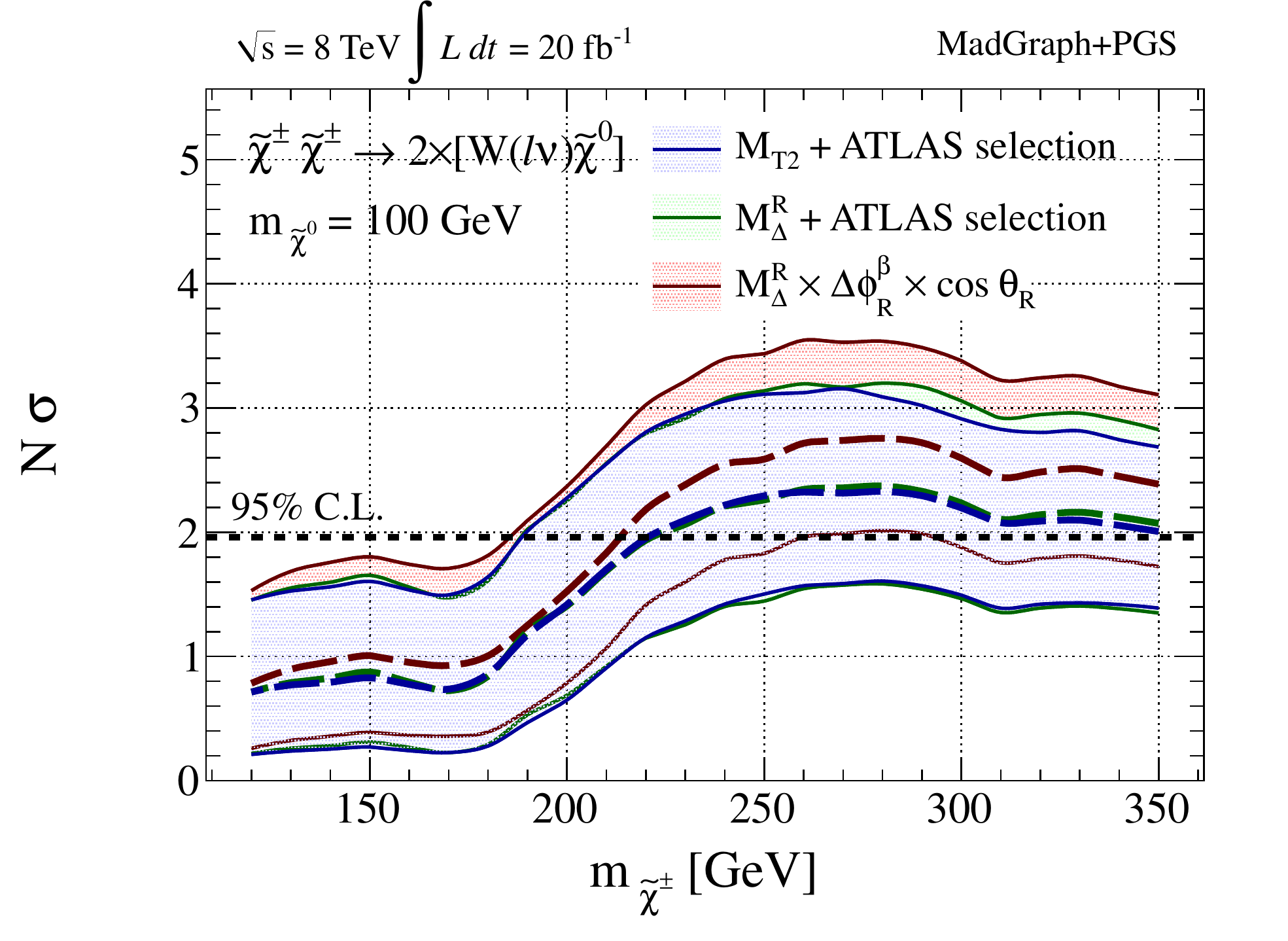}
\includegraphics[width=0.40\columnwidth]{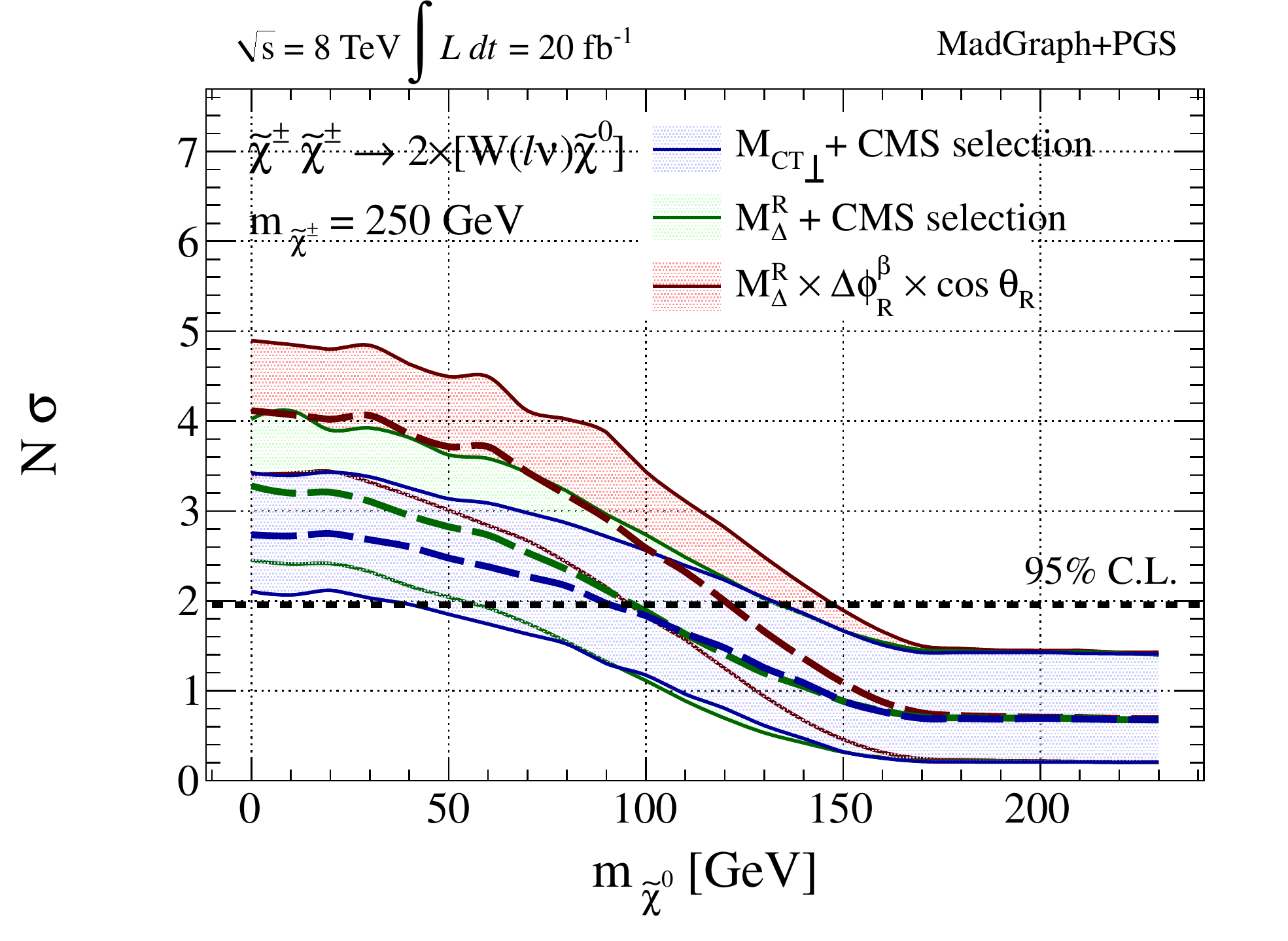}
\includegraphics[width=0.40\columnwidth]{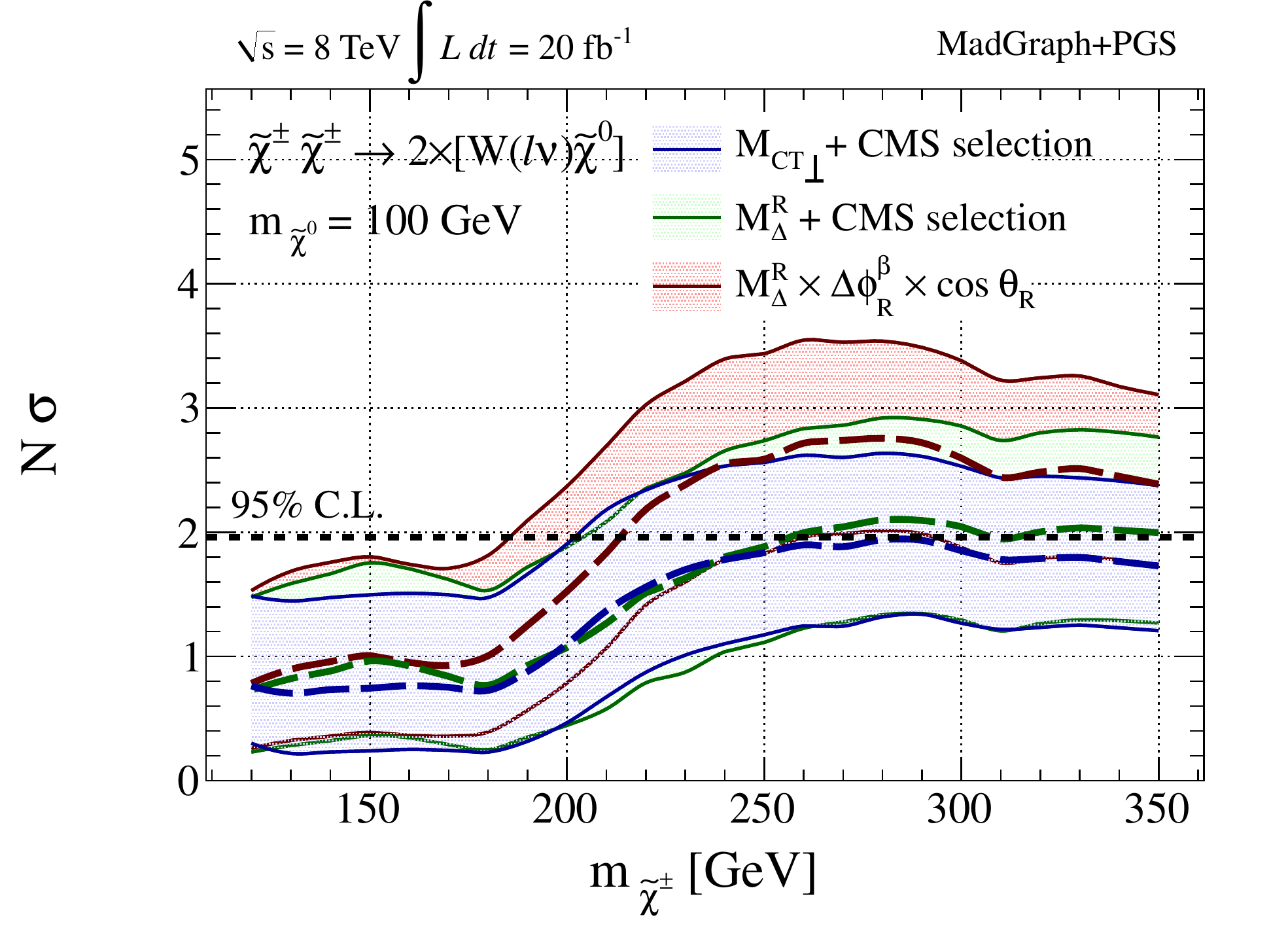}
\caption{Expected exclusion limits (in units of $\sigma$) for charginos decaying to neutralinos and leptonic $W$ bosons using 20~fb$^{-1}$ of 8 TeV data, as a function of neutralino mass with 250 GeV charginos (upper and lower left) or as a function of selectron mass with 100 GeV neutralinos (upper and lower right). Expected limits are shown for our multi-dimensional razor analysis (red), and compared to either ATLAS (upper plots) or CMS (lower plots) mass variables and selection criteria. \label{fig:results_chargino_1D_compare}}
\end{figure}

Moderate improvements over the $M_\Delta^R$ analysis are visible for the selectrons, while the chargino sensitivity is greatly increased in the low-mass degeneracy regime. The relative improvements can be more clearly seen in the Figures~\ref{fig:results_slepton_1D_compare} and \ref{fig:results_chargino_1D_compare}, where we show the exclusion reach for fixed values of selectron/chargino or neutralino masses.

The super-razor improvements in the sensitivity to compressed spectra can be understood from
the additional kinematic information provided by the angles $\Delta\phi_R^\beta$ and $|\cos\theta_{R+1}|$. Recall that the magnitude of the approximate razor boost $\vec{\beta}_R$ is systematically larger than the correct boost $\vec{\beta}^{\, \rm CM}$, because of the the assumption that the energy of the event is evenly split between the visible and invisible systems. This causes a peaking of $\Delta\phi_R^\beta$ at $\pi$, since the sum of the visible momenta tends to be anti-aligned with the boost direction. As the spectrum becomes more and more compressed, this effect is magnified, as seen in Figure \ref{fig:dphi_v_dphi} of Section \ref{sec:variables}. Thus for compressed spectra  $\Delta\phi_R^\beta$ is a particularly good disciminator to appeal to in future searches.

As described in Section \ref{sec:variables}, $|\cos\theta_{R+1}|$ is related to the energy
difference of the leptons in the razor frame $R$, the approximation to the CM frame.
This difference is expected to be small for the Drell-Yans + jets background, and is also
peaked at zero for the $W^-W^+$ background, because of polarization effects.
For signal events the distributions in $|\cos\theta_{R+1}|$ are much flatter; the
polarization effects are absent either because the parent particles are spin zero
(sleptons) or because we have two-step decays (charginos).

Each of the super-razor variables, $M_{\Delta}^{R}$, $\sqrt{\hat{s}}_{R}$, $\vec{\beta}_{R}$, $\vec{\beta}_{R+1}$, $\Delta \phi_{R}^{\beta}$, 
$|\cos\theta_{R+1}|$, and the angle $|\Delta\phi (\vec{p}_{\ell\ell}^{\, \text{lab}},\vec{E}_{T}^\text{miss})|$
used in the super-razor selection,
represents a different piece of information about an event.
The collection can be thought of as a kinematic basis, which raises the question of whether one
can identify an optimal kinematic basis for a particular type of search, e.g. searches for
sleptons with compressed spectra. The answer to this question depends not just on the
kinematic properties of the signal, but also on kinematics of the major backgrounds and
especially on the detector effects that dominate the systematic uncertainties. It seems plausible that in some cases there may be a family of approximately equivalent kinematic bases, such that
more or less the same kinematic information is exploited in different ways but resulting in
approximately equivalent sensitivity.

\begin{acknowledgments}
We acknowledge helpful discussions with Paul Jackson, Maurizio Pierini, Chiu-Tien Yu, Javier Duarte and Avi Yagil.   
JL acknowledges the hospitality and support of the Theoretical Physics Group at SLAC.
Fermilab is operated by Fermi Research Alliance, LLC, under contract DE-AC02-07CH11359 with the United States Department of Energy. MS and CR are funded by the United States Department of Energy under Grant DE-FG02-92-ER40701 and acknowledge the support of the Weston Havens Foundation.  
\end{acknowledgments}
\bibliographystyle{apsrev}
\bibliography{superrazor}

\begin{thebibliography}{58}
\expandafter\ifx\csname natexlab\endcsname\relax\def\natexlab#1{#1}\fi
\expandafter\ifx\csname bibnamefont\endcsname\relax
  \def\bibnamefont#1{#1}\fi
\expandafter\ifx\csname bibfnamefont\endcsname\relax
  \def\bibfnamefont#1{#1}\fi
\expandafter\ifx\csname citenamefont\endcsname\relax
  \def\citenamefont#1{#1}\fi
\expandafter\ifx\csname url\endcsname\relax
  \def\url#1{\texttt{#1}}\fi
\expandafter\ifx\csname urlprefix\endcsname\relax\def\urlprefix{URL }\fi
\providecommand{\bibinfo}[2]{#2}
\providecommand{\eprint}[2][]{\url{#2}}

\bibitem[{\citenamefont{Aad et~al.}(2013{\natexlab{a}})}]{Aad:2012naa}
\bibinfo{author}{\bibfnamefont{G.}~\bibnamefont{Aad}} \bibnamefont{et~al.}
  (\bibinfo{collaboration}{ATLAS Collaboration}),
  \bibinfo{journal}{Eur.Phys.J.} \textbf{\bibinfo{volume}{C73}},
  \bibinfo{pages}{2362} (\bibinfo{year}{2013}{\natexlab{a}}),
  \eprint{1212.6149}.

\bibitem[{\citenamefont{Chatrchyan
  et~al.}(2013{\natexlab{a}})}]{Chatrchyan:2012uea}
\bibinfo{author}{\bibfnamefont{S.}~\bibnamefont{Chatrchyan}}
  \bibnamefont{et~al.} (\bibinfo{collaboration}{CMS Collaboration}),
  \bibinfo{journal}{Phys.Rev.Lett.} \textbf{\bibinfo{volume}{111}},
  \bibinfo{pages}{081802} (\bibinfo{year}{2013}{\natexlab{a}}),
  \eprint{1212.6961}.

\bibitem[{\citenamefont{Chatrchyan
  et~al.}(2013{\natexlab{b}})}]{Chatrchyan:2013fk}
\bibinfo{author}{\bibfnamefont{S.}~\bibnamefont{Chatrchyan}}
  \bibnamefont{et~al.} (\bibinfo{collaboration}{CMS Collaboration})
  (\bibinfo{year}{2013}{\natexlab{b}}), \eprint{1301.2175}.

\bibitem[{\citenamefont{Aad et~al.}(2013{\natexlab{b}})}]{Aad:2013wta}
\bibinfo{author}{\bibfnamefont{G.}~\bibnamefont{Aad}} \bibnamefont{et~al.}
  (\bibinfo{collaboration}{ATLAS Collaboration})
  (\bibinfo{year}{2013}{\natexlab{b}}), \eprint{1308.1841}.

\bibitem[{\citenamefont{Chatrchyan
  et~al.}(2013{\natexlab{c}})}]{CMS-PAS-SUS-13-004}
\bibinfo{author}{\bibfnamefont{S.}~\bibnamefont{Chatrchyan}}
  \bibnamefont{et~al.} (\bibinfo{collaboration}{CMS Collaboration}),
  \bibinfo{type}{Tech. Rep.} \bibinfo{number}{CMS-PAS-SUS-13-004},
  \bibinfo{institution}{CERN}, \bibinfo{address}{Geneva}
  (\bibinfo{year}{2013}{\natexlab{c}}).

\bibitem[{\citenamefont{Carena et~al.}(2008)\citenamefont{Carena, Freitas, and
  Wagner}}]{Carena:2008mj}
\bibinfo{author}{\bibfnamefont{M.}~\bibnamefont{Carena}},
  \bibinfo{author}{\bibfnamefont{A.}~\bibnamefont{Freitas}}, \bibnamefont{and}
  \bibinfo{author}{\bibfnamefont{C.}~\bibnamefont{Wagner}},
  \bibinfo{journal}{JHEP} \textbf{\bibinfo{volume}{0810}}, \bibinfo{pages}{109}
  (\bibinfo{year}{2008}), \eprint{0808.2298}.

\bibitem[{\citenamefont{Bi et~al.}(2012)\citenamefont{Bi, Yan, and
  Yin}}]{Bi:2011ha}
\bibinfo{author}{\bibfnamefont{X.-J.} \bibnamefont{Bi}},
  \bibinfo{author}{\bibfnamefont{Q.-S.} \bibnamefont{Yan}}, \bibnamefont{and}
  \bibinfo{author}{\bibfnamefont{P.-F.} \bibnamefont{Yin}},
  \bibinfo{journal}{Phys.Rev.} \textbf{\bibinfo{volume}{D85}},
  \bibinfo{pages}{035005} (\bibinfo{year}{2012}), \eprint{1111.2250}.

\bibitem[{\citenamefont{Bai et~al.}(2012)\citenamefont{Bai, Cheng, Gallicchio,
  and Gu}}]{Bai:2012gs}
\bibinfo{author}{\bibfnamefont{Y.}~\bibnamefont{Bai}},
  \bibinfo{author}{\bibfnamefont{H.-C.} \bibnamefont{Cheng}},
  \bibinfo{author}{\bibfnamefont{J.}~\bibnamefont{Gallicchio}},
  \bibnamefont{and} \bibinfo{author}{\bibfnamefont{J.}~\bibnamefont{Gu}},
  \bibinfo{journal}{JHEP} \textbf{\bibinfo{volume}{1207}}, \bibinfo{pages}{110}
  (\bibinfo{year}{2012}), \eprint{1203.4813}.

\bibitem[{\citenamefont{Alves et~al.}(2013)\citenamefont{Alves, Buckley, Fox,
  Lykken, and Yu}}]{Alves:2012ft}
\bibinfo{author}{\bibfnamefont{D.~S.} \bibnamefont{Alves}},
  \bibinfo{author}{\bibfnamefont{M.~R.} \bibnamefont{Buckley}},
  \bibinfo{author}{\bibfnamefont{P.~J.} \bibnamefont{Fox}},
  \bibinfo{author}{\bibfnamefont{J.~D.} \bibnamefont{Lykken}},
  \bibnamefont{and} \bibinfo{author}{\bibfnamefont{C.-T.} \bibnamefont{Yu}},
  \bibinfo{journal}{Phys.Rev.} \textbf{\bibinfo{volume}{D87}},
  \bibinfo{pages}{035016} (\bibinfo{year}{2013}), \eprint{1205.5805}.

\bibitem[{\citenamefont{Han et~al.}(2012)\citenamefont{Han, Katz, Krohn, and
  Reece}}]{Han:2012fw}
\bibinfo{author}{\bibfnamefont{Z.}~\bibnamefont{Han}},
  \bibinfo{author}{\bibfnamefont{A.}~\bibnamefont{Katz}},
  \bibinfo{author}{\bibfnamefont{D.}~\bibnamefont{Krohn}}, \bibnamefont{and}
  \bibinfo{author}{\bibfnamefont{M.}~\bibnamefont{Reece}},
  \bibinfo{journal}{JHEP} \textbf{\bibinfo{volume}{1208}}, \bibinfo{pages}{083}
  (\bibinfo{year}{2012}), \eprint{1205.5808}.

\bibitem[{\citenamefont{Bhattacherjee and Ghosh}(2012)}]{Bhattacherjee:2012mz}
\bibinfo{author}{\bibfnamefont{B.}~\bibnamefont{Bhattacherjee}}
  \bibnamefont{and} \bibinfo{author}{\bibfnamefont{K.}~\bibnamefont{Ghosh}}
  (\bibinfo{year}{2012}), \eprint{1207.6289}.

\bibitem[{\citenamefont{Carena et~al.}(2013)\citenamefont{Carena, Gori, Shah,
  Wagner, and Wang}}]{Carena:2013iba}
\bibinfo{author}{\bibfnamefont{M.}~\bibnamefont{Carena}},
  \bibinfo{author}{\bibfnamefont{S.}~\bibnamefont{Gori}},
  \bibinfo{author}{\bibfnamefont{N.~R.} \bibnamefont{Shah}},
  \bibinfo{author}{\bibfnamefont{C.~E.} \bibnamefont{Wagner}},
  \bibnamefont{and} \bibinfo{author}{\bibfnamefont{L.-T.} \bibnamefont{Wang}},
  \bibinfo{journal}{JHEP} \textbf{\bibinfo{volume}{1308}}, \bibinfo{pages}{087}
  (\bibinfo{year}{2013}), \eprint{1303.4414}.

\bibitem[{\citenamefont{Delgado et~al.}(2013)\citenamefont{Delgado, Giudice,
  Isidori, Pierini, and Strumia}}]{Delgado:2012eu}
\bibinfo{author}{\bibfnamefont{A.}~\bibnamefont{Delgado}},
  \bibinfo{author}{\bibfnamefont{G.~F.} \bibnamefont{Giudice}},
  \bibinfo{author}{\bibfnamefont{G.}~\bibnamefont{Isidori}},
  \bibinfo{author}{\bibfnamefont{M.}~\bibnamefont{Pierini}}, \bibnamefont{and}
  \bibinfo{author}{\bibfnamefont{A.}~\bibnamefont{Strumia}},
  \bibinfo{journal}{Eur.Phys.J.} \textbf{\bibinfo{volume}{C73}},
  \bibinfo{pages}{2370} (\bibinfo{year}{2013}), \eprint{1212.6847}.

\bibitem[{\citenamefont{Dutta et~al.}(2012)\citenamefont{Dutta, Kamon, Kolev,
  Sinha, and Wang}}]{Dutta:2012kx}
\bibinfo{author}{\bibfnamefont{B.}~\bibnamefont{Dutta}},
  \bibinfo{author}{\bibfnamefont{T.}~\bibnamefont{Kamon}},
  \bibinfo{author}{\bibfnamefont{N.}~\bibnamefont{Kolev}},
  \bibinfo{author}{\bibfnamefont{K.}~\bibnamefont{Sinha}}, \bibnamefont{and}
  \bibinfo{author}{\bibfnamefont{K.}~\bibnamefont{Wang}},
  \bibinfo{journal}{Phys.Rev.} \textbf{\bibinfo{volume}{D86}},
  \bibinfo{pages}{075004} (\bibinfo{year}{2012}), \eprint{1207.1873}.

\bibitem[{\citenamefont{Evans and Kats}(2013)}]{Evans:2012bf}
\bibinfo{author}{\bibfnamefont{J.~A.} \bibnamefont{Evans}} \bibnamefont{and}
  \bibinfo{author}{\bibfnamefont{Y.}~\bibnamefont{Kats}},
  \bibinfo{journal}{JHEP} \textbf{\bibinfo{volume}{1304}}, \bibinfo{pages}{028}
  (\bibinfo{year}{2013}), \eprint{1209.0764}.

\bibitem[{\citenamefont{Kilic and Tweedie}(2013)}]{Kilic:2012kw}
\bibinfo{author}{\bibfnamefont{C.}~\bibnamefont{Kilic}} \bibnamefont{and}
  \bibinfo{author}{\bibfnamefont{B.}~\bibnamefont{Tweedie}},
  \bibinfo{journal}{JHEP} \textbf{\bibinfo{volume}{1304}}, \bibinfo{pages}{110}
  (\bibinfo{year}{2013}), \eprint{1211.6106}.

\bibitem[{\citenamefont{Buckley
  et~al.}(2013{\natexlab{a}})\citenamefont{Buckley, Plehn, and
  Takeuchi}}]{Buckley:2013lpa}
\bibinfo{author}{\bibfnamefont{M.~R.} \bibnamefont{Buckley}},
  \bibinfo{author}{\bibfnamefont{T.}~\bibnamefont{Plehn}}, \bibnamefont{and}
  \bibinfo{author}{\bibfnamefont{M.}~\bibnamefont{Takeuchi}},
  \bibinfo{journal}{JHEP} \textbf{\bibinfo{volume}{1308}}, \bibinfo{pages}{086}
  (\bibinfo{year}{2013}{\natexlab{a}}), \eprint{1302.6238}.

\bibitem[{\citenamefont{Bai et~al.}(2013)\citenamefont{Bai, Cheng, Gallicchio,
  and Gu}}]{Bai:2013ema}
\bibinfo{author}{\bibfnamefont{Y.}~\bibnamefont{Bai}},
  \bibinfo{author}{\bibfnamefont{H.-C.} \bibnamefont{Cheng}},
  \bibinfo{author}{\bibfnamefont{J.}~\bibnamefont{Gallicchio}},
  \bibnamefont{and} \bibinfo{author}{\bibfnamefont{J.}~\bibnamefont{Gu}},
  \bibinfo{journal}{JHEP} \textbf{\bibinfo{volume}{1308}}, \bibinfo{pages}{085}
  (\bibinfo{year}{2013}), \eprint{1304.3148}.

\bibitem[{\citenamefont{Barbier et~al.}(2005)\citenamefont{Barbier, Berat,
  Besancon, Chemtob, Deandrea et~al.}}]{Barbier:2004ez}
\bibinfo{author}{\bibfnamefont{R.}~\bibnamefont{Barbier}},
  \bibinfo{author}{\bibfnamefont{C.}~\bibnamefont{Berat}},
  \bibinfo{author}{\bibfnamefont{M.}~\bibnamefont{Besancon}},
  \bibinfo{author}{\bibfnamefont{M.}~\bibnamefont{Chemtob}},
  \bibinfo{author}{\bibfnamefont{A.}~\bibnamefont{Deandrea}},
  \bibnamefont{et~al.}, \bibinfo{journal}{Phys.Rept.}
  \textbf{\bibinfo{volume}{420}}, \bibinfo{pages}{1} (\bibinfo{year}{2005}),
  \eprint{hep-ph/0406039}.

\bibitem[{\citenamefont{Csaki et~al.}(2012)\citenamefont{Csaki, Grossman, and
  Heidenreich}}]{Csaki:2011ge}
\bibinfo{author}{\bibfnamefont{C.}~\bibnamefont{Csaki}},
  \bibinfo{author}{\bibfnamefont{Y.}~\bibnamefont{Grossman}}, \bibnamefont{and}
  \bibinfo{author}{\bibfnamefont{B.}~\bibnamefont{Heidenreich}},
  \bibinfo{journal}{Phys.Rev.} \textbf{\bibinfo{volume}{D85}},
  \bibinfo{pages}{095009} (\bibinfo{year}{2012}), \eprint{1111.1239}.

\bibitem[{\citenamefont{Berger et~al.}(2013)\citenamefont{Berger, Perelstein,
  Saelim, and Tanedo}}]{Berger:2013sir}
\bibinfo{author}{\bibfnamefont{J.}~\bibnamefont{Berger}},
  \bibinfo{author}{\bibfnamefont{M.}~\bibnamefont{Perelstein}},
  \bibinfo{author}{\bibfnamefont{M.}~\bibnamefont{Saelim}}, \bibnamefont{and}
  \bibinfo{author}{\bibfnamefont{P.}~\bibnamefont{Tanedo}},
  \bibinfo{journal}{JHEP} \textbf{\bibinfo{volume}{1304}}, \bibinfo{pages}{077}
  (\bibinfo{year}{2013}), \eprint{1302.2146}.

\bibitem[{\citenamefont{Chatrchyan et~al.}(2012{\natexlab{a}})}]{CMS:yut}
\bibinfo{author}{\bibfnamefont{S.}~\bibnamefont{Chatrchyan}}
  \bibnamefont{et~al.} (\bibinfo{collaboration}{CMS Collaboration}),
  \bibinfo{type}{Tech. Rep.} \bibinfo{number}{CMS-PAS-SUS-12-027},
  \bibinfo{institution}{CERN} (\bibinfo{year}{2012}{\natexlab{a}}).

\bibitem[{\citenamefont{Chatrchyan et~al.}(2013{\natexlab{d}})}]{CMS:swa}
\bibinfo{author}{\bibfnamefont{S.}~\bibnamefont{Chatrchyan}}
  \bibnamefont{et~al.} (\bibinfo{collaboration}{CMS Collaboration}),
  \bibinfo{type}{Tech. Rep.} \bibinfo{number}{CMS-PAS-SUS-13-003},
  \bibinfo{institution}{CERN} (\bibinfo{year}{2013}{\natexlab{d}}).

\bibitem[{\citenamefont{Aad et~al.}(2013{\natexlab{c}})}]{Aad:2013txa}
\bibinfo{author}{\bibfnamefont{G.}~\bibnamefont{Aad}} \bibnamefont{et~al.}
  (\bibinfo{collaboration}{ATLAS Collaboration}), \bibinfo{journal}{JINST}
  \textbf{\bibinfo{volume}{1307}}, \bibinfo{pages}{P07015}
  (\bibinfo{year}{2013}{\natexlab{c}}), \eprint{1305.2284}.

\bibitem[{\citenamefont{Chatrchyan
  et~al.}(2013{\natexlab{e}})}]{Chatrchyan:2012jwg}
\bibinfo{author}{\bibfnamefont{S.}~\bibnamefont{Chatrchyan}}
  \bibnamefont{et~al.} (\bibinfo{collaboration}{CMS Collaboration}),
  \bibinfo{journal}{Phys.Lett.} \textbf{\bibinfo{volume}{B722}},
  \bibinfo{pages}{273} (\bibinfo{year}{2013}{\natexlab{e}}),
  \eprint{1212.1838}.

\bibitem[{\citenamefont{Jegerlehner and Nyffeler}(2009)}]{Jegerlehner:2009ry}
\bibinfo{author}{\bibfnamefont{F.}~\bibnamefont{Jegerlehner}} \bibnamefont{and}
  \bibinfo{author}{\bibfnamefont{A.}~\bibnamefont{Nyffeler}},
  \bibinfo{journal}{Phys.Rept.} \textbf{\bibinfo{volume}{477}},
  \bibinfo{pages}{1} (\bibinfo{year}{2009}), \eprint{0902.3360}.

\bibitem[{\citenamefont{Miller et~al.}(2007)\citenamefont{Miller, de~Rafael,
  and Roberts}}]{Miller:2007kk}
\bibinfo{author}{\bibfnamefont{J.~P.} \bibnamefont{Miller}},
  \bibinfo{author}{\bibfnamefont{E.}~\bibnamefont{de~Rafael}},
  \bibnamefont{and} \bibinfo{author}{\bibfnamefont{B.~L.}
  \bibnamefont{Roberts}}, \bibinfo{journal}{Rept.Prog.Phys.}
  \textbf{\bibinfo{volume}{70}}, \bibinfo{pages}{795} (\bibinfo{year}{2007}),
  \eprint{hep-ph/0703049}.

\bibitem[{\citenamefont{Buckley
  et~al.}(2013{\natexlab{b}})\citenamefont{Buckley, Hooper, and
  Kumar}}]{Buckley:2013sca}
\bibinfo{author}{\bibfnamefont{M.~R.} \bibnamefont{Buckley}},
  \bibinfo{author}{\bibfnamefont{D.}~\bibnamefont{Hooper}}, \bibnamefont{and}
  \bibinfo{author}{\bibfnamefont{J.}~\bibnamefont{Kumar}},
  \bibinfo{journal}{Phys.Rev.} \textbf{\bibinfo{volume}{D88}},
  \bibinfo{pages}{063532} (\bibinfo{year}{2013}{\natexlab{b}}),
  \eprint{1307.3561}.

\bibitem[{\citenamefont{Carena et~al.}(2012)\citenamefont{Carena, Gori, Shah,
  Wagner, and Wang}}]{Carena:2012gp}
\bibinfo{author}{\bibfnamefont{M.}~\bibnamefont{Carena}},
  \bibinfo{author}{\bibfnamefont{S.}~\bibnamefont{Gori}},
  \bibinfo{author}{\bibfnamefont{N.~R.} \bibnamefont{Shah}},
  \bibinfo{author}{\bibfnamefont{C.~E.} \bibnamefont{Wagner}},
  \bibnamefont{and} \bibinfo{author}{\bibfnamefont{L.-T.} \bibnamefont{Wang}},
  \bibinfo{journal}{JHEP} \textbf{\bibinfo{volume}{1207}}, \bibinfo{pages}{175}
  (\bibinfo{year}{2012}), \eprint{1205.5842}.

\bibitem[{\citenamefont{Beringer et~al.}(2012)}]{Beringer:1900zz}
\bibinfo{author}{\bibfnamefont{J.}~\bibnamefont{Beringer}} \bibnamefont{et~al.}
  (\bibinfo{collaboration}{Particle Data Group}), \bibinfo{journal}{Phys.Rev.}
  \textbf{\bibinfo{volume}{D86}}, \bibinfo{pages}{010001}
  (\bibinfo{year}{2012}).

\bibitem[{\citenamefont{Aad et~al.}(2012{\natexlab{a}})}]{:2012gg}
\bibinfo{author}{\bibfnamefont{G.}~\bibnamefont{Aad}} \bibnamefont{et~al.}
  (\bibinfo{collaboration}{ATLAS Collaboration})
  (\bibinfo{year}{2012}{\natexlab{a}}), \eprint{1208.2884}.

\bibitem[{\citenamefont{ATLAS}(2013{\natexlab{a}})}]{ATLAS-CONF-2013-049}
\bibinfo{author}{\bibnamefont{ATLAS}} (\bibinfo{collaboration}{ATLAS
  Collaboration}), \bibinfo{type}{Tech. Rep.}
  \bibinfo{number}{ATLAS-CONF-2013-049}, \bibinfo{institution}{CERN},
  \bibinfo{address}{Geneva} (\bibinfo{year}{2013}{\natexlab{a}}).

\bibitem[{\citenamefont{Chatrchyan
  et~al.}(2012{\natexlab{b}})}]{CMS-PAS-SUS-12-022}
\bibinfo{author}{\bibfnamefont{S.}~\bibnamefont{Chatrchyan}}
  \bibnamefont{et~al.} (\bibinfo{collaboration}{CMS Collaboration}),
  \bibinfo{type}{Tech. Rep.} \bibinfo{number}{CMS-PAS-SUS-12-022},
  \bibinfo{institution}{CERN} (\bibinfo{year}{2012}{\natexlab{b}}).

\bibitem[{\citenamefont{Chatrchyan
  et~al.}(2013{\natexlab{f}})}]{CMS-PAS-SUS-13-006}
\bibinfo{author}{\bibfnamefont{S.}~\bibnamefont{Chatrchyan}}
  \bibnamefont{et~al.} (\bibinfo{collaboration}{CMS Collaboration}),
  \bibinfo{type}{Tech. Rep.} \bibinfo{number}{CMS-PAS-SUS-13-006},
  \bibinfo{institution}{CERN}, \bibinfo{address}{Geneva}
  (\bibinfo{year}{2013}{\natexlab{f}}).

\bibitem[{\citenamefont{Aad et~al.}(2012{\natexlab{b}})}]{Aad:2012ku}
\bibinfo{author}{\bibfnamefont{G.}~\bibnamefont{Aad}} \bibnamefont{et~al.}
  (\bibinfo{collaboration}{ATLAS Collaboration})
  (\bibinfo{year}{2012}{\natexlab{b}}), \eprint{1208.3144}.

\bibitem[{\citenamefont{Aad et~al.}(2013{\natexlab{d}})}]{Aad:2012hba}
\bibinfo{author}{\bibfnamefont{G.}~\bibnamefont{Aad}} \bibnamefont{et~al.}
  (\bibinfo{collaboration}{ATLAS Collaboration}), \bibinfo{journal}{Phys.Lett.}
  \textbf{\bibinfo{volume}{B718}}, \bibinfo{pages}{841}
  (\bibinfo{year}{2013}{\natexlab{d}}), \eprint{1208.3144}.

\bibitem[{\citenamefont{Aad et~al.}(2013{\natexlab{e}})}]{ATLAS-CONF-2013-036}
\bibinfo{author}{\bibfnamefont{G.}~\bibnamefont{Aad}} \bibnamefont{et~al.}
  (\bibinfo{collaboration}{ATLAS Collaboration}), \bibinfo{type}{Tech. Rep.}
  \bibinfo{number}{ATLAS-CONF-2013-036}, \bibinfo{institution}{CERN},
  \bibinfo{address}{Geneva} (\bibinfo{year}{2013}{\natexlab{e}}).

\bibitem[{\citenamefont{ATLAS}(2013{\natexlab{b}})}]{ATLAS-CONF-2013-035}
\bibinfo{author}{\bibnamefont{ATLAS}} (\bibinfo{collaboration}{ATLAS
  Collaboration}), \bibinfo{type}{Tech. Rep.}
  \bibinfo{number}{ATLAS-CONF-2013-035}, \bibinfo{institution}{CERN},
  \bibinfo{address}{Geneva} (\bibinfo{year}{2013}{\natexlab{b}}).

\bibitem[{\citenamefont{Rogan}(2010)}]{Rogan:2010kb}
\bibinfo{author}{\bibfnamefont{C.}~\bibnamefont{Rogan}} (\bibinfo{year}{2010}),
  \eprint{1006.2727}.

\bibitem[{\citenamefont{Chatrchyan
  et~al.}(2012{\natexlab{c}})}]{Chatrchyan:2011ek}
\bibinfo{author}{\bibfnamefont{S.}~\bibnamefont{Chatrchyan}}
  \bibnamefont{et~al.} (\bibinfo{collaboration}{CMS Collaboration}),
  \bibinfo{journal}{Phys.Rev.} \textbf{\bibinfo{volume}{D85}},
  \bibinfo{pages}{012004} (\bibinfo{year}{2012}{\natexlab{c}}),
  \eprint{1107.1279}.

\bibitem[{\citenamefont{Fox et~al.}(2012)\citenamefont{Fox, Harnik, Primulando,
  and Yu}}]{Fox:2012ee}
\bibinfo{author}{\bibfnamefont{P.~J.} \bibnamefont{Fox}},
  \bibinfo{author}{\bibfnamefont{R.}~\bibnamefont{Harnik}},
  \bibinfo{author}{\bibfnamefont{R.}~\bibnamefont{Primulando}},
  \bibnamefont{and} \bibinfo{author}{\bibfnamefont{C.-T.} \bibnamefont{Yu}},
  \bibinfo{journal}{Phys.Rev.} \textbf{\bibinfo{volume}{D86}},
  \bibinfo{pages}{015010} (\bibinfo{year}{2012}), \eprint{1203.1662}.

\bibitem[{\citenamefont{Chatrchyan et~al.}(2012{\natexlab{d}})}]{CMS:2012dwa}
\bibinfo{author}{\bibfnamefont{S.}~\bibnamefont{Chatrchyan}}
  \bibnamefont{et~al.} (\bibinfo{collaboration}{CMS Collaboration}),
  \bibinfo{type}{Tech. Rep.} \bibinfo{number}{CMS-PAS-SUS-12-009},
  \bibinfo{institution}{CERN}, \bibinfo{address}{Geneva}
  (\bibinfo{year}{2012}{\natexlab{d}}).

\bibitem[{\citenamefont{Chatrchyan
  et~al.}(2012{\natexlab{e}})}]{Chatrchyan:2012gq}
\bibinfo{author}{\bibfnamefont{S.}~\bibnamefont{Chatrchyan}}
  \bibnamefont{et~al.} (\bibinfo{collaboration}{CMS Collaboration})
  (\bibinfo{year}{2012}{\natexlab{e}}), \eprint{1212.6961}.

\bibitem[{\citenamefont{Lester and Summers}(1999)}]{Lester:1999tx}
\bibinfo{author}{\bibfnamefont{C.}~\bibnamefont{Lester}} \bibnamefont{and}
  \bibinfo{author}{\bibfnamefont{D.}~\bibnamefont{Summers}},
  \bibinfo{journal}{Phys.Lett.} \textbf{\bibinfo{volume}{B463}},
  \bibinfo{pages}{99} (\bibinfo{year}{1999}), \eprint{hep-ph/9906349}.

\bibitem[{\citenamefont{Barr et~al.}(2003)\citenamefont{Barr, Lester, and
  Stephens}}]{Barr:2003rg}
\bibinfo{author}{\bibfnamefont{A.}~\bibnamefont{Barr}},
  \bibinfo{author}{\bibfnamefont{C.}~\bibnamefont{Lester}}, \bibnamefont{and}
  \bibinfo{author}{\bibfnamefont{P.}~\bibnamefont{Stephens}},
  \bibinfo{journal}{J.Phys.} \textbf{\bibinfo{volume}{G29}},
  \bibinfo{pages}{2343} (\bibinfo{year}{2003}), \eprint{hep-ph/0304226}.

\bibitem[{\citenamefont{Matchev and Park}(2011)}]{Matchev:2009ad}
\bibinfo{author}{\bibfnamefont{K.~T.} \bibnamefont{Matchev}} \bibnamefont{and}
  \bibinfo{author}{\bibfnamefont{M.}~\bibnamefont{Park}},
  \bibinfo{journal}{Phys.Rev.Lett.} \textbf{\bibinfo{volume}{107}},
  \bibinfo{pages}{061801} (\bibinfo{year}{2011}), \eprint{0910.1584}.

\bibitem[{\citenamefont{Tovey}(2008)}]{Tovey:2008ui}
\bibinfo{author}{\bibfnamefont{D.~R.} \bibnamefont{Tovey}},
  \bibinfo{journal}{JHEP} \textbf{\bibinfo{volume}{0804}}, \bibinfo{pages}{034}
  (\bibinfo{year}{2008}), \eprint{0802.2879}.

\bibitem[{\citenamefont{Chatrchyan
  et~al.}(2012{\natexlab{f}})}]{Chatrchyan:2012jx}
\bibinfo{author}{\bibfnamefont{S.}~\bibnamefont{Chatrchyan}}
  \bibnamefont{et~al.} (\bibinfo{collaboration}{CMS Collaboration}),
  \bibinfo{journal}{JHEP} \textbf{\bibinfo{volume}{1210}}, \bibinfo{pages}{018}
  (\bibinfo{year}{2012}{\natexlab{f}}), \eprint{1207.1798}.

\bibitem[{\citenamefont{Aaltonen et~al.}(2010)}]{Aaltonen:2009rm}
\bibinfo{author}{\bibfnamefont{T.}~\bibnamefont{Aaltonen}} \bibnamefont{et~al.}
  (\bibinfo{collaboration}{CDF Collaboration}), \bibinfo{journal}{Phys.Rev.}
  \textbf{\bibinfo{volume}{D81}}, \bibinfo{pages}{031102}
  (\bibinfo{year}{2010}), \eprint{0911.2956}.

\bibitem[{\citenamefont{Rogan}(2013)}]{roganthesis}
\bibinfo{author}{\bibfnamefont{C.}~\bibnamefont{Rogan}}, Ph.D. thesis,
  \bibinfo{school}{California Institute of Technology} (\bibinfo{year}{2013}).

\bibitem[{\citenamefont{Alwall et~al.}(2011)\citenamefont{Alwall, Herquet,
  Maltoni, Mattelaer, and Stelzer}}]{Alwall:2011uj}
\bibinfo{author}{\bibfnamefont{J.}~\bibnamefont{Alwall}},
  \bibinfo{author}{\bibfnamefont{M.}~\bibnamefont{Herquet}},
  \bibinfo{author}{\bibfnamefont{F.}~\bibnamefont{Maltoni}},
  \bibinfo{author}{\bibfnamefont{O.}~\bibnamefont{Mattelaer}},
  \bibnamefont{and} \bibinfo{author}{\bibfnamefont{T.}~\bibnamefont{Stelzer}},
  \bibinfo{journal}{JHEP} \textbf{\bibinfo{volume}{1106}}, \bibinfo{pages}{128}
  (\bibinfo{year}{2011}), \eprint{1106.0522}.

\bibitem[{\citenamefont{Sjostrand et~al.}(2006)\citenamefont{Sjostrand, Mrenna,
  and Skands}}]{Sjostrand:2006za}
\bibinfo{author}{\bibfnamefont{T.}~\bibnamefont{Sjostrand}},
  \bibinfo{author}{\bibfnamefont{S.}~\bibnamefont{Mrenna}}, \bibnamefont{and}
  \bibinfo{author}{\bibfnamefont{P.~Z.} \bibnamefont{Skands}},
  \bibinfo{journal}{JHEP} \textbf{\bibinfo{volume}{0605}}, \bibinfo{pages}{026}
  (\bibinfo{year}{2006}), \eprint{hep-ph/0603175}.

\bibitem[{\citenamefont{Rainwater and Zeppenfeld}(1999)}]{Rainwater:1999sd}
\bibinfo{author}{\bibfnamefont{D.~L.} \bibnamefont{Rainwater}}
  \bibnamefont{and}
  \bibinfo{author}{\bibfnamefont{D.}~\bibnamefont{Zeppenfeld}},
  \bibinfo{journal}{Phys.Rev.} \textbf{\bibinfo{volume}{D60}},
  \bibinfo{pages}{113004} (\bibinfo{year}{1999}), \eprint{hep-ph/9906218}.

\bibitem[{\citenamefont{Beenakker et~al.}(1999)\citenamefont{Beenakker, Klasen,
  Kramer, Plehn, Spira et~al.}}]{Beenakker:1999xh}
\bibinfo{author}{\bibfnamefont{W.}~\bibnamefont{Beenakker}},
  \bibinfo{author}{\bibfnamefont{M.}~\bibnamefont{Klasen}},
  \bibinfo{author}{\bibfnamefont{M.}~\bibnamefont{Kramer}},
  \bibinfo{author}{\bibfnamefont{T.}~\bibnamefont{Plehn}},
  \bibinfo{author}{\bibfnamefont{M.}~\bibnamefont{Spira}},
  \bibnamefont{et~al.}, \bibinfo{journal}{Phys.Rev.Lett.}
  \textbf{\bibinfo{volume}{83}}, \bibinfo{pages}{3780} (\bibinfo{year}{1999}),
  \eprint{hep-ph/9906298}.

\bibitem[{\citenamefont{Cacciari et~al.}(2012)\citenamefont{Cacciari, Salam,
  and Soyez}}]{Cacciari:2011ma}
\bibinfo{author}{\bibfnamefont{M.}~\bibnamefont{Cacciari}},
  \bibinfo{author}{\bibfnamefont{G.~P.} \bibnamefont{Salam}}, \bibnamefont{and}
  \bibinfo{author}{\bibfnamefont{G.}~\bibnamefont{Soyez}},
  \bibinfo{journal}{Eur.Phys.J.} \textbf{\bibinfo{volume}{C72}},
  \bibinfo{pages}{1896} (\bibinfo{year}{2012}), \eprint{1111.6097}.

\bibitem[{\citenamefont{Cacciari et~al.}(2008)\citenamefont{Cacciari, Salam,
  and Soyez}}]{Cacciari:2008gp}
\bibinfo{author}{\bibfnamefont{M.}~\bibnamefont{Cacciari}},
  \bibinfo{author}{\bibfnamefont{G.~P.} \bibnamefont{Salam}}, \bibnamefont{and}
  \bibinfo{author}{\bibfnamefont{G.}~\bibnamefont{Soyez}},
  \bibinfo{journal}{JHEP} \textbf{\bibinfo{volume}{0804}}, \bibinfo{pages}{063}
  (\bibinfo{year}{2008}), \eprint{0802.1189}.

\bibitem[{\citenamefont{Junk}(1999)}]{Junk:1999kv}
\bibinfo{author}{\bibfnamefont{T.}~\bibnamefont{Junk}},
  \bibinfo{journal}{Nucl.Instrum.Meth.} \textbf{\bibinfo{volume}{A434}},
  \bibinfo{pages}{435} (\bibinfo{year}{1999}), \eprint{hep-ex/9902006}.

\bibitem[{\citenamefont{Read}(2002)}]{Read:2002hq}
\bibinfo{author}{\bibfnamefont{A.~L.} \bibnamefont{Read}},
  \bibinfo{journal}{J.Phys.} \textbf{\bibinfo{volume}{G28}},
  \bibinfo{pages}{2693} (\bibinfo{year}{2002}).

\end{thebibliography}

\end{document}